PhD Thesis

# Caustics and catastrophes in strong-field physics

## Picard–Lefschetz theory as a universal approach to saddle-point methods in attosecond science

by

Anne Weber

*First Supervisor:* Dr. Emilio Pisanty
*Second Supervisor:* Dr. Amelle Zaïr

*submitted on 31st January 2026*

**Examiners**
Prof. Nirit Dudovich
Prof. Raphael Assier
— many thanks for the thorough feedback on my work and the inspiring viva!

**Oral examination**
13th May 2026

**Degree award**
1st June 2026

This version incorporates minor typographical corrections made prior to digital archiving.

**Digital version**
https://arxiv.org/abs/2605.03794 (October 14, 2026)

# Abstract

The creation of the shortest flashes of light — laser pulses with durations in the order of attoseconds — is typically achieved via high-order harmonic generation (HHG). In this highly nonlinear optical effect, the interaction of atoms with ultra-strong light fields causes the emission of a broad spectrum of harmonic frequencies. Ever since its discovery, HHG has been described in terms of a 'quantum orbits' model based on several interfering electron trajectories, thereby incorporating both quantum-mechanical effects and an intuitive picture of classical dynamics. By tuning the parameters of the driving laser field, the interplay between these trajectories can be controlled, shaping the emitted light. Mathematically, this quantum orbits model is based on the expression of the harmonic response in terms of a highly-oscillatory integral for the generated dipole moment. Applying saddle-point methods to this integral allows it to be decomposed into contributions from distinct saddle points of the semi-classical action — analogous to Feynman's path integral formalism. Therewith, saddle-point methods provide the essential link between the fully quantum-mechanical description of the process and the intuitive picture in terms of classical electron trajectories. However, a rigorous framework for applying saddle-point methods to highly-oscillatory integrals with algorithms that work uniformly for arbitrary configurations and laser drivers, has previously not been established.

In this thesis, we introduce the key ideas of Picard–Lefschetz theory — the foundation of all saddle-point approximations — and develop practical numerical methods for its application. These methods allow us to evaluate generic one- and two-dimensional highly-oscillatory integrals across scans of external parameters, and to identify the relevant contributions from individual critical points of the integrand.

We apply these techniques to both strong-field ionisation and high-harmonic generation. In particular, we can use Picard–Lefschetz methods to uncover and analyse caustics. Just as caustics appear in everyday life, when rays of light are reflected from curved surfaces, they also arise in nonlinear optical processes, where they can produce prominent enhancement features in HHG spectra. In these situations, multiple semi-classical trajectories coalesce, causing traditional saddle-point approximations to break down and rendering existing heuristics insufficient. Our techniques, however, work uniformly across such caustic patterns. We showcase a wide range of parameter scans in which they emerge and provide insight into features that could not previously be understood within the standard framework. By enabling a deeper understanding of strong-field quantum-orbit dynamics, this thesis lays the foundation for controlling light-matter interactions on ultrafast timescales, bridging theoretical understanding with the versatility of modern experimental setups.

# Contents

# 1

# Introduction

The interaction of light and matter lies at the heart of a vast range of physical phenomena and has long played a central role in scientific research. From its earliest beginnings, a recurring ambition has been to 'take pictures' of nature — to directly observe and resolve the microscopic processes that govern the behaviour of matter. At the most fundamental level, these processes are driven by the motion of electrons: chemical reactions, bond formation, and charge transport all arise from the dynamics of electronic charge and the resulting forces imparted on its surroundings. Consequently, if we wish to resolve *how* and *why* chemical reactions are happening, we must be able to resolve the motion of electrons in real time.

The natural timescale of electronic motion is extremely short, typically on the order of attoseconds ($10^{-18}$s). To observe such dynamics directly — and ultimately to construct a 'movie' of electrons moving within atoms and molecules — one must therefore probe matter with light pulses of comparable duration. The development of attosecond science has been driven precisely by this challenge: the generation, characterisation, and application of ultrashort flashes of light capable of resolving electron motion on its natural timescale.

Attosecond light pulses can be produced in large-scale facilities with free-electron lasers, where relativistic electron bunches are accelerated and compressed to generate extremely short radiation bursts. Alternatively, attosecond pulses can also be generated in table-top laboratory setups through the process of high-order harmonic generation (HHG). In this approach, intense laser pulses — with peak intensities in the order of $10^{14}\,\mathrm{W/cm^2}$ — are focused into a target medium such as a noble gas, a solid, or even a liquid, shown schematically in Fig. 1.1(a). Under these extreme conditions, the interaction between light and matter becomes highly nonlinear, giving rise to the emission of harmonic radiation at multiples of the driving laser's frequency.

The resulting harmonic spectrum of a gaseous target medium, as in Fig. 1.1(b), typically exhibits a broad plateau of nearly constant intensity. This plateau is followed by a sharp drop in intensity at the so-called high-harmonic cutoff. The Fourier transform of such a spectrum corresponds to an ultrashort burst of radiation, with durations in the order of attoseconds. The properties of these attosecond pulses, e.g., their temporal structure, polarisation and bandwidth, are determined by the spectral amplitude and phase of the generated harmonic frequencies. Over the past decades, extensive experimental and theoretical efforts have therefore focused on understanding and controlling the HHG process, both as a source of attosecond pulses and as a probe of the ultrafast electron dynamics that constitute the process itself.

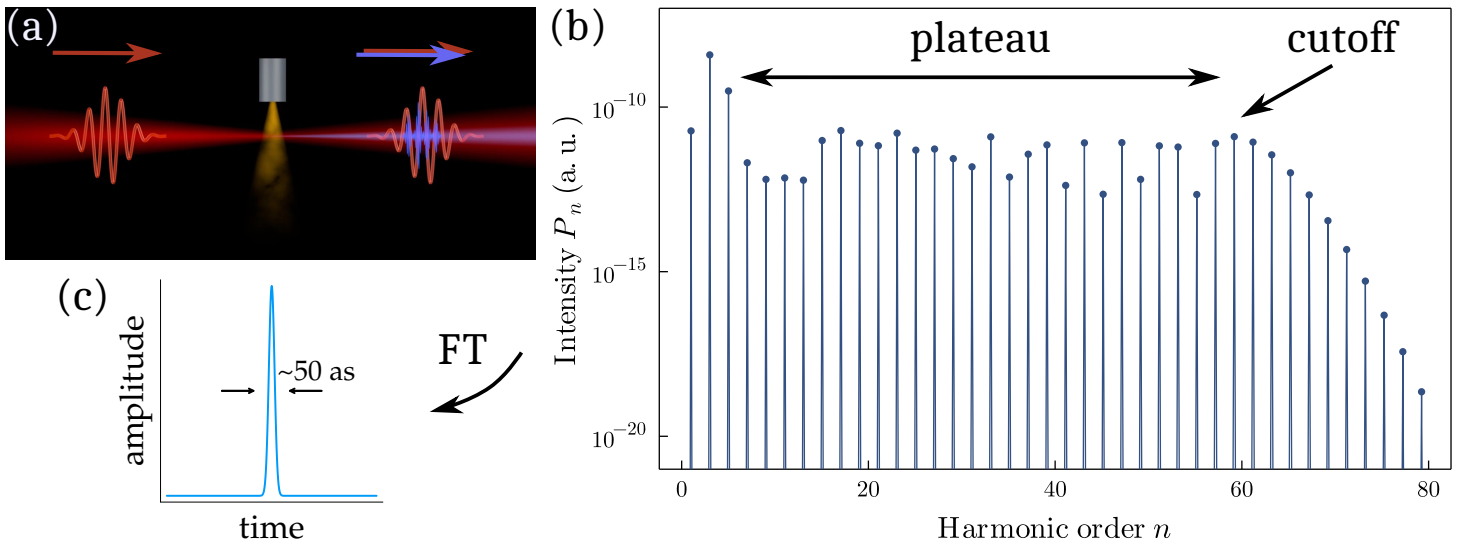

Figure 1.1: Creating attosecond pulses using the generation high-order harmonics. (a) A strong laser pulse (red) is focused into a gas jet medium, and produces a train of HHG pulses (blue).[1] (b) On a microscopic scale, the generated radiation spectrum typically consists of a long plateau followed by a sharp drop in intensity at the cutoff. (c) Conceptually, via a Fourier-transform of the spectrum the broad-band HHG radiation creates a train of ultrashort light pulses, with durations in the order of attoseconds.

On a microscopic level, the interaction between the strong incoming laser field and the individual atoms of the gaseous target is commonly interpreted within the framework of strong-field physics. Therein, the process is captured by an intuitive three-step model: first, an electron is ionised from the atom by the strong laser field; secondly, it propagates freely under the influence of only the oscillating field; and third, it may return to its parent ion. Upon recombination, the acquired energy is released in form of a high-energy photon whose frequency is an integer multiple of the driver's frequency. This semi-classical picture provides a remarkably successful interpretation of HHG and directly connects the emitted radiation to the underlying electron dynamics.

The analytical theory that naturally incorporates this picture is the strong-field approximation (SFA). Within the SFA, the atomic response to a strong laser field can be expressed in terms of integrals over the interaction time between the laser field and the atom. These integrals are highly oscillatory and are commonly evaluated using saddle-point methods, whereby the dominant contributions arise from stationary points of the complex-valued action. Each stationary point corresponds to a semi-classical electron trajectory, characterised by an ionisation time and a recombination time, thereby drawing a direct analogy between the three-step model and Feynman's path-integral formulation of quantum mechanics.

In general, several such trajectories contribute simultaneously to the harmonic emission, and their coherent superposition leads to quantum-path interference — a key feature that may be observed in a myriad of ways. By varying the properties of the driving laser field and/or its interaction geometry with the target, these interference patterns can be modified, offering a powerful means of probing and controlling electron dynamics. Traditionally, theoretical treatments have relied on a set of heuristics to determine which saddle-point solutions correspond to physically relevant trajectories. For simple driving fields, such as monochromatic linearly polarised lasers, this approach is well established, leading to a popular classification into ‘short’ and ‘long’ trajectories that interfere throughout the plateau, and where the short trajectory has to be discarded after the cutoff.

[1] Illustration taken from www.laserfocusworld.com/lasers-sources/article/14169549/high-harmonic-generation-sources-enable-extreme-ultraviolet-lensless-imaging, on 24/01/2026.

In the past few decades, however, the use of increasingly complex driving laser fields has become an indispensable tool in studying strong-field effects. By combining laser fields of different frequencies, amplitudes, polarisation states, and relative phase delays, it is possible to engineer tailored optical waveforms whose characteristics can be directly imprinted onto the HHG process. Such waveform control provides a powerful means to manipulate electron trajectories and to probe the underlying strong-field dynamics. Beyond classical field shaping, recent advances have also explored the use of non-classical light, such as squeezed states, thereby introducing concepts from quantum information science into the study of nonlinear light-matter interaction. These complex driving fields offer unprecedented control over electron motion, but they also pose a significant theoretical challenge: the established heuristic rules often fail to determine which of the many stationary points of the action constitute the relevant electron trajectories for the HHG process. As a result, theoretical understanding increasingly lags behind experimental capability.

This lack of a thorough understanding of the link between the semi-classical trajectory picture, the full quantum-mechanical description and the experimental observations becomes particularly evident in the presence of caustics. Caustics are ubiquitous phenomena, familiar from everyday life as the rippled light patterns at the bottom of a swimming pool or the bright structure atop a coffee cup. More generally, they describe phenomena that arise when interfering classical trajectories coalesce and produce a pronounced feature in the observable. At such points, the mathematical description of the total integral in terms of separate contributions breaks down.

The appearance of caustics is typically understood within the mathematical framework of catastrophe theory, which studies the generic behaviour of critical points under smooth variations of external parameters. Catastrophe theory provides a hierarchical classification of the way in which critical points can merge at the eponymous catastrophe points, and predicts the caustic patterns associated with each type of coalescence in terms of canonical diffraction integrals.

While the application of catastrophe theory to attosecond science is still in its infancy, both experimental and theoretical examples do exist. In this context, the interfering trajectories correspond to electron trajectories, and the observable quantity is the HHG radiation dipole which creates the harmonic spectrum. The most pertinent example of a catastrophe in HHG is the high-harmonic cutoff, which constitutes a caustic, as the two trajectories (almost) coalesce and thereby produce a distinct feature in the spectrum. A rigorous analysis of the HHG cutoff using the framework of catastrophe theory has only been provided a few years ago. Further examples of caustics appearing in attosecond science experiments are certainly intriguing, and provide abundant motivation to study the underlying mathematical link in more depth.

The mathematical framework that links highly-oscillatory integrals with their description in terms of separate contributions (associable with semi-classical trajectories) is Picard–Lefschetz theory. It rigorously proves how a deformation of the integration contour into the complex domain yields an efficient means of evaluating the integral exactly. This ultimately allows us to express the integral as a sum of contributions from a subset of critical points. Crucially, Picard–Lefschetz theory thereupon provides an unambiguous criterion for determining which stationary points contribute to this integral representation. Importantly, the contour deformation does not change the value of the integral, such that the final expression in terms of individual contributions remains exact. As a consequence, Picard–Lefschetz theory offers the tools required to calculate contributions across continuous ranges of external parameters and thereby to resolve the emergence of caustics.

## Motivation and outline of this thesis

In this thesis, we introduce and develop the methods of Picard–Lefschetz theory for applications in attosecond science. We formulate practical techniques for both one- and two-dimensional integrals and apply them to the strong-field ionisation amplitude and the HHG dipole within the SFA framework. In doing so, we demonstrate how Picard–Lefschetz theory enables a rigorous identification of the relevant quantum trajectories in regimes that lie well beyond the reach of established heuristic approaches. A particular focus of this work lies on strongly polychromatic driving fields. We introduce the concept of a colour switchover — a smooth transition between a monochromatic and a strongly bichromatic driving field — which encapsulates the fundamental questions regarding the trajectory interplay in HHG driven by complex laser waveforms. Through this approach, we illustrate how the powerful toolbox of Picard–Lefschetz theory provides new insight into the emergence of caustics and the underlying electron dynamics in modern attosecond experiments.

This thesis is structured as follows. We begin by providing the theoretical background required for the concepts used throughout this work. In Chapter 2, we give a brief overview of the theoretical approaches that give rise to the quantum-orbit formalism within attosecond science. We focus on strong-field tunnelling ionisation and high-harmonic generation, as these processes naturally lead to one- and two-dimensional oscillatory integrals. Subsequently, Chapter 3 introduces the mathematical framework of catastrophe theory and provides a brief review of its applications in attosecond science. In Chapter 4, we present the fundamental ideas of Picard–Lefschetz theory.

Following this introductory part, Part II presents the methods we developed in this thesis. In Chapter 5, we introduce techniques derived from Picard–Lefschetz theory for generic one- and two-dimensional integrals. Their application to the integrals arising in strong-field tunnelling ionisation and HHG is demonstrated in Chapter 6. We then introduce the colour switchover scheme in Chapter 7 as a technique that essentially encapsulates the open questions of quantum-orbit approaches in attosecond science.

These methods are then brought together, such that in Part III we show a range of results. In order to analyse the complex (in both meanings of the word) integration methods for the two-dimensional time integral that describes the dipole response for HHG, it is indispensable that we fully understand a one-dimensional integral first. Accordingly, Chapter 8 investigates the transition amplitude for strong-field tunnel ionisation throughout the colour switchover. This analysis reveals a particularly intricate and unexpected phenomenon: the existence of a tunnel-ionisation event occurring at a time when the electric field vanishes, and hence when no tunnelling barrier is present. This finding is discussed in detail in Chapter 9. We then proceed to the two-dimensional integral describing the HHG dipole response. In Chapter 10, Picard–Lefschetz methods are applied to HHG during the colour switchover, illustrating how new trajectories emerge in the transition from a fundamental driving field to its second harmonic. We subsequently consider a different, more established type of parameter scan within the attosecond-science community, namely the variation of the relative phase between the two field components, in Chapter 11. In Chapter 12, all previously introduced techniques are combined to study caustics and catastrophes arising in the various parameter scans.

Finally, Chapter 13 summarises the results of this thesis and provides an outlook, including perspectives on our future work.

## Publications

Parts of the results presented in this thesis are published in the following publications:

[1] A. Weber, M. Khokhlova, and E. Pisanty, "Quantum tunneling without a barrier", *Phys. Rev. A* **111**, 043103 (2025).

[2] A. Weber, J. Feldbrugge, and E. Pisanty, "Universal approach to saddle-point methods in attosecond science", *Phys. Rev. A* **113**, 063111 (2026).

[3] X. Zou*, L. Jurkoviča*, A. Weber*, et al., "2D quantum-path interference in high-harmonic generation driven by highly-bichromatic fields", *New J. Phys.* (accepted, 2026) *equal contribution

# Part I

## Theoretical Background

# 2

# Quantum-orbit approaches in attosecond physics

In this Chapter, we outline the theoretical models commonly employed to describe the interaction of intense laser fields with atomic targets. We begin with an overview of the fundamental concepts of strong-field physics, including some phenomenological explanations that provide physical intuition. The discussion then centres around the strong-field approximation (SFA), which is the workhorse framework for both qualitative and quantitative description of strong-field light-matter interaction. Following that, we detail the specific theoretical approach and provide key results for tunnelling ionisation in strong laser fields, as well as the application of the SFA to high-order harmonic generation. Note that all of the material presented here reviews established concepts and methods that are well-known within the (theoretical) attosecond-science community.

## 2.1 Introduction to strong-field physics

Historically, the interaction of light and matter has successfully been described within the framework of linear optics and time-dependent perturbation theory, where the energy of the electromagnetic waves is treated as a weak perturbation in addition to the intrinsic energy structure of the target. The advent of the laser and the subsequent development of coherent and more intense light sources enabled nonlinear optical phenomena such as multiphoton absorption and harmonic generation. For the description of these phenomena higher-order terms in the perturbative expansion become relevant. However, as laser intensities continued to increase, the perturbative description itself ultimately broke down: when the electric field strength becomes comparable to the Coulomb binding force in atoms and solids, qualitatively new phenomena emerged that could no longer be captured perturbatively. This so-called 'strong-field' regime is understood via Keldysh theory, which provides a non-perturbative description of ionisation and unifies the multiphoton and tunnelling limits, thereby laying the foundation for modern strong-field and attosecond physics. The pioneering theoretical approach to how atomic targets respond to the interaction with such intense external electric fields was provided by Keldysh [4], and then later refined by Faisal [5] and Reiss [6] and is nowadays known as the KFR theory. In parallel, alternative formulations were developed by Peremelomov, Popov and Terent'ev [7–9], known as PPT theory, as well as Ammosov, Delone and Krainov [10] who derived analytic ionisation rates in the tunnelling regime, known as ADK rates. These theoretical frameworks established the foundations of strong-field physics, by including the intense laser fields into a quantum-mechanical approach in a non-perturbative way. Following the works on strong-field ionisation and experimental observations of radiation of higher frequencies, high-order harmonic generation (HHG) was explained within a similar framework by including the possibility of recombination with the parent ion [11]. The more quantitative description was then provided in Ref. [12], building on earlier experimental insights [13]. This formulation, now known as the strong-field approximation (SFA), introduced a fully quantum-mechanical treatment, capable of describing the ionisation, propagation and recombination of an electron in a strong-laser field — naturally including tunnelling, wave packet dispersion and interference effects. In the PPT formulation, the tunnelling problem is expressed as a time integral over the interaction between the laser field and the atomic target. This representation naturally entailed the insight that the integral can be evaluated using saddle-point methods. For the case of HHG, where time integration includes ionisation and recombination times, these would turn the highly-oscillatory integral into a sum of contributions from several distinct quantum paths — a nice synergy between mathematical techniques and physical intuition. Back then, a major point of interest was the estimation of the maximum energy of the produced harmonic radiation — the HHG cutoff — which was found to depend on the parameters of the driving field and the type of target atom. Because these early studies were mostly employing linearly, monochromatic laser fields to drive the process, the respective theoretical models where developed for those configurations. Subsequent work has generalised the SFA to more complex driving field configurations and incorporated additional effects such as the atomic binding potential. By combining computational simplicity with a powerful intuitive picture of the underlying electron dynamics the SFA has not only shaped our understanding of strong-field phenomena but also remains the central workhorse for theoretical simulations in attosecond science.

### 2.1.1 Parameter regimes and Keldysh theory

Strong-field physics specifically addresses the extraordinary processes that happen in regimes where the strength of the electromagnetic radiation is comparable to the atomic and electronic binding forces of the targets. For the simplest atom, Hydrogen, the electron experiences an attractive Coulomb field of roughly $E_{\text{atom}} = e/a_B^2 = 5.1 \times 10^9 \text{V/cm}$, with the Bohr radius $a_B$ of the $1s$ ground state orbital and the charge of the electron $e$. In comparison, the amplitude of an electromagnetic wave is given by $E_0 = \sqrt{8\pi I_0/c}$ with the light speed $c$ and laser intensity $I_0$. Laser with intensities in the order of $I_0 \approx 4\times10^{14} \text{W/cm}^2$ therefore create field strength of approximately $E_0 \approx 5.5 \times 10^8 \text{V/cm}$. That is, for these light fields a perturbative approach will fail to capture the relevant physics.

In order to describe the physics in a meaningful way without having to worry about the large range of orders of magnitudes and various involved natural constants, we will use atomic units (a.u.) throughout this thesis, if not stated otherwise. That is, as a unit length we use the radius of the lowest energy orbital of the Hydrogen atom, the Bohr radius $a_B$. Unit charge and mass are those of an electron, $e$ and $m_e$ respectively. The reduced Planck's constant as well as the vacuum permittivity are hence set to unity, $\hbar = 4\pi\epsilon_0 = 1$, and the speed of light $c = 1/\alpha$ with the fine-structure constant $\alpha = e^2/(4\pi\epsilon_0\hbar c) \approx 1/137$.

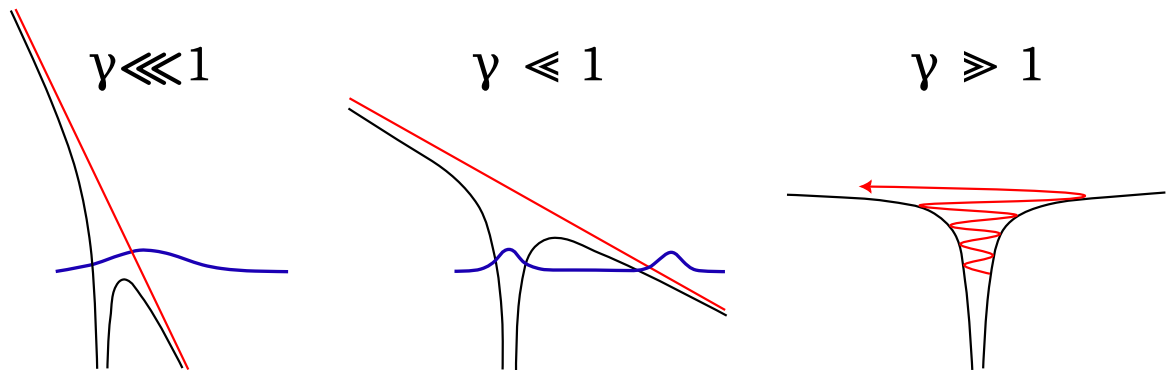

Figure 2.1: Regimes of ionisation for a distorted Coulomb barrier (black) with an electronic wave function (blue) upon the interaction with a strong laser field (red), resulting in different Keldysh parameters $\gamma$. For $\gamma \lll 1$: over-the-barrier ionisation and depletion of the ground state (left); for $\gamma \ll 1$: tunnel ionisation (centre); for $\gamma \gg 1$: multi-photon ionisation (right).

If the laser field is of comparable energy with the binding force the electron can escape the atomic bound state. The parameter that distinguishes different mechanisms of ionisation is the Keldysh adiabaticity parameter $\gamma$, which was originally derived from the ratio of the tunnelling time to the cycle period of the laser field [4]. Nowadays it is often simply expressed as

$$\gamma = \sqrt{\frac{\mathcal{I}_\text{p}}{2U_\text{p}}} \tag{2.1}$$

where $\mathcal{I}_\text{p}$ is the ionisation potential of the atom. $U_\text{p}$ is the ponderomotive energy of the laser field, given by the kinetic (wiggle) energy of the electron in the laser field, averaged over one cycle

$$U_\text{p} = \langle A^2(t)\rangle_t \tag{2.2}$$

with the vector potential of the laser field $A(t) = -\int E(t)\mathrm{d}t$. For a simple monochromatic laser field this yields $U_\mathrm{p} = E_0^2/(4\omega^2)$. The Keldysh parameter then distinguishes different mechanism of ionisation, shown schematically in Fig. 2.1:

- Over-the-barrier ionisation: In the case of $\gamma \lll 1$, due to a very intense laser the field is simply so strong that the barrier is suppressed below the ground state, causing an over the barrier ionisation and a quick depletion of the ground state.
- Tunnel ionisation: For $\gamma \ll 1$, the strong electric field changes so slowly that it effectively creates a short-enough barrier that allows the electron to tunnel out. Intuitively, this means the tunnelling is more likely when the laser field frequency is decreased and the strength of the field is increased.
- Multiphoton ionisation: If $\gamma \gg 1$, the laser oscillates so quickly that the electron can absorb multiple photons to overcome the ionisation barrier.

Note that this distinction only provides an intuition for the relative likelihood of the several processes to occur. That is, we can still calculate tunnel ionisation rates for laser field configurations that corresponds to $\gamma > 1$, but they will not be as meaningful because in that regime the dominating ionisation process will be multiphoton ionisation. Furthermore, if $\gamma > 1$ means that the barrier possibly changes quite significantly during the tunnelling process and we won't accurately capture what happens inside the barrier.

### 2.1.2 Above-threshold ionisation

Within this thesis, we focus on the mechanism of above-threshold ionisation via tunnelling (ATI). The probability for tunnel ionisation at a time $t$ can be derived in terms of an instantaneous tunnelling rate [10]. A central assumption within this derivation is that the tunnelling barrier is static and the tunnelling process happens adiabatically. That is, the laser field is assumed to change so slowly that the atom sees a static, tilted Coulomb potential. This assumption becomes less appropriate when the laser field changes more rapidly, and the Keldysh parameter increases. A more accurate description is therefore given in terms of the PPT ionisation amplitude, which captures the nonadiabaticity of the process by assuming a 'moving' barrier. The electron can therefore acquire energy during the tunnelling. In a later Chapter 9 we will demonstrate that taking the nonadiabaticity of the tunnelling ionisation into account becomes particularly relevant for non-monochromatic driving fields.

The associated observable with strong-field ionisation are the momentum distributions of the released electrons, recorded at a detector. For a linearly polarised driving field electrons can be observed over a large range of frequencies, and the spectrum (proportional to the parallel component of the momentum) has a typical shape, shown in Fig. 2.2 [14]. Upon (or even during) ionisation electrons can follow different types of orbits which produce final momenta up to certain energies. Conversely, the respective ATI spectrum has two cutoffs that can be associated with different ionisation orbits: electrons that are ionised and immediately fly towards the detector have energies up to $2U_\mathrm{p}$. These are called *direct electrons.* Electrons that elastically rescatter off the core before they are recorded at the detector gain frequencies up to $10U_\mathrm{p}$. For the remainder of this thesis, when we mention ATI and/or strong-field ionisation we will most likely be referring to the direct electrons.

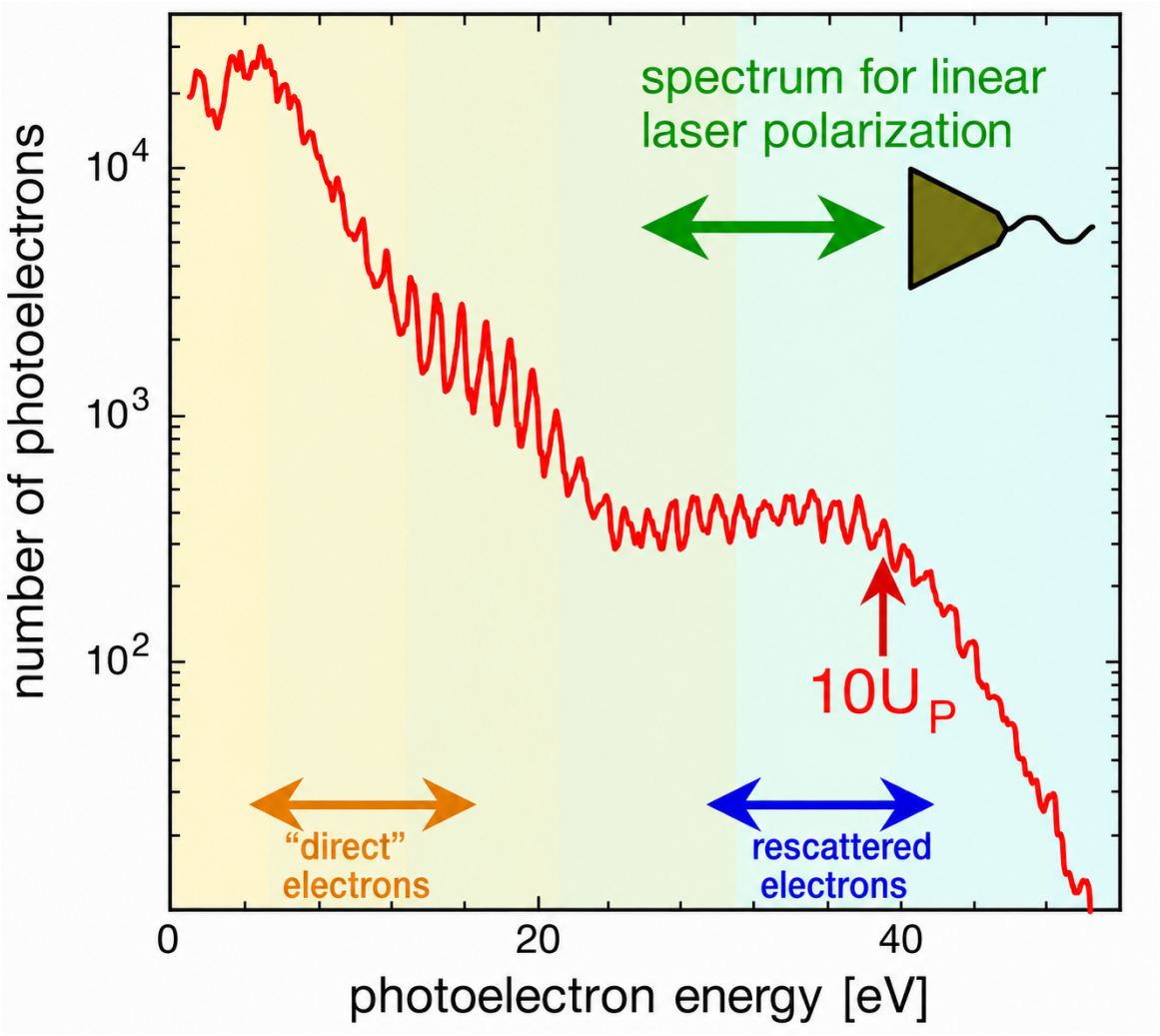

Figure 2.2: Schematic of a typical spectrum of electron ionised above threshold, showing contributions of direct electrons up to energies of $\approx 2U_\mathrm{p}$ and electrons that rescatter at the core, reaching up to $\approx 10U_\mathrm{p}$. From [15].

The spectrum in Fig. 2.2 shows the total energy of the detected electrons in the direction of the laser polarisation. Furthermore resolving the photoelectron momenta spatially yields an intricate structure that promises to contain information about the ionised target. For example, certain asymmetries in the electron distributions have recently been attributed to the chirality of the molecular target [16, 17]. In general, depending on the driving laser field the photoelectron momentum distributions (PEMDs) consist of a myriad of interference patterns between various pathways. Those interfering pathways are different pathways the electron may take after appearing in the continuum, but also different trajectories to escape the Coulomb potential to begin with. Resolving those interference patterns by refining the theoretical models of strong-field ionisation is an active field of research, called electron holography [18–20]. For example, caustics can be found in specific scenarios where the several orbits coalesce [21–23]. Moreover, in order to match the experimental observations and/or the simulations of the time-dependent Schrödinger equation (TDSE) it can be necessary to include the effects of the Coulomb potential during the propagation in the continuum [20, 21, 24, 25]. This is of course particularly relevant for orbits in the vicinity of the core.

### 2.1.3 Classical picture of recolliding electrons: The simple-man model

Following the seminal works on strong-field ionisation, as well as experimental observations of radiation of higher frequencies, the generation of these high-order harmonics was explained within a similar framework by including the possibility of recombination with the parent ion. The first explanation of HHG was provided by Corkum [11] and Kulander [26] in terms of the three-step model where the photons with harmonic frequencies are emitted upon recollision or rescattering, also referred to as the simple-man model, sketched in Fig. 2.3.

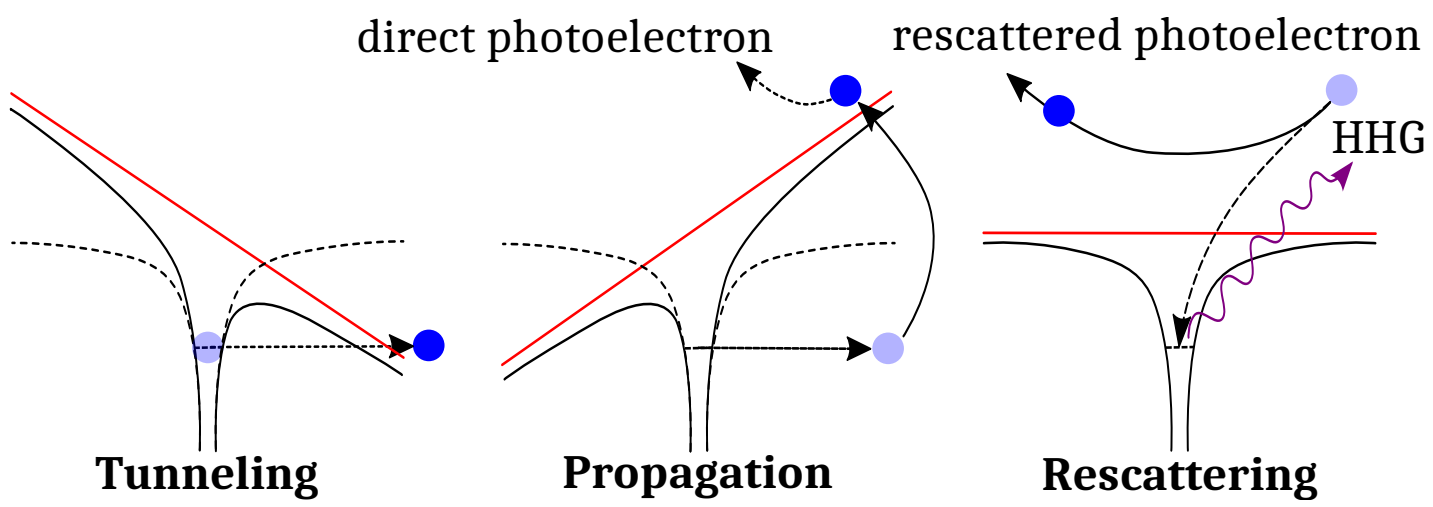

Figure 2.3: Sketch of the three-step model of HHG. The electric field of the laser (red) distorts the Coulomb barrier such that the electron can tunnel out. In the continuum the electron propagates as a free particle and can be detected as a direct photoelectron. Alternatively, it may rescatter, or recombine with the parent ion to emit a high-energetic photon (HHG).

- Step one: Tunnel ionisation.
  An electron tunnels out of a quasi-static barrier, created by the combination of the strong electric field and the atom's Coulomb binding potential, i.e., in the tunnelling regime where $\gamma < 1$. This initial tunnelling step is most likely when the field is maximal as then the distortion of the Coulomb potential is strongest.

- Step two: Propagation.
  Once the electron appears in the continuum it propagates as a free particle accelerated (read: gaining energy) by the electric field.

- Step three: Recombination.
  As the electric field changes direction, the electron is driven back to the atomic core where it recollides with its parent ion. During the time spent in the continuum it has gathered additional energy. Therefore it recombines on a higher energy level, and releases the acquired energy in the form of a photon.

In fact, step three allows for alternative processes as well. Instead of recombining and emitting a HHG photon, the electron may rescatter elastically from the core without energy exchange. Those rescattered electrons contribute to the ATI spectrum up to energies of $10U_\mathrm{p}$ as mentioned above. Alternatively, the returning electron may scatter inelastically such that a second electron is released, a process called non-sequential double-ionisation [27]. Or, the roaming electron can be captured in an auto-ionising state before emitting a photon — a process described within a 'four-step model' and leading to resonant HHG.

Within this simple man's picture it is instructive to do some basic classical calculations to gain intuition about the process of HHG. For this, we assume that the electrons are born into the continuum across a given time at some rate and then propagate like a classical particle accelerated the laser's electric field, fulfilling Newton's equations of motion. The velocity $\mathbf{v}(t)$ and position $\mathbf{x}(t)$ can easily found by integrating the acceleration $\mathbf{a}(t)$ over time:

$$\begin{aligned}
\mathbf{a}(t) &= -\mathbf{E}(t) \\
\mathbf{v}(t,t_0) &= \mathbf{v}_0 + \int_{t_0}^{t} \mathbf{a}(t')\mathrm{d}t' = \mathbf{v}_0 + \mathbf{A}(t) - \mathbf{A}(t_0) \\
\mathbf{x}(t,t_0) &= \int_{t_0}^{t} \mathbf{v}(t'+t_0)\mathrm{d}t' = (\mathbf{v}_0 - \mathbf{A}(t_0))(t-t_0) + \int_{t_0}^{t} \mathbf{A}(t')\mathrm{d}t'
\end{aligned} \tag{2.3}$$

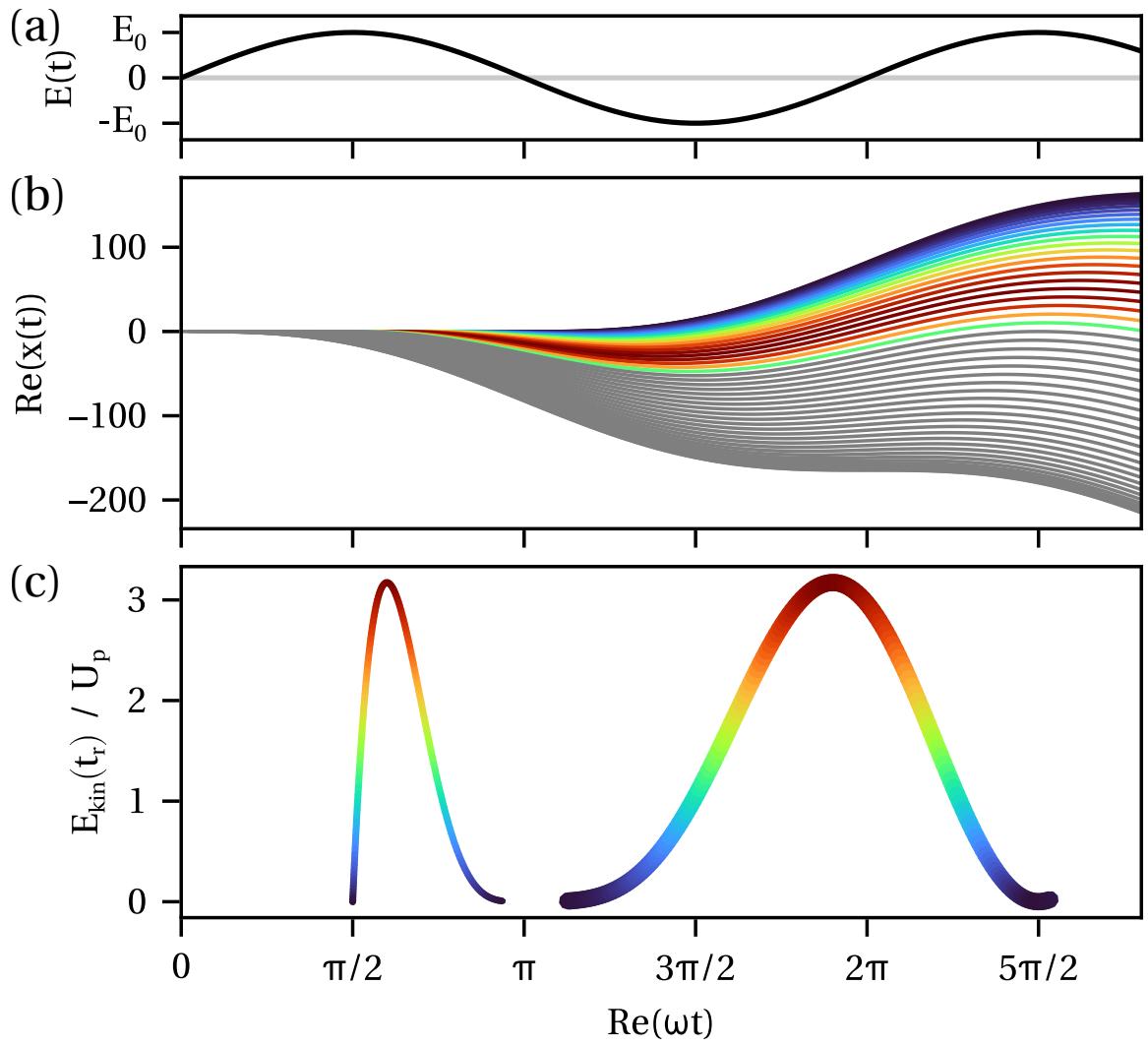

Figure 2.4: Classical electron trajectories (shown in (b)) in a linearly polarised monochromatic laser field (electric field shown in (a)). The trajectories are released into the continuum with a constant rate with zero initial velocity, and propagate under the influence of the monochromatic driving field. Trajectories released before the field maximum will not return to the core (drawn in grey). Returning trajectories are drawn in colour, indicating their kinetic energy. Those kinetic energies at the first return time $t_r$, normalised to the ponderomotive energy $U_\mathrm{p}$, are shown for the ionisation time (left thin line), and the recollision time (right heavy line). The maximum return energy is $E_\mathrm{kin}(t_r) \simeq 3.17 U_\mathrm{p}$.

In this purely classical treatment we ignore the tunnelling step and assume that the electron begins its journey at time $t = t_0$ exactly at the core at the position $\mathbf{x}(t = t_0) = 0$ with zero initial velocity $\mathbf{v}(t = t_0) = \mathbf{v}_0 = 0$. For the case of a monochromatic driving laser field, where $\mathbf{E}(t) = \sin(\omega t)\mathbf{e}_x$, we show a family of such classical trajectories in Fig. 2.4. These trajectories reveal that not all electrons return back to the core, but only those which are released after the field maximum (drawn in colour in panel (b)). Other trajectories simply drift away and do not recombine (drawn in grey). The electron's energy upon recombination is given as a classical kinetic energy

$$E_\mathrm{kin}(t_\mathrm{r}) = \tfrac{1}{2}|\mathbf{v}(t_\mathrm{r})|^2 \tag{2.4}$$

with the return time $t_r$ defined here by $\mathbf{x}(t = t_r) = 0$. For electrons in a monochromatic field the maximum return energy can numerically be found to be $E_{\text{kin,max}} \simeq 3.17U_p$, as visualised in Fig. 2.4(c). Assuming that all this energy is transferred into the emitted photon the maximum frequency is given by the classical harmonic cutoff law

$$\omega q_{\text{max}} = \mathcal{I}_p + 3.17U_p \tag{2.5}$$

According to this classical argument, no harmonic radiation should be generated with frequencies larger than this [11].

A further insight from the classical trajectories is that for any given return energy (i.e., colour in the figure, harmonic order) below this cutoff there exist two possible ionisation times that lead to the same recombination energy — corresponding to the so-called short and long trajectories. This hints at another quantum-mechanical feature of HHG: the interference of multiple electron pathways to contribute to the same emitted harmonic radiation. In conclusion, while the classical model is certainly helpful to convey the central ideas of the strong-field light-matter interaction, it omits other important features of the HHG process, such as the tunnelling ionisation, the generation of harmonics beyond the classical cutoff, and the interference pattern throughout the harmonic plateau. Nevertheless, it provides a valuable intuitive framework for understanding the qualitative structure of high-harmonic emission and the resulting spectrum.

### 2.1.4 Semi-classical framework and the strong-field approximation

The theoretical framework that incorporates this rather intuitive model of the strong-field light-matter interaction into a fully quantum-mechanical description of the interaction is provided by the strong-field approximation (SFA) [12]. This semi-classical treatment truly involves a set of approximations that are typically made such as to obtain tractable quantities for the observables associated with the various processes. In the following we want to briefly sketch the main ideas. The key assumptions in the SFA are:

- dipole approximation: any spatial dependency of the laser's vector potential is neglected and hence we neglect magnetic fields of the laser driver
- single active electron approximation: we consider the dynamics of one electron only, and ignore any interactions with other electron charges
- strong-field approximation: while the electron is in the continuum the atomic potential is neglected

How these assumptions lead to an expression for the time-dependent radiation dipole that can be used to calculate the HHG spectrum will be detailed in Sec. 2.4 below.

### 2.1.5 Beyond the single-atom SFA

The general framework of the SFA paved the way for a meaningful interpretation of other strong-field effects, such as high-order above-threshold ionisation [28, 29], and non-sequential double-ionisation in atoms [29, 30, 30, 31], as well as HHG from molecules [32], and from solids.

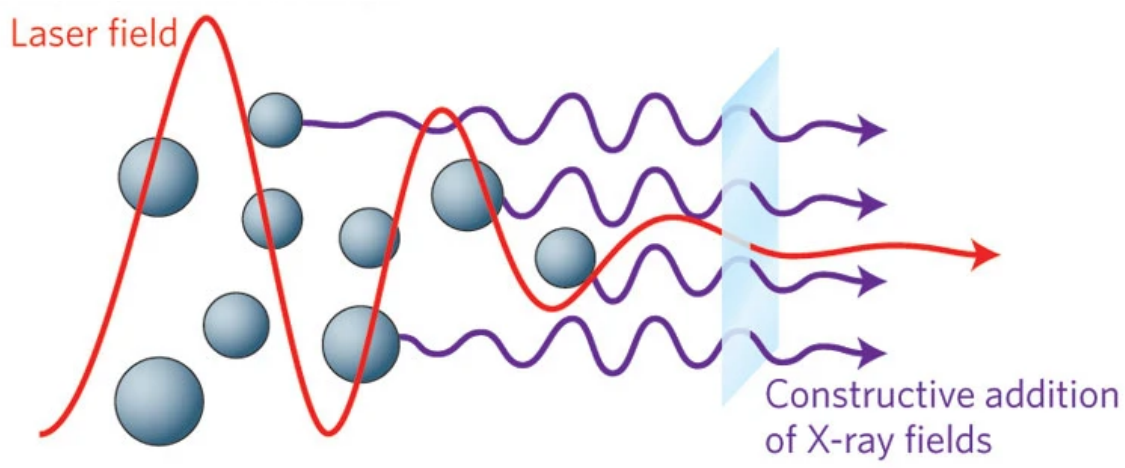

Figure 2.5: Harmonic signal emitted from each atomic target needs to phase-match and constructively interfere in order to be detected as an attosecond pulse. From [33].

A dismissed feature of the vanilla SFA approach is that the ionisation process is assumed to happen somewhat instantaneously. That is, it is assumed that the barrier created by the laser field remains fixed throughout the tunnelling process. However, numerous studies have highlighted the non-adiabatic features of the tunnelling process. The most prominent example is the debate around how long the electron spends inside the tunnel, fuelled by various definitions of 'tunnelling time'. Experimental observations are provided by the attoclock scheme, but other approaches to answer this question exist as well. In Chapter 9 we will show a scenario that emphasizes the nonadabaticity of the tunnelling step, because we find an ionisation event at a time when — within the non-adiabatic picture — there is no barrier.

## 2.2 Description of the driving laser fields

**Macroscopic structure of the driving beams**

The light sources used to probe strong-field ionisation and to generate high-order harmonics are typically pulsed laser beams which are tightly focused into a gas jet in order to reach the high intensities required for HHG. Such laser pulses offer a wide range of tunable properties, including their spatial intensity profile, temporal envelope, and polarisation state. By varying fundamental parameters like wavelength, intensity, and polarisation, or by combining multiple fields, it is possible to create a myriad of different polarisation states. These include beams with spatially structured intensity profiles, such as Laguerre–Gaussian modes with characteristic doughnut-shaped profiles, as well as fields carrying angular momentum. In particular, suitable field combinations allow for the generation of orbital angular momentum, spin-orbit-coupled states, and more exotic angular momentum structures like torus knots [34–36].

The systematic variation of these laser parameters offers powerful experimental control over the HHG process and the properties of the resulting attosecond pulses. In this thesis, however, the focus lies on the theoretical description of the microscopic strong-field response, specifically on the calculation and interpretation of the highly oscillatory integrals that describe the interaction of strong laser fields with a single atomic target.

**Microscopic field shapes at the target**

For the microscopic response, we work in the dipole approximation and assume that the length scales of our considered processes are much smaller than any changes in the electromagnetic field shape. That means that for the vector potential of the laser field we only consider the time dependency and assume that any spatial structure can be absorbed into a prefactor.

The temporal structure of laser fields often exhibits dynamical symmetries, defined as the combination of temporal translations and spatial rotations. This symmetry is present in the most simplest waveforms, for example a linearly polarised monochromatic field: the second half-cycle reproduces the first, rotated by 180°. In the various applications discussed in the Chapters later on, we will consider a variety of field configurations and systematically vary their parameters. At this stage, however, we provide a brief overview to illustrate the versatility of these approaches and to establish a common foundation for later reference.

While the approaches we present are generic to all driving laser fields, most of the showcased applications in this thesis are based on two-colour fields. That is, the combination of two driving fields with different frequencies — most often we will use a combination of $\omega$ and its second harmonic, $2\omega$. This frequency ratio is particular popular, because the creation of the second harmonic frequency using a Barium borate (BBO) crystal is well-understood and controllable. For example, the intensity ratio between the two fields can be tuned by changing the conversion efficiency by rotating the BBO crystal.

On the theoretical side, let us briefly show how the specific parameter changes affect the respective field shapes. A generic description of a two-colour field is

$$\mathbf{E}(t) = E_{01}\sin(r\omega t)\mathbf{e}_1 + E_{02}\sin\big(s\omega t+\varphi\big)\mathbf{e}_2 \tag{2.6}$$

with the polarisation vectors $\mathbf{e}_1$ and $\mathbf{e}_2$, individual field amplitudes $E_{01}$ and $E_{02}$, frequencies $r\omega$ and $s\omega$, and the two-colour phase shift $\varphi$.

Within this thesis we mostly focus on two representative configurations: collinear fields with $\mathbf{e}_1 = \mathbf{e}_2 = \mathbf{e}_x$, and co-orthogonal fields where $\mathbf{e}_1 \perp \mathbf{e}_2$, both realised for the frequency combination of $r = 1$ and $s = 2$, i.e., $\omega - 2\omega$-fields. Below we examine how changes in the amplitudes $E_{01}$, $E_{02}$ and in the relative phase shift $\varphi$ influence the resulting field shapes. In Fig. 2.6 we show examples of a co-linear $\omega - 2\omega$ driving field,

$$\mathbf{E}(t) = E_{01}\sin(\omega t)\mathbf{e}_x + E_{02}\sin\big(2\omega t+\varphi\big)\mathbf{e}_x \tag{2.7}$$

for a range of different parameters. Each panel shows the two constituent fields $E_1(t) = E_{01}\sin(\omega t)$ (red dashed) and $E_2(t) = E_{02}\sin\big(2\omega t+\varphi\big)$ (blue dotted), respectively, as well as the combined field (black). Across the columns we vary the phase shift $\varphi$. Along the rows we vary the relative amplitude ratio $R$ by using

$$E_{01} = \sqrt{1-R}E_0 \qquad \text{and} \qquad E_{02} = \sqrt{R}E_0 \tag{2.8}$$

such that the total intensity $I_0$ remains constant [37], as

$$R = \frac{I_{2\omega}}{I_{2\omega}+I_\omega} = \frac{I_{2\omega}}{I_0}\,. \tag{2.9}$$

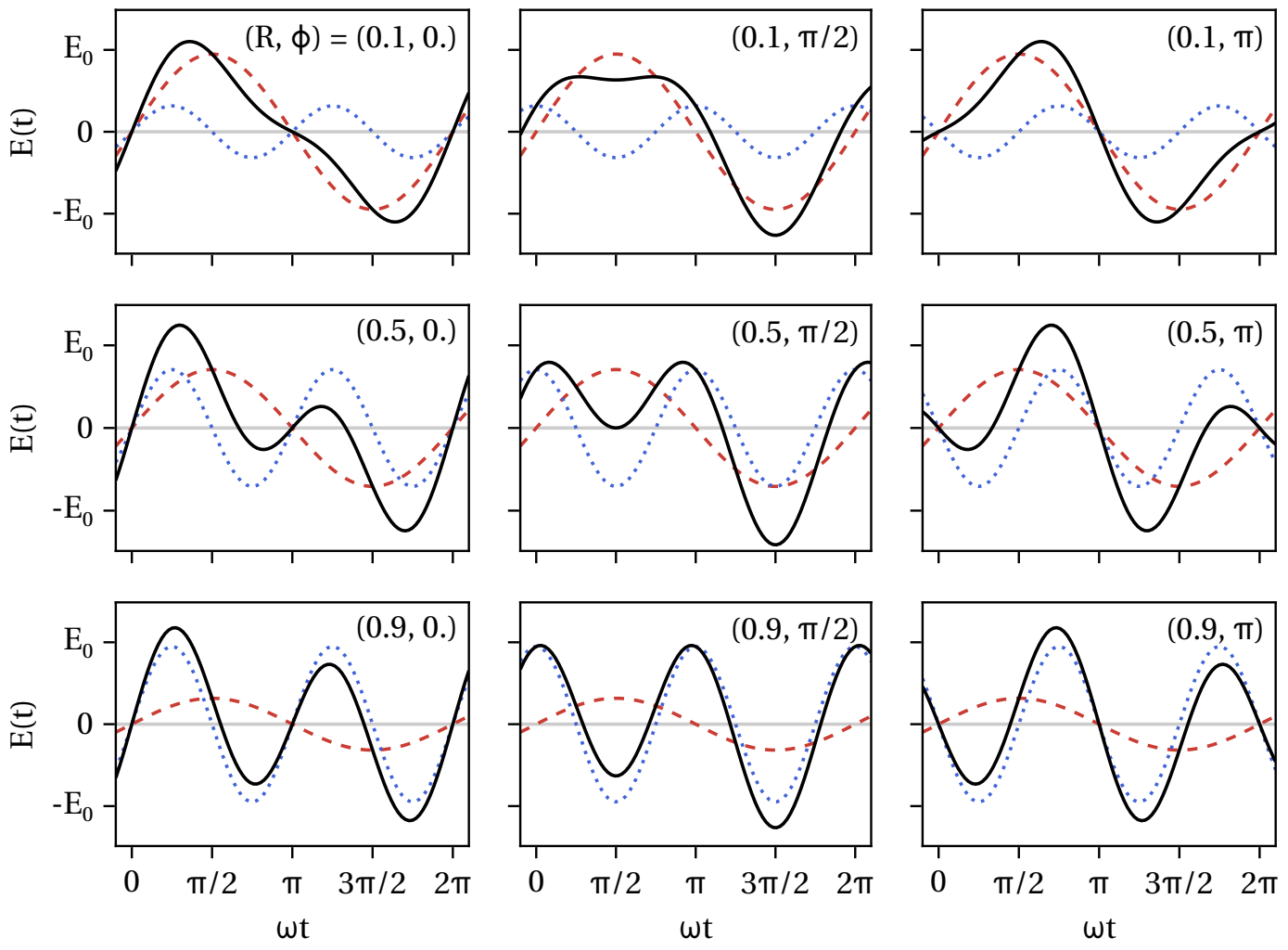

Figure 2.6: Co-linear two-colour field amplitude (black line) composed of a fundamental $\omega$-field (red dashed) and a $2\omega$-field (blue dotted), both are linearly polarised in the $x$-direction. The several panels show the two-colour field for different amplitude ratios $R$ (rows) and phase shifts $\varphi$ (columns) between the two fields.

Note that within the literature and throughout this thesis there exist several definitions of (a continuous change of) the amplitude ratio between two fields.

The combination of two fields with orthogonal polarisation axes creates non-trivial polarisation states. The various field shapes and temporal waveforms that are created from the superposition of two linearly polarised fields with orthogonal polarisation axes are shown in Fig. 2.7. The generic expression for the electric field is

$$\mathbf{E}(t) = E_{01}\sin(\omega t)\mathbf{e}_x + E_{02}\sin\big(2\omega t + \varphi\big)\mathbf{e}_y\,. \tag{2.10}$$

Analogously to Fig. 2.6, in each panel of Fig. 2.7 we show the two component fields over time as well as the field amplitude $|\mathbf{E}(t)|$ in black. Furthermore we show the Lissajous figure as an inset, where the arrows indicate the field and its temporal evolution at $t = 0$. Along the rows of Fig. 2.7 the fields are superimposed with varying phase shifts $\varphi$, and across the columns we vary the amplitude ratio in the same fashion as above.

For completeness, we also briefly mention bicircular driving fields. These are typically generated by superimposing two counter-rotating circularly polarised fields at frequencies $\omega$ and $2\omega$. Equal amplitudes of the two components give rise to the characteristic trefoil-shaped Lissajous figure exhibiting threefold rotational symmetry. Variations in the relative amplitudes or in the phase shift between the fields allow for a wide range of polarisation states and may break the underlying dynamical symmetry.

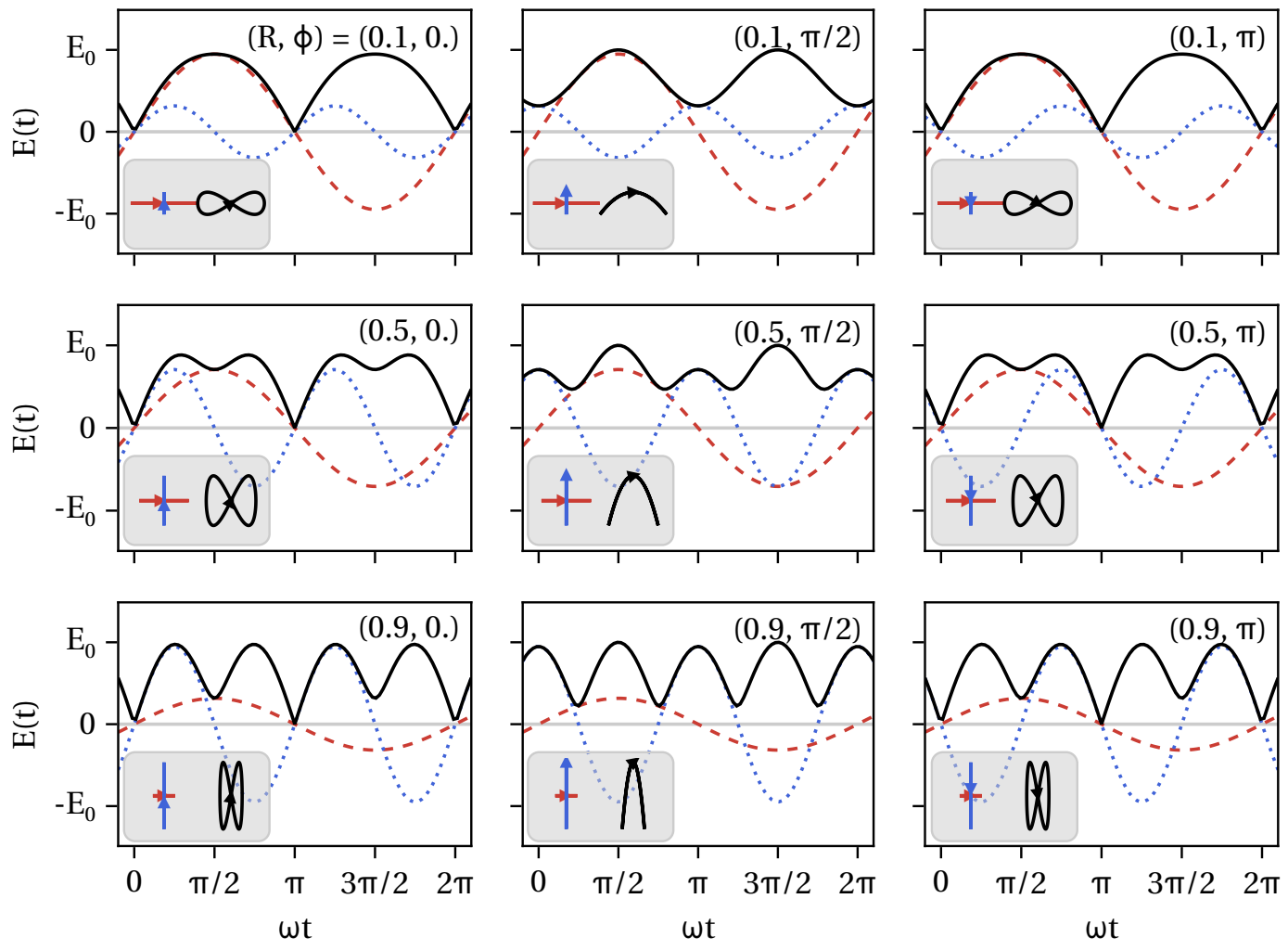

Figure 2.7: Co-orthogonal two-colour fields — similar as Fig. 2.6, but the two field components are polarised in orthogonal directions, hence the black line shows the total field magnitude. The respective Lissajous figures are shown as an inset within each panel.

# 2.3 Strong-field ionisation within the SFA

## 2.3.1 The tunnel ionisation amplitude in the SFA

In the following we derive the transition amplitude that describes the probability amplitude to detect a photoelectron with momentum $p$ at time $t$ at the detector. For momenta up to $2U_\mathrm{p}$ those will mostly be electrons that have tunnelled out of the Coulomb potential and reached the detector on a direct pathway, i.e., the direct photoelectrons mentioned above. A majority of the work (both theoretical and experimental) ATI goes into the various electron pathways and their interference as briefly discussed above. Here, we focus only on the direct electrons for the simple reason that the transition amplitude is structurally similar to the HHG integral we will derive and study later on. The following derivation largely follows the presentation in Refs. [12, 38, 39].

Let us begin by finding a suitable wave function under the Hamiltonian that captures the physics we are seeking to describe.

**The Hamiltonian**

We consider an electron subjected to a strong electromagnetic field $\mathbf{E}(t)$, such that the Schrödinger equation can be written[1] as

$$\mathrm{i}\frac{\partial}{\partial t}|\Psi(\mathbf{r},t)\rangle = \left(-\frac{1}{2}\nabla^2 + V(\mathbf{r}) + V_L(t)\right)|\Psi(\mathbf{r},t)\rangle \tag{2.11}$$

where $V(\mathbf{r})$ is the atomic binding potential and $V_L(t)$ describes the interaction with the laser field. In the velocity gauge, we take $V_L(t) = \mathbf{k}\cdot\mathbf{A}(t)$, where $\mathbf{k}$ is the instantaneous, kinematic momentum and $\mathbf{A}$ is the vector potential of the electric field, $\mathbf{A}(t) = -\int \mathbf{E}(t')\,\mathrm{d}t'$. In the length gauge the interaction is described as $V_L(t) = \hat{\mathbf{r}}\cdot\mathbf{E}(t)$ with the position operator $\hat{\mathbf{r}}$. While generally the observables of a quantum mechanical system should remain independent of the gauge, within the SFA there are so many approximations that the choice of gauge yields different results [40]. For the semi-classical description of HHG from a single atom within the SFA the length gauge is the preferred choice [41].

This Hamiltonian can be approached separately, in terms of the field-free Hamiltonian $H_0$, and the 'laser-only' Hamiltonian $H_L$. The field-free Hamiltonian

$$H_0 = -\frac{1}{2}\nabla^2 + V(\mathbf{r}) \tag{2.12}$$

defines the energetic ground state of the atom, where we use the Coulomb potential for Hydrogen-like atoms, $V(\mathbf{r}) = -\frac{1}{|\mathbf{r}|}$, such that

$$H_0|\Psi_0\rangle = -\mathcal{I}_\mathrm{p}|\Psi_0\rangle \tag{2.13}$$

with the ionisation potential $\mathcal{I}_\mathrm{p}$. The respective wave function is given as the temporal evolution of the spatial distribution of the ground state:

$$|\Psi(t)\rangle = |\Psi_0\rangle\,\mathrm{e}^{\mathrm{i}\mathcal{I}_\mathrm{p}t}. \tag{2.14}$$

As this ground state we typically assume a Hydrogen-like $s$-orbital, with $\alpha = 2\mathcal{I}_\mathrm{p}$, given by

$$\left|\Psi_{0,1s}\right\rangle = \frac{\alpha^{3/4}}{\sqrt{\pi}}\mathrm{e}^{-\sqrt{\alpha}|\mathbf{r}|}. \tag{2.15}$$

For Hydrogen, $\mathcal{I}_\mathrm{p} = 0.5\,\mathrm{a.u.} \approx 13.6\,\mathrm{eV}$, and the extent of the electron wave packet is in the order of magnitude of the Bohr radius, $a_B = 5.2\times10^{-2}\mathrm{nm}$. This is small compared to the typical wavelength taken for HHG experiments of 800 nm. Hence the *electric dipole approximation* is used, i.e., we neglect the spatial dependence of the laser field's vector potential and assume $\mathbf{A}(\mathbf{r},t) \approx \mathbf{A}(t)$. This implies that the magnetic field of the laser beam is ignored and the wave function is solely dependent on the laser field's time-dependent electric field $\mathbf{E}(t)$.

[1] We use bra-ket notation for the inner product on a complex Hilbert space, where $|\phi\rangle$ denotes a vector, $\langle\psi|$ its dual, and $\langle\psi|\phi\rangle$ their inner product.

**The electron in the continuum: Volkov states**
To model the electronic wave function in the continuum, we consider the Hamiltonian of an electron that propagates freely in the electric field of a laser, in the length gauge:

$$H_L(t) = -\frac{1}{2}\nabla^2 + \hat{\mathbf{r}} \cdot \mathbf{E}(t)\,. \tag{2.16}$$

Solutions to the TDSE for this Hamiltonian are the Volkov states

$$\left|\Psi_{\mathbf{p}}(t)\right\rangle = \exp\left(-\mathrm{i}S_{\mathrm{V}}(\mathbf{p},t,t_0)\right)\left|\mathbf{p}+\mathbf{A}(t)\right\rangle, \tag{2.17}$$

with the so-called Volkov phase for momentum $\mathbf{p}$

$$S_{\mathrm{V}}(\mathbf{p},t,t_0) = \int_{t_0}^{t} \frac{1}{2}\left(\mathbf{p}+\mathbf{A}(t')\right)^2 \mathrm{d}t'\,. \tag{2.18}$$

Their spatial distribution is

$$\left\langle \mathbf{r}\middle|\mathbf{p}+\mathbf{A}(t)\right\rangle = \frac{1}{(2\pi)^{3/2}}\mathrm{e}^{\mathrm{i}(\mathbf{p}+\mathbf{A}(t))\cdot\mathbf{r}}\,. \tag{2.19}$$

The Volkov states are specified by their kinematic momentum $\mathbf{k} = \mathbf{p}+\mathbf{A}(t)$, constituting the velocity, and can be understood as simple plane waves for the laser field. The canonical momentum $\mathbf{p}$ is the 'intrinsic' drift momentum of the electron which will ultimately be recorded at the detector, and which is a conserved quantity (as long as we ignore the effects of the Coulomb potential).

**The SFA ansatz for the wave function**
The most important step of the SFA is now to make the actual *strong-field approximation* and write the wave function as a combination of the ground state wave function and these Volkov states:

$$|\Psi(t)\rangle = a(t)\mathrm{e}^{\mathrm{i}\mathcal{I}_{\mathrm{p}}t}|\Psi_0\rangle + \int \mathrm{d}^3\mathbf{p}\, b(\mathbf{p},t)\left|\Psi_{\mathbf{p}}\right\rangle \tag{2.20}$$

with the ground state amplitude $a(t)$ and the amplitudes of the respective Volkov states in the continuum $b(\mathbf{p},t)$. By making this ansatz, we assume that the electron is *either* bound in its atomic ground state *or* freely propagating in the continuum, only driven by the laser field. That is, we ignore any influence of the ionic core's Coulomb potential on the dynamics in the continuum. And we ignore any resonances with any intermediate excited energy states. Moreover, by writing Eq. (2.20) we assume that the system can be described in terms of a single wave function, an approximation known as the *single active electron approximation.*

**The ionisation amplitude**
To obtain an expression for the transition amplitude $b(\mathbf{p},t)$ the TDSE (2.11) is solved with the ansatz Eq. (2.20) such that we obtain [12, 38, 42]

$$\begin{aligned} b(\mathbf{p},t) &= -\mathrm{i}\int_{-\infty}^{t} \mathrm{d}t'\, \left\langle \Psi_{\mathbf{p}}\middle|\mathbf{r}\cdot\mathbf{E}(t')\middle|\Psi_0\right\rangle \\ &= -\mathrm{i}\int_{-\infty}^{t} \mathrm{d}t'\, a(t')\mathbf{E}(t')\cdot\mathbf{d}(\mathbf{p}+\mathbf{A}(t'))\exp\left(\mathrm{i}\mathcal{I}_{\mathrm{p}}t' + \mathrm{i}S_{\mathrm{V}}(\mathbf{p},t',t_0)\right) \end{aligned} \tag{2.21}$$

where the integration runs over the (past) interaction time of the laser field with the atom, $t'$ and we have used the dipole transition matrix element (also known as a form factor)

$$\mathbf{d}(\mathbf{k}) = \langle \mathbf{k} | \mathbf{r} | \Psi_0 \rangle \,. \tag{2.22}$$

For the hydrogenic ground state Eq. (2.15) this matrix element evaluates to[2]

$$\mathbf{d}(\mathbf{k}) = -\mathrm{i}\frac{2^{7/2}}{\pi}\frac{\alpha^{(5/4)}\mathbf{k}}{(\mathbf{k}^2+\alpha)^3}\,. \tag{2.23}$$

The amplitude $b(\mathbf{p}, t)$ is the probability to detect a photoelectron with momentum $\mathbf{p}$ at any given time $t$. To obtain the *total* ionisation amplitude $\Psi(\mathbf{p})$ we consider the full interaction time between the laser pulse and the target, i.e., $t \to \infty$, where $|\mathbf{A}(t \to \pm\infty)| = 0$, and neglect the depletion of the ground state by setting $a(t) = 1$ as only few atoms will be ionised anyway. The total ionisation amplitude for a given drift momentum $\mathbf{p}$ then reads

$$\Psi(\mathbf{p}) = \int_{-\infty}^{+\infty} P(\mathbf{p}+\mathbf{A}(t))\mathrm{e}^{-\mathrm{i}S_{\mathrm{ATI}}(\mathbf{p},t)}\,\mathrm{d}t \tag{2.24}$$

with the semi-classical Volkov action for ATI

$$S_{\mathrm{ATI}}(\mathbf{p}, t) = \mathcal{I}_{\mathrm{p}}t + \frac{1}{2}\int_{-\infty}^{t}(\mathbf{p}+\mathbf{A}(t'))^2\,\mathrm{d}t' \tag{2.25}$$

and the prefactor $P(\mathbf{k})$ carrying the information about the transition amplitude between the ground state orbital and the respective Volkov state.

In the tunnelling limit, the kinematic momentum at time of ionisation is roughly in the order of the ionisation potential, such that $|\mathbf{k}| \approx \sqrt{\alpha}$ and for the denominator in Eq. (2.23) we can approximate $(\mathbf{k}^2+\alpha)^3 \approx (2\alpha)^3$. The magnitude of the dipole transition matrix element therefore simplifies considerably and hence, the prefactor in Eq. (2.24) is assumed as

$$P(\mathbf{k}) = \frac{\mathrm{i}}{\sqrt{\pi}}\left(2\mathcal{I}_{\mathrm{p}}\right)^{1/4}\,. \tag{2.26}$$

## 2.3.2 Saddle-point approach and ionisation events

The integral Eq. (2.24) derived above is highly oscillatory and can therefore be solved using saddle-point methods. Of course, these will be discussed in detail below as their 'extension' to Picard–Lefschetz methods are the main subject of this thesis. For now, however, we shall briefly sketch the idea.

**Saddle-point approximation (brief preview)**

Integrals in the shape of

$$\int g(\mathbf{x})\exp\left(\mathrm{i}kf(\mathbf{x})\right)\mathrm{d}\mathbf{x} \tag{2.27}$$

with a slowly varying prefactor $g(\mathbf{x})$ and large scalar factor $k$ can be approximated as a sum over contributions associated with the stationary points of the exponentiated phase function $f(\mathbf{x})$, where $f:\mathbb{R}^N \to \mathbb{R}$ and must be analytic [43]. For that, we assume that neither $g(\mathbf{x})$ nor $f(\mathbf{x})$ have singularities [44], and we can use the Taylor expansion for $f(\mathbf{x})$ around the stationary points $\mathbf{x}_s$, which are defined by

$$\nabla_{\mathbf{x}} f(\mathbf{x})|_{\mathbf{x}=\mathbf{x}_s} = 0\,. \tag{2.28}$$

[2] Note that here $\mathbf{k}^2 = \mathbf{k}^T\mathbf{k} \neq |\mathbf{k}|^2 = \mathbf{k}^H\mathbf{k}$.

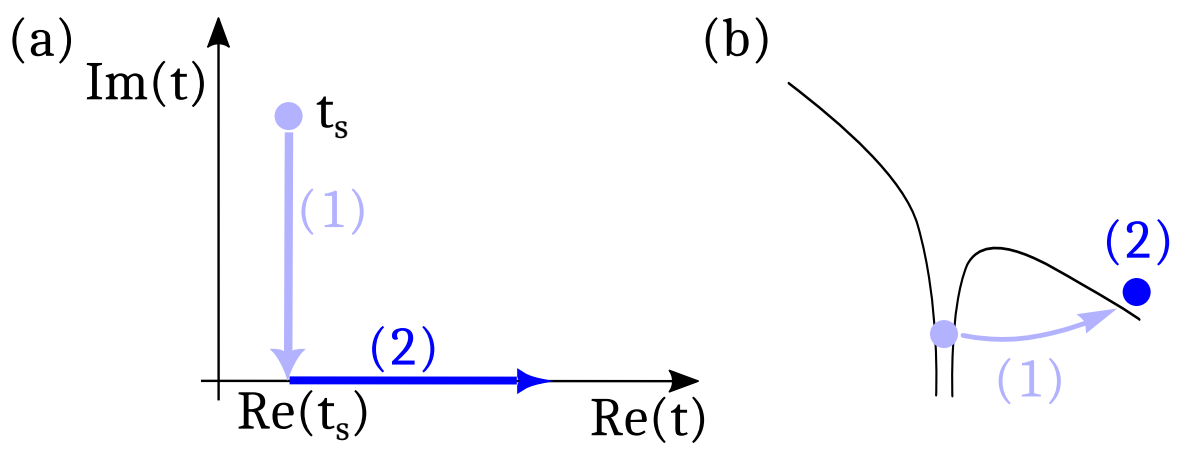

Figure 2.8: Schematic depiction of (a) how time passes in the complex plane during the two-step process of tunnel ionisation (b). First, imaginary time traverses inside the tunnel, and subsequently, upon appearance in the continuum, along the real axis.

Terminating the Taylor expansion after the second order results in a trivial integral of Gaussian shape, that can be solved analytically. Therefore

$$\int g(\mathbf{x}) \exp\left(\mathrm{i}kf(\mathbf{x})\right) \mathrm{d}\mathbf{x} \approx \sum_{\mathbf{x}_s} \sqrt{\frac{(2\pi\mathrm{i})^N}{k \det\left(f''(\mathbf{x}_s)\right)}}\, g(\mathbf{x}_s) \exp\left(\mathrm{i}f(\mathbf{x}_s)\right) \tag{2.29}$$

where the $\mathbf{x}_s$ are the stationary points, and the Hessian is

$$f''(\mathbf{x}) = \left(\frac{\partial^2 f(\mathbf{x})}{\partial x_i \partial x_j}\right)_{i,j=1...N} \tag{2.30}$$

which we assume to have a non-zero determinant at the stationary point, $\det\left(f''(\mathbf{x}_s)\right) \neq 0$, such that it is non-degenerate and well-separated from other stationary points. This approximation becomes more accurate for larger $k$. An often overlooked fact is that the summation does not run over *all* stationary points that fulfil Eq. (2.28), but only those that are part of the deformed steepest-descent integration contour, a property that shall be discussed in detail later.

**Saddle points of strong-field ionisation**

For the case of the strong-field ionisation amplitude Eq. (2.24), the phase function is the semi-classical action $S_{\mathrm{ATI}}(\mathbf{p}, t)$ with the first derivative

$$\frac{\partial S_{\mathrm{ATI}}}{\partial t} = \mathcal{I}_{\mathrm{p}} + \frac{1}{2}\left(\mathbf{p} + \mathbf{A}(t)\right)^2\,. \tag{2.31}$$

Stationary points thereof are defined as

$$\left.\frac{\partial S_{\mathrm{ATI}}}{\partial t}\right|_{t=t_s} = 0 \tag{2.32}$$

and are generally complex-valued numbers (because $\mathcal{I}_\mathrm{p} > 0$) and correspond to discrete ionisation events. The imaginary part of the ionisation time $t_s$ is commonly interpreted as the time the electron spends in the tunnelling barrier, as the “non-classical” part of the process [38]. This is shown schematically in Fig. 2.8. We assume that as a first step, the imaginary part of $t_s$ passes, and then the electron appears in the continuum at $t = \mathrm{Re}\, t_s$. Subsequently, as a second step, time passes on the real axis while the electron propagates in the continuum.

For the simplest case of a monochromatic driving field the saddle point solutions are shown in Fig. 2.9. We use a stereotypical driver with $I_0 = 4\times 10^{14}\,\mathrm{W/cm^2} = E_0^2$, $\omega = 0.057\,\mathrm{a.u.}$ ($\lambda = 800\,\mathrm{nm}$) with the electric field $E(t) = E_0 \sin(\omega t)$, shown in panel (a). The ponderomotive energy for this driver is $U_\mathrm{p} \approx 0.88\,\mathrm{a.u.}$ In panel (b) we show the solutions to Eq. (2.32), with $\mathcal{I}_\mathrm{p} = 0.58\,\mathrm{a.u.}$, in the complex time plane for momenta $p = p_x$ in the range $-2U_\mathrm{p} \lesssim p \lesssim 2U_\mathrm{p}$ (negative momenta in red, positive momenta in blue). As expected, the solutions are symmetrically centred around the maxima of the driving field, and periodic with respect thereto. Panel (c) shows the corresponding semi-classical electron trajectories upon appearance in the continuum, i.e., only for the second leg of the time contour in Fig. 2.8. For visual clarity, trajectories stemming from the second half cycle are drawn dashed.

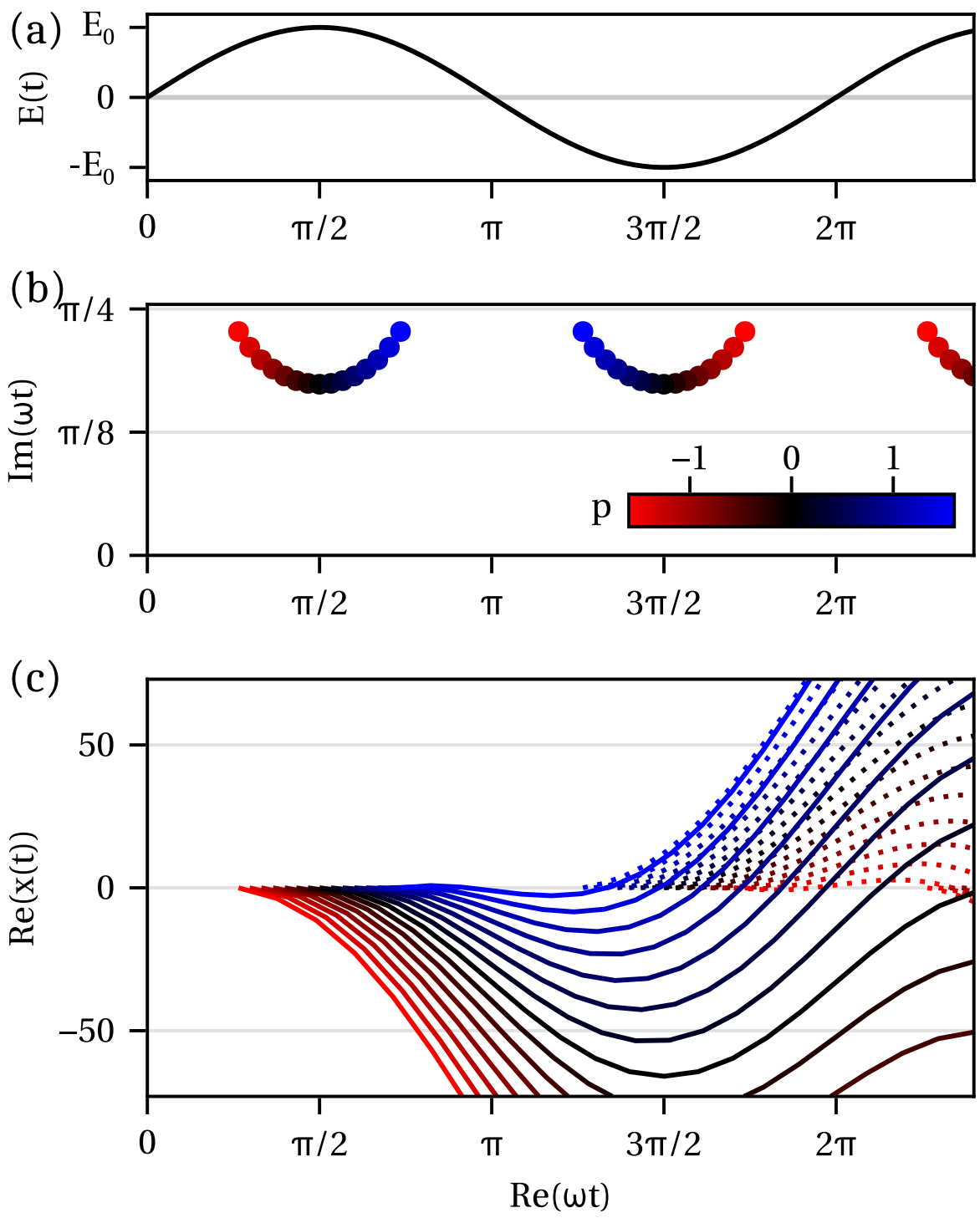

Figure 2.9: (a) Monochromatic electric field, (b) saddle points in the complex plane for a range of momenta $p$ (in colour), (c) semi-classical electron trajectories for different momenta in the first (solid lines) and second (dotted lines) half cycle, evaluated for the second leg of the temporal contour of Fig. 2.8(a).

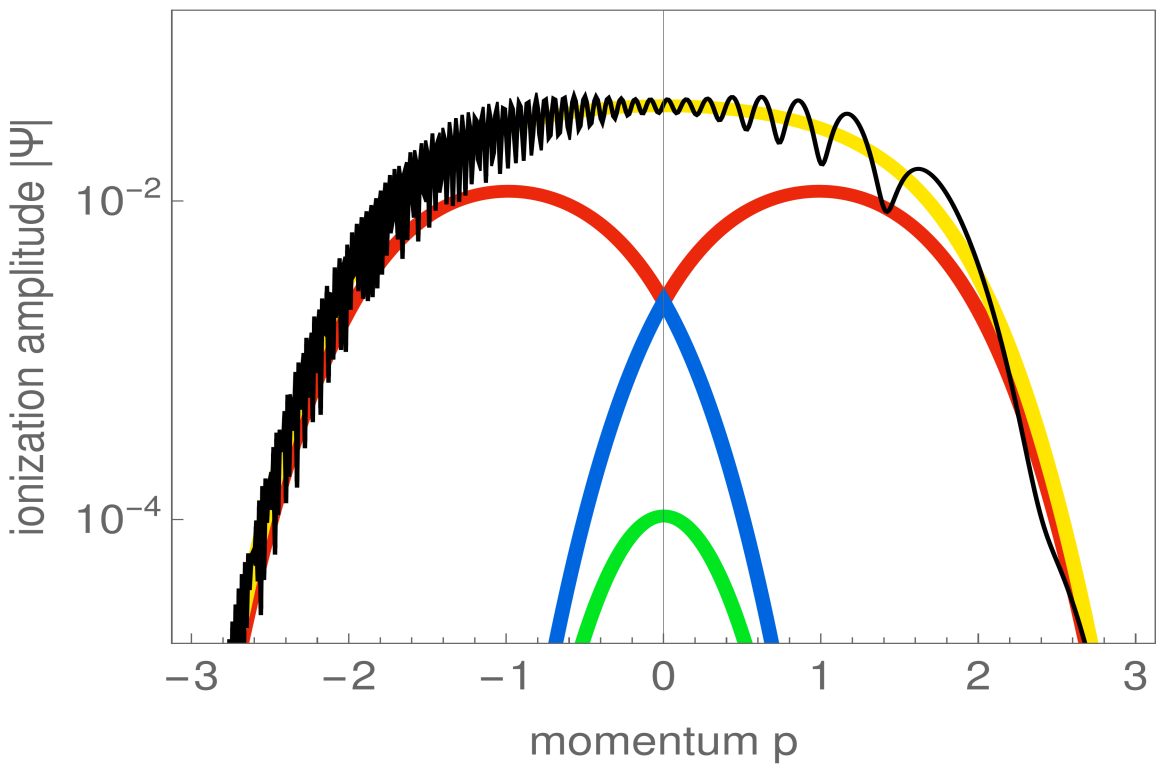

Figure 2.10: Spectrum of the ionisation amplitude of direct photoelectron from a two-colour driving field with equal amplitudes. Contributions from separate ionisation events are shown in separate colours, as well the total amplitude in black.

### 2.3.3 The ATI spectrum of direct photoelectrons

In order to compute the full ionisation amplitude Eq. (2.24) in the saddle-point approximation for a given momentum $\mathbf{p}$, we can now sum over the contributions of each of the relevant ionisation events:

$$\Psi(\mathbf{p}) \approx \sum_s \sqrt{\frac{2\pi \mathrm{i}}{S''_{\mathrm{ATI}}(\mathbf{p}, t_s)}} P(\mathbf{p}+\mathbf{A}(t_s)) \mathrm{e}^{-\mathrm{i}S_{\mathrm{ATI}}(\mathbf{p},t_s)} \tag{2.33}$$

with the second derivative

$$S''_{\mathrm{ATI}}(\mathbf{p}, t) = -\mathbf{E}(t)\cdot\mathbf{p} + \mathbf{E}(t)\cdot\mathbf{A}(t)\,. \tag{2.34}$$

How to rigorously identify which of the solutions to Eq. (2.32) are relevant to the summation will be discussed in detail later on. So far, the understanding in the community is that the saddle points must be part of a valid integration path. In most cases, however, some simple heuristics are sufficient to determine the relevant saddle points. For instance, those saddle points with negative imaginary parts will be discarded because they yield unphysical contributions.

As an example, the total ionisation amplitude $|\Psi(\mathbf{p})|$ of direct photoelectrons for a co-linear $\omega - 2\omega$ driving laser field with equal amplitudes for the two field components across a whole spectrum of momenta is shown in Fig. 2.10. Therein, we have plotted the individual contributions of four discrete ionisation events in different colours, corresponding to the magnitudes of the different summands of Eq. (2.33). The total amplitude (in black) shows distinct ripples, present in the sketch Fig. 2.2 as well. They are the fringes from the intracycle interference between the several involved trajectories. These peaks are equally spaced in distances of $\Delta U_\mathrm{p} = \hbar\omega$. Here we plot in units of $U_\mathrm{p}$, such that the relative spacing is proportional to $\hbar\omega/U_\mathrm{p}$, and hence, shrinking with increasing $U_\mathrm{p}$ (viz. towards the right-hand side).

## 2.4 High-order harmonic generation in the SFA

### 2.4.1 Theoretical derivation of the harmonic dipole (Lewenstein integral)

The following derivation largely follows Refs. [12] and [38], and is standard knowledge in the community of (theoretical) attosecond science. We include it here nevertheless, just to demonstrate the origin of the integrals whose evaluation we will study in detail later.

**Calculating the time-dependent dipole**
The goal is to find an expression for the intensity of all the possible frequencies that be produced from the HHG process, i.e., the HHG spectrum. Since this harmonic radiation originates from the induced dipole at a single atom, in the following we will derive an expression for this time-dependent dipole moment $\mathbf{D}(t)$. The dipole moment of a wave function $|\Psi(t)\rangle$ is given as the expectation value of the position operator

$$\mathbf{D}(t) = \langle\Psi(t)|\hat{\mathbf{r}}|\Psi(t)\rangle\,. \tag{2.35}$$

To calculate the dipole induced at time $t$ we use the SFA ansatz for the wave function Eq. (2.20) derived above, such that

$$\begin{aligned}\mathbf{D}(t) &= \langle\Psi(t)|\hat{\mathbf{r}}|\Psi(t)\rangle \\ &= a^2(t)\,\langle\Psi_0|\hat{\mathbf{r}}|\Psi_0\rangle + \int \mathrm{d}^3\mathbf{p}\, a^\dagger(t) b(\mathbf{p},t)\,\langle\Psi_0|\hat{\mathbf{r}}|\mathbf{p}\rangle \\ &+ \int \mathrm{d}^3\mathbf{p}\, a(t) b(\mathbf{p},t)^\dagger\,\langle\mathbf{p}|\hat{\mathbf{r}}|\Psi_0\rangle + \int \mathrm{d}^3\mathbf{p} \int \mathrm{d}^3\mathbf{p}'\, b(\mathbf{p}',t')^\dagger b(\mathbf{p},t)\,\langle\mathbf{p}'|\hat{\mathbf{r}}|\mathbf{p}\rangle\,.\end{aligned} \tag{2.36}$$

Due to the spherical symmetry of the hydrogenic $1s$ ground state $|\Psi_0\rangle$ considered here, the first term vanishes, $\langle\Psi_0|\hat{\mathbf{r}}|\Psi_0\rangle = 0$. Furthermore, the last term describes transitions between different continuum states — we neglect those too, assuming no resonances contribute to the process of HHG. This allows us to write the dipole moment in more compact form, as

$$\mathbf{D}(t) = \int \mathrm{d}^3\mathbf{p}\, a^\dagger(t) b(\mathbf{p},t)\,\langle\Psi_0|\hat{\mathbf{r}}|\mathbf{p}\rangle^* + c.c. \tag{2.37}$$

where we use the transition amplitude $b(\mathbf{p},t)$ derived above, $(\cdot)^*$ denotes complex conjugation, $(\cdot)^\dagger$ Hermitian conjugation, and $c.c.$ represents the complex-conjugate of the previous term. Introducing the scalar function

$$\Upsilon(\mathbf{p},t) = \mathbf{E}(t)\cdot\mathbf{d}(\mathbf{p}+\mathbf{A}(t)) \tag{2.38}$$

finally yields a full expression for the dipole moment, known as the Lewenstein integral:

$$\mathbf{D}(t) = \mathrm{i}\int_{-\infty}^{t} \mathrm{d}t' \int \mathrm{d}^3\mathbf{p}\,\mathbf{d}^*(\mathbf{p}+\mathbf{A}(t))\Upsilon(\mathbf{p},t') \times \exp\left(-\mathrm{i}S_\mathrm{V}(\mathbf{p},t,t')\right) + c.c. \tag{2.39}$$

This expression for the dipole moment is the central result of the SFA approach to HHG and can be interpreted in terms of the three-step model mentioned above: The factor $\Upsilon(\mathbf{p},t')\exp\left(\mathrm{i}\mathcal{I}_\mathrm{p} t'\right)$ denotes the ionisation step at time $t'$, when the electron transitions from the ground state into the continuum (step one). During the propagation in the continuum (step two), the electron accumulates the phase $\exp\left(-\mathrm{i}\int_{t'}^{t}\frac{1}{2}\left(\mathbf{p}+\mathbf{A}(\tau)\right)^2\mathrm{d}\tau\right)$. The term $\exp\left(-\mathrm{i}\mathcal{I}_\mathrm{p}t\right)\mathbf{d}^*\left(\mathbf{p}+\mathbf{A}(t)\right)$ on the other hand describes the recombination (step three), which coincides with the emission of the harmonic radiation at time $t$.

**The HHG dipole in spectral domain**

As mentioned earlier, the intensity at the harmonic frequency $q\omega$, where $q$ is the harmonic order[3] and $\omega$ the fundamental frequency of the driver, is given as the Fourier transform $\mathcal{F}$ of the time-dependent dipole acceleration. That is, $I(q\omega)\propto\left|\mathcal{F}(\ddot{\mathbf{D}}(t))(q\omega)\right|^2$, and because $\mathcal{F}\left(\ddot{\mathbf{D}}(t)\right)=-(q\omega)^2\mathcal{F}(\mathbf{D}(t))$ for $\lim_{t\to\pm\infty}\mathbf{D}(t)=0$, we commonly use

$$I(q\omega)\propto|\mathcal{F}(\mathbf{D}(t))|^2=(q\omega)^4\left|\mathbf{D}(q\omega)\right|^2 \tag{2.40}$$

with the time-dependent dipole moment $\mathbf{D}(t)$ and its frequency-dependent analogue, notated as $\mathbf{D}(q\omega)$. The Lewenstein integral Eq. (2.39) in spectral domain is given by

$$\mathbf{D}(q\omega)=\mathrm{i}\int_{-\infty}^{\infty}\mathrm{d}t\int_{-\infty}^{t}\mathrm{d}t'\int\mathrm{d}^3\mathbf{p}\,\mathbf{d}^*\left(\mathbf{p}+\mathbf{A}(t)\right)\Upsilon(\mathbf{p},t')\times\exp\left(-\mathrm{i}S(\mathbf{p},t,t')\right)+c.c. \tag{2.41}$$

with the full semi-classical action

$$\begin{aligned}S(\mathbf{p},t,t')&=-\mathrm{i}S_\mathrm{V}(\mathbf{p},t,t')+\mathrm{i}q\omega t\\&=-\mathrm{i}\int_{t'}^{t}\mathrm{d}\tau\left[\frac{1}{2}\left(\mathbf{p}+\mathbf{A}(\tau)\right)^2+\mathcal{I}_\mathrm{p}\right]+\mathrm{i}q\omega t.\end{aligned} \tag{2.42}$$

The five-dimensional integral Eq. (2.41) can be evaluated using quadrature rules, but will require serious computational efforts.

**Saddle-point approximation for the momentum integral**

During its journey in the continuum, the electron's classical canonical momentum $\mathbf{p}$ appears as a conserved quantity because we neglect the influence of the parent ion. The three-dimensional integral over all possible momenta $\mathbf{p}$ is often simplified using a saddle-point approximation.

The application of the saddle-point approximation in this case is well-justified as the dipole matrix transition elements $\mathbf{d}(\mathbf{p})$ vary smoothly over large energy ranges of tens of eV. The phase factor $S(\mathbf{p},t,t')$ on the other hand changes rapidly within one cycle of the driving laser field as $|\mathbf{p}|^2/2\approx 1/(t-t')$ which amounts to less than 1 eV for drivers with wavelength 800 nm.

The saddle points for $\mathbf{p}$ are given by

$$\begin{aligned}\nabla_\mathbf{p}S(\mathbf{p},t,t')&=\nabla_\mathbf{p}\int_{t'}^{t}\mathrm{d}\tau\left[\frac{1}{2}\left(\mathbf{p}+\mathbf{A}(\tau)\right)^2+\mathcal{I}_\mathrm{p}\right]\\&=\int_{t'}^{t}\mathrm{d}\tau\nabla_\mathbf{p}\frac{1}{2}\left(\mathbf{p}+\mathbf{A}(\tau)\right)\cdot\left(\mathbf{p}+\mathbf{A}(\tau)\right)\\&=\int_{t'}^{t}\mathrm{d}\tau\left(\mathbf{p}+\mathbf{A}(\tau)\right)=0\end{aligned} \tag{2.43}$$

[3] Note that $q$ does not have to be an integer.

which we can solve to yield the saddle-point momentum explicitly:

$$\mathbf{p}_s = \mathbf{p}_s(t,t') = -\frac{1}{t-t'}\int_{t'}^{t} \mathrm{d}\tau\, \mathbf{A}(\tau) \tag{2.44}$$

The Hessian w.r.t. $\mathbf{p}$ is given as $\nabla^2_{\mathbf{p}} S(\mathbf{p},t,t') = (t-t')\mathbb{1}$, with the $3\times 3$ unit matrix $\mathbb{1}$, making the determinant simply $(t-t')^3$. Physically, the resulting prefactor $\left(\frac{1}{t-t'}\right)^{3/2}$ then denotes the diffusion of the quantum wave packet during the time it spends in the continuum. The dipole moment $\mathbf{D}(q\omega)$ with saddle-point approximation for the momentum integration then reads

$$\mathbf{D}(q\omega) = \mathrm{i}\int_{-\infty}^{\infty}\mathrm{d}t\int_{-\infty}^{t}\mathrm{d}t'\left(\frac{(2\pi\mathrm{i})^3}{(t-t')-\mathrm{i}\epsilon}\right)^{3/2}\mathbf{d}^*(\mathbf{p}_s+\mathbf{A}(t))\Upsilon(\mathbf{p}_s,t')\times\exp\left(-\mathrm{i}S(\mathbf{p}_s,t,t')\right)+c.c. \tag{2.45}$$

where a small regularisation constant $\epsilon$ is introduced to avoid the divergence at the boundary of the integration limit where $t \to t'$.

### 2.4.2 Saddle-point approach and quantum orbits

That the Lewenstein integral Eq. (2.39) is highly oscillatory in its time-dependency and can hence be solved using saddle-point methods has been recognised from the very outset [12]. In the following we will focus rather on the *phenomenological* aspects of applying the saddle-point approximation to the time-integration, as the mathematically rigorous way of doing it is arguably the purpose of this thesis. That is, we give a brief overview of the so-called quantum orbit approach, which summarises the idea of writing the dipole response as a sum of contributions from several quantum trajectories. We will explain those fundamental insights mostly using the example of a monochromatic driving field but also use this section to introduce the variety of approaches that are used to study HHG and which we will later on apply.

To identify the stationary points of the exponent of Eq. (2.45), we simultaneously solve the first derivatives of the action w.r.t. the two times. For convenience, we slightly re-write the action (2.42), simply in order to have the explicit time-dependency on the ionisation time $t_\mathrm{i}$ and recombination time $t_\mathrm{r}$, and use the saddle-point approximation for the momentum $\mathbf{p}_s$:

$$\begin{aligned} S_\mathrm{HHG}(t_\mathrm{i},t_\mathrm{r}) &= S_\mathrm{V}(\mathbf{p}_s,t_\mathrm{r},t_\mathrm{i}) + (t_\mathrm{r}-t_\mathrm{i})\mathcal{I}_\mathrm{p} - q\omega t_\mathrm{r} \\ &= \frac{1}{2}\int_{t_\mathrm{i}}^{t_\mathrm{r}}\left(\mathbf{p}_s(t_\mathrm{i},t_\mathrm{r})+\mathbf{A}(t)\right)^2\mathrm{d}t + (t_\mathrm{r}-t_\mathrm{i})\mathcal{I}_\mathrm{p} - q\omega t_\mathrm{r}\,. \end{aligned} \tag{2.46}$$

We find the stationary points by setting the respective derivatives to zero:

$$\begin{aligned} \frac{\partial S_\mathrm{HHG}}{\partial t_\mathrm{i}} &= \frac{1}{2}\left(\mathbf{p}_s(t_\mathrm{i},t_\mathrm{r})+\mathbf{A}(t_\mathrm{i})\right)^2 + \mathcal{I}_\mathrm{p} = 0 \\ \frac{\partial S_\mathrm{HHG}}{\partial t_\mathrm{r}} &= \frac{1}{2}\left(\mathbf{p}_s(t_\mathrm{i},t_\mathrm{r})+\mathbf{A}(t_\mathrm{r})\right)^2 + \mathcal{I}_\mathrm{p} - q\omega = 0\,. \end{aligned} \tag{2.47}$$

Solutions to this are pairs of complex numbers, $(t_\mathrm{i}, t_\mathrm{r})$, representing saddle points in the complex planes, and can usually be found numerically or — for simple driving fields — even analytically. Each of these time pairs correspond to a quantum trajectory, uniquely defined by its ionisation and recombination time. As in the case of the strong-field ionisation saddle points, generally, there are many (periodic) solutions to Eq. (2.47) and not all of them will be relevant. Firstly, we limit our considerations to the solutions originating from one period of the driving field, such that $0 \leq \mathrm{Re}(t_\mathrm{i}) < T$. Secondly, we will only consider those solutions where recombination happens *after* ionisation. And, following that, we mostly neglect those trajectories for which the travel time exceeds the cycle period of the laser field. Hence, we only consider solutions for which $0 < \mathrm{Re}(t_\mathrm{r} - t_\mathrm{i}) < T$. Lastly, we select the solutions for which the tunnelling time is positive, i.e., $\mathrm{Im}(t_\mathrm{i}) > 0$.

For the case of a monochromatic driving field where $\mathbf{E}(t) = E_0 \sin(\omega t)\mathbf{e}_x$ and $\mathbf{A}(t) = \frac{E_0}{\omega}\cos(\omega t)\mathbf{e}_x$ these saddle point solutions are shown in Fig. 2.11 for a range of harmonic orders $q$, denoted by colour. Here and below we use $E_0 = 0.05\,\mathrm{a.u.}$ (intensity $I_0 = E_0^2 = 0.92 \times 10^{14}\mathrm{W/cm^2}$), $\omega = 0.044\,\mathrm{a.u.}(\lambda = 1030\,\mathrm{nm})$ and $\mathcal{I}_\mathrm{p} = 0.58\,\mathrm{a.u.} = 15.76\,\mathrm{eV}$ (Argon). In Fig. 2.11 we recognise a familiar observation: The ionisation times of the saddle points are centred around the maximum of the electric field, as indicated by the grey line in panel (a). Analogously to the case of ATI, this is when the distortion of the atomic Coulomb potential is strongest and hence, tunnel ionisation is most likely. Recombination on the other hand, indicated as a grey line on the right-hand side of the figure, happens across a much wider time range.

### 2.4.3 Classification of the solutions

**Motivation**

Generally, and as seen in Fig. 2.11, the saddle points depend smoothly on the harmonic order $q$ and often we can readily identify different 'branches' of solutions, which we will call different 'types' of trajectories. In the well-known case of a monochromatic driver for example, one can identify 'short and 'long' trajectories, referring to the time they spend roaming in the continuum, and the resulting distance they travel. As we will see below, for the harmonic frequencies in the HHG plateau, both types of trajectories need to be taken into account when calculating the HHG dipole response from a single atom emitter.

In an HHG experiment however, the gaseous target consists of billions of atoms and the *macroscopic* response (i.e., calculating the induced dipole at each atom and then propagating this harmonic radiation) can differ between the different types of trajectories. For example, in experiments, the general understanding is that the short trajectories have better phase matching properties. From our viewpoint of microscopic simulations this can be understood as follows. If we assume the target to be a plane sheet of gas atoms, then across this target the laser beam has a specific spatial intensity profile (most typically a Gaussian) such that every atom on the target plane 'sees' slightly different field parameters. The different types of trajectories may depend differently on the change of those external parameters. An example study of this is detailed in the later Sec. 9.5. In order to optimise the interference between the various quantum paths, the macroscopic parameters can be modified so as to gain control over the selected quantum paths.

Apart from this, a classification of solutions has proven desirable within the theoretical approach to the atomic response. Firstly, the various types of trajectories might have quite different physical properties — e.g., trajectories with very long travel times can be classified as higher-order returns. Secondly, as we will discuss in extensive detail below, the identification of the high-order harmonic cutoff frequency goes hand in hand with the classification of the solutions. That is, in the traditional approach the definition of the HHG cutoff is based on a purely geometrical argument as the point of closest vicinity between the two solutions of the short and long trajectories. This, of course, implicitly assumes the consistent classification between the two. This approach becomes particularly challenging in cases where the two solutions approach each other very closely, as shall be seen in Fig. 10.3. In a more recent and more rigorous approach it is the other way around: based on the definition of the cutoff harmonic point in the complex plane we can infer the classification of solutions.

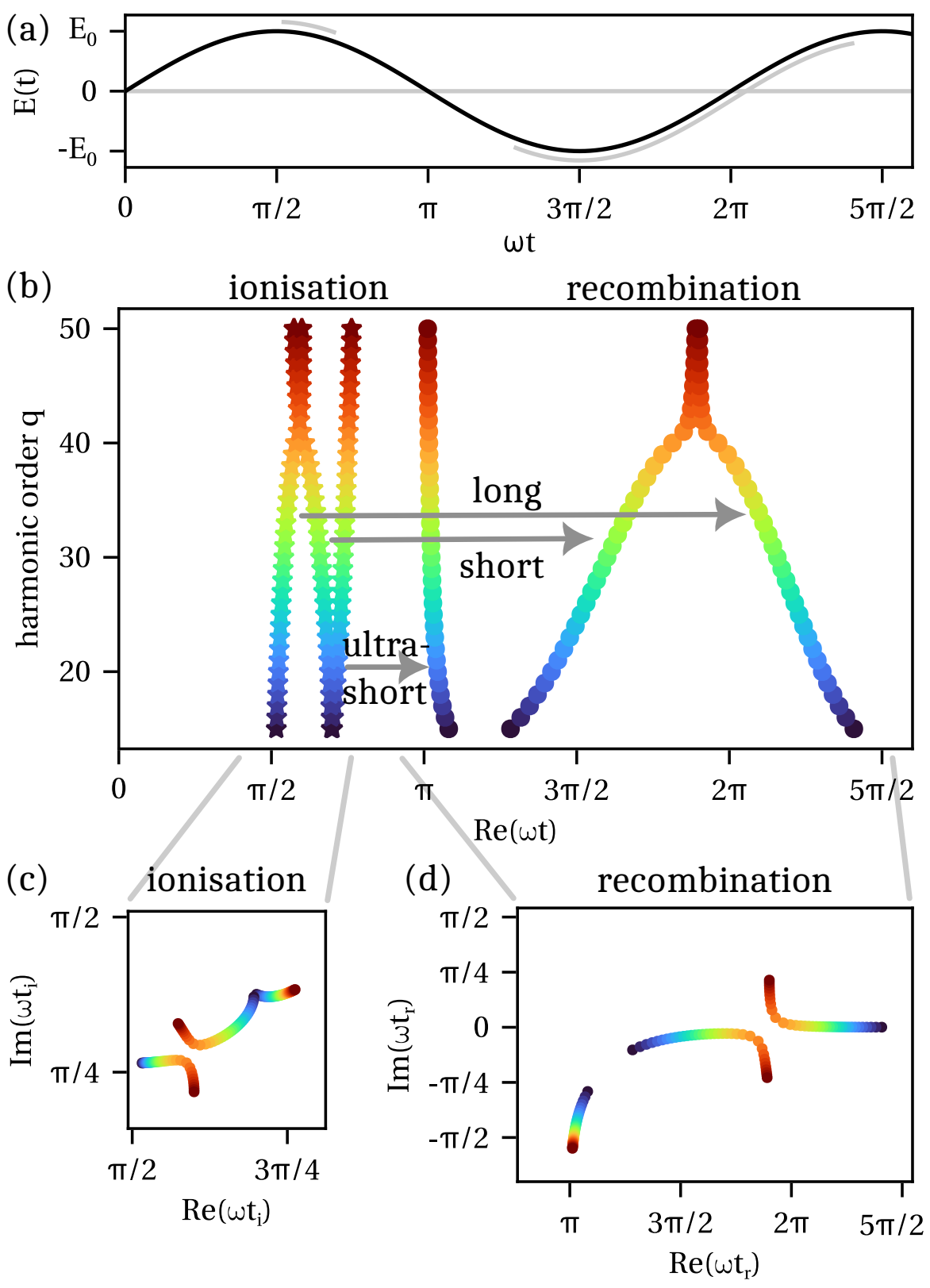

Figure 2.11: Saddle points for a monochromatic driving field (panel (a), ionisation and recombination time windows indicated in grey) over real time (in panel (b)) for a range of harmonic orders. The solutions can be classified as ultrashort, short and long trajectories, depending on the time between ionisation and recombination. The markers are coloured depending on the harmonic order as shown in panel (b). In the complex time planes of ionisation and recombination (panel (c) and (d) respectively) they trace clear branches of solutions where the short and long solutions perform a missed approach for increasing harmonic orders.

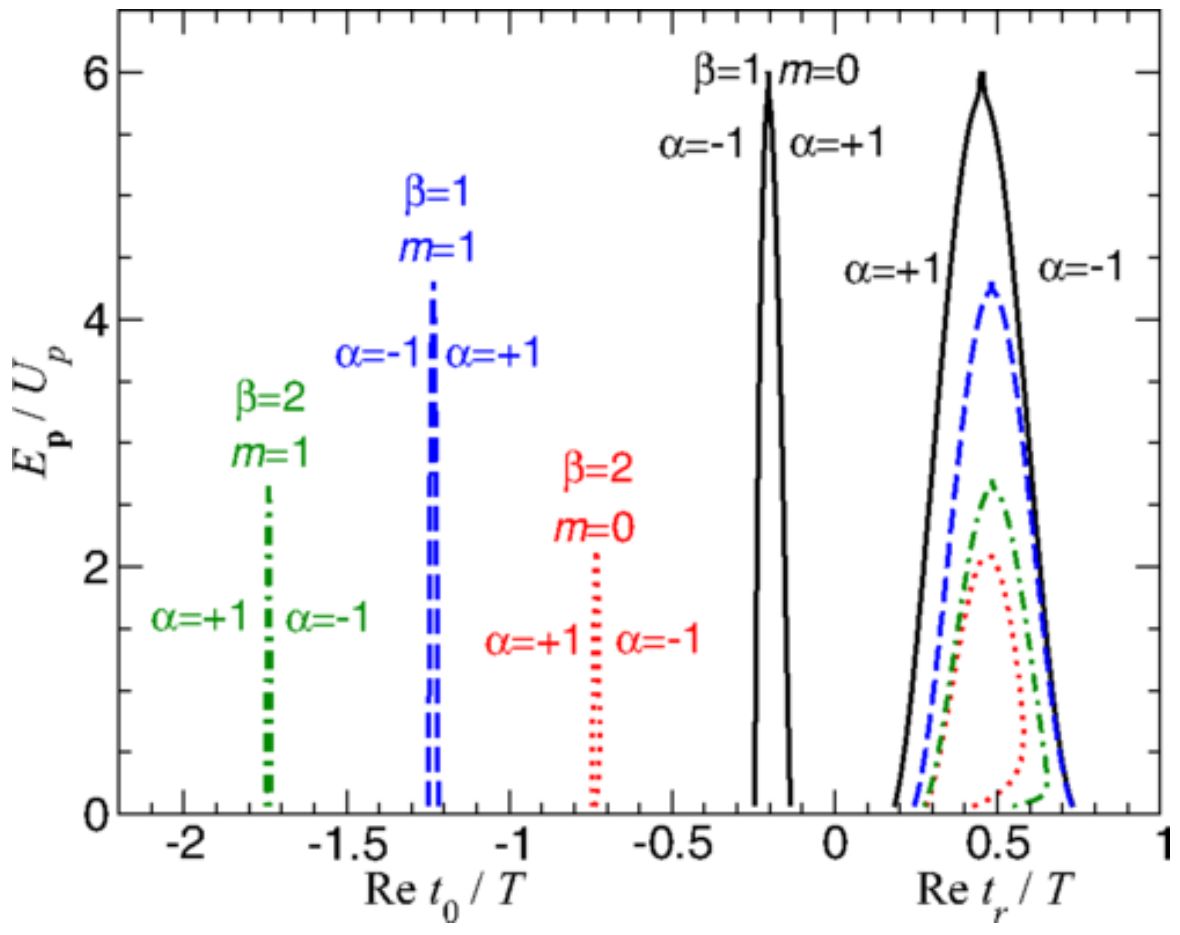

Figure 2.12: Example of the $(\alpha, \beta, m)$ classification scheme, here shown for rescattered ATI photoelectrons, driven by a monochromatic driving field. The scheme is identical to the one used for HHG saddle points. Taken from [45].

To conclude, the classification of the different solutions to Eq. (2.47) provides a meaningful tool to study both the underlying dynamics of the quantum process of HHG, as well as mathematical insight into the topology of the full saddle point landscape.

**Classification for drivers with dynamical symmetry**

To classify the set of solutions obtained from Eq. (2.47) it is most instructive to view the solutions in the complex planes for ionisation time $t_\mathrm{i}$ and recombination time $t_\mathrm{r}$. As shown in Fig. 2.11, for a monochromatic driving field the solutions trace two clear 'branches', performing a missed approach for increasing harmonic orders. The harmonic order for which they are close, gives a fair estimate for the harmonic cutoff order $q_\mathrm{max}$, in Fig. 2.11 we find $q_\mathrm{max} \approx 42$. A rigorous discussion (and definition) of this within the framework of catastrophe theory will follow in the later Sec. 12.1. Furthermore there is a type of ultra-short trajectory but we will see later that they are not particularly relevant.

More generally, for driving fields with dynamical symmetries a classification scheme has been proposed and used by Milošević and co-workers in a multitude of publications, e.g. [45–47], and is shown in Fig. 2.12. In the scheme each branch of solutions is classified in terms of the multi-index $(\alpha, \beta, m)$, where $\alpha$ distinguishes short ($\alpha = +1$) and long ($\alpha = -1$) trajectories, $\beta$ denotes the pairs of solutions within one cycle of the laser field, and $m$ is the approximate travel time in units of the fundamental laser cycle. That is, the first short trajectory in a monochromatic laser field (as shown in Fig. 2.11) would be labelled $(+1, 1, 0)$.

This classification scheme is helpful for a given driving field configuration. For modifications to the field configuration that change the number of ionisation bursts, however, the scheme fails. Furthermore, it does not provide any further insights apart from the distinction of branches itself because the classification is done "empirically", especially for the distinction between long and short orbits around the cutoff.[4] For scans of the phase delay between two field components, or a scan of intensity ratios, this classification scheme can be misleading as the number (and the ordering!) of ionisation bursts might change, as we will see later on in Sects. 10 and 11. Nonetheless, in general, this classification scheme gives a good first intuition for the number and general temporal structure of saddle point solutions, and is perceived as helpful in the attosecond community.

**Classification for arbitrary polychromatic drivers**

For generic combinations of driving fields with arbitrary polarisations, frequencies and amplitudes, the saddle point solutions will still trace branches in the complex plane, but not necessarily with such an obvious structure as for a monochromatic driver. As an example, we show the saddle-point solutions for the two-colour field given by

$$\mathbf{E}(t) = E_0 \cos(\omega t)\mathbf{e}_x + RE_0 \cos\big(2\omega t + \varphi\big)\mathbf{e}_x \tag{2.48}$$

with $R = 0.45$ and $\varphi = 0.75$ in Fig. 2.13. For this somewhat arbitrarily chosen two-colour driving field, without any dynamical symmetries, the saddle-point solutions can be attributed to four distinct ionisation windows, as indicated by the labelling A through D, and marked on the electric field in panel (a). In Fig. 2.13(c) we show the corresponding saddle points in the complex plane in the same colour scheme as in Fig. 2.11. For visual clarity we show the saddle points for the four ionisation windows in separate rows, and plot the complex ionisation time plane in closer detail on the left-hand side (panel (b)). For ionisation bursts A, B and C, we find similar structures to the monochromatic case: pairwise missed approaches for increasing harmonic orders. The structure for the saddle points from ionisation window D however, is slightly different as several branches of the solutions appear to connect. In this scenario, there is no immediately obvious classification heuristic.

In general, for most cases, a classification of saddle points into several types could manually be done on a case-by-case basis. But it will be rather tedious, especially given structures like the one in ionisation window D. Furthermore, the structure might change upon the tuning of an external parameter, as we will see later on in this thesis.

### 2.4.4 Semi-classical electron trajectories

As the saddle points define ionisation and recombination time, we can calculate the physical trajectories, i.e., the displacement from the nucleus, for each solution. According to the three-step model we describe the motion of the electron in the laser field classically between ionisation and recombination time, $t_{\mathrm{i},s}$ and $t_{\mathrm{r},s}$ respectively, with a conserved canonical momentum $\mathbf{p} = \mathbf{p}_s$ as defined in Eq. (2.44). The displacement of the electron is therefore given as the time-integral of the kinematic velocity, such that

$$\mathbf{x}(t) = (t - t_{\mathrm{i},s})\,\mathbf{p}_s + \int_{t_{\mathrm{i},s}}^{t} \mathbf{A}(t')\,\mathrm{d}t'. \tag{2.49}$$

[4] The decision which solution shall be discarded, and hence, which shall be called the short trajectory, is based on a comparison with the result of the numerical integration [conversation with D.B. Milošević in April 2024].

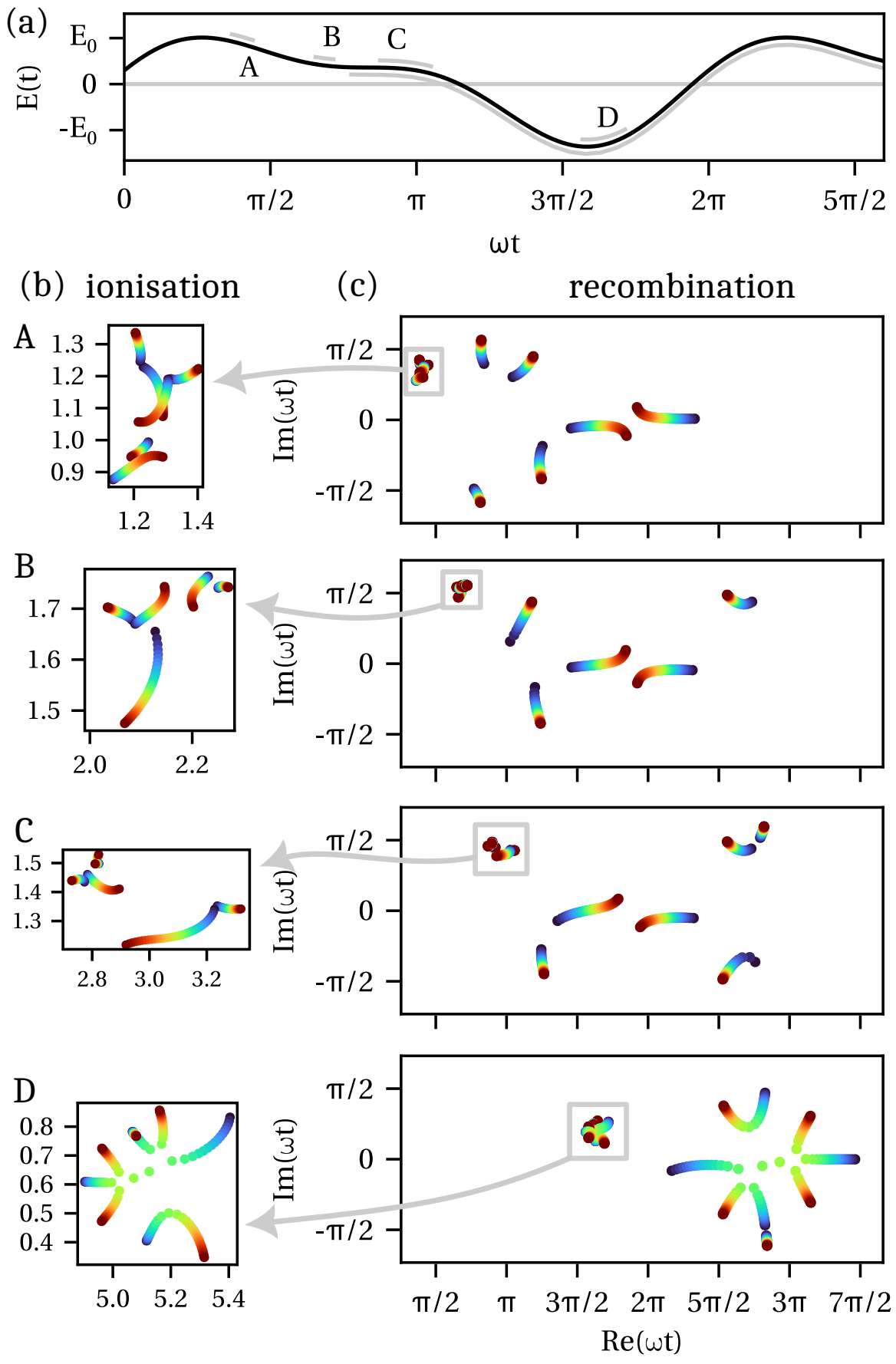

Figure 2.13: For the field shape shown in panel (a) (parameters in the text): saddle points in the complex time planes (right-hand side column, panels in (c)), separated for the four ionisation bursts A, B, C and D (rows) and with close-ups of the ionisation times in the left-hand side column (panels in (b)). The ionisation and recombination windows are indicated as grey lines in panel (a).

Since the start and end time of this trajectory are complex numbers, there is a freedom w.r.t. the integration contour. Typically, the contour is taken in two steps, shown in Fig. 2.14. The first leg of the contour starts at $t_{i,s} = \mathrm{Re}(t_{i,s}) + i\mathrm{Im}(t_{i,s})$ and goes all the way down to $\mathrm{Re}(t_{i,s}) + \mathrm{Im}(t_{r,s})$ to describe the tunnelling process. The second leg of the contour describes the propagation in real time and goes from $\mathrm{Re}(t_{i,s}) + i\mathrm{Im}(t_{r,s})$ all the way to $\mathrm{Re}(t_{r,s}) + i\mathrm{Im}(t_{r,s}) = t_{r,s}$, the recombination time.

The displacement Eq. (2.49) for the shortest three trajectories within one half cycle of a monochromatic driving field are shown in Fig. 2.15, for a range of harmonic orders (denoted by colour). They correspond to the saddle points in Fig. 2.11. The time integration follows the two-legged path shown in Fig. 2.14 and we only plot the real part of the displacement, $\mathrm{Re}(\mathbf{x}(\omega t))$. Hence, the tunnelling step of the process appears simply as a vertical line in Fig. 2.15 at the beginning of each trajectory (i.e., at $\mathrm{Re}(\omega t)) = \mathrm{Re}(\omega t_i)$) because no real time passes. The displacement during the propagation step then reveals the actual trajectories in real space (and time). We can identify the three different types of trajectories where the classification into ultra-short, short and long trajectories now attains meaning in a spatial sense as well. Furthermore, we find the short and long trajectories become more and more similar (and eventually 'overlap') for increasing harmonic order. As the saddle points are closest in the complex plane at the high-harmonic cutoff order, this harmonic order is the one where the respective trajectories are the most similar.

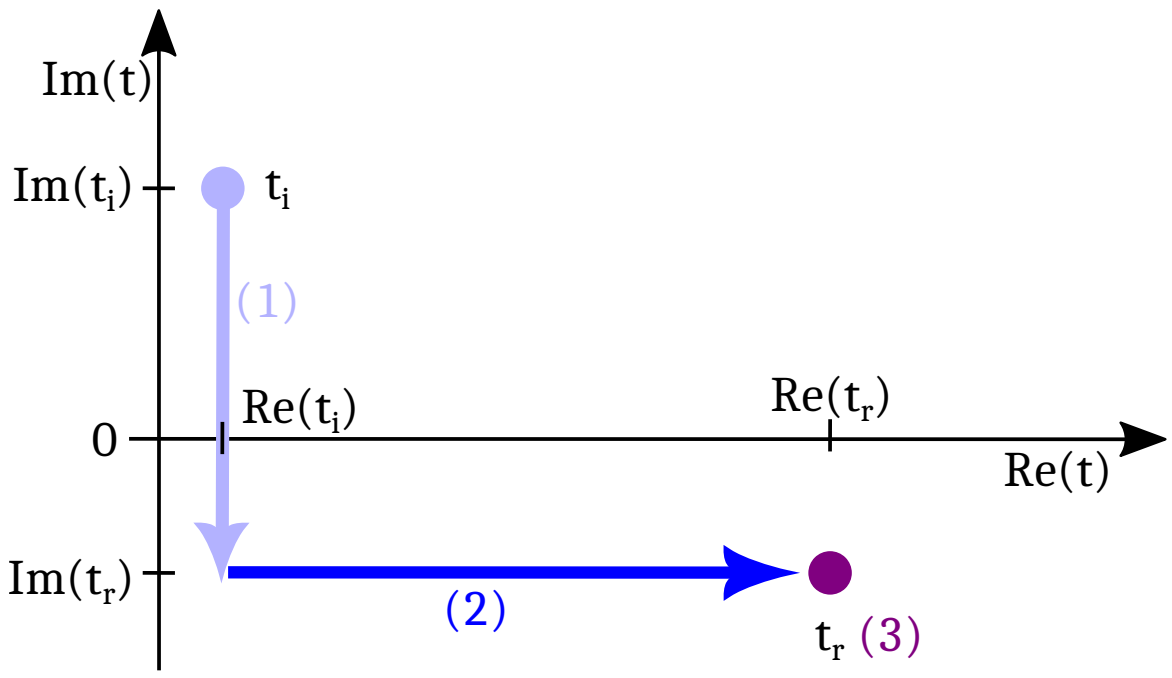

Figure 2.14: The integration between $t_{i,s}$ and $t_{r,s}$ for the integral in Eq. (2.49) follows a two-legged contour in the complex plane.

For a field that is polarised in two dimensions the electron trajectories extend in two dimensions as well. It is therefore instructive to show the trajectories in the $x-y$-plane throughout their journey in real time, i.e., only the second leg of the time contour in Fig. 2.14. We demonstrate this for a bicircular $\omega - 2\omega$ field in Fig. 2.16. This bicircular field is composed of a circularly polarised fundamental field with frequency $\omega$, superimposed with a circularly polarised field of equal amplitude, frequency $2\omega$, which is rotating in opposite direction. The resulting electric field components are shown parametric in time in Fig. 2.16(a) and create a trefoil shape which is characteristic for bicircular fields. In panels (b) and (c) we show the spatial components of the short and long trajectories, respectively. For clarity, we only show the trajectories for the harmonic orders of the HHG plateau. We find that the short trajectories appear in the continuum at a small distance away from the core, which is located at $(0,0)$. They subsequently move away radially, change direction and return back to the core. In comparison, the excursion of the long trajectories follows a triangular shape with two changes in direction before they recombine at the core. Both types of trajectories inherit the clear threefold dynamic symmetry from the driving field. For completeness, the driving field's vector potential is shown in panel (d). The instantaneous velocities for the respective electron trajectories are shown in panels (e) and (f), and follows the shape of the vector potential.

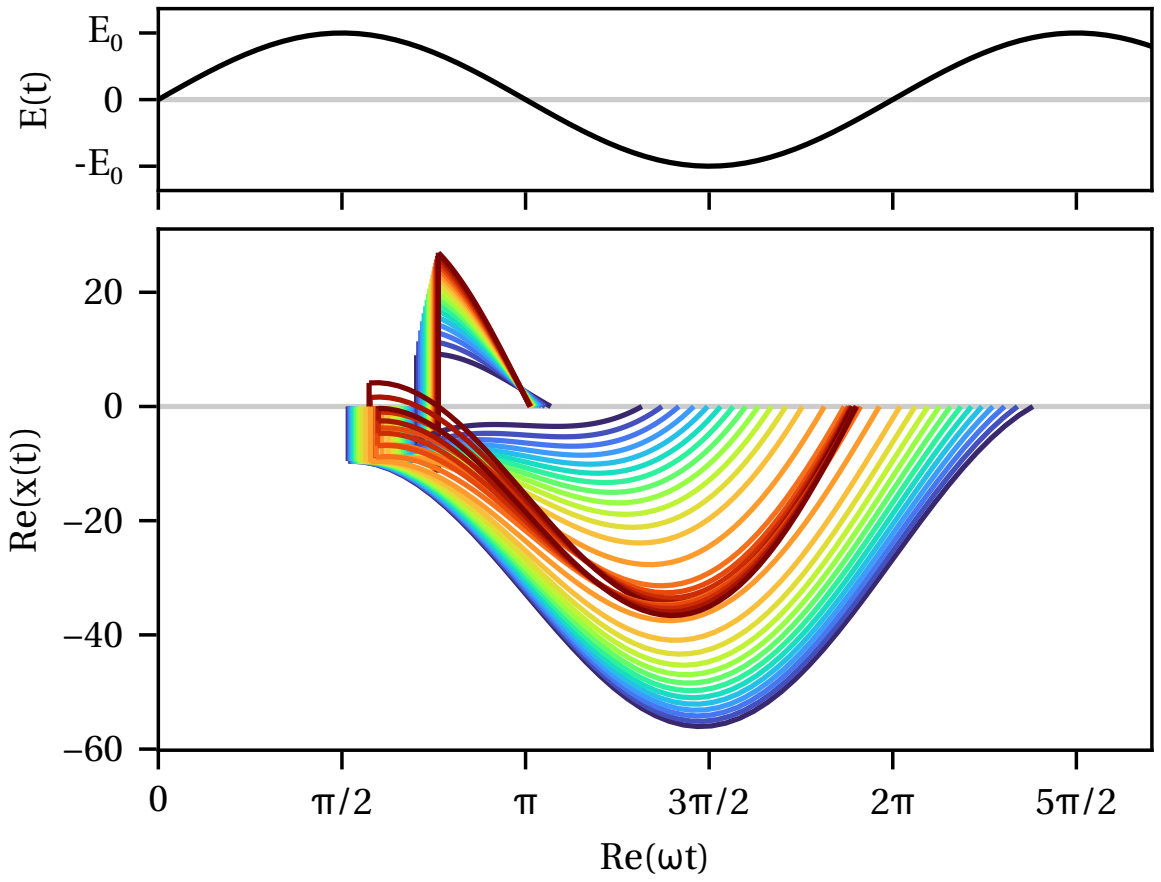

Figure 2.15: The real part of the trajectories according to Eq. (2.49) for the saddle points shown in Fig. 2.11 for a monochromatic electric field (shown on top). The colour indicates the harmonic order, as in Fig. 2.11.

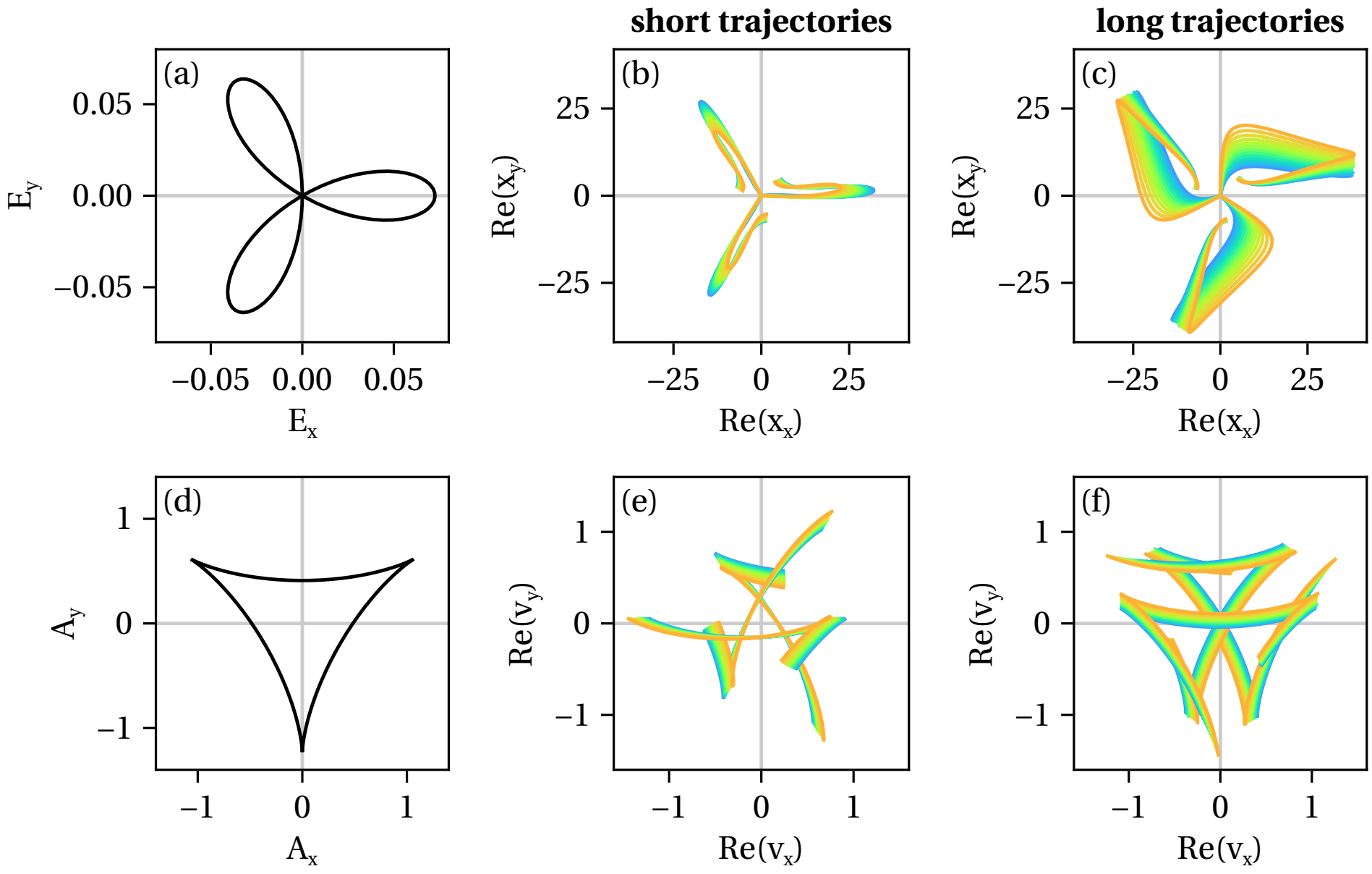

Figure 2.16: Semi-classical electron trajectories in space, for a bicircular driving field. (a) Electric field and (d) vector potential of the bicircular driving field. Electron displacement for the first short and long trajectories is shown in panels (b) and (c), respectively, for a range of harmonic orders (indicated by the colour) in the plateau. The respective instantaneous velocity is shown in panels (e) and (f).

## 2.4.5 The total dipole and the HHG spectrum

**The total dipole as a sum of several trajectories' contributions**
The HHG spectrum emitted from a single atom is calculated from the Fourier-transformed dipole moment, Eq. (2.45). Using the saddle-point approximation for $t_\mathrm{i}$ and $t_\mathrm{r}$, this integral can be written as a sum over contributions from individual trajectories

$$\begin{aligned}\mathbf{D}(q\omega) \approx \sum_s \mathbf{D}_s(q\omega) = \sum_s & \frac{2\pi}{\sqrt{-\det(S''_\mathrm{HHG}(t_{\mathrm{i},s}, t_{\mathrm{r},s}))}} \mathbf{d}\left(\mathbf{p}_s(t_{\mathrm{i},s}, t_{\mathrm{r},s}) + \mathbf{A}(t_{\mathrm{r},s})\right) \\ & \times \Upsilon\left(\mathbf{p}_s(t_{\mathrm{i},s}, t_{\mathrm{r},s}) + \mathbf{A}(t_{\mathrm{i},s})\right) \\ & \times \left(\frac{2\pi}{\mathrm{i}(t_{\mathrm{r},s} - t_{\mathrm{i},s})}\right)^{3/2} \mathrm{e}^{-\mathrm{i}S_\mathrm{HHG}(t_{\mathrm{i},s}, t_{\mathrm{r},s})} .\end{aligned} \tag{2.50}$$

where the sum runs over *all relevant* trajectories. How to define which trajectories are relevant and which are not can be inferred using the insights of Picard–Lefschetz theory — a connection which is the central takeaway of this thesis and which will be detailed later. Here, we shall continue to review the current understanding in the literature. So far, it is a well-known heuristic that the solutions for the short trajectories shall be discarded beyond the Stokes transition near the high-harmonic cutoff. This Stokes transition is the specific harmonic order $q_\mathrm{St}$ at which the real part of the action of the two saddle point solutions is equal:

$$\mathrm{Re}(S_\mathrm{HHG}(t_{\mathrm{i},s1}, t_{\mathrm{r},s1})) = \mathrm{Re}(S_\mathrm{HHG}(t_{\mathrm{i},s2}, t_{\mathrm{r},s2})) . \tag{2.51}$$

Which of the two solutions should be discarded can be decided by checking which of the solutions grows exponentially beyond the cutoff.

In Fig. 2.17(a) we show the intensity of the dipoles from individual trajectories, i.e.,

$$I_s(q\omega) = (q\omega)^4 \left|\mathbf{D}_s(q\omega)\right|^2 \tag{2.52}$$

across a range of harmonic orders for a monochromatic driving field. The contributions come from trajectories ionised within the first half cycle of the driving laser field and empty circles denote contributions that must be discarded. The saddle points and trajectories for the first few sets of solutions, namely the ultra-short (U-S) and the first pair of short and long trajectories (S1 and L1), are shown in Fig. 2.11 and in Fig. 2.15, respectively. In Fig. 2.17(a) we furthermore show the contributions from trajectories with longer travel times as well, labelled according to approximate travel times — S2 and L2 have travel times of around $1.25T$, S3 and L4 around $1.75T$. Those pairs will have saddle-point structures similar to the pair shown in Fig. 2.11. As a result, the contributions of each pair have similar characteristics: each pair has a certain cutoff harmonic order at which the contributions approach each other to form a small peak. This peak is a caustics arising from the close vicinity of the saddle points upon the missed approach, and is more pronounced the closer the saddle points approach each other in the complex plane. After this cutoff, the contribution of one of the trajectories has to be neglected "because" it exponentially increases, while the contribution of the other trajectory drops rapidly. This is the common reasoning in the literature, and — as this thesis will demonstrate — not a mathematically sound argument.

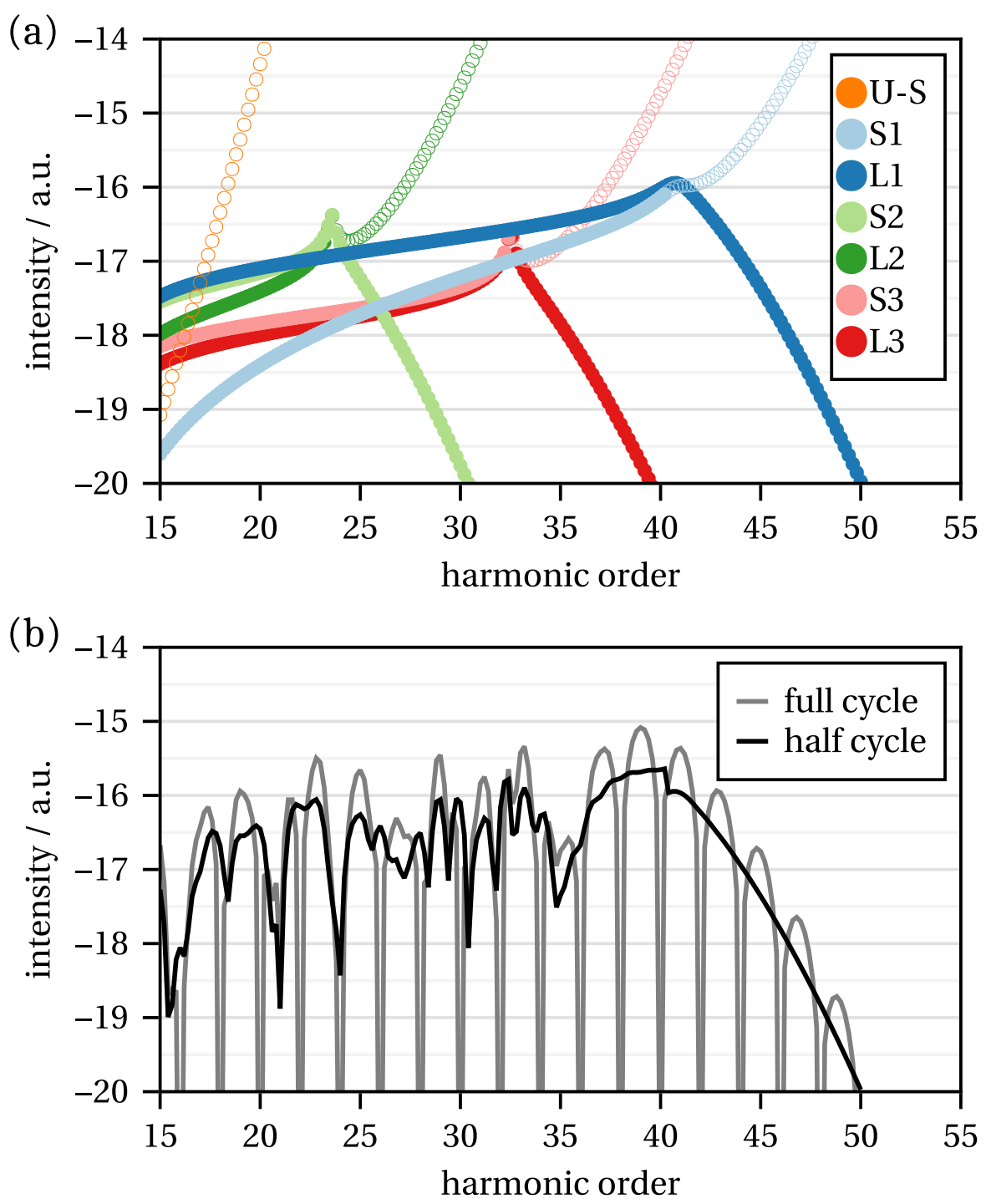

Figure 2.17: HHG spectrum from a monochromatic driving field, in logarithmic scale. Top panel: intensity contributions from individual saddle points Eq. (2.52), classified as short (S) and long (L) trajectories and numbered according to their travel time, as well as the ultra-short (U-S) trajectory. Empty circles denote contributions that will be neglected in the total sum, i.e., the ultra-short trajectory, and trajectories S1, L2 and S3 after their respective cutoffs. This total sum for the trajectories stemming from one half cycle is shown in the bottom panel in black. The total sum including trajectories from both half cycles of the driving field is shown in grey, demonstrating the typical fringe pattern of a HHG spectrum.

Generally, we find the contributions of the individual trajectories depends strongly on their travel times. The ultra-short trajectories are not relevant for harmonic orders beyond the ionisation threshold $q\omega = \mathcal{I}_{\mathrm{p}}$. Throughout the harmonic plateau, the trajectories with travel times less than one cycle (S1 and L1) are dominant. Trajectories with longer travel times contribute orders of magnitude less.

#### The typical shape of an HHG spectrum

In Fig. 2.17(b) we show the intensity from the coherent sum of all the dipoles of the trajectories shown in (a), i.e., the dipole created from trajectories ionised within the first half cycle (black line). We see seemingly unstructured ripples and artificial discontinuities around harmonic order 42.

When including the trajectories from the other half cycle as well (grey line), we find the typical structure of an HHG spectrum: An intensity plateau across a long range of harmonic frequencies up to a certain cutoff frequency at which the intensity drops sharply. Throughout the spectrum we see clear intra-cycle interference fringes and the suppression of the even harmonic orders, which is characteristic for a monochromatic, linearly polarised driver.

### The high-order harmonic cutoff

An approximation for the high-harmonic cutoff frequency can be derived from the classical model reviewed above, see Eq. (2.5). It describes the maximum classical kinetic energy that the electrons can acquire during their propagation in the continuum, assuming they ionised with zero initial velocity. For the configuration of Fig. 2.17 the simple *classical* cutoff frequency is $q_{\mathrm{max}} \approx 37$. From panel (b) we find that this classical limit slightly underestimates the actual cutoff frequency, i.e., the harmonic frequency at which the signal exhibits a sharp drop. A proper definition for the cutoff frequency in terms of a 'hidden' fold catastrophe between the saddle points of the short and the long trajectories has only been given recently [48] and will be introduced further below. Nonetheless, for now it is important to note that the cutoff frequency scales with $U_{\mathrm{p}}$. That is, a larger fundamental intensity will extend the plateau. A limitation to this scaling is given in terms of the Keldysh parameter such as to remain in the tunnelling regime. Moreover, it is well-known that at the cutoff the saddle-point approximation breaks down as the determinant of the Hessian diverges, when the two saddle point solutions approach each other. Typically, this is solved by using an Airy-type integral to model the total dipole, rather than two separate Gaussians for each trajectory. A more in-depth description of the respective procedure will follow in Sec. 12.1.

### Selection rules and quantum path interferences

Generally, the fringe pattern of the total harmonic spectrum can be attributed to the interference of the several quantum paths. On a 'computational' level this is simply because for the total dipole $|\mathbf{D}| = |\mathbf{D}_1 + \mathbf{D}_2 + \ldots| \neq |\mathbf{D}_1| + |\mathbf{D}_2| + \ldots$, as all the dipoles are complex-valued vectors.

More specifically, however, the interference patterns can be traced back to the symmetries of the system and formalised in terms of selection rules. The dynamic symmetries of the driving fields translate throughout the HHG process and can suppress (or enhance) the generation of specific harmonic orders [49–51]. A well-known example is that for a monochromatic driving field the spectrum only shows odd harmonic orders. The even orders are suppressed due to the half-cycle symmetry of the driving field, as seen in Fig. 2.17.

For the case of a bicircular driver the electric field has a characteristic trefoil shape signifying a periodic symmetry. This symmetry can be found in the saddle points and the respective trajectories (shown in Fig. 2.16 above), and ultimately appears in the spectrum as well. In Fig. 2.18(a) we show the contributions from the first two pairs of trajectories stemming from within a third of the laser period, $0 \leq \mathrm{Re}(t_{\mathrm{i},s}) \leq T/3$. As in the case of the monochromatic driving field, solutions come in pairs of 'short' and 'long' trajectories that approach each other in the complex planes for increasing harmonic orders. For each pair, beyond a certain cutoff order, the short trajectory's contribution to the spectrum has to be discarded. Once again, we show the non-relevant contributions as empty circles. The total sum of all relevant contributions (from all three thirds of the cycle) is then drawn in black and we find the typical shape — a long-range plateau followed by a sharp cutoff in intensity. Throughout the spectrum, every third harmonic order is strongly suppressed.

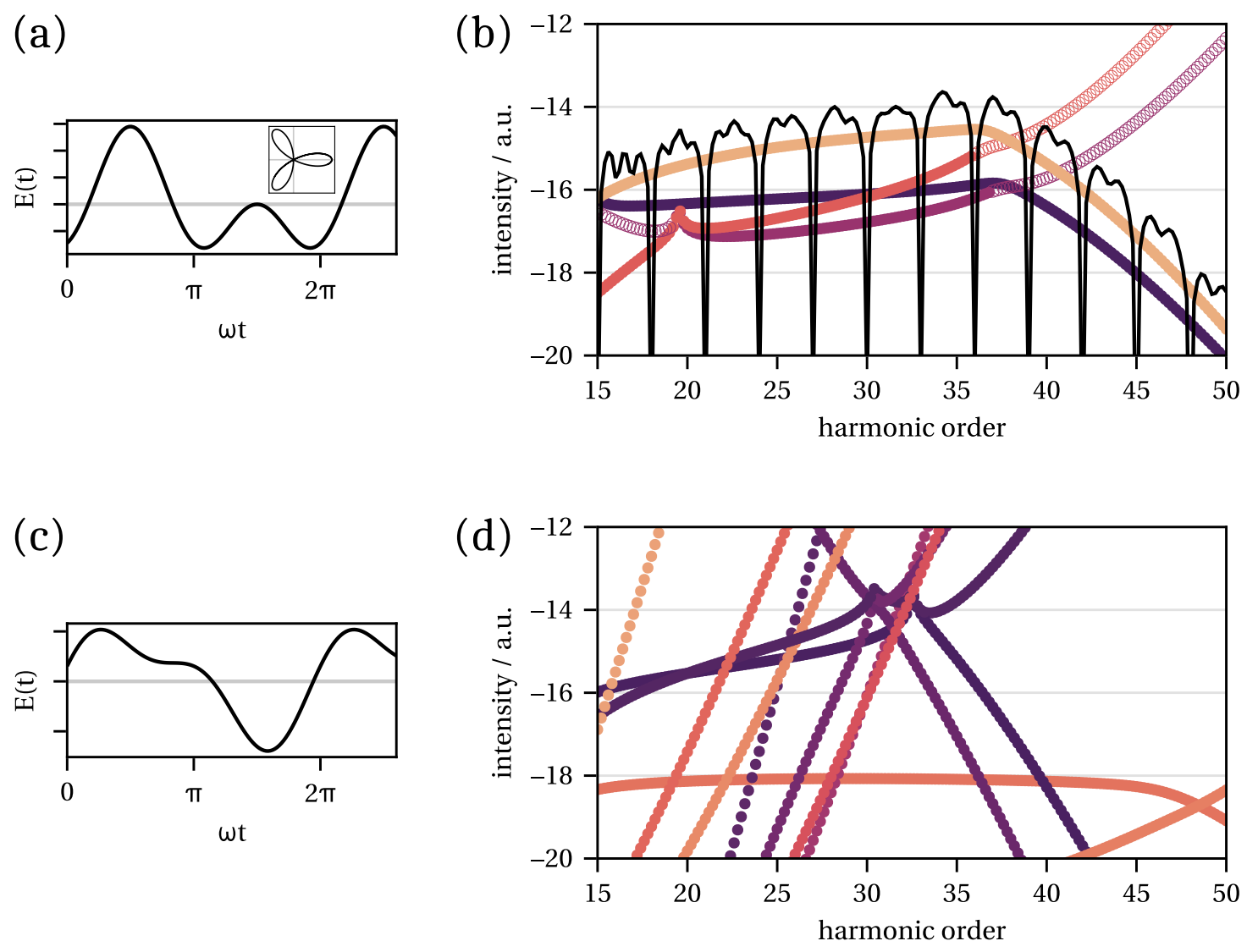

Figure 2.18: (b) HHG spectrum for the bicircular driving field shown in panel (a), with the contributions from several trajectories (in colours) and the resulting total harmonic intensity (black). (d) Contributions from the saddle points shown in Fig. 2.13 from the arbitrary driving field shown in panel (c).

As soon as the periodicity of the driving laser field is broken, all harmonic orders (also non-integers) are generated. As an example of this in Fig. 2.18(b) we show the harmonic spectrum of the arbitrary two-colour field introduced earlier, in Fig. 2.13. In colours, we show the contributions from the several trajectories. For this arbitrary driver, the heuristics about which saddle points are relevant to the total dipole sum fail. Hence, at this point, we are unable to draw the total dipole generated from the relevant subset of trajectories.

Note that the temporal structure of the driving field not only impacts the relative intensity of neighbouring harmonics, but also the overall 'baseline' intensity of the spectrum. For example, the HHG spectrum of a bicircular driver in-/decreases the spectral intensity by orders of magnitude compared to the bilinear driving field.

## 2.4.6 The generated harmonic radiation

The induced atomic dipole generates the electric field

$$\mathbf{E}_{q\omega}(t) = \mathbf{D}(q\omega)\, \mathrm{e}^{\mathrm{i}q\omega t} \tag{2.53}$$

with the afore calculated radiation dipole $\mathbf{D}(q\omega)$. For a given harmonic order this radiation dipole may be calculated as the sum of dipoles associated with the various saddle points, as introduced above. As sketched in Fig. 2.19, the temporal evolution of the electric field $E_{q\omega}(t)$ generally describes a polarisation ellipse and can hence be expressed in term of its major and minor polarisation axes **M** and **N**, and a rectifying phase $\gamma$ [52, 53]:

$$\mathbf{E}_{q\omega}(t) = \mathrm{e}^{\mathrm{i}\gamma}(\mathbf{M} + \mathrm{i}\mathbf{N}) \tag{2.54}$$

Analogously to the intensity of the generated harmonics, also the polarisation can be inherited from the dynamical symmetry of the field. For example, HHG driven from a bilinear driving field will produce only linearly polarised harmonics while HHG from a bicircular driving field produces circularly polarised harmonics, with alternating helicities. Any deviation from such drivers with dynamical symmetries cause not only the selection rules to break, but also may drastically impact the polarisation structure of the harmonics. For example, even only the slightest alteration from a bi-linear driving field can generate harmonics with high ellipticity [54–56].

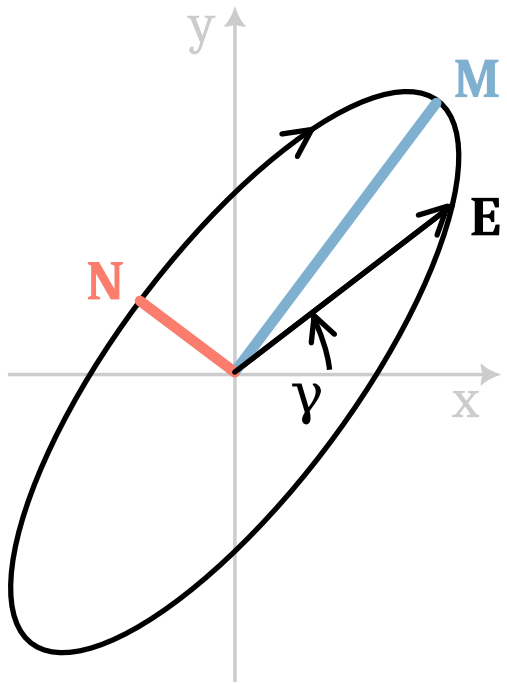

Figure 2.19: Sketch of the polarisation ellipse of the emitted harmonic radiation.

# 3

# Catastrophe theory

This Chapter introduces the theoretical foundations of catastrophe theory and its relevance for wave phenomena governed by interfering classical trajectories. We review the mathematical structure underlying caustics and discuss the canonical catastrophes that are central to the analysis presented in this thesis. Building on this framework, we provide an overview of existing applications of catastrophe theory in attosecond and strong-field physics, highlighting how trajectory coalescences manifest in experimentally observable signals and motivate the approaches developed in the following Chapters.

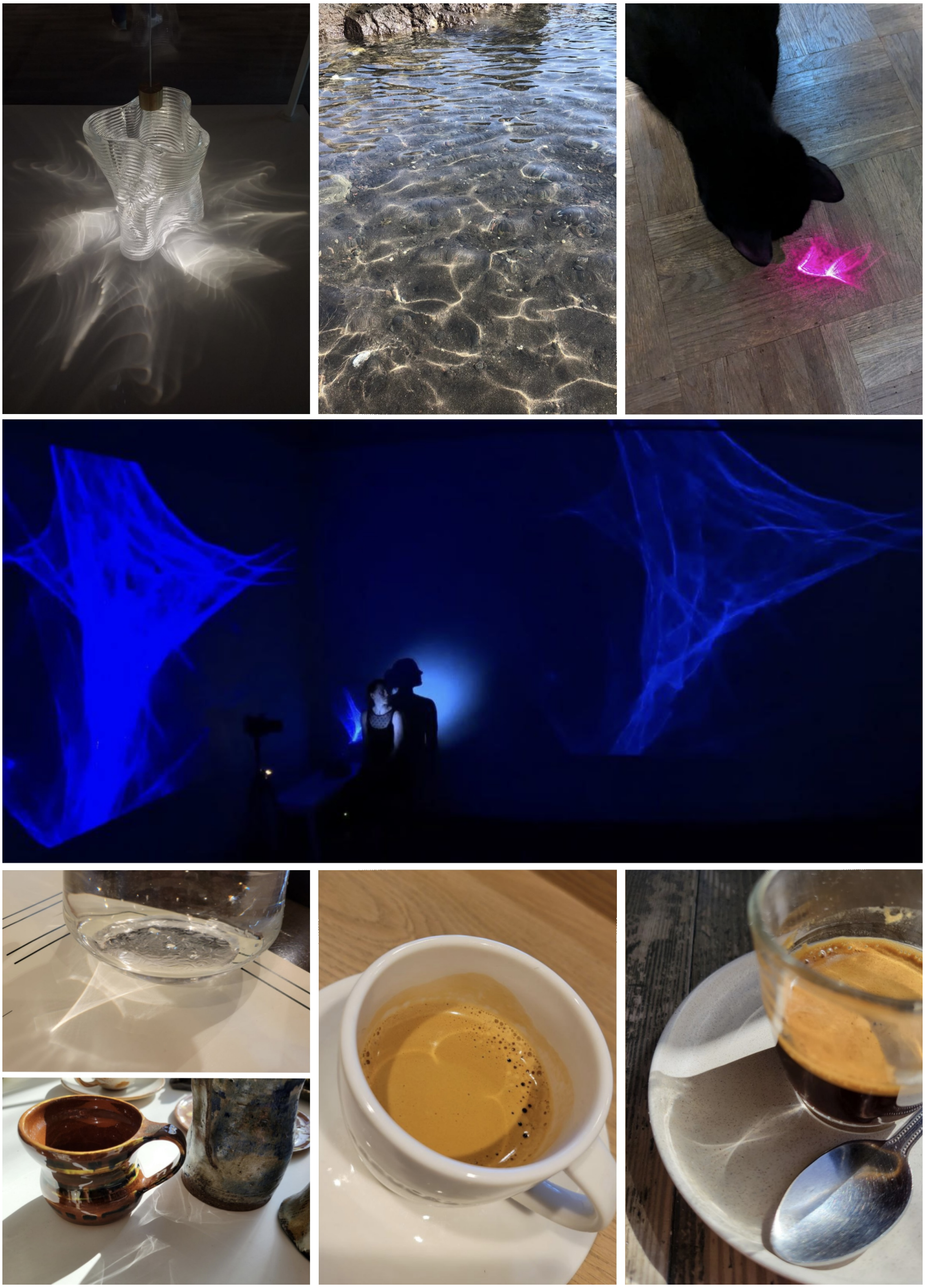

Figure 3.1: Caustics appearing in everyday life.

## 3.1 Introduction

Caustics are an ubiquitous phenomenon in nature and can be observed in everyday life, such as the rippled bright features at the bottom of a swimming pool or the cusp-like structure in a coffee cup, see Fig. 3.1. Beyond these familiar examples, caustics also arise in a wide range of more involved physical contexts, including real-time path integrals [57], quantum many-body dynamics [58, 59], dynamics of Bose–Einstein condensates [60], optical wave front shaping [61], mathematical physics [62, 63], photo-detachment in electromagnetic fields [64, 65], branched flow for wave propagation in periodic potentials [66], lensing (both optical and gravitational) [67–69], the formation of large-scale structures of our universe [70, 71], and, of course, attosecond science, e.g. [37, 72–79].

From a mathematical perspective, caustics are most naturally and rigorously described within the framework of catastrophe theory. The fundamental idea is that any wave phenomenon can be described as the coherent superposition of contributions from multiple classical or semi-classical rays. The form of those rays is governed by external parameters to the studied system. For example, the shape of the coffee cup will determine the trajectories of the classical light rays inside it, and changing the parameters of the driving laser field will change the electron trajectories in an HHG experiment. Excitingly, changes in those external parameters may cause trajectories to merge into each other and coalesce. Catastrophe theory now provides a universal classification of these scenarios and predicts that each type of coalescence gives rise to a characteristic and structurally stable diffraction pattern — the caustics. Consequently, any physical problem that involves the interference of several trajectories and (the control of) external parameters may result in observable caustics when these trajectories coalesce.

In this Chapter we provide an overview of the mathematical framework of catastrophe theory to show the mathematical ideas, and to introduce the canonical types of catastrophes that arise within this thesis. While the application of catastrophe theory to attosecond science is not new, the body of work at this intersection remains sufficiently compact to allow for a comprehensive overview. Accordingly, the second part of this Chapter aims to introduce existing applications in the attosecond and strong-field physics literature.

## 3.2 Canonical catastrophe theory

Catastrophe theory is the study of systems that experience sudden changes in behaviour despite the parameters underlying the system varying slowly and smoothly. As such, it is a branch of the mathematical field of bifurcation theory and singularity theories and “is just another tool which sometimes gives one a little extra insight into a situation” [80].

### 3.2.1 Historical context

In the 1970s the French mathematician René Thom coined the striking term “catastrophe theory” for the relation between geometry and observable discontinuities upon continuous changes of possible environmental causes. The popularity of the theory was fuelled by the experimental demonstrations of Christopher Zeeman. The great fame of the mathematical theory and, in particular, its application to a wide range of sciences led to controversies about the validity and the (false?) claims of its ability to predict behaviour [81]. Albeit the (potentially erroneous and slightly overambitious) application of catastrophe theory in research areas that are not rigorous in the usage of numerical calculus, the mathematical insights are still indubitably helpful to seize our understanding of multivariate problems.

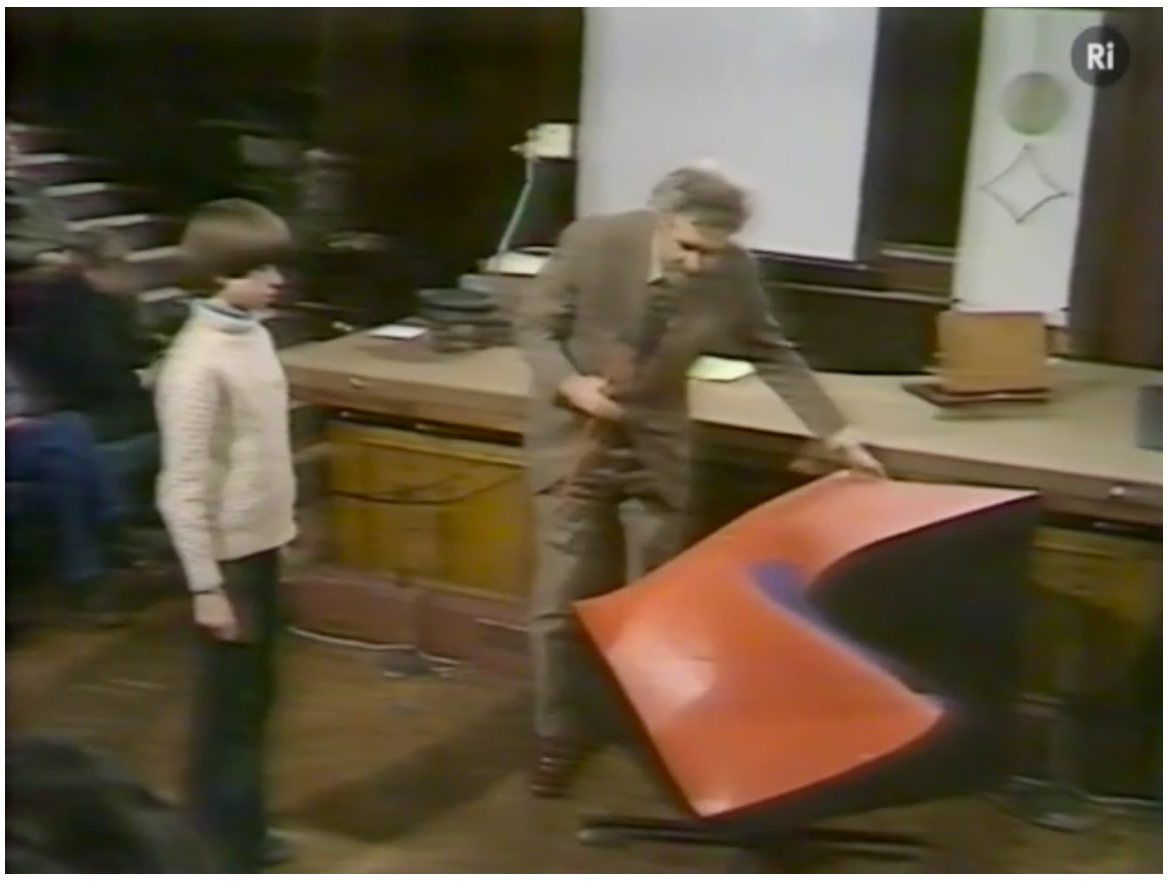

Figure 3.2: Christopher Zeeman demonstrating a folded potential surface to illustrate the fundamental ideas of catastrophe theory in a public lecture, from Ref. [82].

### 3.2.2 Theoretical derivation of the main ideas

In the following sections we provide a brief explanation of the fundamental mathematical concepts that pave the way for the identification of the characteristic set of catastrophes. We largely follow the derivations provided in Refs. [83] and [84], which we recommend as an accessible introduction and for any further details, respectively.

**Critical points and structural stability**

Catastrophe theory is mainly concerned with the behaviour of functions at (and around) their stable critical points, primarily through a classification thereof. Let us consider a real-valued, smooth function $f : \mathbb{R}^N \mapsto \mathbb{R}$. A *critical point* of $f$ is a point $\mathbf{x}_c \in \mathbb{R}^N$ at which the gradient vanishes:

$$\nabla_{\mathbf{x}} f = \frac{\partial f}{\partial \mathbf{x}}\Big|_{\mathbf{x}=\mathbf{x}_c} = 0. \tag{3.1}$$

The nature of the critical point can be studied by finding the Hessian matrix of second derivatives, also called the *deformation tensor*:

$$\mathcal{H} f = \frac{\partial^2 f}{\partial x_i \partial x_j} \qquad \text{for} \quad (i,j) = 1, \ldots, N \tag{3.2}$$

Critical points for which the determinant of this Hessian vanishes, $\det(\mathcal{H}f) = 0$, are called *degenerate*, i.e., the Hessian is singular and not invertible. Analogously, *non-degenerate* critical points are those with a non-vanishing determinant of the Hessian. In the applications in the further Chapters of this thesis we will refer to these as *saddle points*.

In physical applications, critical points often refer to the stationary solution of an energy, action or phase problem, and as such, are of special interest. A naturally arising question of such an equilibrium situation persists under a small perturbation. This notion is captured by the mathematical concept of structural stability. In Fig. 3.3 we show this for some simple polynomials. Considering the function $x^2$ and adding the term $\varepsilon x$ will leave the type of the critical point unchanged. Hence, this critical point (and the whole family of functions, for parameters $\varepsilon$) is structurally stable. Now consider the function $x^3$ with a small additive perturbation $\varepsilon x$. For a small perturbation $\varepsilon > 0$, this function does not have any critical points, for $\varepsilon = 0$ it has one degenerate critical point, and for $\varepsilon < 0$ there are two non-degenerate critical points. Hence, this critical point $x^3$ (and the family of functions) is structurally unstable. Similar examples for higher orders $n$ of the functions $x^n$ are shown in Fig. 3.3, demonstrating how a perturbation can lead to the splitting into $n-1$ critical points.

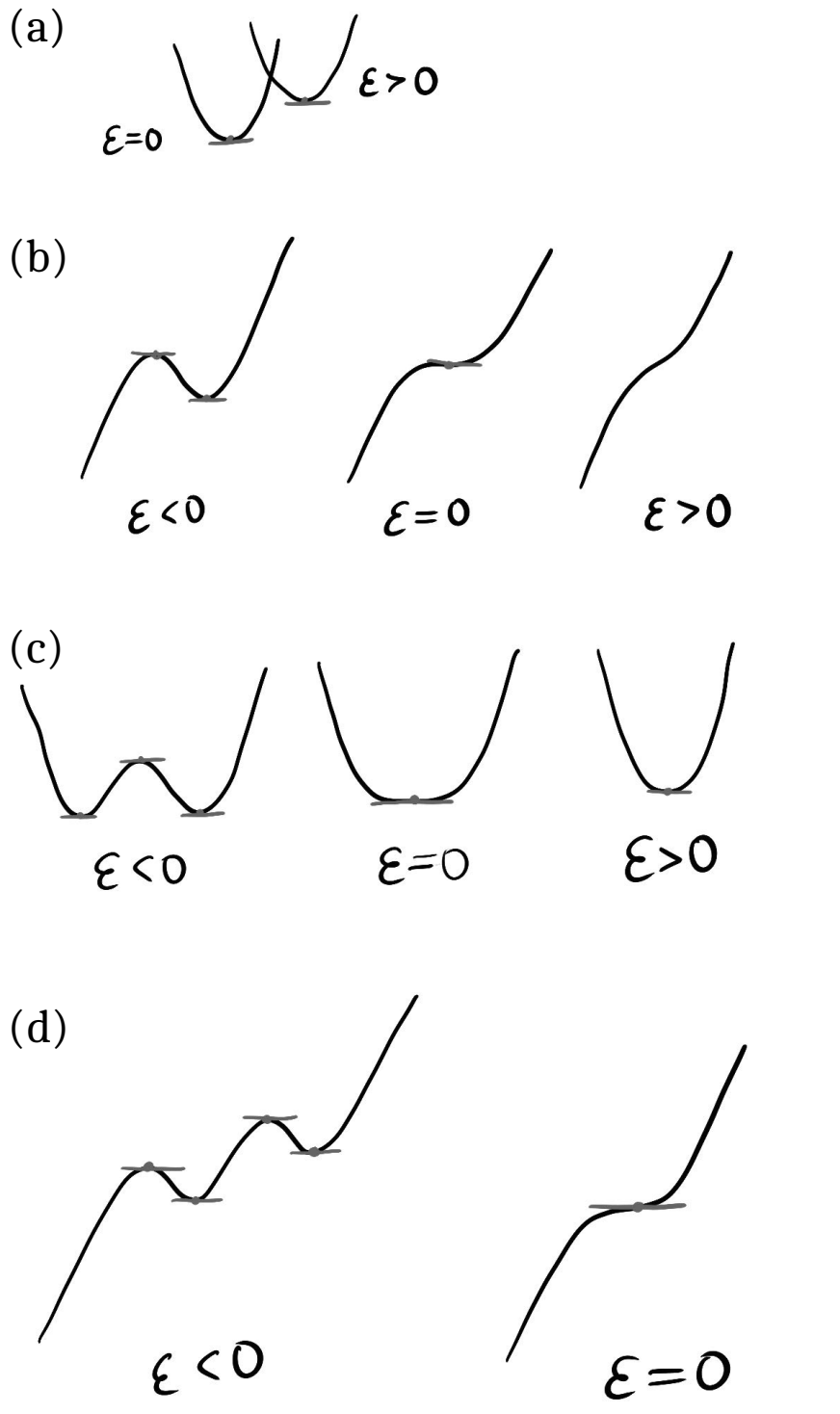

Figure 3.3: The effect of a small perturbation $\varepsilon$ on polynomials of different orders. (a) $x^2+\varepsilon x$, (b) $x^3+\varepsilon x$, (c) $x^4+\varepsilon x^2$, (d) $x^5+\varepsilon x^3$. Adapted from [84].

This property of stability around non-degenerate points of any (!) smooth function is stated in general form as the *Morse Lemma.* In preparation of that, the following Lemma is helpful:

**Lemma 3.2.1** *Let $f : \mathbb{R}^n \to \mathbb{R}$ be smooth in the neighbourhood of $0$, with $f(0) = 0$. Then in a possibly smaller neighbourhood there exist functions $g_i : \mathbb{R}^n \to \mathbb{R}$, such that $f = \sum_{i=1}^n x_i g_i$, where each $g_i$ is smooth and $g_i(0) = \frac{\partial f}{\partial x_i}\big|_0$.*

This means that the topology of a manifold can easily be described by a linear combination of differentiable functions on that manifold. It follows the theorem:

**Theorem 3.2.2 (Morse lemma)** *Let $u$ be a non-degenerate critical point of the smooth function $f : \mathbb{R}^n \to \mathbb{R}$. Then there is a local coordinate system $(y_1, y_2, ..., y_n)$ in a neighbourhood $U$ of $u$, with $y_i(u) = 0$ for all $i$, such that $f = f(u) - y_1^2 - y_2^2 - ... - y_l^2 + y_{l+1}^2 + ... + y_n^2$ for all $y$ in $U$.*

That is, every non-degenerate critical point can be diffeomorphic (think smooth and reversibly differentiable) transformed into a so-called Morse $l$-saddle, where $l = 0$ refers to a minimum and $l = n$ to a maximum, and therefore specifies the 'type' of the critical point. Furthermore, for functions of a single variable ($n = 1$) one can prove that its behaviour near a critical point is completely determined by the first non-vanishing jet, that is, the lowest-order non-zero term in the Taylor expansion about that point. For functions of more than one variable, this property no longer holds.

However, the idea of breaking up the function into parts of structural stability can be recovered using the following theorem:

**Theorem 3.2.3 (Splitting lemma)** *Let $f\colon \mathbb{R}^n \to \mathbb{R}$ be a smooth function with $\nabla f(0) = 0$, and the Hessian at $0$ has rank $r$ and corank $n - r$. Then at $0$, $f$ is equivalent to a function of the form $\pm {x_1}^2 \pm \cdots \pm {x_r}^2 + \hat{f}(x_{r+1}, \ldots, x_n)$, where $\hat{f}\colon \mathbb{R}^{n-r} \to \mathbb{R}$ is smooth.*

The idea is that at a degenerate critical point we can split (hence the name) the function into two parts: A Morse part which is 'inessential and not involved in the degeneracy, and a degenerate part which is 'essential in the sense that it is responsible for the structural instability. The latter lives on a different set of variables, whose dimensionality is the *corank*. We can think of it as 'in how many directions is the critical point degenerate?'.

To summarise, the behaviour of a function near a degenerate critical point is dominated by a function only of a subset of variables. How many of those 'extra variables we need, is the corank of the Hessian. That is, locally we could write $f(x) = f(x_0) + \hat{f}(\hat{x}) + Q(x)$, such that $\hat{x} \in \mathbb{R}^{n-r}$ describes the degenerate part of the function and $Q(x)$ is the non-degenerate quadratic Morse part $Q(x)$. The splitting lemma thereby serves as a starting point for the classification of singularities and catastrophes, as it states the existence of a diffeomorphism that allows the isolation of the essential degrees of freedom responsible for the nontrivial local structure.

**Parameters and codimension**

This notion of dimensionality hints at one of the key concepts of catastrophe theory: the distinction between state and control variables. The smooth potential (or generating) functions that describe the behaviour of a system can be written to depend on two distinct sets of variables. The *state variables* are those that describe the internal 'state of the function, which potentially extremises it. *Control variables*, on the other hand, are all the external variables that we can adjust to influence the overall shape of the potential function. The small perturbation factor $\varepsilon$ in above's examples would be the control variable, and the $x$ is the state variable. When considering a simple parabola $f(x) = ax^2 + b$, then we are interested in the extrema w.r.t. $x$, as the parabola might describe a physical potential energy. Specific coordinates of $x$ therein describe the state of the system. The location of the minimum changes, however, depending on the parameters $a$ and $b$, which may be used to control the stability of the studied system.

A *catastrophe* then occurs when a smooth and gentle change of the control variables causes a qualitative change in the structure of the types of critical points. These sudden changes occur at degenerate saddle points, where multiple stationary points coalesce and which are referred to as the catastrophe points. Around this degenerate critical point, which is structurally unstable, we can add extra terms to the function in order to supply structural stability. The resulting polynomial is called the *unfolding* of the singularity — "It is like the unfolding of a bud to reveal a flower" [83].

The number of variables required to fully 'unfold' the singularity by lifting all degeneracies is the *codimension*. More generally, the codimension of a submanifold is the number of independent equations needed to describe it locally as a subset of an ambient manifold. This number may be different to the number of parameters required to parametrise the submanifold itself.

In the context of catastrophe theory, the codimension quantifies how many independent control parameters are needed to resolve a degenerate critical point into non-degenerate stationary points under small perturbations. Intuitively, the codimension of a catastrophe is the minimum number of control variables that must be tuned in order for the corresponding degenerate critical point to persist. Conversely, a catastrophe of codimension $K$ can be regarded as arising from the coalescence of $K+1$ non-degenerate critical points into a single degenerate catastrophe point (upon the precise tuning of the $K$ parameters).

Equivalently, the codimension measures the degree of instability of the singularity with respect to generic perturbations. As a result, catastrophes with lower codimension are rather generic and robust, as they only require few parameters to be tuned. Higher codimension catastrophes are increasingly fragile and unlikely to be observed without dedicated fine-tuning.

**Canonical unfoldings and bifurcation sets**

The generic form for potential functions with one control variable (corank 1) are called *cuspoids*. Their canonical form reads

$$f_K(x;\boldsymbol{\mu}) = x^{K+2} + \sum_{m=1}^{K} \mu_m x^m \tag{3.3}$$

for the codimension $K$, the state variable $x$ and the control variables $\boldsymbol{\mu} = \{\mu_m\} \in \mathbb{R}^K$. Potential functions with two-dimensional control parameter space, and hence corank 2, are called *umbilics* and have the following canonical forms for up to codimension $K = 3$:

$$f^{(E)}(x,y;\boldsymbol{\mu}) = x^3 - 3xy^2 + \mu_3(x^2+y^2) + \mu_2 y + \mu_1 x \tag{3.4}$$

$$f^{(H)}(x,y;\boldsymbol{\mu}) = x^3 + y^3 + \mu_3 xy + \mu_2 y + \mu_1 x \tag{3.5}$$

which are the elliptic and hyperbolic umbilics, respectively.

The set of all critical points is called the *catastrophe manifold* [84] or equilibrium surface [83]. The set of control-variable values for which critical points are degenerate define the *bifurcation set* $\mathcal{B}$. For a generic function $f: \mathbb{R}^n \mapsto \mathbb{R}$ with state variables $\mathbf{x} \in \mathbb{R}^n$ and control variables $\boldsymbol{\mu} \in \mathbb{R}^K$, that is

$$\mathcal{B} = \left\{\boldsymbol{\mu} \in \mathbb{R}^K \middle| \nabla_x f(x;\boldsymbol{\mu}) = 0 \quad \text{and} \quad \det\left(\mathcal{H}f\right) = 0\right\}. \tag{3.6}$$

Intuitively, the bifurcation set contains all the values of control parameters for which critical points of the function coincide to higher-order singularities. Sometimes, it is therefore also referred to as the singularity set. For parameters outside the bifurcation set, all critical points are non-degenerate, such that their number and type remains stable under small parameter changes. The bifurcation set is a $K$-dimensional manifold and separates regions of the parameter space with qualitatively distinct behaviour. The big achievement of catastrophe theory is the finding that these unfoldings of the catastrophes are universal for each order of singularity, and can in fact be classified depending solely on their dimensionality, which we will explicate below. Before we detail the examples of specific catastrophes, however, let us motivate the usage of catastrophe theory in its application to describe physical phenomena.

### 3.2.3 Wave phenomena and highly-oscillatory integrals

In physics, the simplest way to describe the propagation of light is through classical ray optics, where light is treated as travelling along straight-line trajectories. While this approach captures basic propagation, it fails to describe many common wave phenomena, such as intense focusing, evanescent waves and diffraction patterns. To account for these effects, one must consider the superposition of waves, a description that naturally extends to quantum mechanics and sound waves as well. Consider therefore a path integral that describes the wave field at an observation point $\boldsymbol{\mu}$ in the $N$-dimensional space of $\mathbf{x} \in \mathbb{R}^N$, the Fresnel–Kirchoff diffraction integral

$$\Psi(k;\boldsymbol{\mu}) = \left(\frac{k}{\pi}\right)^{N/2} \int_{\mathbb{R}^N} \mathrm{e}^{\mathrm{i}kf(\mathbf{x};\boldsymbol{\mu})} \mathrm{d}\mathbf{x} \tag{3.7}$$

with the wave number $k$. Any alteration of the optical path, for example by an $N$-dimensional lens, would be contained in the function $f(\mathbf{x};\boldsymbol{\mu})$. Put differently, the light is now considered to accumulate a phase during its propagation along the optical path. This approach is known as semi-classical theory, or WKB method[1] as it captures wave-like phenomena such as diffraction and interference, but still retains the ray-like propagation. The total observed brightness is then given by

$$I(k;\boldsymbol{\mu}) \propto \left|\Psi(k;\boldsymbol{\mu})\right|^2. \tag{3.8}$$

In most regions, the integral Eq. (3.7) can be evaluated using a stationary-phase approach, as introduced briefly in Sec. 2.3.2 already. Each non-degenerate critical point of the exponent contributes as a classical 'ray', resulting in an amplitude that scales with $k^{-N/2}$.

Near regions where classical rays converge, known as *caustics*, the ray approximation breaks down: the density of rays diverges, and singular-like patterns appear. This is ultimately because the critical points of the stationary-phase method coincide and the prefactor diverges. To describe the resulting wave behaviour uniformly across a range of wave numbers, physicists employ *canonical integrals* of the diffraction catastrophes. These are oscillatory integrals of the form

$$\Psi_K(\boldsymbol{\mu}) = \int_{-\infty}^{+\infty} \mathrm{e}^{\mathrm{i}f_K(x,\boldsymbol{\mu})} \mathrm{d}x \tag{3.9}$$

[1] Named after Gregor Wentzel, Hendrik Kramers and Léon Brillouin.

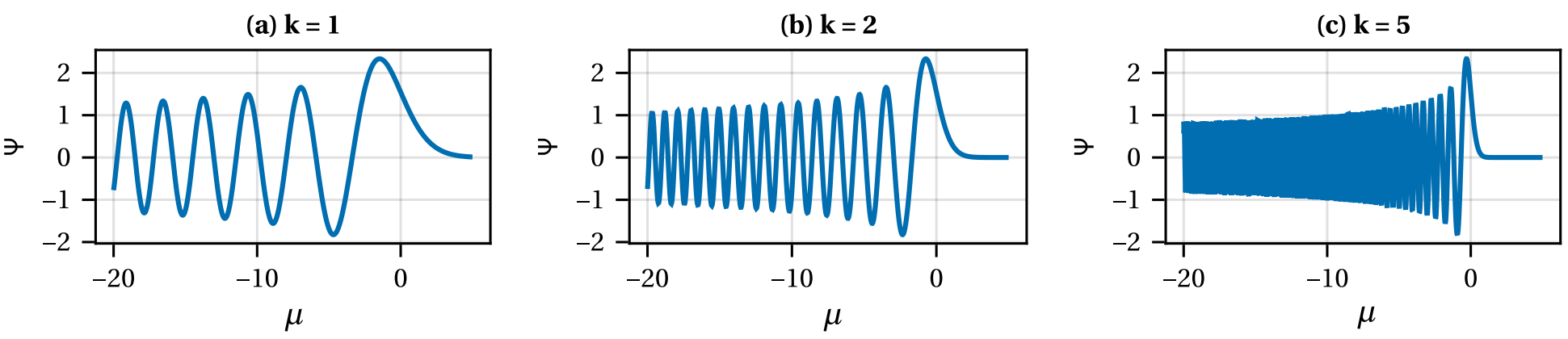

Figure 3.4: The diffraction integral $\Psi(k;\mu)$ for a fold catastrophe ($K = 1$), for different wave numbers $k$.

for cuspoids of codimension $K$, Eq. (3.3), and

$$\Psi^{(U)}(\boldsymbol{\mu}) = \int_{-\infty}^{+\infty}\int_{-\infty}^{+\infty} \mathrm{e}^{\mathrm{i}f^{(U)}(x,y;\boldsymbol{\mu})}\mathrm{d}x\mathrm{d}y \tag{3.10}$$

for the umbilic unfoldings (Eqs. 3.4 and 3.5, respectively). They are conveniently listed online in the DLMF, Chapt. 36 [85]. These integrals can then be adapted to model, for example, the observed light intensity across a spatial domain. The critical points of $f$ correspond to classical rays, and their degeneracies generate catastrophic focusing and universal interference structures. The external parameters (for example the points in space) where they coincide often immediately reveal the bifurcation set of the associated catastrophe point. By classifying these critical points via catastrophe theory, one identifies canonical forms that appear throughout optics, quantum mechanics, fluid dynamics, and gravitational lensing. Studying these critical points and their hierarchy is therefore essential: it predicts caustic structure, guides uniform approximations near singularities, and explains how the geometry of classical trajectories shapes observable wave patterns across physical systems.

For the canonical forms of the cuspoid and umbilic catastrophes, the caustics appear at $\boldsymbol{\mu} = 0$. For increasing frequency values $k$, the intensity of the wavefront diverges and the fringe spacing of the interference patterns shrink. This is demonstrated in Fig. 3.4, where we show the Fresnel–Kirchoff diffraction integral associated with the cuspoid of codimension $K = 1$ (a fold catastrophe) for different frequencies $k$. The integral is calculated in terms of the Airy function across a range of the external parameter $\mu$, and the caustic appears at $\mu = 0$. For different values of the frequency $k$, the overall shape of the integral remains the same. As $k$ increases, the fringes of the oscillatory region for $\mu < 0$ become narrower, the peak intensity at the caustic increases and the transition between the two regions on either side of the caustic becomes sharper.

Specific motivation stems from the enhancement that is observable at the caustic. The ‘brightness’ of the caustic in comparison to its surrounding is quantifiable in terms of the so-called *twinkling exponents* [86]. In the high-frequency regime of large values of $k$ (panel (c)), the maximum intensity at the caustic follows the scaling law

$$I(k;\boldsymbol{\mu} = 0) = I_0 k^{2\beta} \tag{3.11}$$

where $I_0$ is the intensity in the vicinity of the caustics and $\beta$ is the twinkling exponent.

Note that in application to physical problems, the definition of $I_0$ is not necessarily immediately obvious. For the derivation of Eq. (3.11), $I_0$ is taken as the integral of wave amplitudes along a specific ‘path’ in parameter space $\boldsymbol{\mu}$. That is, the scaling with the frequency ultimately arises as a relation between the amplitude integrals along a path that includes the caustic, and the along the path that is far away from it. In the case where $\boldsymbol{\mu}$ denotes the position of the observer, this could just be one line in, say, the $x$-direction. In generic cases, however, it is important to bear in mind that the path is by no means universal and has to be considered carefully depending on the specific application.

On the other hand, the twinkling exponent in Eq. (3.11), related to the singularity index [87, 88], is universal in that it only depends on the classification of the underlying catastrophe. As a result, it provides a quantitative link between catastrophe theory and the diffraction catastrophe integrals, capturing how the local geometry of critical points controls observable wave fluctuations in optics, quantum mechanics, and related wave phenomena.

### 3.2.4 (Some of) The seven elementary catastrophes

As mentioned above, the codimension of critical points quantifies the number of control parameters required to tune a set of non-degenerate critical points to coalesce into one. Higher codimension therefore requires much more precision in the adjustment of the parameters, and, as a result, the appearance of catastrophes becomes less likely with increasing codimensionality. The critical points with codimension $K \leq 4$, however, are ubiquitous in nature and have been classified by René Thom [89]. In the following we review these seven catastrophes and their hierarchical order, and then introduce features of the cuspoids that will appear in this thesis.

**The hierarchy of the seven elementary catastrophes**

The seven elementary catastrophes are characterised by their corank and codimension, i.e., the number of control and state variables. They have characteristic unfoldings in parameter space, which inspired the naming of the catastrophes as ‘fold’, ‘cusp’, ‘swallowtail’, ‘butterfly’ (corank 1) as well as ‘elliptic’, ‘hyperbolic’ and ‘parabolic’ umbilic (corank 2). The full classification is listed in Tab. 3.1. The hierarchy of the catastrophe is ultimately established via Coxeter reflection groups [87, 90, 91], included as the respective symbol in the table. These furthermore imply that the system exhibiting a specific catastrophe always contains the lower-order catastrophes as well. For example, the bifurcation set of the swallowtail catastrophe also inherently incorporates cusp and fold points.

For the canonical forms of the potential functions shown in Tab. 3.1, one can formulate the condition for each of the cuspoid catastrophe points simply as:

$$\frac{\partial^n}{\partial x^n} f = 0 \quad \text{and} \quad \frac{\partial^{K+2}}{\partial x^{K+2}} f \neq 0\,. \tag{3.12}$$

where $n = 1, \ldots, K+1$. However, in most applications, the potential function does not appear in its canonical form and we only know from the generality of the theorems that there exists a diffeomorphism that maps it to these forms. In practice, it is often easy to spot the appearance of a caustic, or a specific catastrophe, but finding the precise mapping between the studied physical system and the canonical catastrophe proves to be challenging. The conditions for the exact catastrophe point therefore need to be formulated more carefully [70, 71, 92, 93]. We will detail some generic conditions, alongside some more phenomenological details for the most relevant catastrophes in the subsequent sections.

Table 3.1: Hierarchy of the seven elementary catastrophes, with $\mathbf{x} = (x, y)$ and corank $r$.

| *Name* | $r$ | $K$ | *Symbol* | $\beta$ | *potential function* $f(\mathbf{x};\boldsymbol{\mu})$ |
|---|---|---|---|---|---|
| Fold | 1 | 1 | $A_2$ | $1/6$ | $\frac{1}{3}x^3 + \mu_1 x$ |
| Cusp | 1 | 2 | $A_3$ | $1/4$ | $\frac{1}{4}x^4 + \frac{1}{2}\mu_2 x^2 + \mu_1 x$ |
| Swallowtail | 1 | 3 | $A_4$ | $3/10$ | $\frac{1}{5}x^5 + \frac{1}{3}\mu_3 x^3 + \frac{1}{2}\mu_2 x^2 + \mu_1 x$ |
| Butterfly | 1 | 4 | $A_5$ | $1/3$ | $\frac{1}{6}x^6 + \frac{1}{4}\mu_4 x^4 + \frac{1}{3}\mu_3 x^3 + \frac{1}{2}\mu_2 x^2 + \mu_1 x$ |
| Elliptic umbilic | 2 | 3 | $D_4^-$ | $1/3$ | $x^3 - 3xy^2 + \mu_3(x^2 + y^2) + \mu_2 y + \mu_1 x$ |
| Hyperbolic umbilic | 2 | 3 | $D_4^+$ | $1/3$ | $x^3 + y^3 + 6xy + \mu_3 xy + \mu_2 y + \mu_1 x$ |
| Parabolic umbilic | 2 | 4 | $E_6$ | $3/8$ | $x^2 y + y^4 + \mu_4 y^2 + \mu_3 x^2 + \mu_2 y + \mu_1 x$ |

**The fold catastrophe**

The fold catastrophe has codimension $K = 1$ and is therefore defined as the coalescence of two critical points. For the canonical potential function of the fold catastrophe,

$$f(x;\mu) = x^3 + \mu x, \tag{3.13}$$

this is fulfilled if the first and second derivative vanish — a requirement that we can evaluate analytically. Setting the first derivative to zero,

$$\frac{\partial f}{\partial x} = 3x2 + \mu = 0, \tag{3.14}$$

yields the two extremal points $x_{1/2} = \pm\sqrt{-\mu/3}$, assuming $\mu \le 0$ to ensure that $x \in \mathbb{R}$. Requesting that the second derivative vanishes too,

$$\frac{\partial^2 f}{\partial x^2} = 6\mu = 0, \tag{3.15}$$

yields $\mu = 0$ as the location of the degenerate critical point in parameter space. That is, the bifurcation set is simply $\mathcal{B} = \{\mu = 0\}$. For any value $\mu \neq 0$ the two critical points $x_1$ and $x_2$ are separate, but at $\mu = 0$ they coalesce into one and the same object.

The canonical diffraction integral for the fold catastrophe is the Airy function:

$$\Psi_1(\mu) = \int_{-\infty}^{+\infty} \mathrm{e}^{\mathrm{i}(x^3+\mu x)}\mathrm{d}x = \frac{2\pi}{3^{1/3}}\mathrm{Ai}\left(\frac{\mu}{3^{1/3}}\right) \tag{3.16}$$

with

$$\mathrm{Ai}\left(\mu\right)=\frac{1}{2\pi}\int_{-\infty}^{+\infty}\mathrm{e}^{\mathrm{i}\left(\frac{t^3}{3}+\mu t\right)}\mathrm{d}t=\frac{1}{\pi}\int_{0}^{\infty}\cos\left(\frac{t^3}{3}+\mu t\right)\mathrm{d}t. \tag{3.17}$$

This diffraction integral is shown in Fig. 3.4(a), where the frequency is unity. The Airy function has two qualitatively distinct regions. For $\mu < 0$, it exhibits oscillations with slowly decreasing amplitudes as $\mu \to -\infty$. In this region, there are two, typically real-valued, stationary points of the phase function which contribute towards the integral and yield this fringed interference pattern. For $\mu > 0$ the Airy function decays rapidly and monotonically towards zero, without any oscillations and no real stationary points. The point $\mu = 0$ is the caustics that marks the maximum amplitude and the breakdown of the standard stationary-phase approximation. In total, the Airy function provides a uniform approximation as it bridges smoothly between the two regimes.

Fold catastrophes can appear in functions with a higher number of control and state parameters. To identify them in these higher-dimensional regimes, the conditions for a fold catastrophe have to be defined more carefully [70]. In general, a fold catastrophe is that, at a critical point, the matrix of second derivatives is singular, meaning the determinant of the Hessian vanishes. That is, a fold catastrophe appears when

$$\nabla_{\mathbf{x}} f = 0, \qquad \det\left(\mathcal{H} f\right) = 0 \tag{3.18}$$

As such, the control parameters that fulfill these equations define the bifurcation set.

**The cusp catastrophe**

The canonical form of the unfolding for the cusp catastrophe with its two-dimensional parameter space $\boldsymbol{\mu} = (\mu_1, \mu_2)$ is

$$f(x;\boldsymbol{\mu}) = \tfrac{x^4}{4} + \tfrac{\mu_2}{2}x^2 + \mu_1 x\,. \tag{3.19}$$

The set of critical points (the catastrophe set) is given by the manifold

$$\frac{\mathrm{d}f}{\mathrm{d}x} = x^3 + \mu_2 x + \mu_1 = 0 \tag{3.20}$$

which is shown in the top of Fig. 3.5 across the parameter space $\mu_1, \mu_2$ (at the bottom). At $\boldsymbol{\mu} = (0,0)$ we find three non-degenerate critical points coalescing into a single cusp point. For a small perturbation, this point will unfold into two fold lines. This bifurcation set is visualised in the bottom projection in Fig. 3.5. The bifurcation set indicates regions of numbers of saddle points: at the cusp point there is one (high-order) critical point, along the fold lines there are two critical points. In the region between the fold lines there are three critical points — seen as three sheets of the catastrophe manifold (often referred to as the potential surface) in Fig. 3.5. Assuming $\boldsymbol{\mu} \in \mathbb{R}^2$, outside that region there is only one critical point.

For the canonical case, the analytical expression for the bifurcation set can be found by setting the second derivative of the potential function to zero, i.e.,

$$\frac{\mathrm{d}^2 f}{\mathrm{d}x^2} = 3x^2 + \mu_2 = 0 \tag{3.21}$$

and solving for $x$. This gives the location of inflection points at $x_{1/2} = \pm\sqrt{-\mu_2/3}$. We can plug this directly into the first derivative Eq. (3.20) and with that eliminate $x$. The bifurcation set of the canonical cusp catastrophe is hence described by

$$4\mu_2^3 + 27\mu_1^2 = 0\,, \tag{3.22}$$

also known as Neile's parabola, sketched in blue at the bottom of Fig. 3.5, and which can be found in everyday situations as seen in Fig. 3.1. That is, parameter combinations that fulfil Eq. (3.22), yield a coalescence of two saddle points. For $\mu_1 = \mu_2 = 0$ we obtain the full cusp catastrophe where all three saddle points coalesce.

The diffraction integral associated with the canonical cusp catastrophe is the Pearcey function [94]

$$\Psi_2(\boldsymbol{\mu}) = P(\mu_2, \mu_1) = \int_{-\infty}^{+\infty} e^{i(x^4+\mu_2 x^2+\mu_1)}\,dx \tag{3.23}$$

shown in Fig. 3.6. The uniform and accurate evaluation of which, across ranges of frequencies, is an active topic of research [95–99].

For a non-canonical form of a generic potential function $f(\mathbf{x}, \boldsymbol{\mu})$, the coalescence of three critical points into a cusp catastrophe is given when two fold lines coincide tangentially. For that to happen, the direction of change of the vanishing eigenvalue along the fold has to become orthogonal to its eigenvector. Put differently, the vanishing eigenvalue has zero gradient along its own eigenvector. Hence, the generic conditions for a cusp catastrophe are

$$\nabla_{\mathbf{x}} f = 0, \qquad \epsilon_i = 0, \qquad \mathbf{v}_i \cdot \nabla \epsilon_i = 0 \tag{3.24}$$

where $\mathbf{v}_i$ is the eigenvector to the vanishing eigenvalue $\epsilon_i$ of the Hessian matrix $\mathcal{H}$ (equivalently, $\det(\mathcal{H}f) = 0$). Intuitively, by working with the eigensystem of the deformation tensor $\mathcal{H}f$ we ensure that the further derivatives are taken in 'canonical' directions of the system.

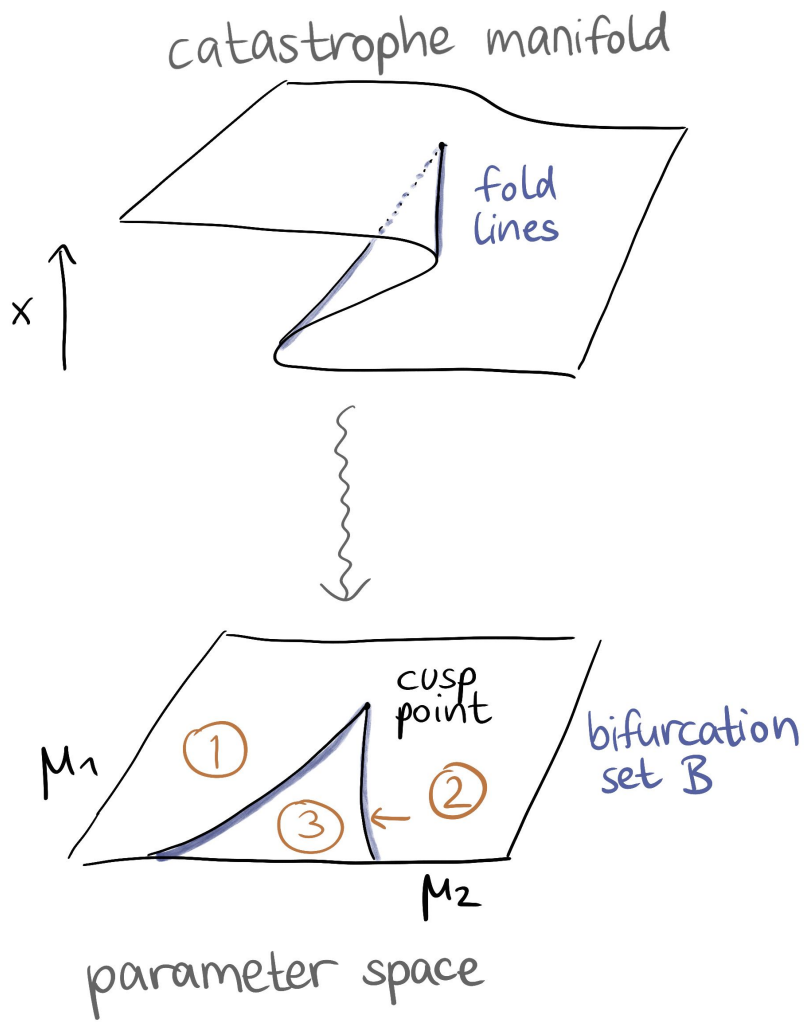

Figure 3.5: The catastrophe manifold (top) and the bifurcation set in parameter space (bottom) for the cusp catastrophe. In parameter space, the bifurcation set (blue) separates region with different numbers of relevant critical points (encircled numbers).

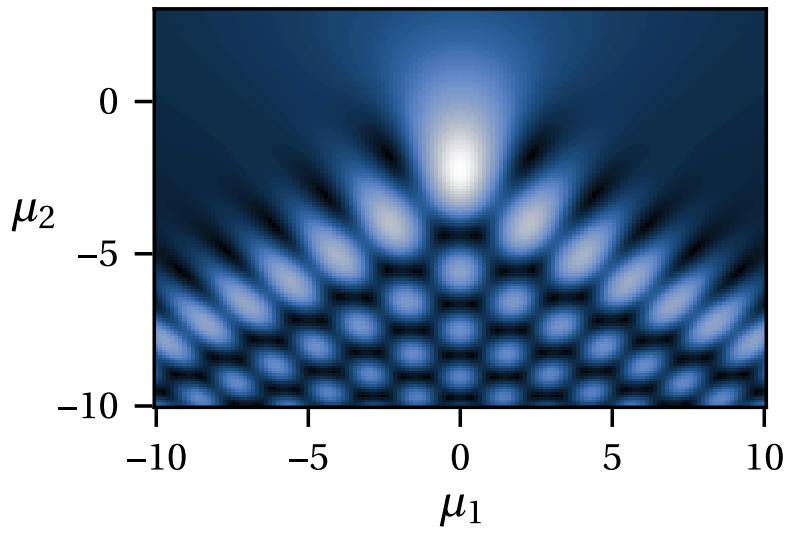

Figure 3.6: The canonical diffraction integral of the cusp catastrophe evaluated across parameter space.

**The swallowtail catastrophe**

For the swallowtail catastrophe the control space is three-dimensional, $\boldsymbol{\mu} = (\mu_1, \mu_2, \mu_3)$, and the respective canonical potential surface overlies itself at most four times at the degenerate catastrophe point. That is, in the vicinity of the swallowtail point, the potential surface unfolds into four separate sheets of non-degenerate critical points. Owing to this high dimensionality, we cannot plot the potential surface directly, but only the bifurcation set. Let us find an analytical expression for the canonical bifurcation set. The canonical unfolding is given as

$$f(x, \boldsymbol{\mu}) = \tfrac{1}{5}x^5 + \tfrac{\mu_3}{3}x^3 + \tfrac{\mu_2}{2}x^2 + \mu_1 x \tag{3.25}$$

and the critical points are given by the vanishing first derivative:

$$\frac{\mathrm{d}f}{\mathrm{d}x} = x^4 + \mu_3 x^2 + \mu_2 x + \mu_1 = 0. \tag{3.26}$$

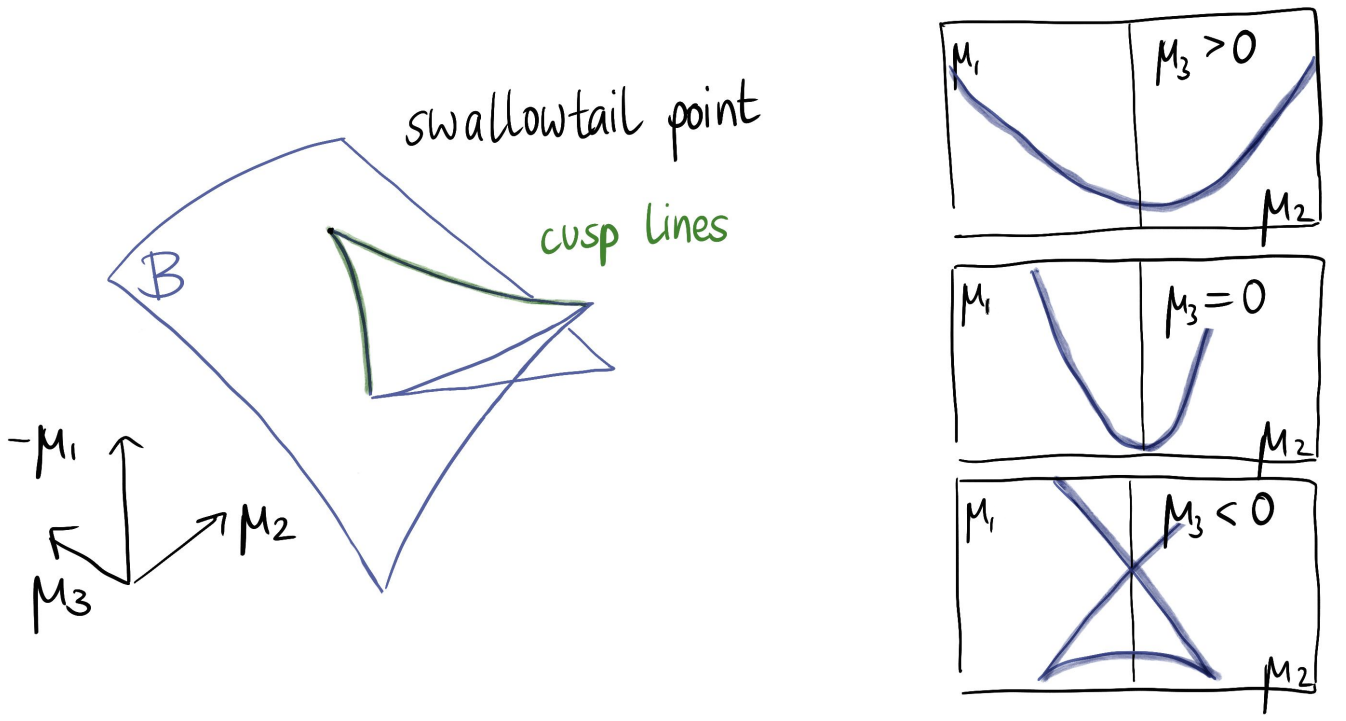

Figure 3.7: Bifurcation set of the swallowtail catastrophe (left) and slices of it at constant $\mu_3$ (right).

Degenerate critical points are those for which the second derivative vanishes as well, i.e.,

$$\frac{\mathrm{d}^2 f}{\mathrm{d}x^2} = 4x^3 + 2\mu_3 x + \mu_2 = 0. \tag{3.27}$$

By rearranging, we find a parametric form of the bifurcation set for fixed values of $\mu_3$:

$$\begin{aligned} \mu_1 &= x^2\left(3x^2 + \mu_3\right) \\ \mu_2 &= -2x\left(2x^2 + \mu_3\right) \qquad \text{for} \quad -\infty < x < \infty \end{aligned} \tag{3.28}$$

which is shown on the left-hand side of Fig. 3.7 and with some imagination has the shape of a swallowtail. Characteristic features of the bifurcation set are the two cusp lines (sometimes referred to as 'ribs'), which can be found by setting the third derivative $\frac{\mathrm{d}^3 f}{\mathrm{d}x^3}$, giving $x^2 = -\frac{1}{6}\mu_3$. Plugging this into the bifurcation set (3.28) yields

$$\mu_1 = -\frac{1}{12}\mu_3 \quad \text{and} \quad \mu_2 = -\frac{8}{27}\mu_3 \quad \text{and} \quad \mu_3 \leq 0\,. \tag{3.29}$$

For clarity, on the right-hand side we show slices through the bifurcation set in separate panels. At the bottom of the shape, for $\mu_3 < 0$ two sheets of the bifurcation set intersect.

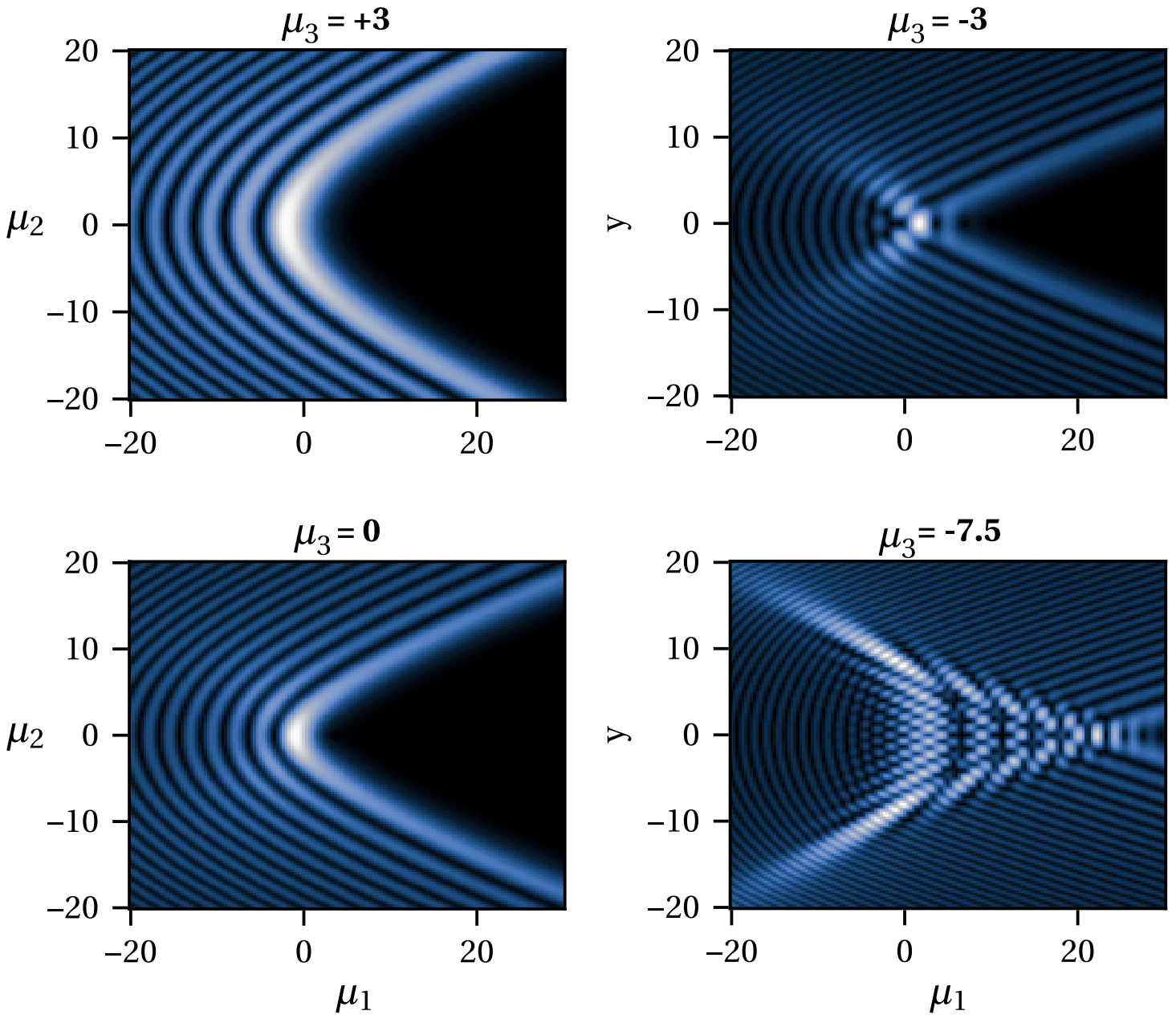

Figure 3.8: Cuts through the diffraction integral of the swallowtail catastrophe, $\Psi_4(\boldsymbol{\mu})$ for several values of $\mu_3$.

That is, there are two separate sets of fold points which peacefully coexist without coalescing. This self-intersection line is found be realising that for $\mu_2 = 0$ two solutions to Eq. (3.28) both appear for the same parameters

$$\mu_1 = -\frac{3}{4}\mu_3^2 \quad \text{and} \quad \mu_2 = 0 \quad \text{and} \quad \mu_3 \leq 0\,. \tag{3.30}$$

Generally, perturbations into the direction of $\mu_3 \leq 0$ allow the swallowtail point to unfold into three different non-degenerate points, whereas the side of $\mu_3 > 0$ only yields one solution.

The canonical diffraction integral for the swallowtail potential is evaluated in the three-dimensional space of control parameters. Therefore, we can only show slices of it, at specific values of $\mu_3$. Characteristic visualisations are shown in Fig. 3.8.

The identification of a swallowtail point in a generic function $f(\mathbf{x},\boldsymbol{\mu})$ is a bit more intricate. Conceptually, we want the critical point to be 'even more degenerate' than a cusp, and take that to mean that the change of eigenvalue is vanishing in one more direction. More precisely, at a swallowtail point all the conditions for a cusp catastrophe must be fulfilled, and additionally we require the eigenvector to be orthogonal to the direction of change of the cusp condition. That leads us to the following conditions:

$$\nabla_{\mathbf{x}} f = 0, \qquad \epsilon_i = 0, \qquad \mathbf{v}_i \cdot \nabla \epsilon_i = 0, \qquad \mathbf{v}_i \cdot \nabla\left(\mathbf{v}_i \cdot \nabla \epsilon_i\right) = 0\,. \tag{3.31}$$

## 3.3 Applications of catastrophe theory in attosecond science

Caustics as a reason for an enhanced signal in strong-field effects is not unknown. In fact, it has been recognised from the very beginning that the cutoff in high-order above-threshold ionisation (ATI) spectra can be attributed to "the classical rainbow effect" [100, 101]. Since then, numerous studies have investigated such spectral cutoffs, both in ATI and in high-order harmonic generation (HHG). Beyond the overall spectral extent, detailed analyses of the resulting interference patterns have provided valuable insight into the role of the atomic binding potential in shaping the electron dynamics. More recently, caustic-induced enhancements in HHG have been identified both theoretically and experimentally, often motivated by the prospect of tailoring the driving field to selectively enhance specific harmonic orders. Despite these advances, a rigorous and systematic connection to catastrophe theory is only rarely established. In the following, we provide a brief review of the most relevant developments and corresponding literature that addresses the application of caustics in attosecond science.

### 3.3.1 Caustics in ATI from the interaction with the Coulomb potential

**The low-energy structure in the ATI spectrum**

The intricate structure of ATI spectra includes an "ionisation surprise" [102]: a peak-like feature at an energy of around $\approx 0.1 U_\mathrm{p}$, which is now commonly referred to as "low-energy structure" (LES) [103–108], as shown on the left-hand side of Fig. 3.9. Its appearance was rather unexpected as it is not predicted within the standard SFA framework — even though it is in the parameter range where the SFA was expected to hold well. A whole line of research has therefore emerged to explain its origin, see e.g. [104–114].

One of the approaches identifies a particular peak as a caustic stemming from the combined influence of the Coulomb potential and the laser field on the electron's dynamics. This is mentioned in [21] and [104], and then specifically addressed in [22]. The authors call this a *dynamical* caustic, which is to be distinguished from the commonly studied *spectral* or *kinematical* caustics which constitute the energy cutoffs in the various spectra. In the publication [22], an approach of classical trajectories is used, where electron trajectories are sampled across the initial conditions ionisation time and velocity, and then follow the equations of motion in the continuum until they are captured at the detector with spatially resolved momenta $(p_x, p_y)$. In Fig. 3.10 those final momenta are shown for several initial velocities (different colours) across ranges of ionisation times, corresponding to the several points. The highlighted box in the left-hand side panel marks the LES area. For very low momenta a typical feature in those mappings is the Coulomb focussing (not shown here), which results in a relative decline of transverse momenta (viz. the attraction towards the core origin). In [21] the authors introduce an inverse effect to describe the LES: Coulomb defocussing. The model includes multiple successive interactions and takes into account that the Coulomb potential will only become relevant during the time intervals when the electron is actually revisiting the ionic core. This more refined approach finds that electrons with low initial momentum exhibit larger final transversal momentum and these particular trajectories are predominately responsible for the formation of LES in the ATI spectrum. A closer look at the mapping of initial conditions to final momenta plane then demonstrates the link to a (hidden) catastrophe and is shown in the right-hand side panel of Fig. 3.10. The thick red lines mark the zeros of the Jacobian, and hence the caustics. The black dots highlight cusps in the final-momentum plane. The authors suggest the similarity to a butterfly catastrophe but do not provide an exact mapping to the canonical form, nor an exploration of the full unfolding. While the identification of a butterfly catastrophe is surely intriguing, no rigorous link supports this suggestion yet.

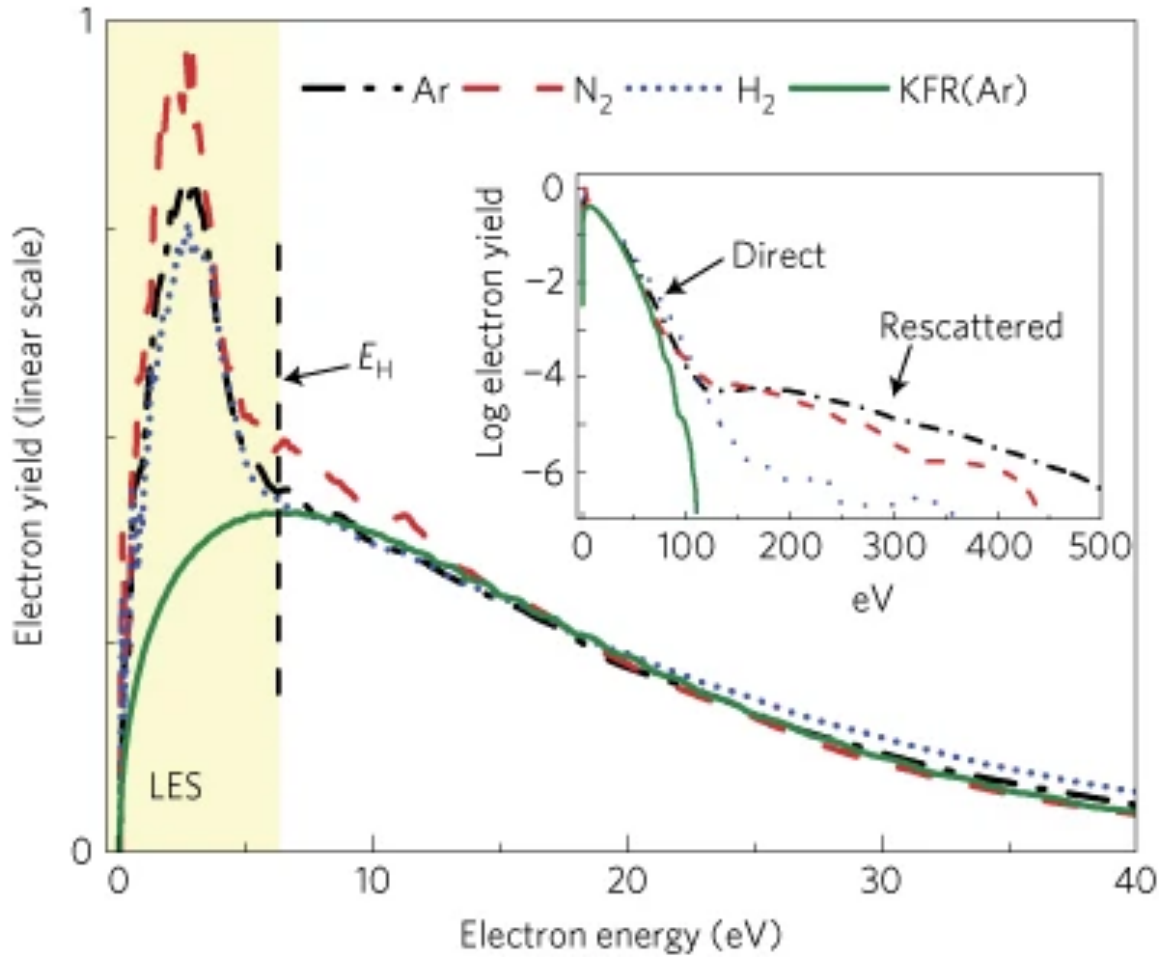

Figure 3.9: Low-energy region of the ATI photoelectron spectrum showing a pronounced peak. Inset: The entire distribution of ATI energies, showing contributions from direct and rescattered pathways. Dashed lines are experimentally measured data from different gases, the solid green line (strongly deviating in the LES regions) is a calculation based of KFR tunnelling theory. Taken from [103].

**Photoelectron holography**

Understanding the overall pattern of photoelectron momentum distributions (PEMDs) of ATI is a complicated task, because they consist of several interferences between a myriad of different electron pathways (see Fig. 3.11 for an example). Each of the pathways may be modelled with different assumptions, which further complicates the identification of their interplay and the resulting caustics, see e.g., Ref. [115] for a review. There are several efforts to disentangle this structure by taking into account the Coulomb effects and classifying the existing orbits, see Fig. 3.12. This work aims at disentangling the intricate interference structure observed in PEMDs and attributing characteristic features to specific pathways and phenomena. This has mainly been done using Monte Carlo methods as mentioned above, and with special attention to existing symmetries of the problem and the influence of taking into account the Coulomb potential [18, 19, 75, 116].

A core feature of the intricate structure of ATI PEMDs is the cutoff, i.e. a sharp drop in intensity, for momenta of around $10U_\text{p}$. This outermost cutoff has been identified early on as a caustic, i.e., a “classical rainbow effect”, from interfering backwards rescattering trajectories [100]. The respective methods to accurately calculate the joint contribu-tion and to prevent artificial discontinuities were developed subsequently [117–119]. In particular, a uniform approximation was developed that allows the calculation of the spectral amplitude across the range of energies that includes the cutoff [117, 119]. It was identified that — in the vicinity of the cutoff — the two saddle points can be modelled as a third-order polynomial such that the resulting amplitude is related to its diffraction integral, namely the Airy function. This high-energy cutoff and its treatment is in direct analogy to the high-order harmonic cutoff for the process of HHG.

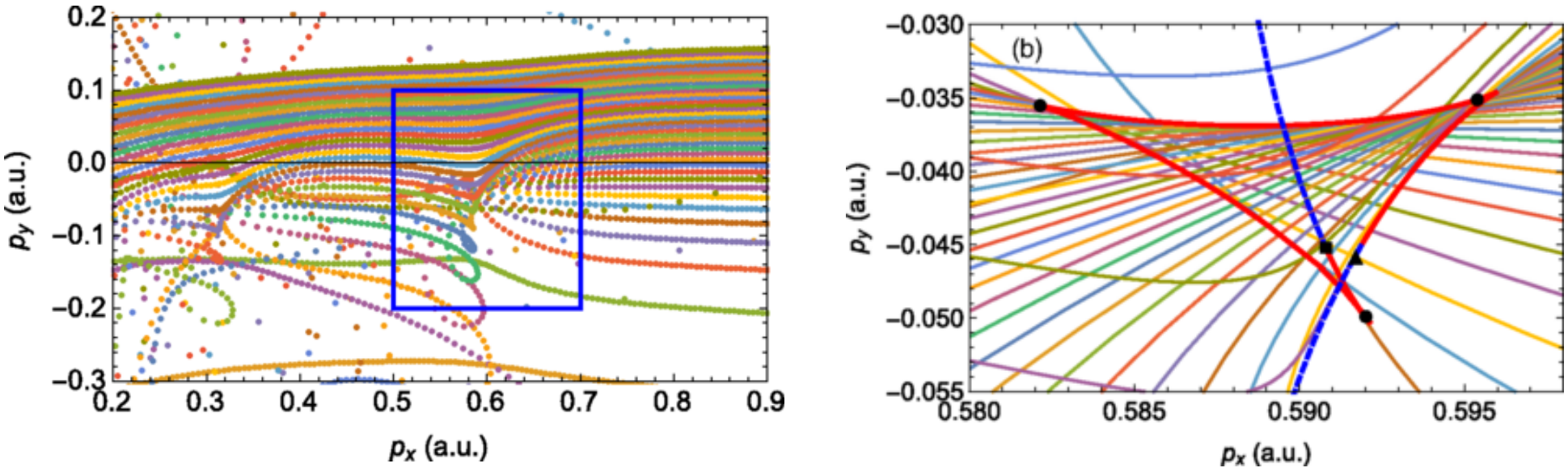

Figure 3.10: Left: Final electron momenta for initial conditions sampled across different initial velocities (colours) and released during a range of ionisation times. The blue window highlights the LES region. Right: Same, but for within the LES. The red lines highlights zeros of the Jacobian, marking the caustics. Black circles denote cusps. Taken from [22].

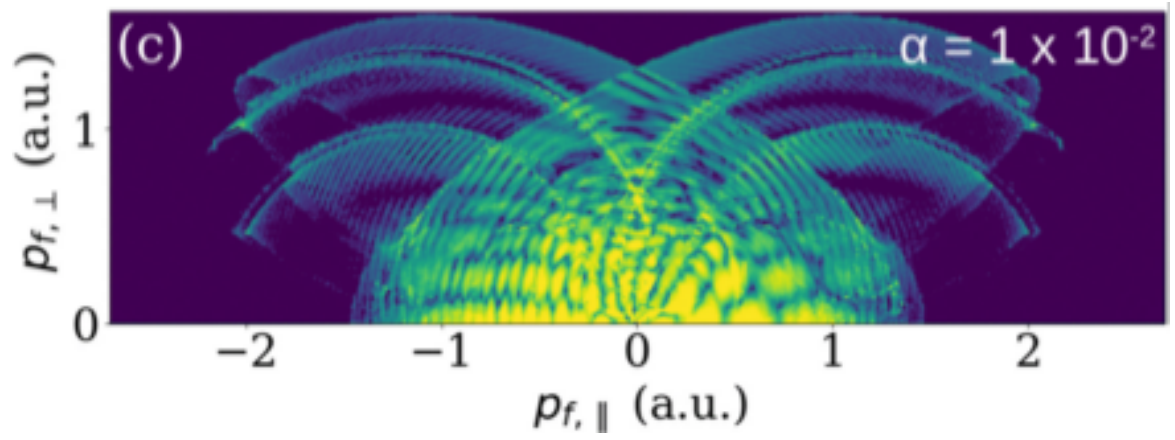

Figure 3.11: Exemplary distributions of photoelectron momenta of ATI driven by a linearly polarised monochromatic field revealing a myriad of interferences. Calculated with a hybrid forward-boundary Coulomb-corrected quantum-orbit SFA approach, taken from [75].

### 3.3.2 The spectral cutoff in HHG: a fold catastrophe

In the very early days of HHG, much research was centred around the maximum kinetic energy the emitted photons could have. Ever since the development of the concept of a trajectory-based model it was recognised that the cutoff marks a caustic feature due to the vanishing Hessian determinant of the prefactor of Eq. (2.50) and the assimilation of the two defining saddle-point solutions for the short and the long trajectories [11]. The artificial discontinuity in the harmonic spectrum can be removed by following a similar approach to the one for the $10U_\text{p}$ cutoff of ATI spectra mentioned above [46]. Interestingly, while many publication identify the HHG cutoff as a caustic stemming from the two interfering saddle points, it was only recently that the underlying fold catastrophe was identified [48, 120]. It is shown that the often-observed missed approach of the two saddle point solutions for a range of harmonic frequencies can be continuously modified into an actual coalescence of the two saddles, by assuming complex-valued harmonic frequencies. This simultaneously provides a unique and rigorous classification scheme for the saddle point solutions around the HHG cutoff(s) in Figs. 3.13(b) and (c), as well as the foundation for the so-called "High-harmonic Cutoff Approximation" — a uniform approximation of the whole spectrum that is generated from just this single coalescence point, see Fig. 3.13(d).

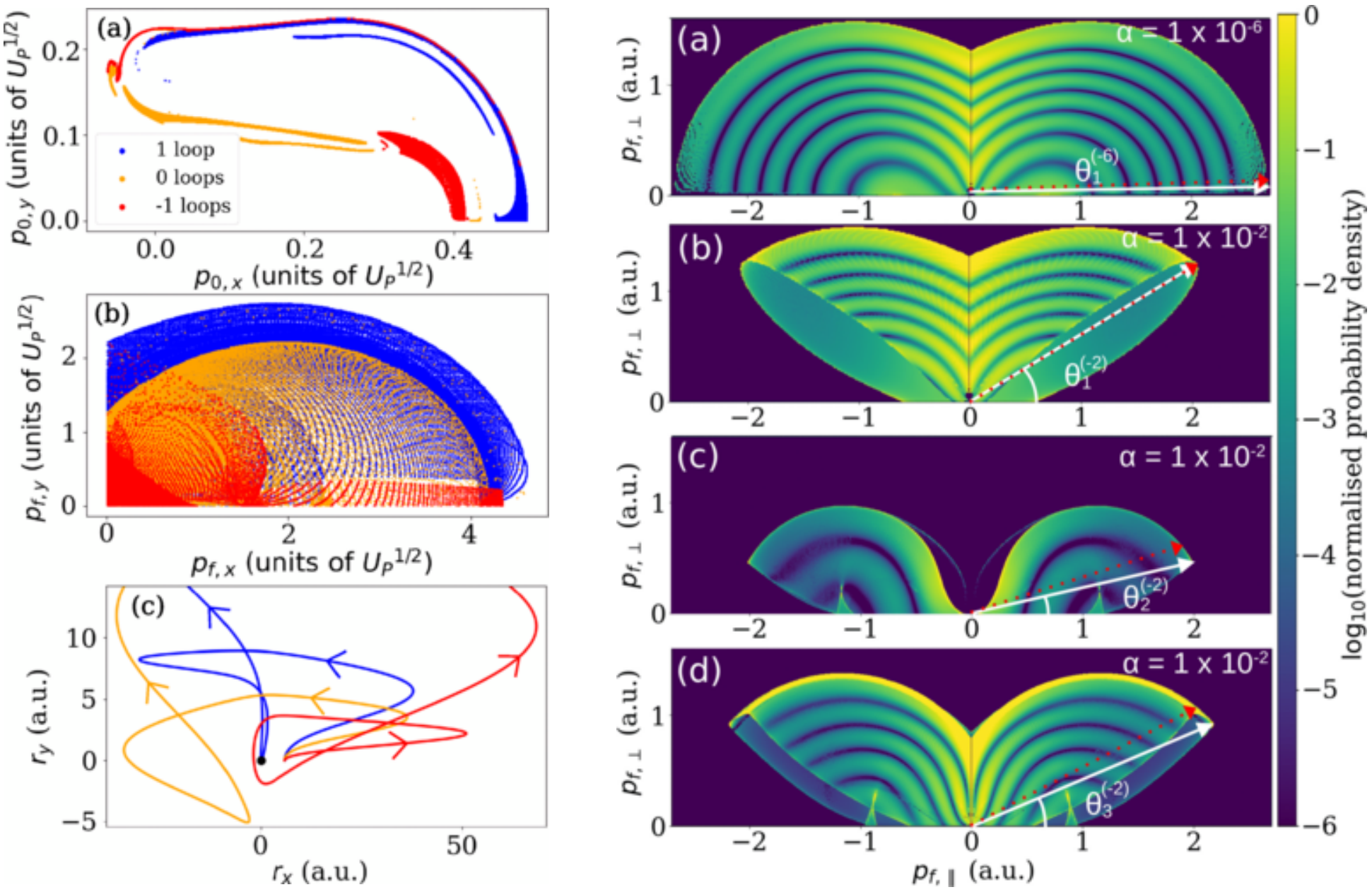

Figure 3.12: Left: Distinct trajectories for rescattering orbits, calculated using the Coulomb-corrected quantum-orbit SFA approach. Shown are their initial (panel (a)) and final (panel (b)) momenta, as well as the orbits in space (panel (c)), taken from [116]. Right: PEMDs with small Coulomb softening parameter $\alpha$ (panel (a)), and individual contributions from the first, second and third pair of backscattered orbits, in panels (b), (c), (d), respectively. Taken from [75].

The details of these findings will be reviewed as part of the results section later on Sec. 12.1, where we will discuss procedures to deal with catastrophe points in HHG in more depth. To put it in context, however, the findings were shown to work for two-colour field in a bicircular configuration as well as for monochromatic fields, but are limited to the case of missed approaches (turned coalescences) of two saddle points strictly.

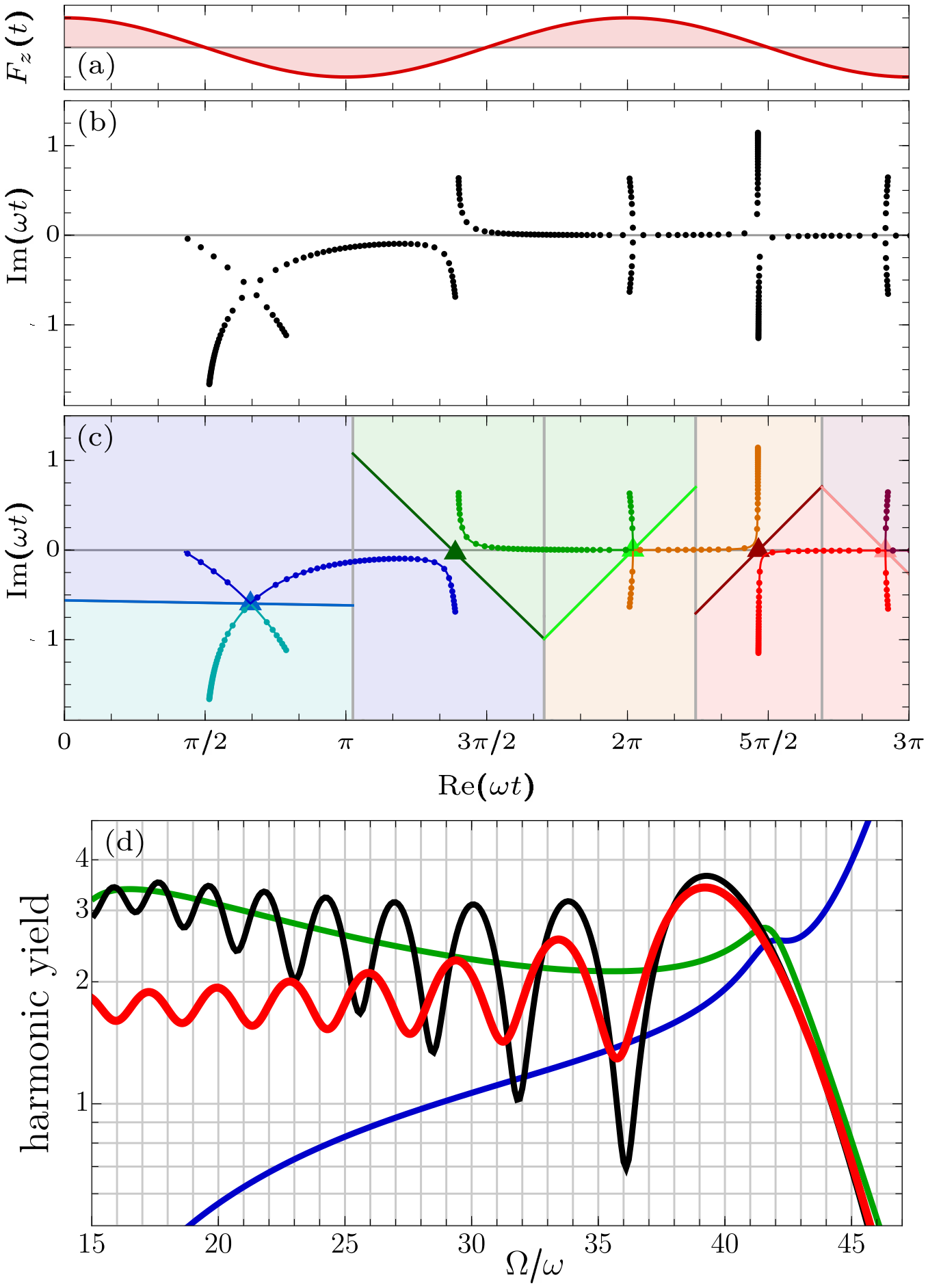

Figure 3.13: Unstructured set of saddle-point solutions for HHG (b) from a monochromatic driver (a). Using the harmonic-cutoff times (triangles) provides a classification scheme to separate solutions into several quantum orbits. (d) The Harmonic-Cutoff Approximation (red line) matches the saddle-point approximation (black line, combining the short (blue) and long (green) trajectories) at the cutoff. Taken from [48].

### 3.3.3 Spectral enhancements in two-colour HHG due to a swallowtail catastrophe

Probably the most prominent and influential work combining catastrophe theory and attosecond sciences stems from HHG experiments driven by a two-colour laser field [72]. By tuning the ratio between the two driving-field components as well as the phase shift between them, Raz et al. identified a spectral enhancement orders of magnitude brighter than the spectral intensity around it, see panel (c) in Fig. 3.14. The publication pioneers the application of the tuning of driving field parameters to study the enhancements in the spectrum. The control parameters are the phase delay $\phi$, amplitude ratio $R$ and the return energy (corresponding to the harmonic order) and therefore allow for a swallowtail catastrophe. The respective bifurcation set is shown in Fig. 3.15.

The experimentally-measured spectral data for scans over the phase delay for a set of different amplitude ratios shows good agreement with (the slices of) the diffraction patterns of a swallowtail catastrophe, see Fig. 3.14. The publication furthermore links the spectral enhancements quantitatively with the twinkling exponents, introduced in Sec. 3.2. A definition of the 'baseline' spectral intensity is however lacking in Ref. [72]. Furthermore, note that the publication works with a simple classical model where the harmonic energy is modelled as a univariate function by using the bijective mapping $t_\mathrm{i} \longmapsto t_\mathrm{i}(t_\mathrm{r})$.

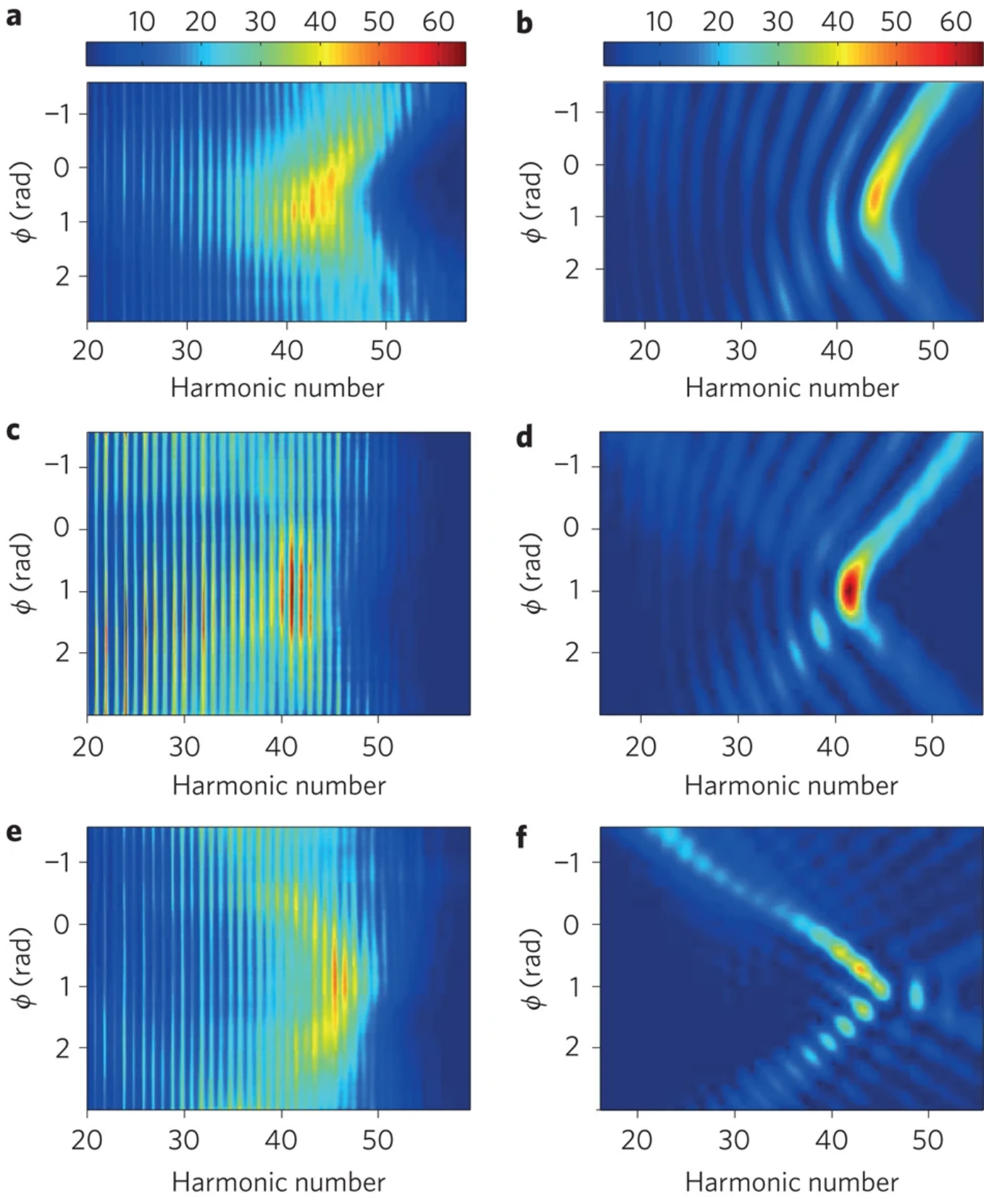

Figure 3.14: Left: Experimentally measured harmonic intensity across a range of phase shifts $\phi$, for the three different values of the amplitude ratio $R$, (a) $R \approx 0.3$, (b) $R \approx 0.45$, (c) $R \approx 0.75$. Right: Calculations of the swallowtail diffraction pattern for the analogous situation, i.e., using a swallowtail unfolding with the parameters $(R, \phi, q)$ as control variables. Taken from [72].

These findings where confirmed by more recent experiments, for a slightly different experimental realisation [37]. There, the coalescence of HHG trajectories was harvested specifically to enhance the signal for each given harmonic order. That is, the experimental study shows how — within a setup of a two-colour driving field and for a fixed harmonic order — the two-colour phase delay together with the amplitude ratio can be varied to achieve the maximum spectral intensity. The classical model of coalescing classical trajectories is used as a theoretical support. Notably, Raab et al. attribute the enhancement to the interaction with the Coulomb potential rather than the kinematical aspects of the HHG process.

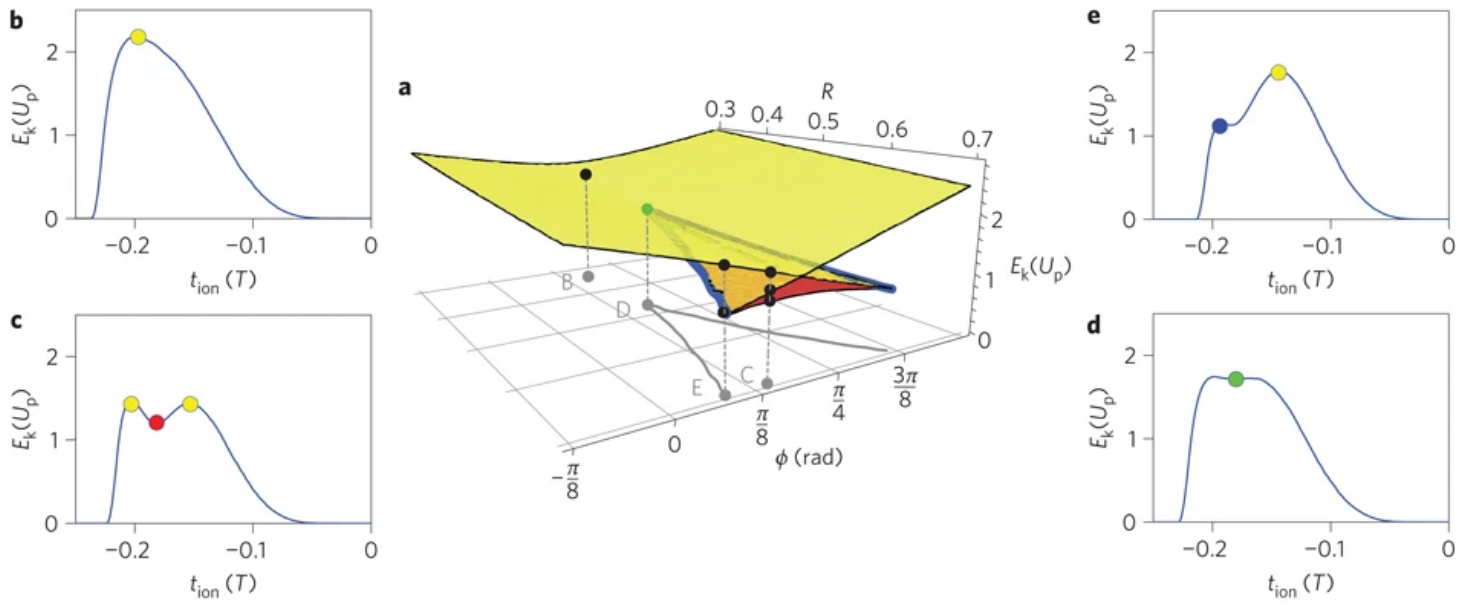

Figure 3.15: The swallowtail bifurcation set as assumed for HHG driven by a two-colour field with amplitude ratio $R$ and phase delay $\phi$, and resulting kinetic photon energy $E_k(U_p)$. Taken from [72].

A purely theoretical study takes a slightly different approach to a spectral enhancement in the HHG spectrum of a two-colour field [73]. By solving a 3D-TDSE it is shown that the harmonic signal is drastically enhanced for specific parameters of the two-colour phase and relative amplitude. The degree of the enhancement is shown to depend on a single parameter: the ratio of the radius of the free electronic oscillation in the laser field to the atomic size. It is then argued that this feature can be used to tailor a quasi-monochromatic XUV source. Interestingly, the enhancement is clearly attributed to a caustic, but there is no mention of a (type of) underlying catastrophe.

To summarise, it has been shown that driving HHG from a two-colour field and using the phase shift and amplitude ratio as ‘free parameters’ allows for the enhancement of specific harmonic orders. Within the three mentioned publications, the definitions of phase shift and intensity ratio differ. For example, the amplitude ratio may be varied by keeping the total intensity constant, or by keeping the intensity of the fundamental field constant. Structurally, however, they define the same catastrophe scenario. The inclusion of the Coulomb potential for the TDSE simulations of Ref. [73] seems certainly intriguing. However, even for that, ultimately a swallowtail catastrophe necessitates the existence of four (eventually coalescing) saddle points. In all the situations above, there are four saddle points in close vicinity such that they can be considered structurally equivalent. We will see our realisation of it, in terms of quantum orbits, in Sec. 12.3.

# 4

# Picard–Lefschetz theory

Integrals of the form $\int e^{i\phi(x)} dx$ arise throughout physics and are notoriously difficult to evaluate due to their highly oscillatory nature and the absence of absolute convergence. A powerful strategy to overcome this challenge is to deform the integration contour into the complex domain, where the oscillations are transformed into exponentially damped contributions and the integral becomes numerically and conceptually tractable. This approach is formalised within the framework of Picard–Lefschetz theory.

In this Chapter, we introduce and derive the central ideas underlying Picard–Lefschetz theory, demonstrating how contour deformation leads to a decomposition of the original integral into a sum of contributions associated with a subset of critical points of the phase function $\phi(x)$. A concise version of this Section is published in

[2] A. Weber, J. Feldbrugge, and E. Pisanty, “Universal approach to saddle-point methods in attosecond science”, *Phys. Rev. A* **113**, 063111 (2026).

In addition, we provide a non-exhaustive overview of applications of Picard–Lefschetz methods across different areas of physics and mathematics, highlighting the broad relevance and versatility of the framework.

## 4.1 Introduction

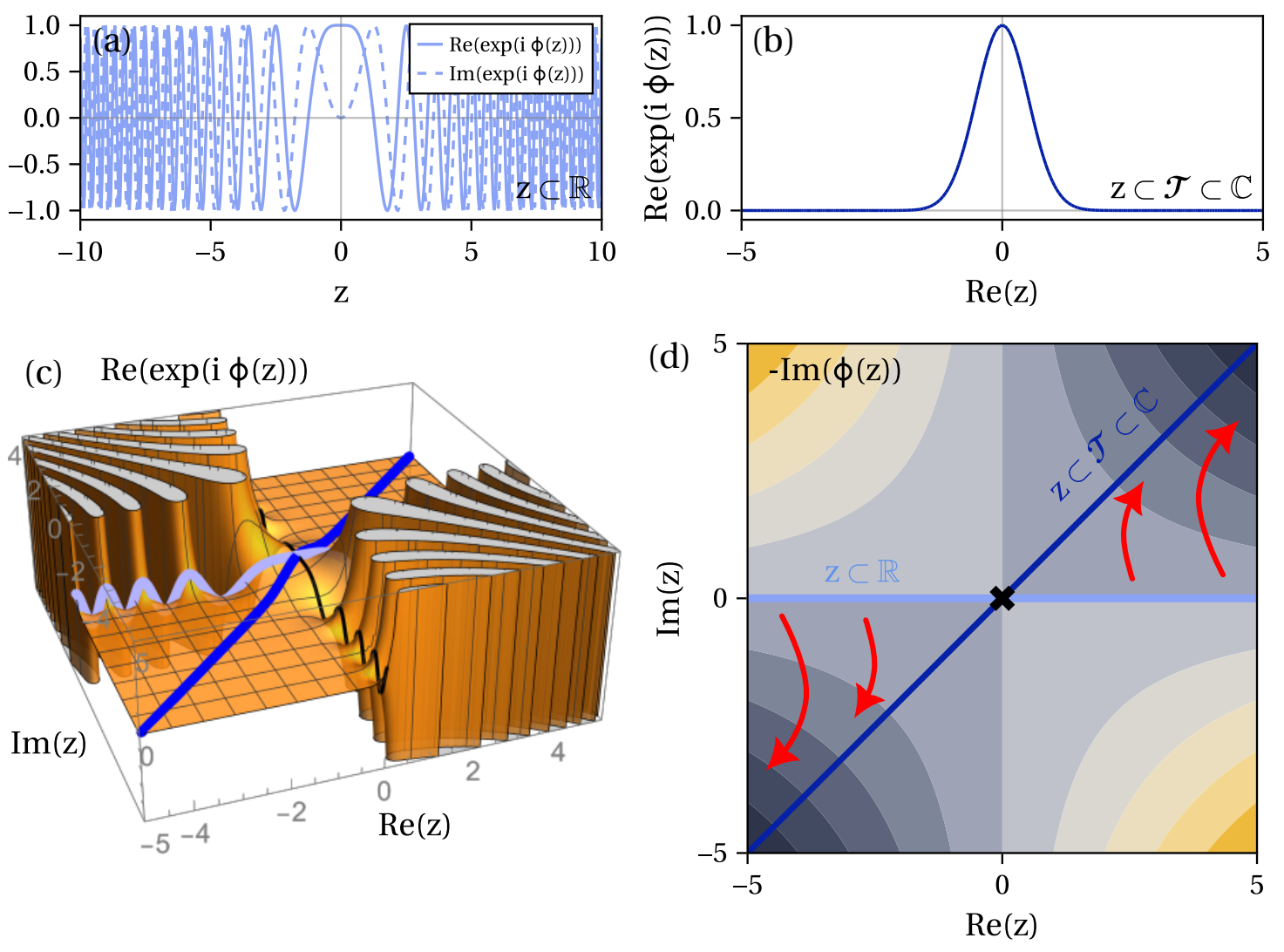

Figure 4.1: Fundamental idea of the integration contour deformation using Picard–Lefschetz theory, shown on the toy model function $\phi(z) = z^2$: The integrand $e^{iz^2}$ is highly oscillatory when evaluated along the real axis (top left panel). The continuation of $z$ into the complex plane (bottom left panel) shows that the oscillations (along the light blue line) vanish if we evaluate the integrand along a different contour (dark blue line). The contour that localises the integrand by minimising the oscillations follows steepest-descent paths of $\mathrm{Im}(z^2)$ (contour plot in the bottom right panel) and is identified by deforming the original integration domain according to the downward flow (red arrows in the bottom right panel) and leads across the saddle point at $z = 0 + 0i$, where $\phi'(z) = 0$. Often, the integrand along the new contour has approximately Gaussian shape (top right panel) and can be evaluated analytically.

Integrals of the form

$$I = \int\limits_{\mathcal{C}_0 \subset \mathbb{R}^N} e^{i\phi(\mathbf{x})/\hbar} \mathbf{dx}, \tag{4.1}$$

evaluated along a path $\mathcal{C}_0$ in $N$-dimensional real space with the real-valued phase function $\phi(\mathbf{x})$ are highly oscillatory and only conditionally convergent. This makes them notoriously difficult to evaluate numerically, especially in the semi-classical limit $\hbar \to 0$ (see Fig. 4.1 for an example). These types of conditionally convergent oscillatory path integrals appear across a vast range of research areas within physics (and beyond!) and each research area has developed different methods to solve them.

The fundamental insight of Picard–Lefschetz theory, shown on a toy model function in Fig. 4.1, is to apply Cauchy's integral theorem, which states that the integral of an analytic function evaluated along a closed loop vanishes. As an immediate consequence follows that the value of a contour integral does not depend on the specific contour.[1] Using this, the integration contour $\mathcal{C}_0 \subset \mathbb{R}^N$ is deformed into the complex space $\mathbb{C}^N$ such that along this new contour the integrand no longer oscillates, the integral converges absolutely, and hence is easier (if not trivial) to evaluate. Picard–Lefschetz theory neatly proves the existence of this optimal contour, proves how to find it, and simultaneously shows that it passes through critical points of the phase function $\phi(x)$. The integral can therefore be expressed as a sum over contributions associated with these critical points. By assuming the critical points to be non-degenerate, and their contribution to be the integral of a Gaussian, this yields an approximation known as the saddle-point method.

Picard–Lefschetz theory originates in the works of Émile Picard and Solomon Lefschetz around 1900, [121] and [122] respectively, whose works were concerned mainly with the idea that one could deform the contour of integration without changing the value of the integral. Picard predominantly explored the algebraic properties of the analytic continuations in the complex plane, while Lefschetz generalised these ideas to higher-dimensional complex manifolds and makes the connection of deformed integration cycles to various homology classes. Thereupon Lefschetz coined the term "thimbles" ("onglets" in his French original) to describe the integration contours that are attached to the critical points of the phase function. Within mathematics, Picard–Lefschetz theory is therefore sometimes known as complexified Morse theory (which is the topological study of a real manifold by inspecting its critical points). Frederic Pham used these insights to study the asymptotic behaviour $\hbar \to 0$ of integrals with polynomial phase functions [123]. He finds that the integration along the thimbles yields Laplace-type integrals on complex-valued integration contours, which are guaranteed to converge — even for asymptotic values of $\hbar$. Furthermore, his work develops the understanding of Lefschetz thimbles as $N$-dimensional manifolds embedded in $\mathbb{C}^N$, as well as the intersection of the dual thimbles with the integration domain. The asymptotic behaviour of the integral is then determined by the vicinity of the critical points that the thimbles are attached to. A possible utilisation of these ideas to applications in physics was foreseen by Pham [123]:

> "I like to dream that the same idea could help understanding the reduction of 'Feynman path integrals' to 'Wiener integrals', a very important challenge put on Mathematics by Theoretical Physics."

This wish came partially true a few decades later when Edward Witten first used the ideas of complex contour deformations in physics, in the context of Chern–Simons quantum field theory [124]. Therein, Picard–Lefschetz theory is employed mainly for the advantage of a proven convergence of the integral, and in some cases for providing solutions to the 'sign problem' of quantum chromodynamics. Unfortunately, however, this did not immediately award Picard–Lefschetz methods the popularity they deserve — given the omnipresence of path integrals in physics.

[1] Intuitively speaking, imagine two different contours $C_1$ and $C_2$ between two points $a$ and $b$, then together these two contours form a loop along which the total integral vanishes, $\int_{C_1+C_2} = 0$. It immediately follows that $\int_{C_1} = -\int_{C_2}$.

## 4.2 Key theoretical concepts

In the following we want to derive the main ideas of Picard–Lefschetz theory. These were briefly sketched in Fig. 4.1 for a toy model function that only has one saddle point. The concepts derived here are particularly relevant for functions with more than one saddle point, as well as for higher dimensional problems. Naturally, we will focus on those aspects that subsequently appear relevant in application to the integrals studied in attosecond science. The realisations of the numerical methods that implement Picard–Lefschetz theory, as well as their specific application to the integrals in attosecond will be detailed in the Methods part of this thesis.

### 4.2.1 Deforming the integration contour into complex space

Assuming that the exponentiated phase function $\phi(\mathbf{z})$ of the integral Eq. (4.1) is a meromorphic function[2] of $\mathbf{z} \in \mathbb{C}^N$, we continuously deform the integration domain towards contours along which the amplitude of the integrand $|\mathrm{e}^{\mathrm{i}\phi}| = \mathrm{e}^{\mathrm{Re}(\mathrm{i}\phi)}$ decreases as rapidly as possible. The direction of the deformation is therefore given by the *downward flow*

$$\frac{\mathrm{d}\mathbf{z}}{\mathrm{d}\lambda} = -\left(\frac{\partial \mathrm{i}\phi}{\partial \mathbf{z}}\right)^* \tag{4.2}$$

which is parametrised by $\lambda$, sometimes referred to as the dimensionless 'flow time'. The flow is initialised along the original integration domain $\mathcal{C}_0 \in \mathbb{R}^N$, such that the resulting integration domain — as a function of the flow parameter $\lambda$ — is the contour

$$\mathcal{T}(\lambda) = \{\mathbf{z}(\lambda, \mathbf{z}_0) \,|\, \mathbf{z}(\lambda = 0, \mathbf{z}_0) = \mathbf{z}_0 \in \mathcal{C}_0\} \,. \tag{4.3}$$

In the limit $\lambda \to \infty$ the deformation of the contour $\mathcal{T}(\lambda)$ converges to yield a set[3] of steepest-descent manifolds $\mathcal{T}_\sigma \subset \mathbb{C}^N$, so-called Lefschetz thimbles:

$$\mathcal{T} = \lim_{\lambda\to\infty} \mathcal{T}(\lambda) = \sum_\sigma n_\sigma \mathcal{T}_\sigma \,. \tag{4.4}$$

Each thimble is attached to a critical point $\mathbf{z}_\sigma$ of the phase function, where $\phi'(\mathbf{z}_\sigma) = 0$ (i.e., saddle points in the complex space) as these are stationary solutions to the downward flow Eq. (4.2) [125].

Let us briefly explain why this works. Given that the exponent of the integrand in Eq. (4.1) is meromorphic, it can be written as

$$\mathrm{i}\phi(\mathbf{z})/\hbar = h(\mathbf{z}) + \mathrm{i}H(\mathbf{z}) \,, \tag{4.5}$$

where $h$ controls the amplitude of the integrand as $|\mathrm{e}^{\mathrm{i}\phi}| = \mathrm{e}^h$, while $H$ controls the oscillations.[4] That is, to localise the integrand we are seeking a contour along which the value

[2] That is, it is locally analytic, i.e., locally complex differentiable, such that it can be approximated by a Taylor series almost everywhere in the complex space. A remark on notation: We are using $\mathbf{z}$ instead of $\mathbf{x}$ to highlight the continuation into the complex space $\mathbb{C}^N$.

[3] We acknowledge that strictly speaking Eq. (4.4) should be a union rather than a sum. However, for consistency with the literature and the *summation* over integral contributions in Eq. (4.9) we use a sum here as well.

[4] We use the notation $h = \mathrm{Re}(\mathrm{i}\phi/\hbar)$ and $H = \mathrm{Im}(\mathrm{i}\phi/\hbar)$ which is standard in the context of Picard–Lefschetz theory.

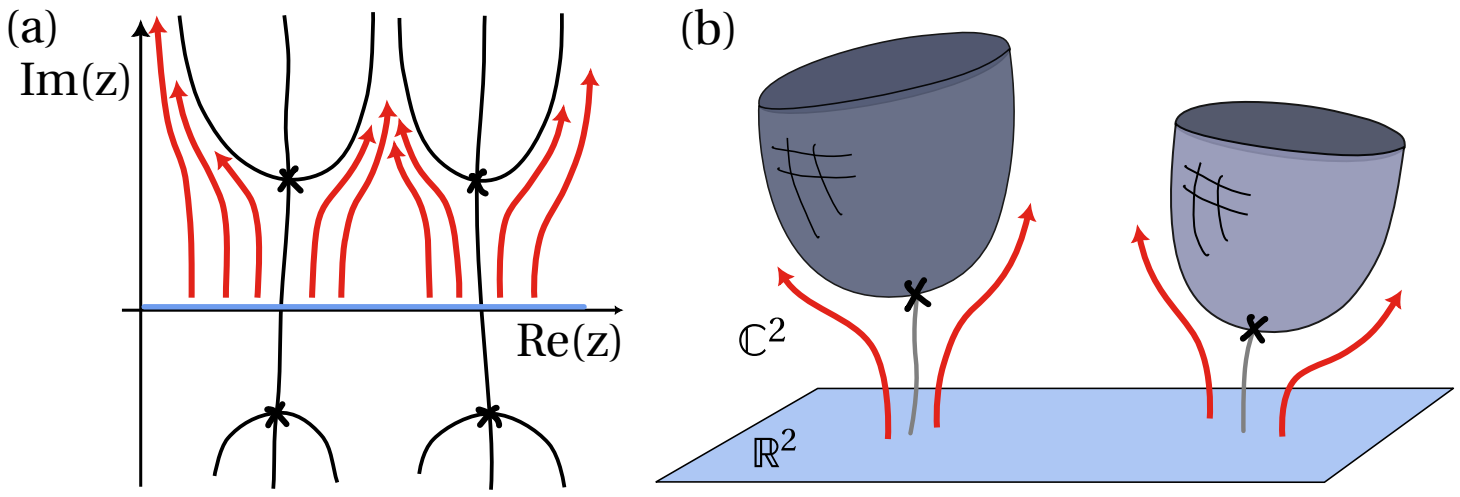

Figure 4.2: For (a) one- and (b) two-dimensional path integrals the downward flow (directions indicated by red arrows) transform the original, real-valued integration domain (light blue) into the complex domain, ultimately towards the steepest descent contours (thimbles, grey) attached to the critical points (cross markers).

of $h$ decreases most rapidly and $H$ is constant. As we analytically continue $\mathrm{i}\phi(\mathbf{z})/\hbar$ into the complex plane, it fulfils the Cauchy–Riemann equations

$$\frac{\partial h}{\partial \mathrm{Re}(\mathbf{z})} = \frac{\partial H}{\partial \mathrm{Im}(\mathbf{z})} \quad \text{and} \quad \frac{\partial h}{\partial \mathrm{Im}(\mathbf{z})} = -\frac{\partial H}{\partial \mathrm{Re}(\mathbf{z})}. \tag{4.6}$$

This is equivalent to writing $\nabla h \cdot \nabla H = 0$, emphasizing that the gradients of $h$ and $H$ are orthogonal. Conversely, contours of constant phase $H$ are those along which $\mathrm{e}^h$ decreases (or increases) most rapidly, i.e., contours of steepest descent (ascent). We can therefore find an optimal integration contour by deforming the integration path into the direction of decreasing $h$, using the downward flow Eq. (4.2) shown above[5] and sketched in Fig. 4.2. Along the flow the value of $H$ remains constant as

$$\frac{\partial \mathrm{i}\phi}{\partial \lambda} = \frac{\partial \mathrm{i}\phi}{\partial \mathbf{z}}\frac{\partial \mathbf{z}}{\partial \lambda} = \frac{\partial \mathrm{i}\phi}{\partial \mathbf{z}}\left(-\frac{\partial \mathrm{i}\phi}{\partial \mathbf{z}}\right)^* = -\left|\frac{\partial \mathrm{i}\phi}{\partial \mathbf{z}}\right|^2 \tag{4.7}$$

which means that indeed $\mathrm{Im}\left(\frac{\partial \mathrm{i}\phi}{\partial \lambda}\right) = \frac{\partial H}{\partial \lambda} = 0$, while $h$ decreases most rapidly: $\mathrm{Re}\left(\frac{\partial \mathrm{i}\phi}{\partial \lambda}\right) = \frac{\partial h}{\partial \lambda} < 0$. Flowing the entire original integration domain $\mathcal{C}_0$ into the complex plane converges the contour to a set of several disconnected[6] Lefschetz thimbles Eq. (4.4). Each thimble is a $N$-dimensional manifold embedded in $\mathbb{C}^N$ (i.e., $2N$ real dimensions) and attached to a critical point $\mathbf{z}_\sigma$, as mentioned above and visualised in Fig. 4.2 for $N = 1$ and $N = 2$ respectively. As $H(\mathbf{z}_\sigma)$ is constant along each of the thimbles $\mathcal{T}_\sigma \subset \mathbb{C}^N$, in the expression for the total integral Eq. (4.9) it acts as a weighting factor for each contribution, while the integration only needs to be carried out across $\mathrm{e}^{h(\mathbf{z})}$:

$$I = \sum_\sigma n_\sigma \, \mathrm{e}^{\mathrm{i}H(\mathbf{z}_\sigma)} \int_{\mathcal{T}_\sigma} \mathrm{e}^{h(\mathbf{z})} \mathrm{d}\mathbf{z}. \tag{4.8}$$

The contribution of each thimble is furthermore governed by the intersection number $n_\sigma \in \mathbb{N}_0$, which will be discussed in more detail below.

[5] Alternative names in other research areas are gradient, Morse or holomorphic flow.

[6] The separation into several thimbles stems from the near-zero integral contribution of regions with low values of $h$. In our numerical implementation this ultimately causes the contour to break up into disconnected thimbles.

Upon the deformation of the contour according to the downward flow, the integral Eq. (4.1) can ultimately be expressed as a sum over contributions from the separate Lefschetz thimbles[7]:

$$I = \int_{\mathcal{C}\subset\mathbb{C}^N} e^{i\phi(\mathbf{z})/\hbar} d\mathbf{z} = \sum_\sigma n_\sigma \int_{\mathcal{T}_\sigma\subset\mathbb{C}^N} e^{i\phi(\mathbf{z})/\hbar}\, d\mathbf{z} \tag{4.9}$$

The integral along each thimble is well-defined, guaranteed to converge, and in some cases almost trivial to evaluate. Importantly, the value of the integral is preserved at any intermediate stage $\lambda$ of the downward flow. The expressions Eq. (4.1) and Eq. (4.9) are thus strict equalities, since all we have done so far is a contour deformation. Furthermore, the deformation of the contour into the contributions of separate thimbles is largely independent of the factor $\hbar$, such that identified Lefschetz thimbles can be 're-used' as integration contours for different values of $\hbar$. Lastly, all the above considerations are independent of the dimension $N$.

### 4.2.2 Dual thimbles and the relevance of critical points

The insight that the integral can be expressed as a sum of separate Lefschetz thimble contributions is incredibly powerful as long as we are able to identify *which* thimbles are relevant contributors — a property which is determined by the intersection number $n_\sigma$. The following section explains how these intersection numbers arise and how they determine the thimbles' contribution.

In analogy to the downward flow, the *upwards flow* is given by

$$\frac{d\mathbf{z}_\lambda}{d\lambda} = +\left(\frac{\partial i\phi}{\partial \mathbf{z}_\lambda}\right)^* \tag{4.10}$$

and shows the direction of steepest *ascent* of $h$, while preserving $H$. The steepest ascent manifold of a saddle point $\mathbf{z}_\sigma$ is known as the *dual* thimble $\mathcal{K}_\sigma \subset \mathbb{C}^N$. Alternative terms are the 'anti' or the 'unstable' thimble [127, 128]. The manifolds $\mathcal{T}_\sigma$ and $\mathcal{K}_\sigma$ intersect (and are locally orthogonal to each other) only in exactly one point: the critical point $\mathbf{z}_\sigma$, as it is a stationary solution to both upwards and downward flow. This is visualised in Fig. 4.3(a) where steepest-descent (blue) and steepest-ascent (green) contours are locally orthogonal lines, intersecting at the saddle point. In the traditional treatment of saddle-point methods, the dual thimbles are often ignored. However, following Picard–Lefschetz theory, the dual thimble $\mathcal{K}_\sigma$ actually governs the relevance (meaning the contribution to the integral Eq. (4.9)) of the thimble $\mathcal{T}_\sigma$ through its intersections with the original integration domain $\mathcal{C}_0 \in \mathbb{R}^N$. Those intersections are counted by the intersection number

$$n_\sigma = \langle \mathcal{K}_\sigma, \mathcal{C}_0 \rangle, \quad n_\sigma \in \mathbb{Z}, \tag{4.11}$$

[7] In this thesis, we focus on the constructive interference of the integrand at the stationary points of exponent $\phi$. However, in general, the sum of thimbles includes both steepest-descent manifolds associated with the stationary points of the exponent and the stationary points of the exponent restricted to the boundary of the original integration domain, i.e. $\phi|_{\mathcal{C}_0}$. These boundary thimbles are always relevant. However, the integral along the first set of thimbles typically dominates over the boundary thimbles. For a systematic investigation of the boundary points in Picard–Lefschetz theory, we refer to [126].

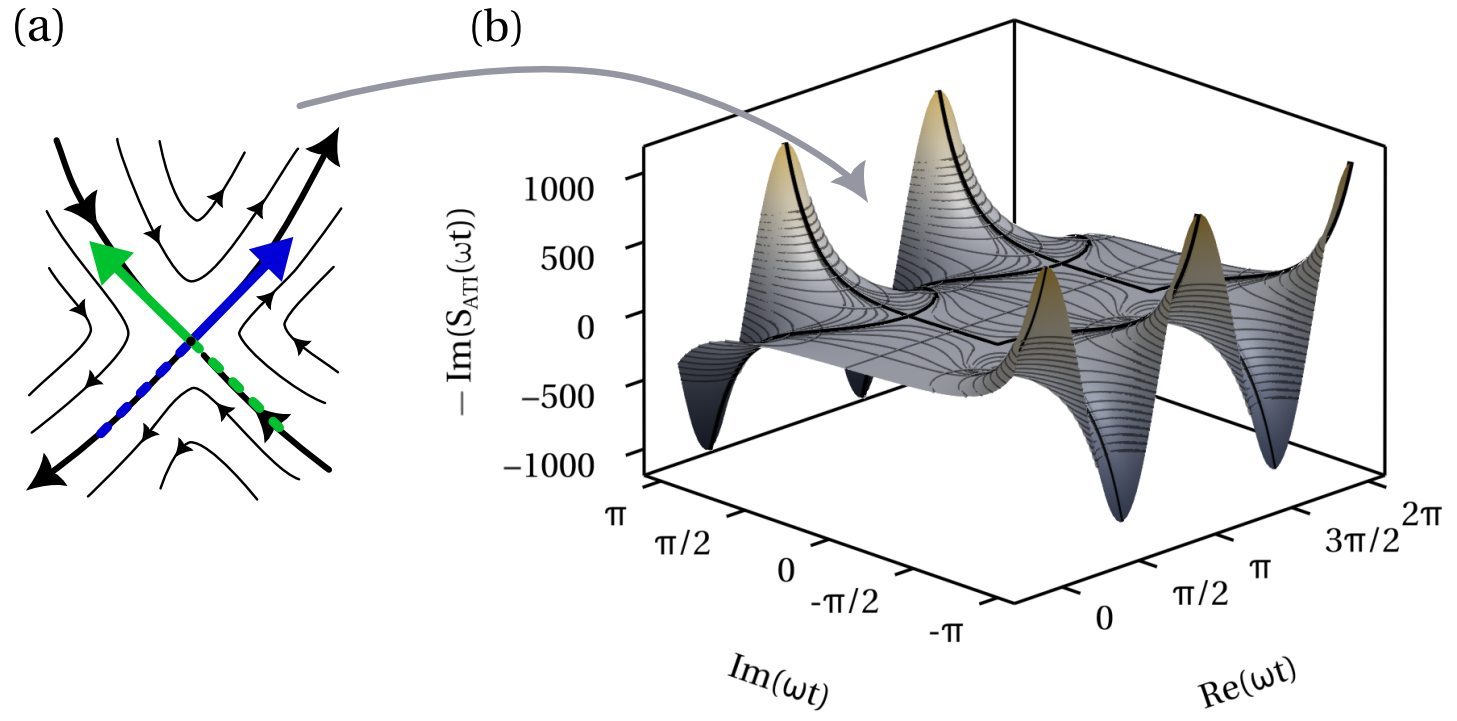

Figure 4.3: For an analytical function zeros of the first derivative constitute saddle points $\omega t_s$ in the complex plane, visible in the contour for $-\mathrm{Im}(S_{\mathrm{ATI}})$ shown in (b). Around the saddle point, level lines of $\mathrm{Im}(S_{\mathrm{ATI}}(\omega t)) = \mathrm{Im}(S_{\mathrm{ATI}}(\omega t_s))$ (black) are locally orthogonal, as shown in (a), and define directions of steepest descent (blue) and steepest ascent (green) of $\mathrm{Re}(S_{\mathrm{ATI}}(\omega t))$.

where the intersection operator $\langle \cdot, \cdot \rangle$ is rigorously defined in relative homology. If the dual thimble $\mathcal{K}_\sigma$ attached to a critical point $\mathbf{z}_\sigma$ intersects the original integration domain ($n_\sigma = \pm 1$), its thimble $\mathcal{T}_\sigma$ is part of the converged integration contour $\mathcal{T}$. If they don't intersect ($n_\sigma = 0$), the respective thimble has to be neglected. That is, the thimble of the critical point $\mathbf{z}_\sigma$ is relevant to the integral iff there exists a point on the original integration domain such that the flow eventually reaches it in the limit $\lambda \to \infty$. Intuitively, because the downward flow defines a continuous, and hence 'unambiguous', contour transformation, we can reverse this procedure: Relevant critical points are therefore those which have the steepest-ascent manifold connecting back to the original integration domain.

### 4.2.3 Saddle-point approximation

If the critical point $\mathbf{z}_\sigma$ is a simple saddle point (i.e., with a non-zero determinant of the matrix of second derivatives Eq. (3.2)), well separated from other critical points, then the integral along its attached thimble can be approximated using the standard saddle-point approximation. For that we terminate the Taylor-expansion of the phase function around $\mathbf{z}_\sigma$ after the second order such that the integrand evaluated along the steepest-descent contour $\mathcal{T}_\sigma \subset \mathbb{C}^N$ resembles a Gaussian, as in Fig. 4.1(c). This Gaussian can then be integrated analytically [43]

$$\int_{\mathcal{T}_\sigma \subset \mathbb{C}^N} \mathrm{e}^{\mathrm{i}\phi(\mathbf{z})/\hbar} \mathrm{d}\mathbf{z} = \int \mathrm{e}^{\mathrm{i}(\phi(\mathbf{z}_\sigma) + \frac{1}{2}\phi''(\mathbf{z}_\sigma)(\mathbf{z}-\mathbf{z}_\sigma)^2 + \ldots)/\hbar} \mathrm{d}\mathbf{z}$$

$$\approx \sqrt{\frac{(\mathrm{i}2\pi\hbar)^N}{\det\left(\phi''(\mathbf{z}_\sigma)\right)}} \mathrm{e}^{\mathrm{i}\phi(\mathbf{z}_\sigma)/\hbar} \tag{4.12}$$

where $\det\left(\phi''(\mathbf{z}_\sigma)\right)$ is the determinant of the Hessian matrix of second derivatives of $\phi(\mathbf{z})$, evaluated at the saddle point $\mathbf{z}_\sigma$.

If the phase function has multiple (well-separated) saddle points, the total integral can then be written as a sum over weighted Gaussian integrals centred around the relevant saddle points of the phase functions:

$$I \approx \sum_\sigma n_\sigma \sqrt{\frac{(\mathrm{i}2\pi\hbar)^N}{\det\big(\phi''(\mathbf{z}_\sigma)\big)}}\,\mathrm{e}^{\mathrm{i}\phi(\mathbf{z}_\sigma)/\hbar}. \tag{4.13}$$

With the prior knowledge of Picard–Lefschetz theory it now is obvious that the intersection number $n_\sigma$ appears in Eq. (4.13) to dictate which of saddle points have to be taken into account and which saddle points are to be neglected. In most textbooks, however, this connection is dismissed and saddle-point approximations are introduced to include "a subset of saddle points", without further specification.

Note that all methods described in the previous sections still constituted an *exact* representation of the integral $I$. The Eq. (4.12) is the first instance where an approximation is made, namely the quadratic expansion around the saddle point. The accuracy of this approximation, and/or alternative ways of estimating the value of the integral along a given contour that crosses a critical point, upon scaling with the 'large parameter' $1/\hbar$, is actively studied within the context of asymptotic analysis. Generally speaking, however, the approximation becomes more accurate for decreasing values of $\hbar$.

Beyond this, the major limitation of the saddle-point approximation is — as the name suggests — that it provides a reliable description only when the critical point is an isolated, non-degenerate saddle point. In situations involving higher-order (degenerate) critical points, or when several saddle points lie in close proximity, the approximation rapidly loses accuracy or breaks down altogether as the determinant of the Hessian in the denominator of Eq. (4.12) vanishes, making the integral diverge.

In such cases, the combined contribution of the nearby or merging critical points must be captured through higher-order or uniform approximation methods, which remain well-behaved in the presence of degeneracies [43, 129]. These scenarios are typically associated with topological changes of the integration contour — such as Stokes transitions — or with the coalescence of multiple saddles. The treatment of these phenomena, and the tools required to address them, will be discussed in the following section.

### 4.2.4 Parametric Picard–Lefschetz theory and topology changes

**Stokes transitions**

Equipped with tools to solve integrals like Eq. (4.1) with arbitrary phase functions $\phi(\mathbf{z})$, we can study how they depend on external parameters. Upon a continuous and smooth scan over such external parameters, the saddle points $\mathbf{z}_\sigma$ of $\phi(\mathbf{z})$ vary smoothly in the complex $\mathbf{z}$ space. Correspondingly, the value of the integral evaluated along associated Lefschetz thimble $\mathcal{T}_\sigma$ varies smoothly. Whether this value contributes towards the total integral, however, is decided by the intersection number $n_\sigma$, which may changes abruptly at so-called *Stokes transitions*. With that, the total number of contributing saddle points (or rather, thimbles) changes, and there must be a topological change in the course of the steepest-descent integration contour. [46, 48, 119, 130, 131]. This is often caused by two (or more) critical points in close proximity, as shown in Fig. 4.4 for an integration contour depending on the external parameter $q$. In the left-hand panel where $q < q_{\mathrm{St}}$ the two critical points are both part of the converged deformed integration contour $\mathcal{T}$ (heavy blue line) and contribute separately to the integral via their thimbles $\mathcal{T}_{\sigma 1}$ and $\mathcal{T}_{\sigma 2}$. In the centre panel, at $q = q_{\mathrm{St}}$, their steepest-descent contours coincide and the deformed contour contains both critical points — this value $q_{\mathrm{St}}$ of the external parameter marks the Stokes transition. For a further increase of the external parameter where $q > q_{\mathrm{St}}$, the steepest-descent contours separate, and $\mathcal{T}$ only contains one of the saddle points.

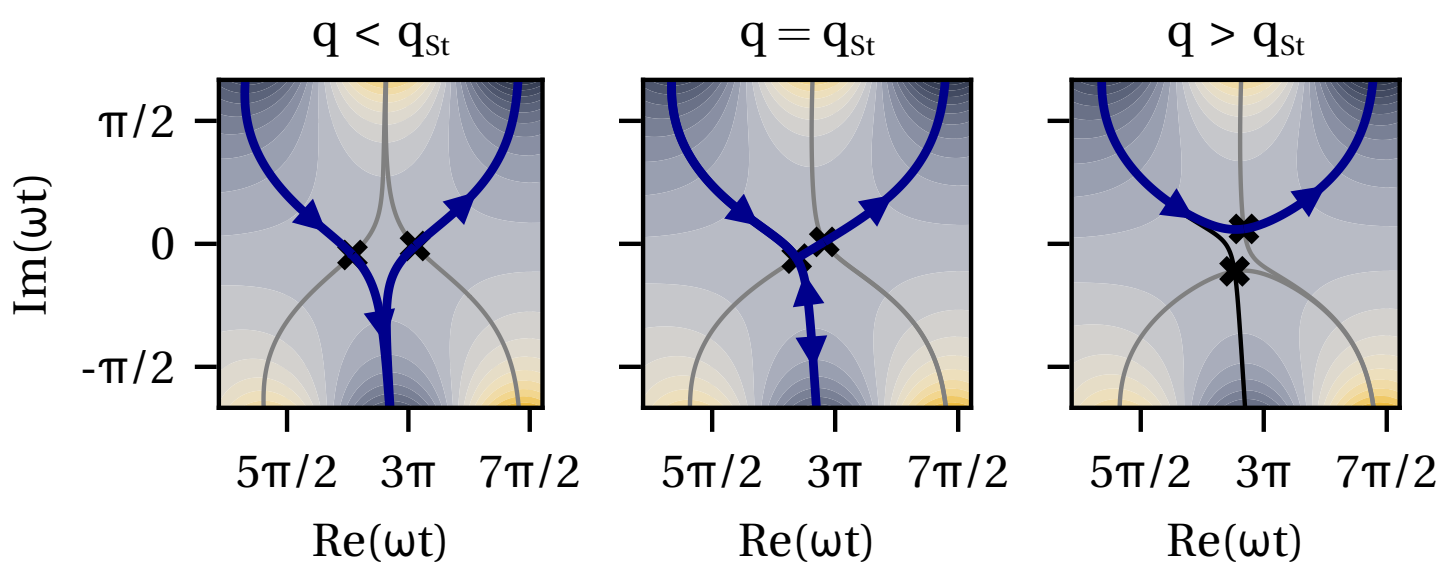

Figure 4.4: Topological change of steepest-descent contours (black) and the resulting integration contour (heavy dark blue line) around a Stokes transition (at $q = q_{\mathrm{St}}$) between two saddle points upon changing an external parameter $q$ (left to right panel).

A necessary condition for the Stokes transition between two critical points $\mathbf{z}_{\sigma 1}$ and $\mathbf{z}_{\sigma 2}$ is that $H(\mathbf{z}_{\sigma 1}) = H(\mathbf{z}_{\sigma 2})$, such that their steepest-descent contours may connect directly to each other. In simple examples, e.g., when $\phi(\mathbf{z})$ is a polynomial with two external parameters, Stokes transitions can be analytically solved for, and yield lines in parameter space [132–134]. For more complicated $\phi(\mathbf{z})$, where there is no closed-form expression for $\mathbf{z}_\sigma$, 'candidate' Stokes transitions can be found numerically by identifying where in parameter space pairs of critical points assume the same value of $H$. If sweeping across this parameter changes the intersection number of one of the saddle points, then we truly found a Stokes transition. Generally, for a phase function with $K$ external control parameters, the Stokes transitions are $(K-1)$-dimensional manifolds in the $K$ dimensional parameter space. They are topological features of this parameter space, as any change of number of relevant saddle points is indubitably linked to a Stokes transition. Vice versa, Stokes transitions define boundaries between regions in parameter space with a certain number of contributors to the integral. Consequently, if they can be calculated a priori, it is unnecessary to calculate each critical point's relevance individually.

While the Stokes transition marks the topological change in the course of the integration contour, the physically more relevant parameter point is the so-called *anti-Stokes transition*. It is defined as the configuration of external parameters for which the value of the real part of the exponent coincides for the two involved saddle points, i.e., $h(\mathbf{z}_{\sigma 1}) = h(\mathbf{z}_{\sigma 2})$. As $h$ governs the amplitude of the saddles' contribution, the anti-Stokes transition marks where the relative dominance of the two saddles changes[8]. That is, in many applications the amplitude of the to-be-discarded thimble increases exponen-tially from the anti-Stokes transition onwards. These two transitions generally happen at distinct points in parameter space.

[8]Note that in different contexts the definition of Stokes and anti-Stokes transition are defined reverse.

**Resolving caustics**
More generally, as we evaluate the total integral across ranges of external parameters, we find not only Stokes transitions but also the related caustics. The latter are the pronounced features in the integral value that arise whenever multiple saddle points are in close proximity and ultimately coalesce. As touched upon above, the analytic description of the integral based on the (order of) critical points and their distance is a delicate subject. The downward flow, in conjunction with a (possibly numerical) evaluation of the integrand along the thimble, elegantly circumvents these problems as it is agnostic to the critical points. With that, it offers the unique capability to evaluate the integral Eq. (4.1) exactly[9] across parameters ranges and to naturally resolve the appearing caustic structures. Moreover, the downward flow allows to identify the direction of the suitable integration contour across a higher-order singularity, which would otherwise require a deliberate, well-informed choice.

## 4.3 Applications in other disciplines

In the following we offer a brief (and certainly incomplete) overview of other research areas that use Picard–Lefschetz theory, with the purpose of demonstrating that the ideas presented in the previous sections provide a versatile and powerful toolbox whose strengths manifest differently across disciplines.

### 4.3.1 (Quantum) Field theory

The first application of Picard–Lefschetz theory was in the context of Chern–Simons gauge theory in order to analytically continue the usual integration contour into the complex space. Chern–Simons theory is a topological 3D quantum field theory. Topological quantum field theories are developed to unite the laws of quantum mechanics with the classical field theories that describe special relativity, without relying on the geometry of the considered space. Within these, the Chern–Simons theory is a pivotal example in three dimensional space [135]. As part of the theory there appears a Feynman path integral (in a form similar to Eq. (4.1)) that provides a topological invariant, but the path integral is taken over all gauge fields on a 3D manifold — which means this is an integral over infinitely many dimensions. Less than two decades ago, Witten suggested to employ the analytic continuation of such path integrals into the complex space and deform the integration contour such that it passes through (complex) critical points of the exponentiated Chern–Simons functional. This reduces the infinite-dimensional integral into a union of tractable contour integrals, and provides some additional topological insights regarding the space of gauge fields and relation to other field theories [124].

Since its introduction to physics, Picard–Lefschetz theory was used to advance other research topics within field theories as well. For example, a so-called 'sign problem' ap-pears in lattice field theories (and the related studies of quantum many-body systems) where an integral over the (gridded) configuration space is taken to define the expectation values of a physical observable. Sampling the configuration space according to its probability measure (the 'weight', the term $e^{i\beta H}$) allows to estimate this integral, which is called the Monte Carlo importance sampling technique. For a high number of fermionic particles this exponent may assume complex values rendering the interpretation (and numerical evaluation) as a probability measure to fail. This is ultimately because $H$ is highly oscillatory such that with 'unfortunate' sampling several contributions may cancel out. Deforming the integration contour was proposed to regularise a lattice field theory on the Lefschetz thimble, and hence to minimise the sign problem on this newly identified (or rather: sampled) integration space for the complex scalar field theory for relativistic Bose gas [136, 137]. This approach has proven to be effective to tackle the sign problem for other field theories [128, 138–141], as well as to address questions beyond that, mostly regarding symmetry breaking in specific bosonic models [142–144].

[9]'Exact' up to uncertainties stemming from the algorithmic realisation and the respective discretisation of the method of course.

### 4.3.2 Real-time path integrals

Furthermore, Picard–Lefschetz theory was recently applied to path integrals, which have been instrumental to quantum theory ever since Feynman's work. For a few, specific problems in quantum mechanics, like the harmonic oscillator, the path integral can be solved in closed form. In scattering theory, the path integral is approximated using Feynman diagram expansions. Lattice quantum field theory typically evaluates the integral using a Wick rotation into the complex plane by replacing the real time $t$ with the complex variable $-i\tau$, and then integrating over $\tau$ using Monte Carlo methods. However, the fundamental real-time path integrals as originally proposed by Feynman have remained inaccessible due to their highly-oscillatory nature, and, moreover, have for a long time lacked a mathematically rigorous definition.

Picard–Lefschetz theory solves these problems. The idea to complexify the configuration space of paths rather than the time parameter makes Feynman's path integrals tractable. Firstly, these ideas were applied to the real-time dynamics of a particle in a double well [125]. In analogy to the methods of lattice quantum field theory the corresponding path integrals were solved using Monte Carlo approaches. More recently, this has led to the further development of a rigorous definition of the real-time path integral [145], and efficient techniques for its evaluation in quantum mechanics [57, 146–148]. A new method explicit for the calculation of the intersection number was proposed very recently as well, and exemplified on the example of a double-well potential [149].

### 4.3.3 Cosmology

On the very other end of the time and length scales of quantum tunnelling problems, Picard–Lefschetz theory also finds application in cosmology. On the quest of solving questions about the origin and composition of the universe as well as the evolution of space and time, the typically-employed mathematical methods are solving differential equations and the quantisation of quantum fields on a curved space time [150]. The usage of Picard–Lefschetz theory extends this toolset and has recently been applied to solve a series of questions. In particular, Picard–Lefschetz theory has invigorated the field of quantum cosmology, a quantum-mechanical approach to questions about the beginning of the universe. Therein, the "no-boundary" proposal assumes an unbounded initial quantum state and is then formulated as a path integral over metrics, analogously to a (Wick-rotated) Euclidean quantum field theory, to define "the wave function of the universe" [151]. However, the rotation of the respective path integral into the complex plane leads to an unbounded, non-positive definite action and hence a non-convergent integral — a problem known as the "conformal factor problem". The deformation of the integration contour into the complex plane solves this problem and makes the integral absolutely convergent [68, 152]. This approach simultaneously removes the necessity of imposing boundary conditions on the superspace (i.e., the configuration space of all possible spatial metrics on a three-dimensional Riemannian manifold), and the arising semi-classical factor provides a meaningful positive cosmological constant, subscribing to an accelerated expansion of the universe [68, 152].

Within experimental and observational cosmology, Picard–Lefschetz theory finds application in the analysis of wave-optical lensing phenomena. Astronomy has so far almost exclusively relied on geometrical optics. This is generally a valid approximation as the wavelength of visible light is short compared to the other length scales, and the studied radiation is incoherent. However, with the advent of gravitational waves and the fast radio bursts in radio astronomy the situation is now drastically different. Their signals have been difficult to distinguish from other frequency content, as astrophysical plasmas act as a dispersive lens between the source of radiation and observers on earth, which needs to be accounted for in a fully wave-optical fashion (rather than just geometric rays). In the resulting Fresnel–Kirchoff lensing integrals, the wavelength typically enters as a scaling parameter in the exponentiated phase function. However, deforming the integration contour according to the downward flow of Picard–Lefschetz theory happens independent of it such that the optimal integration contour can be 're-used' for a range of different wavelengths. These insights makes the evaluation of those integrals and the observable caustic structures computationally feasible, and allows to identify the interference patterns emerging from the lensing of radio waves and gravitational waves [67, 69, 134].

### 4.3.4 Asymptotic and numerical analysis

Apart from the applications in physics, the evaluation of highly oscillatory integrals like Eq. (4.1) is of interest within mathematics as well. For example, they appear in the context of asymptotic analysis, where the value of the integral is studied w.r.t. its asymptotic behaviour upon $\hbar \to 0$, as well as in the adjacent numerical analysis (with ties to computer science) to study the accuracy of discretised evaluation schemes. A comprehensive and nicely written overview is given in [153]. A central motivation is to exploit the peculiarities of highly oscillatory integrals in order to develop both accurate and efficient computation schemes. For analytic phase functions a popular (and most related) method is the method of steepest descents, where the integral is evaluated along the respective contour in the complex plane. In that case, the problem is reduced to the numerical evaluation of an exponentially fast decaying Laplace type integral (i.e., $\int_0^\infty f(t) \mathrm{e}^{-\omega t} \mathrm{d}t$ ) that can be well approximated with a Gauss–Laguerre quadrature. In practical applications however, the identification of the relevant steepest-descent contours, especially in scenarios of multiple saddle points in close vicinity, presents another hurdle. Moreover, for small values of $\omega$, the method becomes less precise.

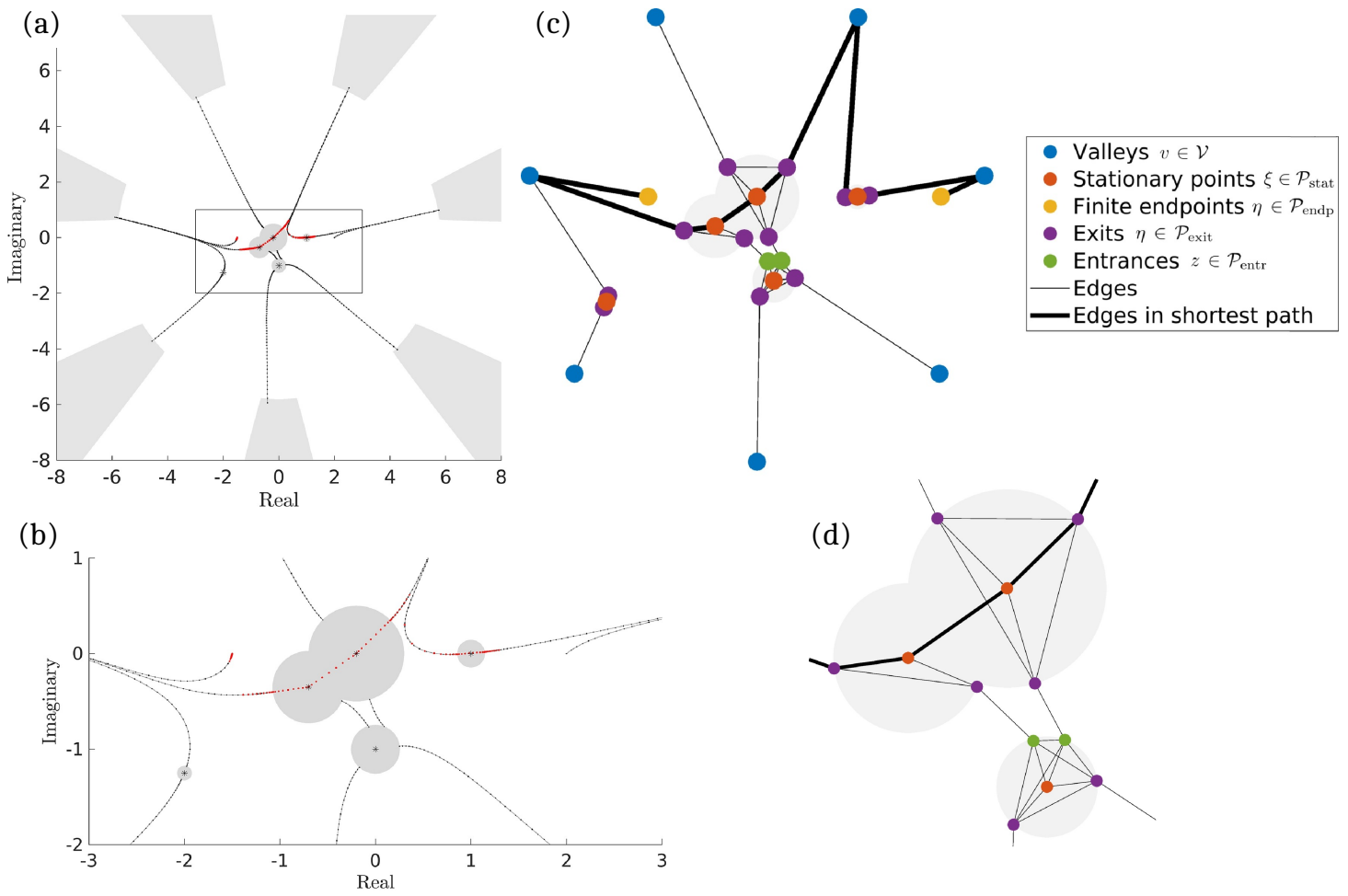

Figure 4.5: Output and methods of the PathFinder algorithm [154]. (a) sampled steepest-descent contours (black spotty lines) passing through a region with stationary points, defined by the balls (grey shaded). (b) A zoom into that area shows the required sampling points for a desired accuracy. (c) Graph and (d) zoom into the graph for the construction of the steepest-descent integration path. The thick line shows the shortest path between the considered endpoints.

To solve these problems, as well as to circumvent the need for fine-tuning of discretisation parameters, a graph-based method has been proposed recently [154]. The scheme, shown in Fig. 4.5, addresses the identification of a 'quasi steepest-descent' integration path for integrals like Eq. (4.1) in one dimension, with polynomial phase functions. The algorithm relies on the identification of 'non-oscillatory regions' around (clusters of) stationary points within which the number of oscillations is limited and hence, integration can be performed to high accuracy with a low number of quadrature points. The connection between several of such regions, valleys at infinity, and ultimately the two endpoints of the original integration contour is traced by means of a shortest-path graph. The scheme proves to be exceptionally powerful in studying scenarios with (nearly) coalescing stationary points. In most methods, those scenarios have to be treated with special attention and e.g. those degenerate critical points need to be identified. Here, thanks to the definition of non-oscillatory regions, the number and nature of saddle points, whether they perform a Stokes transition or coalesce, is irrelevant. While this approach is certainly intriguing, further research is needed to extend it to higher-dimensional integrals. Moreover, whether the idea generalises to non-polynomial phase functions remains an open question.

### 4.3.5 Applied mathematics

Yet another application of contour deformation techniques appears in the context of diffraction of waves from (or in) generic geometries. One example is the study of transient phenomena in generic waveguides [155]. For the propagation of waves in non-trivial layered material only specific modes are allowed. To model the field at a given observation point, the total wave is modelled as a sum of contributing (complex) modes, each of them fulfilling the dispersion relation of the structure. Each contribution has a shape similar to Eq. (4.1) and involves the Fourier transform of the profile of the wave source as a prefactor in the integrand. Oftentimes, there is a large number of propagating modes, such that the direct evaluation of all the integrals cannot be performed directly. An approach informed by the works of Ref. [123] is to analytically continue the frequency into complex space and identify the full Riemann surface of propagating modes. Furthermore, the insights regarding the relevancy of saddle points are applied to transient processes in waveguides, and specifically in application to acoustic waves in car tires [156, 157]. Therein the identification of the 'carcass of the dispersion diagram', essentially the map of relevant saddle points in parameter space, is used to identify meaningful contributions to the summation over contour integrals. By then applying the saddle-point approximation to the relevant saddle point solutions the contour integrals become tractable over the large parameter space and allows for an asymptotic expansion of the pulse response for, e.g., a car tire.

Another example is the diffraction of waves on surfaces of simple geometries like a quarter-plane or a wedge [44, 158–161]. Therein, the difficulty is the analytical continuation of the integration domain into the complex space itself. The configuration space gives rise to several branches in the complex space and special attention is required to treat potential branch cuts and jumps accordingly. These works are embedded into the more broader research on complex contour deformation techniques, tackling challenges like the possible contributions stemming from the boundary of the integration domain or singularities of the integrand [44, 162, 163].

# Part II

## Methods

# 5

# Numerical realisation of Picard–Lefschetz methods

As we have seen in Chapter 4, Picard–Lefschetz theory provides a beautiful framework for evaluating highly-oscillatory integrals by deforming the original integration contour into the complex domain, where the integral becomes convergent and numerically tractable. Upon this contour deformation, the integral can be expressed as a sum of contributions stemming from Lefschetz thimbles attached to critical points of the phase function. Conversely, when the critical points are known, their relevance can be determined by constructing the associated dual thimbles and evaluating their intersection with the original integration domain.

This Chapter focuses on the numerical implementation of these central concepts for one- and two-dimensional integrals of the generic form $\int_{\mathcal{C}_0} e^{i\phi(\mathbf{z})}\,d\mathbf{z}$. We detail the discretised implementation of the downward flow of the original integration domain $\mathcal{C}_0$, in both one and two dimensions. In addition, we introduce the necklace algorithm, a robust numerical procedure that guarantees the identification of the intersection numbers associated with individual critical points.

The methods presented here were developed in close collaboration with Job Feldbrugge. A concise description of the implementation is published in Ref. [2], and an open-source version is available as a julia package [164].

## 5.1 The downward flow

The goal of the downward flow method is to modify the integration contour according to the downward flow Eq. (4.2), such that it can subsequently be integrated numerically. The fundamental algorithm (and its description) are based on Job Feldbrugge's open-source implementation in C++, available at [165]. The implementations are made for generic, meromorphic phase functions $\phi(\mathbf{z})$ of one- or two-dimensional arguments $\mathbf{z}$.

In the following, we outline the general procedure of the algorithm alongside its implementation for the one-dimensional case (where the integration domain is a line). The same general procedure is used for the two-dimensional case, in which the domain is a surface; only the implementation details differ. These specific aspects will be discussed in the subsequent section.

### 5.1.1 General procedure, and its implementation for one-dimensional integrals

The generic algorithm for the downward flow procedure follows a routine as simple as shown in Code 5.1.

Code 5.1: DOWNWARD FLOW PROCEDURE

```
points, simplices = initialise_domain()
while termination_condition
	flow!(points)
	subdivide!(points, simplices)
	clean!(points)
end
```

The initial step is the discretisation of the (original) integration domain, which is a finite interval of the real axis. We discretise this real axis as a list of points, which are connected to line segments. Then, we iteratively apply the downward flow to each of the points, moving it into the complex plane using a first-order Euler method

$$z \mapsto z - \delta_{\text{flow}} \left( \frac{\partial h}{\partial z} \right)^* \tag{5.1}$$

with the small parameter $\delta_{\text{flow}}$. Note that it is sufficient to consider the gradient of $h$ (rather than $\mathrm{i}\phi$) as $H$ remains constant along the flow anyway (see Eq. (4.7)).

We use an adaptive grid in the sense that as soon as two neighbouring points are further than a threshold distance $l_{\text{thresh}}$ apart, we insert a new point in the middle (see on the right-hand side in Fig. 5.1). Furthermore, points are turned 'inactive' (i.e., they are not moved any more, `clean!()`) as soon as their $h$ value drops below a certain threshold, say $h_{\text{thresh}}$, indicated as grey regions and empty points in Fig. 5.1. This will eventually break up the integration contour into disconnected parts, as e.g., in Fig. 4.2(a). The deformation of the integration contour converges to the Lefschetz thimble as $\frac{\partial h}{\partial z}$ vanishes on the thimble. To avoid 'overshooting' this zero-gradient contour of steepest descent, we normalise the gradient as soon as its magnitude drops below a certain threshold.

The `termination condition` in Code 5.1 depends on the purpose of applying the contour deformation. As mentioned in Chapter 4, the value of the integral remains exact upon the deformation. If the aim was simply to obtain a means to calculate the integral with fewer evaluation points, then the algorithm could be terminated after a fixed number of flow steps $i = i_{\max}$. Further, one could evaluate the integral at every iteration step and terminate the procedure as soon as it converges and the relative change upon an iteration step does not exceed a certain accuracy limit. In our application to the one-dimensional integral of the strong-field ionisation amplitude this would typically happen after a few flow steps, as we will see below in Sec. 6.1.1.

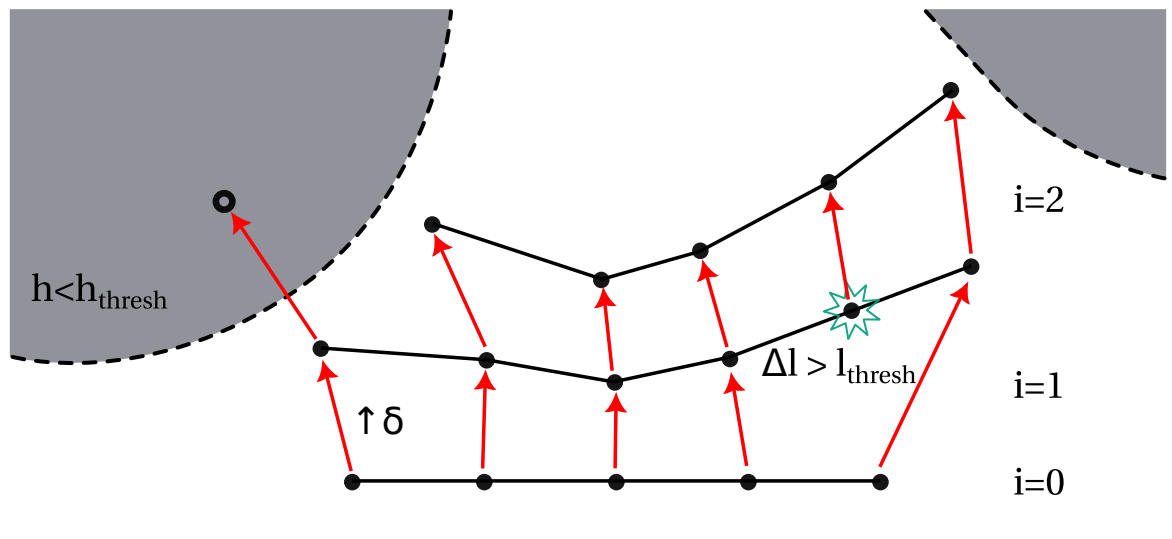

Figure 5.1: Scheme of the numerical downward flow procedure in 1D. The contour is discretised as a set of connected points that are iteratively (iteration step $i$) flowed, possibly subdivided (newly inserted point on the right-hand side) and discarded (empty circle on the left-hand side).

If, however, one is interested in the actual shape of the Lefschetz thimble, then the termination condition needs to be considered more carefully. For example, upon convergence to the steepest-descent contour, the number of (active) points remains roughly constant, and the shape of the contour does not change for subsequent iteration steps. In practice, these criteria seem less convenient because the discretised nature of the flow causes points to be turned inactive and 'drop into the valley' while new points are included via the subdivision routine. That is, the points continue to move downwards in every iteration step. How the several discretisation parameters determine this behaviour is a topic of ongoing research. For our applications an arbitrarily high set number of flow steps is assumed to guarantee the convergence of the downward flow to the Lefschetz thimble.

When the downward flow procedure is terminated, the contour is given as a list of line segments. Along this new contour, the integrand should be less oscillatory and can hence be evaluated following a suitable numerical integration scheme.

### 5.1.2 Implementation for two-dimensional integrals

In the case of a two-dimensional integral the downward flow algorithm technically follows the same procedure. However, each point now has two coordinates (each of them being a complex number) and the integration contour is a surface, embedded in four real dimensions. That is, rather than using line segments we have now discretised our integration domain into quadrilaterals. For the subdivision of a quadrilateral we use the routine sketched in Fig. 5.2: we insert midpoints on the two longest edges (points 5 and 6 in the sketch) and compare the length of the two possible 'dividing cuts'. We pick the shorter length (here between points 2 and 5) and divide the hexagon respectively into two quadrilaterals (here (1,6,2,5) and (2,3,4,5)). We perform this subdivision iteratively until no quadrilateral's edge is longer than the threshold $l_{\text{thresh}}$.

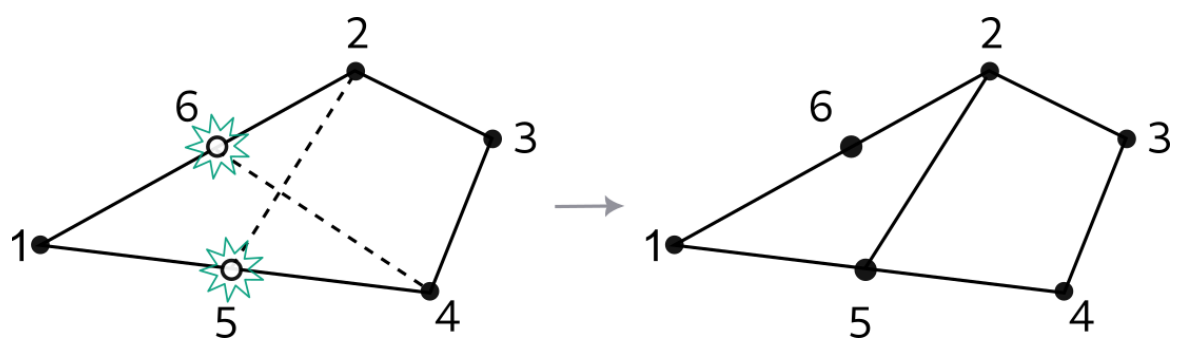

Figure 5.2: Sketch of the subdivision routine for quadrilaterals. We insert new midpoints (here 5 and 6) on the two longest edges, and then subdivide the polygon along the shorter of the new diagonals into two quadrilaterals.

For the integration of the surface discretised by the quadrilaterals we map a Gauss–Legendre quadrature on a reference square to each quadrilateral. Adjusting the number of nodes for the quadrature provides additional control over the speed and accuracy of the integration procedure.

In practice, while this subdivision routine has proven to work well for the integration of the total integral, a conceptual flaw is that the subdivision routine always considers midpoints on the two longest edges of the quadrilateral. This is problematic in cases where only one side actually exceeds the length threshold. In these cases the new point is still also inserted on the second longest corner, such that — over the course of multiple iterations of the subdivision routine — the quadrilateral ends up as a long narrow strip.

An alternative implementation of the downward flow procedure in the two dimensional space uses triangles to discretise the (deformed) integration surface. For the subdivision we use the routine sketched in Fig. 5.3. Each triangle is considered in its plane. For edges exceeding $l_{\text{thresh}}$, we insert $\lfloor \Delta l / l_{\text{thresh}} \rfloor$ new points and then mesh the original triangle using a Delaunay triangulation[1] of all points. Ultimately, for the evaluation of the integral we use a numerical quadrature of the obtained meshed surface [166].

The discretisation of the surface in terms of triangles rather than quadrilaterals is in so far preferred as that they are the universal geometric building blocks of any polygonal surface. This becomes particularly important in the present work, where the surface is embedded in the real four-dimensional space. This makes the development of suitable numerical algorithms and the analysis of the underlying topological structure, for example to determine whether the surface consists of several disconnected components, intrinsically challenging. Using triangles simplifies the problem in so far as that each triangle can be fully viewed in its planar projection.

[1] Coincidentally, the eponymous Boris Delaunay is the father of Nikolai Delone that gave the "D" in the Ammosov–Delone–Krainov (ADK) ionization rates in strong-field physics.

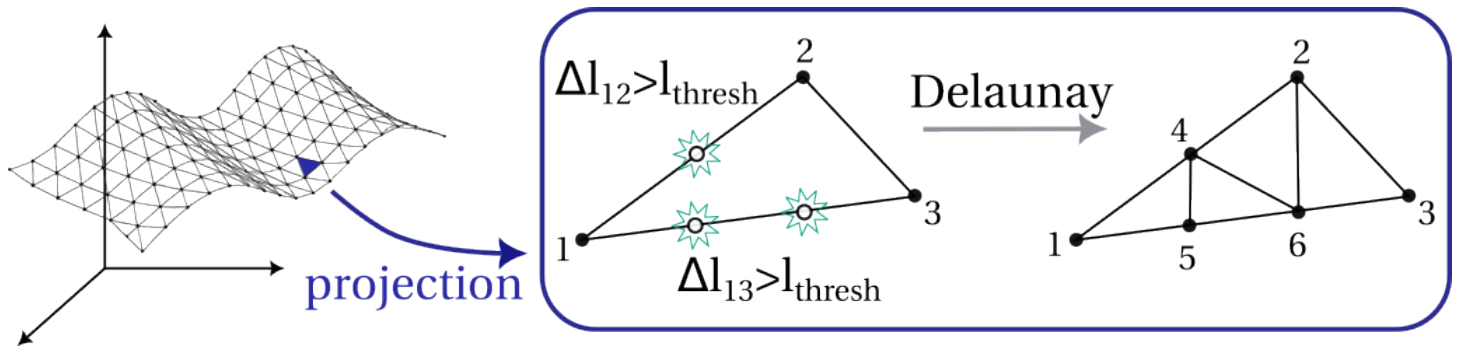

Figure 5.3: Sketch of the discretisation of the surface using triangles, and their respective subdivision routine. For the subdivision each triangle is considered in its projected plane, vertices are subdivided and the set of old and newly inserted points are connected to triangles by Delaunay triangulation.

## 5.2 Saddle-point based approaches

In many applications of Picard–Lefschetz theory, the saddle points are not necessarily of primary interest, but more a 'nice feature' that can possibly simplify calculations. In our applications in attosecond science, however, the saddle points correspond to specific electron pathways as introduced in Chapter 2. We are therefore interested in the behaviour of the individual saddle points, not only how much or whether they contribute to the total integral, but also how they depend on external parameters. As a result, we are particularly interested in the Picard–Lefschetz methods that allow us to answer those questions. In the following we detail our numerical implementations of the methods that assume given saddle points and are aimed at identifying their contribution to the total integral. If not otherwise specified, by 'saddle point' we will here refer to critical points in general. That is, including folds and higher order critical points.

### 5.2.1 Finding saddle points

For the integrals appearing in attosecond science the phase function is given by the semi-classical action $S_{\mathrm{ATI}}$ (for ATI, the 1D case) and $S_{\mathrm{HHG}}$ (for HHG, the 2D case). For both of these, finding saddle points by solving their first derivatives (Eqs. (2.32) and (2.47), respectively) can easily be done using any available numerical root solver, e.g., employing Newton's method. For root solvers that take initial guesses, we need to ensure that all saddle points are found even if they are in close proximity. For that, we typically provide in the order of hundreds of initial guesses, randomly distributed over the complex plane(s), and then filter out duplicate solutions. A more thrifty approach would be to actually identify sophisticated guesses for where saddle point solutions might be located and provide these as initial guesses [167]. However, in our application, *finding* saddle points has never been the bottleneck of the calculations.

### 5.2.2 Identifying and integrating the contribution of relevant saddle points in one-dimensional integrals

Whether the thimble $\mathcal{T}_\sigma$ of a given saddle point $z_\sigma$ contributes to Eq. (4.9) or not is dictated by the intersection number $n_\sigma$ which counts the intersections between the dual thimble $\mathcal{K}_\sigma$ (i.e., the steepest-ascent manifold attached to the critical point) and the original integration domain $\mathcal{C}_0$.

As the value of $H$ is constant along the steepest-ascent manifold, for a one-dimensional integral ($N = 1$) finding the thimble and dual thimble attached to the critical point $z_\sigma \in \mathbb{C}$ corresponds to finding the respective contour level lines where $H(z) = H(z_\sigma)$, drawn as heavy black lines in Fig. 4.3. This can easily be done numerically with, e.g., a marching squares algorithm. The arrows on the level lines in the inset panel in Fig. 4.3 indicate the direction of descending $h$. At the saddle point, these directions of maximised gradient are given by the (orthogonal) eigenvectors of the Hessian, one pointing in direction of steepest ascent (green), and one points in direction of steepest descent (blue). To find the dual thimble we therefore simply pick the contour level lines along which $h$ increases away from the saddle point in the steepest-ascent direction. Locally, these lines coincide with the one of the eigenvectors and its inverse (dotted green vectors) of the Hessian. That is, if $h$ is ascending along a level line away from $z_\sigma$ and eventually connects to the real axis (the original integral domain), then $n_\sigma = 1$ and the critical point contributes to the integral. Note that in order to identify the possible intersection we need to follow the steepest-ascent lines in both directions. That is, in Fig. 4.3 this refers to the solid green line and its counterpart, the dashed green line. We will see how this choice is more complicated for the case of 2D integrals, and hence motivates the development of a dedicated algorithm below.

In most of our applications, the level lines have structures similar to those in Fig. 4.2. There, for the saddle points with a positive imaginary part the steepest-ascent lines intersect the real axis, whereas for those saddle points with a negative imaginary part they don't. This can also be concluded immediately from the fact that for saddle points with $\mathrm{Im}(z_\sigma) < 0$ we have $h(z_\sigma) > 0$, such that there is no way 'uphill' from $z_\sigma$ to the real axis where $h = \mathrm{Re}(\mathrm{i}\phi(\mathrm{Re}(z))) = 0$. This means that generally, if $h(z_\sigma) > 0$ we can immediately conclude that $n_\sigma = 0$ such that $z_\sigma$ is not relevant to the total integral.

A small note regarding the implementation: Marching squares algorithms — which are the common tool for most libraries to compute contour lines — typically operate on a given two-dimensional discretised (grid-like) region. When identifying the steepest-ascent (or -descent) lines one needs to (a) ensure that the chosen region is definitely large enough to contain the possible intersection point with the original integration domain (see Fig. 5.4(a)). And (b), from the found level lines pick those that are actually crossing through the saddle point, taking into account the provided discretisation of the region. In our experience, unexpected results could in most cases be traced back to either of these issues. It might therefore be advisable to construct the dual thimble by initialising two points very near the saddle point and flowing them upwards. This traces out the dual thimble regardless of discretisation and domain limitations.

Once we determined which critical points constitute relevant thimbles, finding their contribution to the integral is rather straightforward. Either we apply the standard saddle-point method and approximate the integral across the thimble to be of Gaussian shape as shown in Eq. (4.13). This literally only requires the evaluation of Eq. (4.12) at the given saddle point. Or, if we aim for an (analytically) exact representation of the integral, we find the Lefschetz thimble $\mathcal{T}_\sigma$ attached to each critical point. Numerically, in the case of one-dimensional integral, the thimble can easily be identified to be the level lines at $H(z) = H(z_\sigma)$, which are attached to the saddle point, and which are descending in $h$ away from it. The value of the integral is then obtained from a numerical quadrature of the integrand along this contour line.

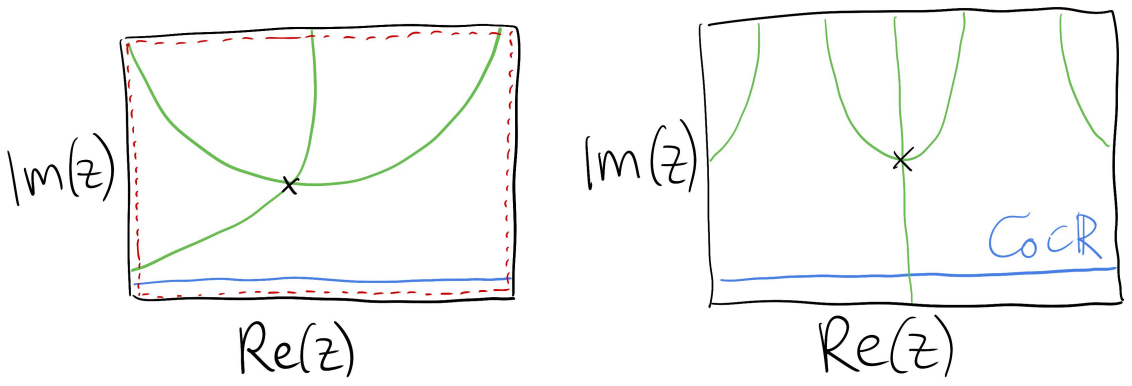

Figure 5.4: Common implementation oversights in the identification of relevant saddle points via level lines of $H$: (a) bounding region (red) does not include the intersection point, (b) there are level lines that are not going through the saddle point.

### 5.2.3 Identifying relevant saddle points in two-dimensional integrals — the necklace algorithm

For the case of a two-dimensional integral, finding the dual thimbles is more complicated. As mentioned above, they are now two-dimensional manifolds (i.e., surfaces) embedded in the four-dimensional space $(\mathrm{Re}(z_1), \mathrm{Im}(z_1), \mathrm{Re}(z_2), \mathrm{Im}(z_2))$. That means, tracing the contour levels $H(\mathbf{z}) = H(\mathbf{z}_\sigma)$ for a critical point $\mathbf{z}_\sigma$ will now yield contour level *surfaces* embedded in 4D which is computationally more advanced.

Here we present a novel technique — which we call the 'necklace algorithm' — to determine the intersection number of a given saddle point for a two-dimensional integral, a problem which has so far remained open,[2] and for which tentative general solutions were only proposed very recently [149]. The basic idea, shown in Fig. 5.6, is to initialise the 'tip' of the dual thimble in the closest vicinity of the saddle point, and then use the upward flow for its further construction 'slice by slice'. We terminate the upward flow as soon as each point reaches $h = 0$, and then check for the intersection with the original integration domain. As we are only interested in this (potential) intersection, it is sufficient to consider the 'brim' of the thimble, which — as it is a discretised closed loop — we dub the necklace. Theoretically, this approach is guaranteed to identify *all* possible intersections of the steepest-ascent manifold with the original integration domain.

Let us briefly explain the procedure on the one-dimensional example. In the vicinity of the saddle point the direction of steepest descent and ascent can be found by linearising the flow. That is, we calculate the second derivatives w.r.t. to both real and imaginary part of the integration variable $z$ and identify the $2 \times 2$ real-valued Hessian matrix, the eigenvectors of which pointing in the direction of maximised gradient. The eigenvector corresponding to the negative eigenvalue points in direction of decreasing $h$ away from the saddle point (drawn as a blue vector in Fig. 4.3), whereas the eigenvector corresponding to the positive eigenvalue increases $h$ away from the saddle point (drawn in green). To obtain the full steepest-ascent manifold (a line), we flow the end point of the steepest-ascent eigenvector and its inverse (drawn as a dashed vector in the opposite direction) until it eventually reaches $h = 0$. If either of the end points of the line hits the real axis, the intersection number $n_\sigma$ counts $+1$. For the one-dimensional integral, of course, this was not necessary as we simply picked the respective contour line.

[2]A similar approach has been used in [138, 168] to approximate the (steepest-descent) thimbles.

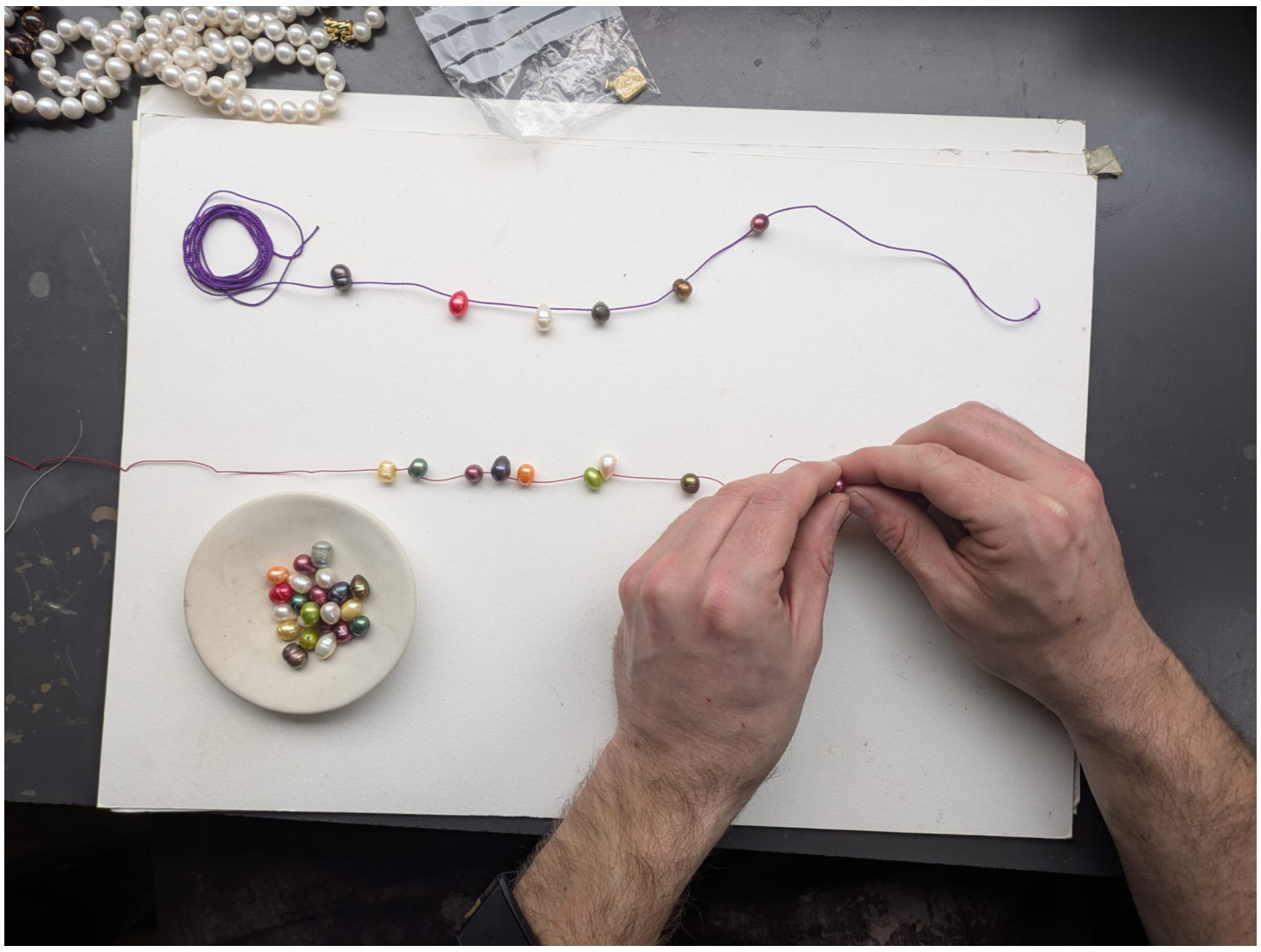

Figure 5.5: Making a necklace by replenishing beads on a thread. By Hannah Upritchard and Jacob Donaghy-Sutton.

Now, for the two-dimensional integral we follow the same procedure. We assume that the upward flow Eq. (4.10) in a small region around the critical point $\mathbf{z}_\sigma$ is linear in $h$ w.r.t. each of the four real-valued dimensions. That is, we use $\mathbf{z} = (\mathrm{Re}(z_1), \mathrm{Im}(z_1), \mathrm{Re}(z_2), \mathrm{Im}(z_2))$ and write

$$\begin{aligned}\frac{dz_\alpha}{d\lambda} &= \left(\frac{\partial^2 h}{\partial z_\alpha \partial z_\beta}\right)^* \Big|_{\mathbf{z}=\mathbf{z}_\sigma} ({z_\beta}^* - {z_{\sigma,\beta}}^*) \\ &= \mathcal{H}_{\alpha\beta}({z_\beta}^* - {z_{\sigma,\beta}}^*) \end{aligned} \tag{5.2}$$

where $\alpha, \beta = 1 \ldots 4$, such that $\mathcal{H}$ is the real-valued, symmetric $4 \times 4$ Hessian of $h$. The solution to the respective eigensystem

$$\mathcal{H}\tilde{\mathbf{v}} = \xi\tilde{\mathbf{v}} \tag{5.3}$$

yields four eigenvalues, coming in pairs, where $\xi_2 = -\xi_1$ and $\xi_4 = -\xi_3$. Analogously, for the corresponding eigenvectors we find $\tilde{\mathbf{v}}_2 = -\mathrm{i}\tilde{\mathbf{v}}_1$ and $\tilde{\mathbf{v}}_4 = -\mathrm{i}\tilde{\mathbf{v}}_3$. Their linear combination solves Eq. (5.2) and hence, defines the directions of constant $H$ around the saddle point $\mathbf{z}_\sigma$. The two vectors $\tilde{\mathbf{v}}$ with the smaller eigenvalues point towards the steepest descent of $h$, and the two vectors with larger eigenvalues point in the direction of steepest ascent of $h$, and they are drawn as blue and green vectors in Fig. 5.6(a) respectively. By re-writing the eigenvectors into complex form as in $\mathbf{v}_\alpha = (\tilde{v}_{\alpha,1} + \mathrm{i}\tilde{v}_{\alpha,2}, \tilde{v}_{\alpha,3} + \mathrm{i}\tilde{v}_{\alpha,4})$, and assuming $\xi_1$ and $\xi_3$ to be the two positive eigenvalues, we can therefore define the directions of steepest ascent,

$$\mathbf{z}(\lambda) - \mathbf{z}_\sigma = a_1 \mathbf{v}_1 \mathrm{e}^{\xi_1 \lambda} + a_3 \mathbf{v}_3 \mathrm{e}^{\xi_3 \lambda}. \tag{5.4}$$

with arbitrary coefficients $a_1$ and $a_3$. We initialise the dual thimble's brim by constructing a vanishing cycle (viz. a 'loop' of constant $H$) around the saddle point. For that we use the two steepest-ascent vectors and draw the circle

$$\mathbf{z}(\gamma, \lambda = 0) = \mathbf{z}_\sigma + \epsilon \left( \cos\gamma \, \mathbf{v_1} + \sin\gamma \, \mathbf{v_3} \right) \tag{5.5}$$

for angles $\gamma \in [0, 2\pi)$ and a small value $\epsilon$, as shown in Fig. 5.6(a).

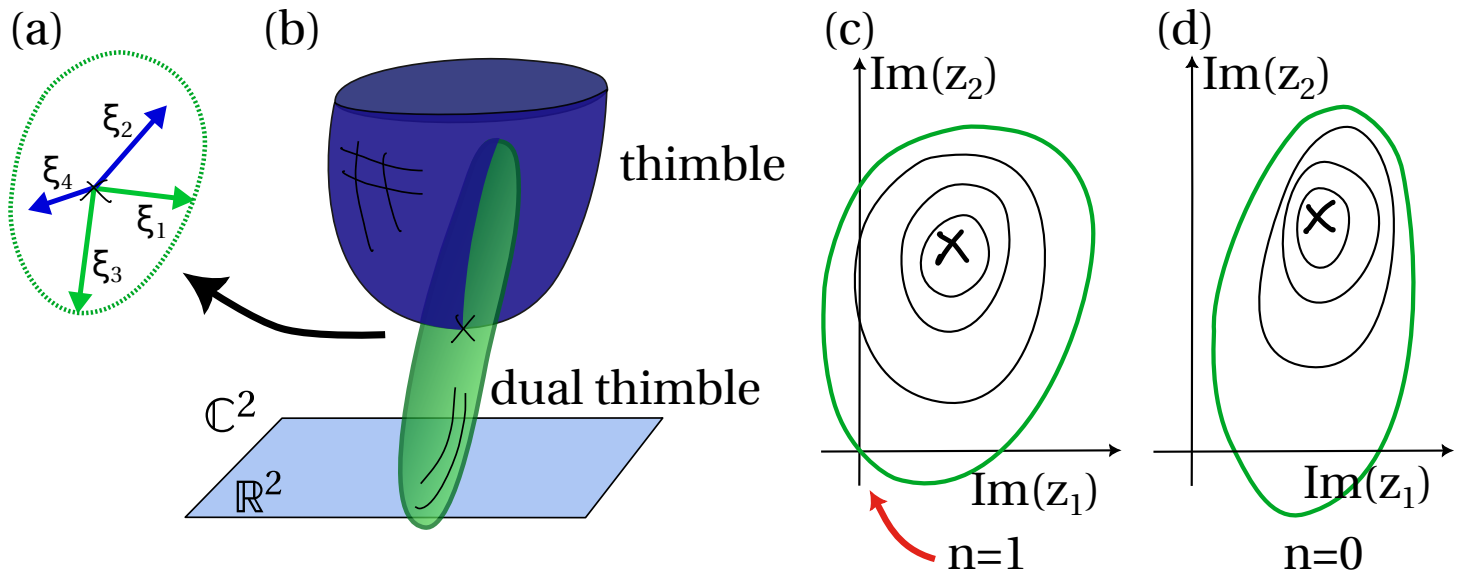

Figure 5.6: The necklace algorithm: Thimble and dual thimble are surfaces embedded in 4D (illustrative 3D projection in panel (b)), locally spanned by the eigenvectors $\xi_\alpha$ of the Hessian at the saddle point (panel (a)). The brim of the dual thimble — the necklace — is initialised as a circle using the steepest-ascent eigenvectors (green), and then flowed upwards. (c) and (d) show projections of example necklaces (intermediate steps in black, necklace for $h = 0$ in green) in the $(\mathrm{Im}(z_1), \mathrm{Im}(z_2))$ projection, where an intersection with $(0,0)$ implies an intersection with the real plane.

Once the necklace has been initialised as this closed list of beads, we apply the upward flow Eq. (4.10) to each bead. We make use of the procedure described for the downward flow in one dimension and/to ensure the necklace remains closed such as to not converge to local minima. With that we construct the dual thimble 'slice by slice' (or rather 'ring by ring') until all beads reach $h = 0$. Finally, it remains indeed to check whether the found brim of the dual thimble intersects the original integration domain $\mathcal{C}_0$.

This part of the necklace algorithm to identify the intersection number is shown schematically in Fig. 5.6(c) and (d), where we show the necklace in the $\mathrm{Im}(\mathbf{z})$ projection as it is constructed (black rings) for two different saddle points (cross markers). The (final) necklace for $h = 0$ is shown in green, and in the case in panel (c) it intersects the real axis (at $\mathrm{Im}((z_1, z_2)) = (0,0)$), resulting in intersection number $n_\sigma = 1$. For the case in panel (d) the necklace does not pass through the real axis, making $n_\sigma = 0$.

As the necklace algorithm essentially traces contours of constant $H$ value, limitations naturally arise in cases where two neighbouring critical points have similar $H$ values. The upward flow then might accidentally 'slip' into (parts of) the dual thimble brim of the other critical point. A schematic illustration of this is attempted in Fig. 5.7, where we show both two projections of an erroneous construction of the necklace. The necklace for $h = 0$ (the outermost ring, not highlighted here) intersects the real axis, but from the rather irregular shape we conclude that it has probably slipped into tracing the dual thimble of a nearby saddle point with similar $H$ value. In practice, a decision about the relevance of the individual critical points can then be made by identifying the exact parameters for the Stokes transitions.

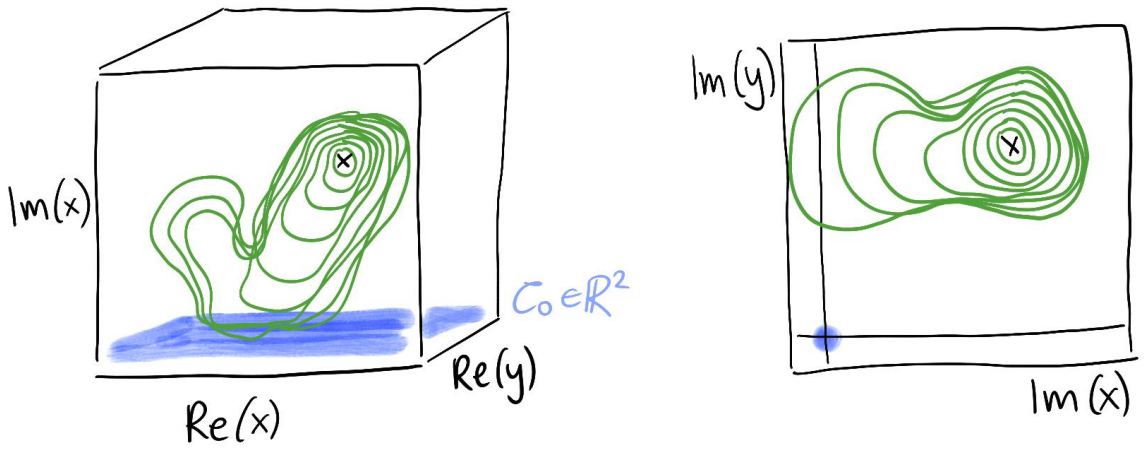

Figure 5.7: Projections of the necklace near a Stokes transition. Its shape deviating from a smooth, regular shape hints at a nearby saddle point with similar $H$ value and hence, a possible Stokes transition.

So far, we have been concerned with determining whether the thimble attached to a given saddle point is relevant or not by means of determining a possible intersection of the steepest-ascent manifold with the original integration domain. In fact, as mentioned above, the intersection number does have a sign. That is, the contribution of a thimble sometimes has to be counted with factor $-1$. For the one-dimensional integrals we study in the remainder of this thesis, however, we assume the intersection number is always positive. For the case of two-dimensional integrals, in lack of a rigorous implementation of the orientation of the dual thimble's manifold, we determine the sign of the contribution of a thimble by comparison with the sign of the Gaussian contribution associated with the respective saddle point.

### 5.2.4 Steepest-descent thimbles for two-dimensional integrals

To calculate the theoretically *exact* contribution of a critical point to the integral, we have to identify the steepest-descent manifold. Note that this is particularly helpful in cases where saddle points are in close vicinity or coalesce to higher order critical points. In those cases the determinant of the Hessian in the denominator of the Gaussian approximation Eq. (4.12) vanishes and the integral contribution diverges.

In the case of one-dimensional integrals the steepest-descent thimbles are the level lines of $H = H(z_\sigma)$ that are descending away from the critical point $z_\sigma$. In the two-dimensional case, however, we have to construct the steepest-descent thimble 'ring by ring'. For that, we use the inverse version of the necklace algorithm (i.e., replacing the upwards flow with the downward flow) and connect flowed points in every iteration step with their previous position to quadrilaterals. Similarly to the downward flow routine, we may remove points as 'inactive' if they reach a certain $h_{\text{thresh}}$. Analogously, the construction procedure can be terminated after a fixed number of iteration steps, or as soon as the calculated integral converges.

# 6

# Applying Picard–Lefschetz methods to SFA integrals

In the previous Chapter, we developed numerical methods for the evaluation of highly oscillatory integrals by deforming the integration contour into the complex domain using the downward flow, and by determining the intersection numbers associated with the critical points of the phase function. In the following Chapter, we explicitly apply these methods to integrals arising in strong-field physics. First, we employ the one-dimensional techniques to the SFA integral over ionisation time, describing direct photoelectrons originating from tunnel ionisation in above-threshold ionisation. Second, we apply the methods developed for two-dimensional integrals to the SFA integral over ionisation and recombination time to calculate the HHG dipole response.

A concise version of this section is published in Ref. [2].

## 6.1 The one-dimensional integral for strong-field ionisation

The ionisation amplitude Eq. (2.24) is a one-dimensional integral over time, as derived in Sec. 2.3.1, and of the shape of a generic path integral. The phase function, which is exponentiated in the integrand, is the semi-classical action $S_{\mathrm{ATI}}(t)$ of the electron on its journey in the field-dressed continuum. The original integration domain is simply the real time axis, representing the interaction time between laser and atom. Having established this link, the application of the downward flow presented in Sec. 5.1.1 is straightforward.

### 6.1.1 The downward flow

We identify the downward flow as

$$\frac{\mathrm{d}t}{\mathrm{d}\lambda} = -\left(\frac{\partial S_{\mathrm{ATI}}}{\partial t}\right)^{*} \tag{6.1}$$

where

$$\frac{\partial S_{\mathrm{ATI}}}{\partial t} = \mathcal{I}_{\mathrm{p}} + \frac{1}{2}\left(\mathbf{p} + \mathbf{A}(t)\right)^2 \tag{6.2}$$

and $\lambda$ is a real-valued scalar that parametrises the flow procedure, and $\mathbf{p}$ and $\mathbf{A}$ are the drift momentum and the driving field's vector potential, respectively. In Fig. 6.1 we show several steps (from left to right, corresponding to different values of $\lambda$) of this flow procedure for the driving laser field shown in Fig. 6.3(b) below. In the top row we show the value of the imaginary part of the integrand, evaluated along the integration contour highlighted in blue on the respective bottom panel. For the original integration contour, i.e., flow step $i = 0$, where the contour is simply the real time axis, the integrand oscillates rapidly. Flowing the integration contour into the complex plane, according to 6.1, soothes the oscillations. For a large number of flow steps (right-hand side panels) the integration contour eventually converges to the steepest-descent contours attached to (two of) the critical points. At this point, the integrand is fully localised to two separate single oscillations, both of almost Gaussian shape, which can easily be integrated.

Note that up to panel (c) the integration contour is one connected component, but in panel (d) the thimble has broken up into two separate components. This is due to the asymptotic valley of the action landscape. At which iteration step of the procedure this split happens, is of course heavily dependant on the chosen discretisation parameters, and in particular, the choice of $h_{\mathrm{thresh}}$ (here a value of $\mathrm{Re}(\mathrm{i}S_{\mathrm{ATI}}(t))$. For Fig. 6.1, the chosen parameters are $\delta_{\mathrm{flow}} = 1$, $\Delta = 4$, $h_{\mathrm{thresh}} = -700$ and the gradient of the action is normalised as soon as its norm exceeds 0.5.

Very briefly, we want to furthermore comment on how the choice of discretisation parameters affects the value of calculated integral. The algorithm presented in Sec. 5.1.1 includes several numerical parameters, that are to be set by the user, and which are described above. Here we want to show how the value of the total integral converges after a very few number of steps of the flow algorithm. Theoretically, the value of the integral remains fixed, no matter how the contour is deformed. For the evaluation along the discretised contour, however, the value depends on how well we can resolve the oscillations of the integrand (or rather, how the oscillations are minimal, and hence easy to resolve, along the deformed contour). In Fig. 6.2 we show how the integral converges for an increasing number of steps of the flow procedure. The integral is calculated as a numerical quadrature along the obtained line segments. We compare different values of the flow step scaling factor $\delta_{\text{flow}}$ (in different shades of blue). If $\delta_{\text{flow}} = 1$, then the step taken is exactly the magnitude of the gradient at the point. We also show how changing the threshold for the subdivision of the line segments $l_{\text{thresh}} = \Delta$ affects the convergence behaviour. The maximum distance two points can flow apart within one iteration step depends on the flow step factor $\delta$, hence we choose the subdivision threshold as a factor of it ($\Delta = 4\delta_{\text{flow}}$ for solid lines, $\Delta = 20\delta_{\text{flow}}$ for dashed lines). For comparison, the result of the Gaussian saddle-point approximation is drawn as a black line (SPM). Let us first briefly examine the value of the integral evaluated along the real axis, i.e, for zero flow steps. The value of the integral is overestimated by several orders of magnitude because the oscillations of the integrand cannot be resolved properly. This overestimation is worse with increased grid spacing $\Delta$. Upon flowing the contour ($i > 0$) the grid spacing becomes less relevant for the value of integral. Generally, we find it is remarkable that the integral value converges after only a few flow steps, while the contour has to flow for much longer to actually assimilate the steepest-descent paths. That is, in our example Fig. 6.1, where we used $\delta_{\text{flow}} = 1.$ and $\Delta = 4.$, the contour converges to the steepest-descent paths after around $i = 100$ flow steps (panel (d)). The value of the integral, however, has already converged after about $i = 5$ steps (see Fig. 6.2). Hence, if one was interested merely in the value of the integral, a few flow steps would be sufficient.

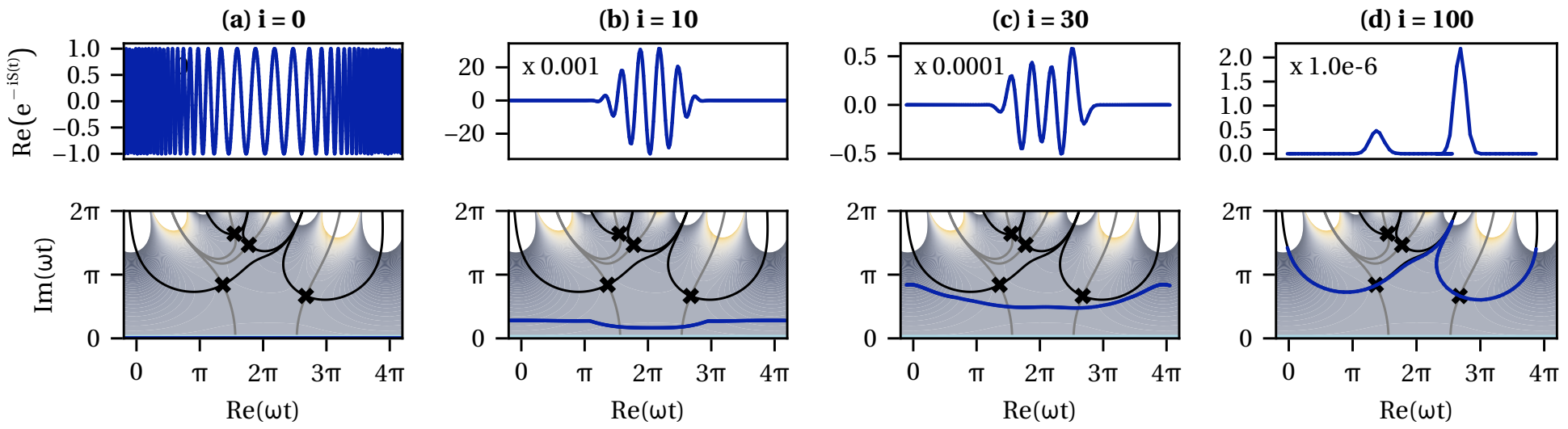

Figure 6.1: Flowing the integration contour (dark blue) according to the downward flow 6.1 for the same scenario as in Fig. 6.3(d) for discretised flow steps $i$ to minimise the oscillations of the integrand. The integrand evaluated along the contour is shown in the top rows.

Overall, this rudimentary benchmark demonstrates that the downward flow procedure provides a rather robust means of evaluating the integral, largely independent of the specific parameter choices. Nevertheless, a more systematic and sophisticated approach to their particular choice would be a worthwhile direction for future work, as their optimal values depend strongly on the intended application. If the primary goal is just a quick evaluation of the integral, relatively large flow step factors and only a small number of flow steps into the complex plane may be sufficient. In contrast, reliably identifying and separating individual thimbles requires more iteration steps, finer grid spacing, and careful consideration of the threshold parameter $h_{\text{thresh}}$. A detailed investigation of these numerical choices is therefore deferred to future studies.

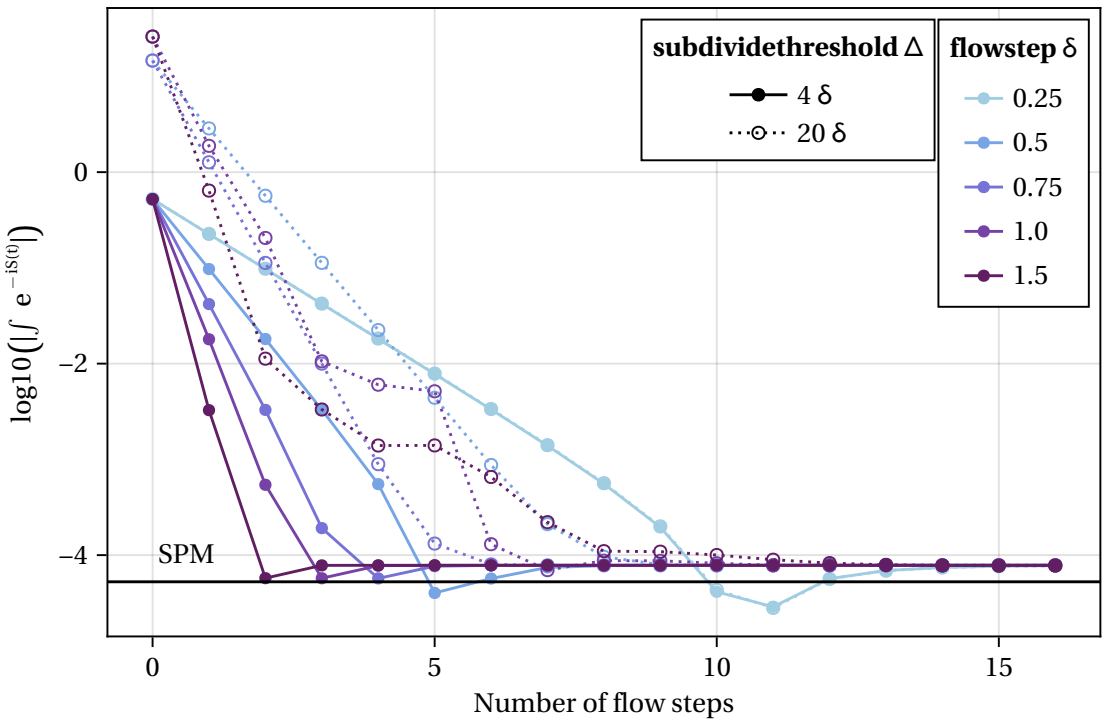

Figure 6.2: The value of the integral along the flow into the complex plane, i.e., for different number of iteration steps $i$, as visualised in Fig. 6.1. Thereby, the downward flow procedure was done with different thresholds for subdivision of line segments $\Delta$ (solid and dashed lines), and different factors for the flow step $\delta$ (shades of blue). The integral was performed by a simple numerical quadrature along the obtain line segments. For comparison, the value of the integral as obtained from a Gaussian approximation around the relevant saddle points (SPM) is drawn as a black line.

### 6.1.2 Determining relevant saddle points

The above-mentioned downward flow works independently of the critical points. However, in our application the saddle points correspond to discrete ionisation events. Theoretically, we could determine which saddle points are relevant parts of the final thimble. That is, comparing with Fig. 6.1, we can identify that saddle points A and D are part of the converged integration contour of panel (c), whereas saddle points B and C are not. However, in practice, this approach is rather unreliable because the convergence of the integration contour to the steepest-descent contours cannot be rigorously defined. This is rooted in the nature of the flow: Remember that the flow is stationary only at the critical points. At any other point, it will direct towards the steepest descent. As a result, for a set of points describing the integration contour, applying the flow will move them all into the direction of the steepest descent, a.k.a. the steepest way down into the valley of the landscape.

Therefore, the more reliable method to identify which saddle points are relevant is finding the steepest-ascent contour. As explained in Sec. 5.2.2 in the one-dimensional case this can be done by finding the contour level lines of $\mathrm{Im}(-\mathrm{i}S_{\mathrm{ATI}}(t_s))$ across $\mathrm{Im}(-\mathrm{i}S_{\mathrm{ATI}}(t))$ evaluated in the complex plane $t \in \mathbb{C}$ for all saddle points $t_s$. In practice, this can be done with any library, e.g., using a marching squares algorithm. The algorithm will typically return a set of lines, from which one needs to choose (a) those lines that actually intersect the saddle point, because $\mathrm{Im}(-\mathrm{i}S_{\mathrm{ATI}}(t))$ could assume the respective value $\mathrm{Im}(-\mathrm{i}S_{\mathrm{ATI}}(t_s))$ elsewhere in the complex plane as well. And (b), those lines that are steepest-ascent lines, i.e., ascending in $\mathrm{Re}(-\mathrm{i}S_{\mathrm{ATI}}(t))$ as they leave the saddle point. Once the steepest-ascent lines attached to the critical point are found, it is trivial to check whether they intersect the real axis, and hence, the saddle point is a relevant contributor to the integral.

In Fig. 6.3 this is shown for two examples. The steepest-descent contours are drawn in black, the steepest-ascent contours are drawn in grey. For the case of a monochromatic driving laser field (electric field shown in panel (a)) both of the two saddle points are relevant as for both of them a grey lines leads down to the real axis. This confirms our knowledge about two relevant ionisation events per cycle of a monochromatic driving field. More interesting is therefore the other case. In panel (d) we show the landscape for an arbitrary two-colour driving field (shown in panel (b)) a momentum $p = 1.2\,\mathrm{a.u.}$ From existing literature there would be no heuristic nor clear expectations as to which saddle points are relevant to the process. Inspecting the steepest-ascent routes here shows unambiguously that for saddle A and D they connect downwards to the real plane. Whereas for saddles B and C the steepest-ascent lines both go 'uphill' towards higher imaginary parts. That they will never 'return' is proven in Ref. [154] and guaranteed by the fact that our phase function is analytic. That is, saddles A and D are relevant to the integral and saddles B and C are not.

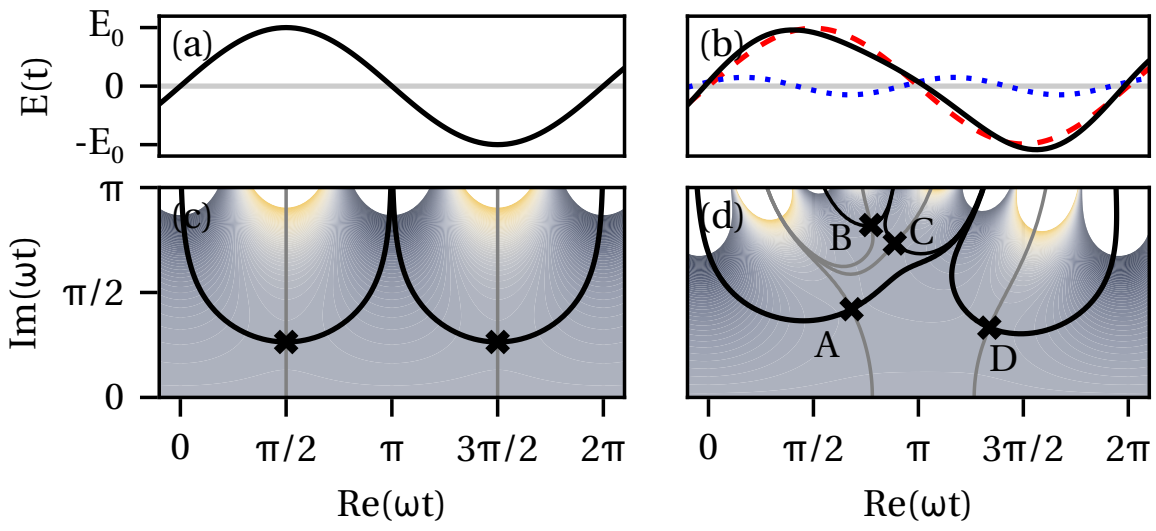

Figure 6.3: Bottom row: Contour plots of $\mathrm{Im}(S_{\mathrm{ATI}}(t))$ for the complex $t$ plane for the two driving fields shown on top (panels (a) and (b)) and drift momenta $p = 0$ and $p = 1.2\,\mathrm{a.u.}$ respectively. Steepest-descent and steepest-ascent contour lines (black and grey lines, respectively) are attached to saddle points Eq. (2.32) (black markers, labelled in (d) for convenience), with the resulting integration path drawn as a heavy black line. The electric field in (b) is composed of the two constituent fields of frequency $\omega$ (red dashed) $2\omega$ (blue dotted) with phase shift $\varphi = 0.5$ and amplitude ratio $E_2/E_1 = 0.15$ according to Eq. (6.5).

To obtain the ionisation amplitude *per saddle point* we can then either numerically integrate their respective steepest-descent contours or make a Gaussian approximation. For the former, we simply pick the contour lines attached to the saddle points that are descending in $\mathrm{Re}(-\mathrm{i}S_{\mathrm{ATI}}(t))$ away from it, and numerically integrate the integrand along this line — analogously as integrating the thimble determined from the downward flow. Note that the contribution of this integral towards the total integral is multiplied with the intersection number $n_s$. It has been shown that for one-dimensional integrals the intersection number can only be $\pm 1$ or 0 [156]. Here, through the intersection with the real domain we have found the magnitude already. We can set the sign of the intersection number simply by ensuring to take the steepest-descent contour oriented in the same direction as the original integration domain (making $n_s = +1$). That is, looking at the action landscapes in Fig. 6.3, we ensure to integrate the steepest-descent lines from left to right.

As a quick alternative, and helpful especially in regimes where saddle points are far apart, we can approximate the integral to be of Gaussian shape around the saddle point. The total integral then results as a summation over all the contributions from the relevant saddle points.

## 6.2 The two-dimensional HHG integral

The harmonic response of an atom subjected to a strong laser field can (within the SFA framework [11, 12]) be calculated in terms of the two-dimensional integral Eq. (2.45) over ionisation and recombination times of the involved electronic wave packet, $t_\mathrm{i}$ and $t_\mathrm{r}$, respectively. This double integral is often rewritten in terms of contributions of separate quantum orbits [169, 170], i.e. semi-classical electron paths defined by a discrete ionisation and recombination time. This allows for an intuitive understanding of the process as the associated trajectories have different properties in e.g. the spatial divergences in the far-field [171, 172]. The quantum orbits are pairs $(t_\mathrm{i}, t_\mathrm{r})$ for which the semi-classical action $S_\mathrm{HHG}(t_\mathrm{i}, t_\mathrm{r})$ is stationary, i.e., saddle points in the complex plane defined by Eq. (2.47). Notably, there are typically far more solutions to Eq. (2.47) than relevant quantum orbits to the process. So far, the existing heuristics to decide whether a given saddle point solution is a relevant quantum orbit rely on a classification of the solutions and dynamic symmetries of the driving field.

For generic driving fields, however, those heuristics fail. In the following we demonstrate how the methods of Picard–Lefschetz theory described in Secs. 5.1.2 and 5.2.3 can be utilised to compute the harmonic dipole integral Eq. (2.45). The central insight is that the integral can be evaluated along a different contour $\mathcal{C}$ in the complex time planes and ultimately expressed as a sum over contributions from separate thimbles $\mathcal{T}_s$:

$$\begin{aligned}\mathbf{D}(q\omega) &= \int_{\mathcal{C}_0} \ldots \mathrm{d}\mathbf{t} = \int_{\mathcal{C}\subset\mathbb{C}^2} \ldots \mathrm{d}\mathbf{t} \\ &= \mathrm{i}\sum_s n_s \int\limits_{\mathcal{T}_s\subset\mathbb{C}^2} \mathrm{d}\mathbf{t}\, \mathbf{d}\left(\mathbf{p}_s(t_\mathrm{i}, t_\mathrm{r}) + \mathbf{A}(t_\mathrm{r})\right) \Upsilon\left(\mathbf{p}_s(t_\mathrm{i}, t_\mathrm{r}) + \mathbf{A}(t_\mathrm{i})\right) \\ &\quad \times \left(\frac{2\pi}{\mathrm{i}(t_\mathrm{r} - t_\mathrm{i})}\right)^{3/2} \mathrm{e}^{-\mathrm{i}S_\mathrm{HHG}(t_\mathrm{i}, t_\mathrm{r})}. \end{aligned} \tag{6.3}$$

Here, we notate $\mathbf{t} = (t_\mathrm{i}, t_\mathrm{r})$ and the original integration domain is $\mathcal{C}_0 = \{(t_\mathrm{i}, t_\mathrm{r}) \in \mathbb{R}^2 \mid t_\mathrm{r} > t_\mathrm{i}\}$. Each thimble $\mathcal{T}_s$ is attached to a critical point $(t_{\mathrm{i},s}, t_{\mathrm{r},s})$ defined by Eq. (2.47), and only contributes for non-vanishing intersection numbers $n_s$. The key difference to and striking advantage over Eq. (2.50) is, that here we still have an equality, as we have not made any assumption on the shape of the integrand around the critical points. If the critical points are well-separated saddle points we can assume the integrand along the thimble to be of Gaussian shape, which is the well-known saddle-point approximation. In that case, the harmonic dipole integral can be approximated as

$$\mathbf{D}(q\omega) \approx \sum_s n_s \mathbf{D}_s(q\omega). \tag{6.4}$$

with the contribution $\mathbf{D}_s(q\omega)$ from each saddle point as in Eq. (2.50).[1]

Note that the presented integration methods 'only' address the two-dimensional temporal integration and therefore hold for any definition of prefactors, waveforms etc.

[1] So the only difference between this formulation and the earlier Eq. (2.50) is the intersection number $n_s$.

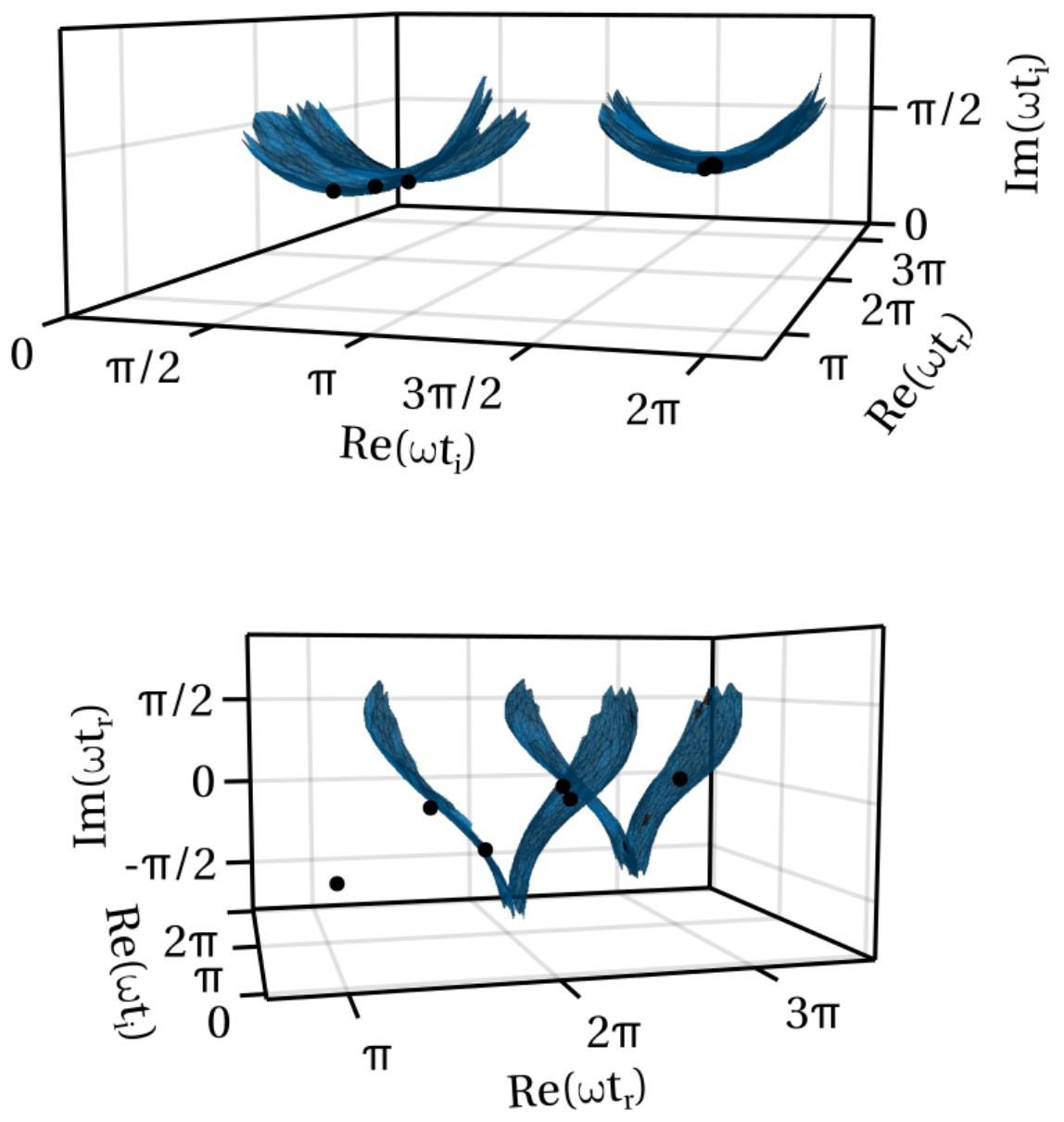

Figure 6.4: Two projections of the HHG thimble for a monochromatic driving field and $q = 25$: Ionisation-time projection $(\mathrm{Re}(t_\mathrm{i}), \mathrm{Im}(t_\mathrm{i}), \mathrm{Re}(t_\mathrm{r}))$ and recombination-time projection $(\mathrm{Re}(t_\mathrm{r}), \mathrm{Im}(t_\mathrm{r}), \mathrm{Re}(t_\mathrm{i}))$, top and bottom respectively. Saddle points are marked as black dots.

With that, the exponentiated phase factor is $-\mathrm{i}S_\mathrm{HHG}$, which we consider in its dependency on ionisation and recombination times only, $S_\mathrm{HHG} = S_\mathrm{HHG}(t_\mathrm{i}, t_\mathrm{r})$ as in Eq. (2.46).

As an example, within this Section we choose to consider HHG driven by a collinear and co-polarised two-colour field that consists of a fundamental laser field with frequency $\omega$, superimposed with its second harmonic (frequency $2\omega$). A generic expression for the electric field then reads

$$\mathbf{E}(t) = E_1 \cos(\omega t)\mathbf{e}_x + E_2 \cos\big(2\omega t + \varphi\big)\mathbf{e}_x \tag{6.5}$$

with the field amplitudes $E_1$ and $E_2$ and the phase delay $\varphi$ between the two field components. For pulses longer than a few cycles, it is a good approximation to restrict our considerations to one cycle of the fundamental frequency, the period $T = 2\pi/\omega$. These types of driving fields are ubiquitous in attosecond science, in both experiment and theory. They allow to probe the inner workings of the process of strong-field light-matter interaction itself, as well as to tailor the properties of the harmonic spectrum and/or the created attosecond pulse [37, 173–178]. The following results are obtained using $\mathcal{I}_\mathrm{p} = 15.8\,\mathrm{eV}$ and $I_0 = E_0^2 = 0.92 \times 10^{14}\mathrm{W/cm^2}$ ($E_0 = 0.05\,\mathrm{a.u.}$), $\lambda = 1030\,\mathrm{nm}$ ($\omega = 0.044\,\mathrm{a.u.}$).

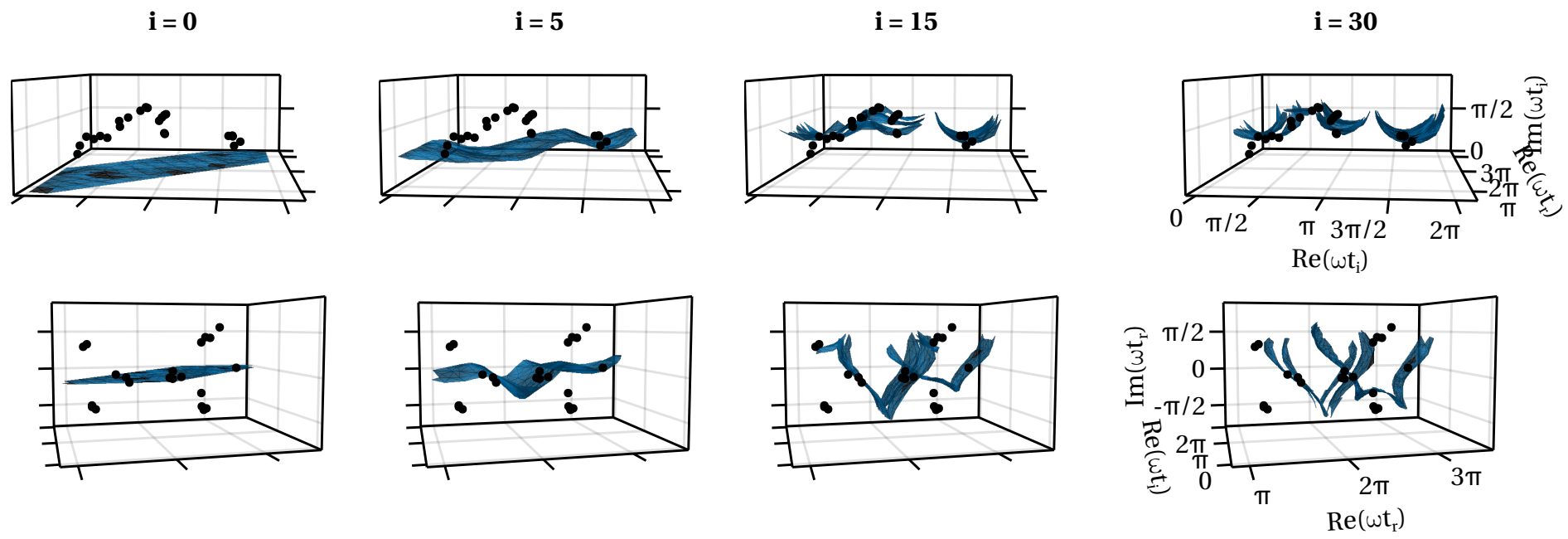

Figure 6.5: Several steps of the deformation of the integration domain towards the Lefschetz thimble, for HHG driven by the two-colour field shown in Fig. 6.3(b), harmonic order $q = 25$.

### 6.2.1 Using the downward flow to deform the integration contour towards Lefschetz thimbles

In the previous Chapter we showed that there exists a continuous deformation of the original integration domain into a contour that minimises the integrand oscillations, which then allows for a more efficient numerical evaluation of the integrand along that new contour. We demonstrate how this downward flow method is applied to the integration contour of the HHG integral Eq. (2.45).

We restrict ionisation to one cycle, so $0 \leq \mathrm{Re}(t_\mathrm{i}) < T$, and recombination to happen after that, with travel times limited to one cycle as contributions from orbits with longer travel times are lower: $0 < \mathrm{Re}(t_\mathrm{r} - t_\mathrm{i}) < T$. This original integration domain is then deformed into the complex domain for both $t_\mathrm{i}$ and $t_\mathrm{r}$, i.e., into the four-dimensional space $(\mathrm{Re}(t_\mathrm{i}), \mathrm{Im}(t_\mathrm{i}), \mathrm{Re}(t_\mathrm{r}), \mathrm{Im}(t_\mathrm{r}))$. As described in Sec. 5.1.1, the deformation of the domain follows a simple first-order scheme for either variable,

$$t_\mathrm{i} \longmapsto t_\mathrm{i} + \delta_\mathrm{flow} \frac{\mathrm{d}t_\mathrm{i}}{\mathrm{d}\lambda} \quad \text{and} \quad t_\mathrm{r} \longmapsto t_\mathrm{r} + \delta_\mathrm{flow} \frac{\mathrm{d}t_\mathrm{r}}{\mathrm{d}\lambda}, \tag{6.6}$$

with a small factor $\delta_\mathrm{flow}$. The direction for every flow step $\lambda$ is dictated by the downward flow Eq. (4.2), recast for the case of HHG:[2]

$$\frac{\mathrm{d}t_\mathrm{i}}{\mathrm{d}\lambda} = -\left(\frac{\partial S_\mathrm{HHG}}{\partial t_\mathrm{i}}\right)^* \quad \text{and} \quad \frac{\mathrm{d}t_\mathrm{r}}{\mathrm{d}\lambda} = -\left(\frac{\partial S_\mathrm{HHG}}{\partial t_\mathrm{r}}\right)^*. \tag{6.7}$$

This routine continuously deforms the (discretised) original integration domain into a two-dimensional steepest-descent surface embedded in 4D space and converges to the Lefschetz thimbles.

The resulting thimbles for the simple case of a monochromatic driver $E(t) = E_0 \cos(\omega t)$ (shown in Fig. 6.3(a)) and harmonic order $q = 25$ are shown in Fig. 6.4 in two projections, as well as the saddle point solutions (black markers). In the ionisation projection

[2]We have dropped the $\lambda$ indices for better readability.

$(\mathrm{Re}(t_\mathrm{i}), \mathrm{Im}(t_\mathrm{i}), \mathrm{Re}(t_\mathrm{r}))$ (top), we observe the deformation into four disconnected surfaces corresponding to the ionisation windows around each maximum of the electric field. In the recombination projection $(\mathrm{Re}(t_\mathrm{r}), \mathrm{Im}(t_\mathrm{r}), \mathrm{Re}(t_\mathrm{i}))$ (bottom), we identify four separate surfaces, corresponding to the expected two pairs of 'short' and 'long' quantum orbits within each half cycle. Each surface is the steepest-descent manifold (thimble) of a relevant saddle point. Saddle points that are not included in the surface are irrelevant.

For comparison, and in order to aid the understanding of the flow method, we show a range of intermediate steps of the continuous deformation for a more complicated situation in Fig. 6.5. The driving field is the two-colour field as in Fig. 6.3(b), and we show how the downward flow deforms the integration domain towards the Lefschetz thimbles (increasing iteration steps from left to right) for harmonic order 25, using the same projections described above. From the ionisation-time projection (top row) we can make out separate ionisation windows, albeit not as distinct as in the monochromatic case.

Finally, the harmonic dipole Eq. (2.45) can be calculated with a simple quadrature along these discretised Lefschetz thimbles. This can be done at any intermediate step of the deformation (as the integral remains unchanged for just a change of contour, see Eq. (4.9)). Note that this deformation of the integration domain is not an efficient method for the detection of all relevant saddle points.

### 6.2.2 Using the necklace algorithm to determine relevant quantum orbits

Saddle points of the action correspond to the quantum orbits that interfere when creating the harmonic dipole response. The properties of those several electron trajectories, e.g. the spread of the wave packet and the recollision angle, imprint on the dipole and therewith on the properties of the emitted radiation [173, 179–183]. Moreover, the contributions of the various quantum paths behave differently upon propagation and give rise to distinct patterns in the far-field spectra measurement [184–187]. Phenomenologically, it is therefore interesting to understand which quantum orbits are at play for the creation of a certain dipole, i.e., to understand which saddle points are relevant contributors to the sum Eq. (6.3). A given saddle point is a relevant contributor if and only if its attached steepest-ascent contour (the dual thimble) connects back to the original integration domain. We find this possible intersection by propagating the brim of the dual thimble upwards until $h = 0$ and then checking for an intersection of this brim with the real plane; this is the necklace algorithm introduced in Sec. 5.2.3. The necklace around the saddle point is initialised as a small circle in the plane of the two eigenvectors corresponding to the largest eigenvalues of the matrix

$$\mathcal{H}_{\alpha,\beta} = \frac{\partial^2}{\partial t_\alpha \partial t_\beta} S_\mathrm{HHG}(t_1 + t_2\mathrm{i}, t_3 + t_4\mathrm{i}) \tag{6.8}$$

where $\alpha, \beta = 1, \ldots, 4$, and where we have taken $\mathbf{t} = (\mathrm{Re}(t_\mathrm{i}), \mathrm{Im}(t_\mathrm{i}), \mathrm{Re}(t_\mathrm{r}), \mathrm{Im}(t_\mathrm{r}))$. Each bead of the (discretised) necklace then flows upwards in $h$, following

$$\frac{\mathrm{d}t_\mathrm{i}}{\mathrm{d}\lambda} = +\left(\frac{\partial S_\mathrm{HHG}}{\partial t_\mathrm{i}}\right)^* \quad \text{and} \quad \frac{\mathrm{d}t_\mathrm{r}}{\mathrm{d}\lambda} = +\left(\frac{\partial S_\mathrm{HHG}}{\partial t_\mathrm{r}}\right)^* \tag{6.9}$$

until $h = 0$ and $S_{\mathrm{HHG}}$ becomes real. If the converged necklace intersects the original integration domain, the given saddle point is relevant.

### 6.2.3 Comparison of the downward flow and the saddle-point based approach across a spectrum

The resulting values of the integrals, in the form of spectral intensities Eq. (2.40) for a range of harmonic orders $q$, i.e., harmonic spectra, are shown in Fig. 6.6 for the monochromatic driving field (top panel) and the two-colour field (bottom panel) as in Fig. 2.11(a) and (b) respectively. We show the Gaussian contribution from each saddle point in coloured markers (for relevant saddle points; light grey for non-contributing saddles). The coherent summation of relevant saddles' contribution is shown in black markers (saddle point-method, SPM, Eq. (6.4)), which is compared to the quadrature along the deformed integration contour (Picard–Lefschetz downward flow, PLDF, Eq. (6.3)) in blue. For the monochromatic driving field (in panel (a)) we recognise the familiar structure of a typical HHG spectrum exhibiting quantum-path interference [184]. Throughout the spectrum there are two types of relevant contributions: from short and long trajectories (red and pink respectively), of which the former become non-relevant at the high-harmonic cutoff at $q_c = 42$. Note small deviations between the two integration approaches only occur around this Stokes transition where the two saddle points are in close vicinity and their contribution should not be modelled as Gaussian, but rather as an Airy-type integral [46, 48, 119].

For the complicated two-colour field (in panel (b)), the harmonic spectrum exhibits a more interesting structure, as we find more than only two types of trajectories. Throughout the first plateau (harmonic orders 15 to 32) we observe the expected interference structure from the two dominant trajectories marked in red and pink. However, we can identify a more interesting feature of the spectrum that we can now attribute to individual trajectories, and which is shown enlarged in the inset. Around the first harmonic cutoff there are two other trajectories (yellow and orange) which contribute significantly to the integral and yield an overall spectral enhancement. This enhancement stems from the 'cluster' of saddle points shown in Fig. 2.13(e) panel D and signifies the appearance of a caustic, which we address in more detail in the following Section. In that region of the spectrum, the SPM exhibits both an artificial discontinuity and also the largest deviation from the PLDF-based signal. Again, this is expected, as for saddle points in close vicinity the integrand along the steepest-descent contour does not resemble a Gaussian. For a proper analytical treatment within saddle-point based methods we would require a uniform approximation to smoothen out the discontinuities.

Within the PLDF method the integrand is directly evaluated along the thimble and agnostic of any saddle (or higher order critical) points. As a result, the evaluated integral is naturally smooth throughout the spectrum and eliminates the need for a carefully constructed uniform approximation that connects different regimes of relevant saddle points. To this end, the PLDF provides a unique tool that captures the exact value of the integral while still allowing for a separation into distinct contributions from the disconnected components of the thimble, where each of them may be identified with a specific electron trajectory.

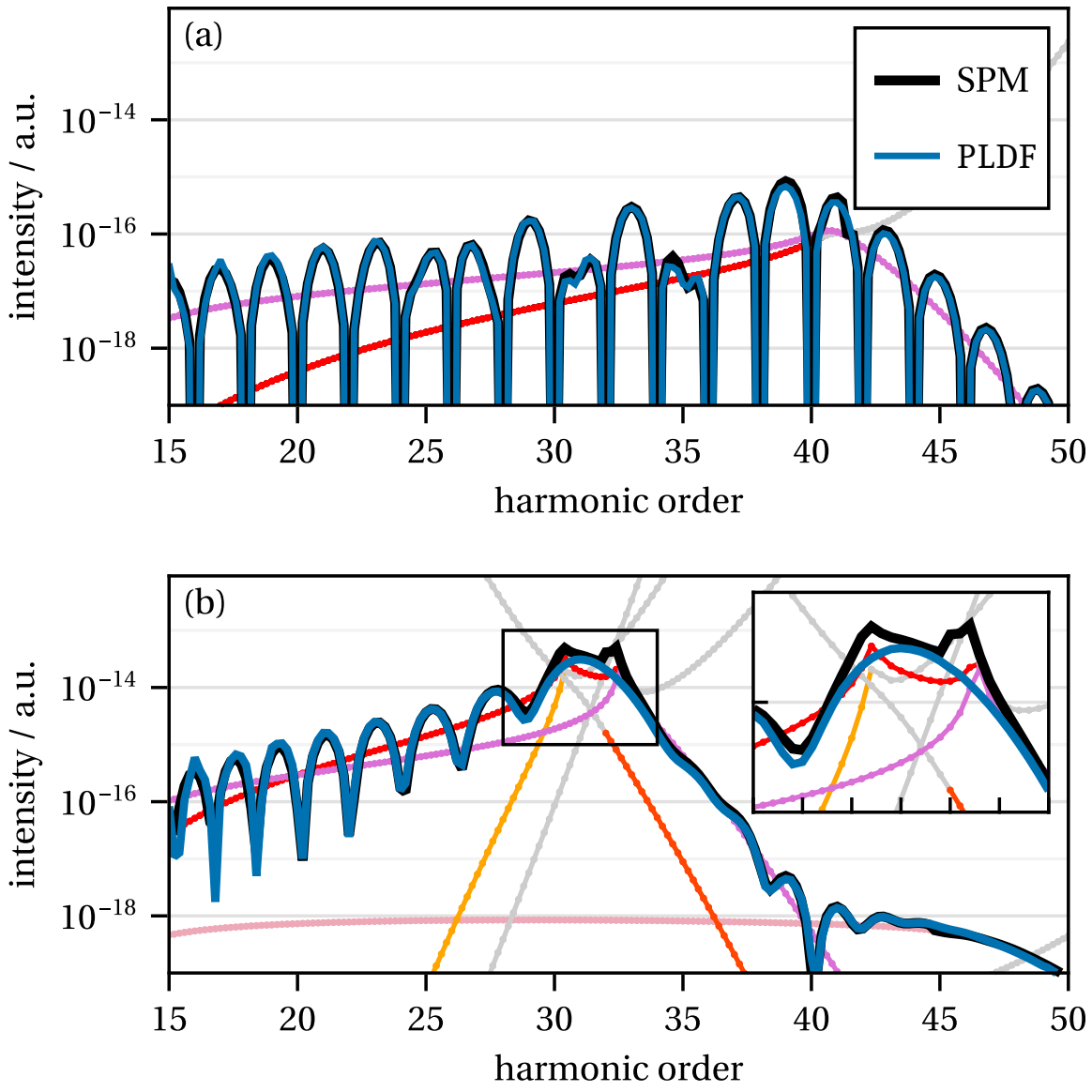

Figure 6.6: HHG spectrum (a) for the monochromatic and (b) for the two-colour field shown in Fig. 6.3 respectively, calculated as a sum of Gaussian contributions from relevant saddle points (SPM, black line), and as a quadrature of the deformed integration domain (PLDF, blue line). Individual saddle points' contributions are shown as markers (coloured for contributing saddles, grey for non-contributing saddles). The bottom panel features an inset for the artificial discontinuity that arises for the SPM when saddle points are in close vicinity.

# 7

# The colour switchover

The *colour switchover* refers to a driving field configuration in which a laser field with fundamental frequency $\omega$ is gradually replaced by a second colour component, often the second harmonic $2\omega$. By continuously varying the relative amplitudes of the two fields this transition includes the full range of amplitude ratio between two field components — thereby connecting the perturbative two-colour field regime with that of fully bichromatic drivers. Either of those are commonly employed in attosecond science. Here we present the colour switchover as a technique that probes the link between them. We have generated this ultrashort Chapter to introduce this powerful technique and the questions arising from it.

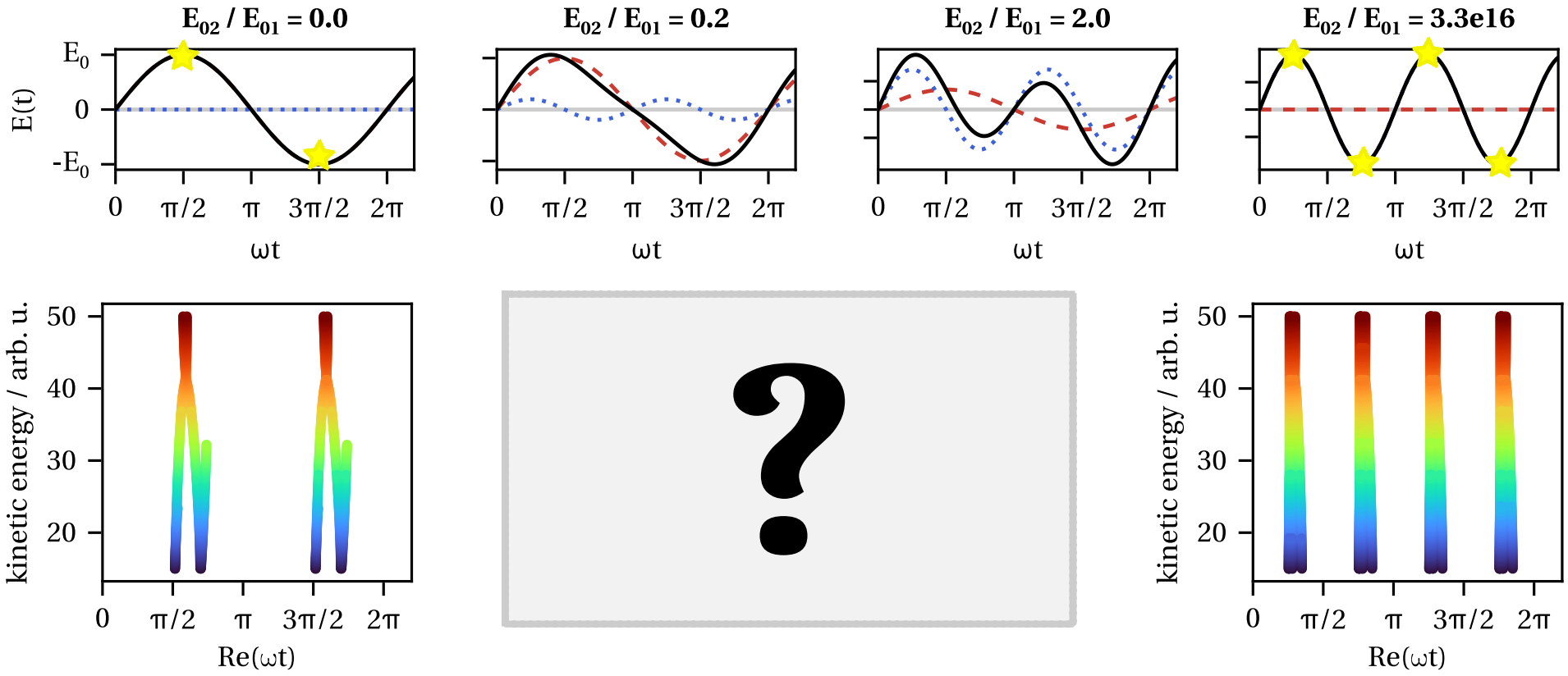

Figure 7.1: Top: Total field (black) composed of a fundamental (red dashed) and a second harmonic (blue dotted) field, for increasing amplitude ratio $E_{02}/E_{01}$ (left to right). Ionisation (highlighted as yellow stars, and saddle point ionisation times shown in the bottom figures for a range of return energies) occurs around the extrema of the electric driving field — for an almost monochromatic $\omega$ field (left) that means two ionisation windows per optical cycle, for a $2\omega$ field (right) there are four ionisation windows per cycle. How the transition between them happens is exactly the question posed by the colour switchover.

With the technological advances of recent years the generation of high harmonics is explored in a plethora of experimental conditions. From a pragmatic point of view, with the increased conversion efficiency and better control over the experimental setups it becomes more and more feasible to conduct experiments where the driving laser field is a combination of several colour fields. This is simply because the generation of the second harmonic is experimentally so ubiquitous that the resulting two-colour field might as well be used to drive HHG 'again'. Commonly studied scenarios are the *perturbative* regime, where the amplitude of the second colour field is in the order of a few percent of the fundamental driving field, and the *fully bichromatic* field, where both fields are of equal amplitude.

Quantum-orbit approaches for the perturbative regime often rely on the fact that the weak second colour field only slightly perturbs the saddle points (now meaning both ionisation events in the case of direct photoelectrons of ATI, and ionisation and recombination times for the trajectories in the case of HHG). In particular, saddle-point solutions stemming from the second colour field are either fully ignored or consciously declared to be negligible. For bichromatic fields, of course, the second field contributes relevant saddle points as well. In those cases quantum orbits are often studied alongside symmetry considerations of the driving field. Of course, the combination of two (or more!) fields of different frequencies opens up a huge space of free parameters [188]: each frequency component can be elliptically polarised with colinear or co-orthogonal polarisation axes, the commensurate fields can have a phase shift between them, we may modify their amplitude ratio, their frequency ratio etc. These parameters vastly impact the overall field shape and, as a result, the location and relevancy of saddle points and the corresponding quantum orbits.

Within the scope of this thesis we are interested in the structure of saddle points and — in particular — cases where they are in close vicinity and potentially coalesce, as well as transitions between regions of various numbers of saddle points. We therefore introduce the *colour switchover* as a technique to study exactly this behaviour in particular.

The idea of the scheme is simple: in a colinear configuration (i.e., where both fields are linearly polarised along the same direction) we replace a field of fundamental frequency $\omega$ with a field of frequency $2\omega$. That is, this replacement constitutes a two-colour field of slowly increasing amplitude ratio. Technically, this replacement can be realised in a couple of different ways. One method is to simply add the second field and increasing its amplitude. This will increase the total field amplitude. Other methods are to gradually increase the amplitude ratio $E_2/E_1$ while keeping the total intensity, or the total amplitude, or the total ponderomotive energy constant.

The fundamental questions, however, remain the same. Considering that in a monochromatic driver ionisation is most likely to happen around the maxima, this corresponds to two ionisation events per cycle $T = 2\pi/\omega$ for a $\omega$ field, and to four ionisation events in $T$ for the $2\omega$ field, as shown in Fig. 7.1. The gradual replacement from $\omega$ to $2\omega$ therefore raises the following questions:

- Where do the new ionisation events come from?
- Where do they go?
- When do they start being relevant?
- Which of the old ionisation events are the new ones?

We aim to answer all of these questions by studying the colour switchover. Since ionisation events arise from saddle points of the underlying action, and constitute topologically protected objects, these questions naturally call for a rigorous mathematical explorations. The observation that a region containing two saddle points evolves into containing four saddle points, under a smooth variation of an external control parameter, immediately brings us into the realms of catastrophe theory. In this context, the colour switchover serves as a 'toy model' for applying the concepts of catastrophe to attosecond science.

# Part III

## Results

# 8

# Strong-field ionisation during the colour switchover

In the following Chapter we use the methods explained in Sec. 5.2.2 to determine the relevant saddle points of the semi-classical action for the strong-field ionisation integral. We use these methods to study the ionisation events during the colour switchover from a monochromatic driving field to its second harmonic. In detail, for a range of drift momenta, we show how new ionisation events emerge and from which stage in the colour switchover on they are relevant. We furthermore identify the fold catastrophe point that serves as an organising centre for the whole set of critical points in the parameter space spanned by the switchover and the electron momenta.

This Chapter ultimately answers the questions posed by the colour switchover for the process of strong-field tunnel ionisation, and provides a solid foundation to study the colour switchover for the process of high-harmonic generation.

## 8.1 Saddle point solutions across the parameter range

We consider the colour switchover where the driving field is defined as

$$\mathbf{E}(t) = E_1 \cos(\omega t)\mathbf{e}_x - E_2 \cos(2\omega t)\mathbf{e}_x \tag{8.1}$$

with $E_1 = E_0 \cos\theta$ and $E_2 = E_0 \sin\theta$.[1] That is, tuning the mixing angle $\theta$ from 0° to 90° performs a gradual replacement of the fundamental field with frequency $\omega$ by its second harmonic, a field with frequency $2\omega$. The respective field shapes are shown on the left-hand side of Fig. 8.1. Apart from this figure, within this chapter most of the results are reported by specifying the amplitude ratio $R = E_2/E_1 = \tan\theta$, for which the full colour switchover corresponds to the transition from $R = 0$ to $\infty$. For now, we use a standard strong-field configuration with $I_0 = 4 \times 10^{14}\mathrm{W/cm^2}$, $\lambda = 800\,\mathrm{nm}$ ($\omega = 0.057\,\mathrm{a.u.}$), and $\mathcal{I}_\mathrm{p} = 0.5$, such that the Keldysh parameter is $\gamma = 0.67$.

We are now interested in the discrete ionisation events, i.e., saddle points of the semi-classical action, for the range of field shapes created by the colour switchover, and the range of electron momenta $p$. For a given amplitude ratio $R$ we numerically find these solutions $t_s$ to Eq. (2.32) and plot them in the complex plane, shown in Fig. 8.2, across discrete values of momentum $p = p\mathbf{e}_x$ indicated by colour (red: $p > 0$, blue: $p < 0$). We choose the momentum magnitudes $|p|$ up to $\approx 2U_\mathrm{p}$, because this is the maximum energy of direct photoelectrons, as laid out in Sec. 2.3.1. First, we observe that a clear separation of two ionisation bands remains present throughout the colour switchover. Within our chosen parameter range, the boundary between the two bands can be drawn at $\omega t \approx \frac{2}{3}\pi$. For increased momentum amplitudes the solutions asymptotically approach each other, but do not coalesce. Secondly, we clearly see that the saddle point solutions depend smoothly (and symmetrically) on the momentum, forming 'lines' in the complex plane. At the beginning of the switchover, i.e., for low values of $R$, we find the two broad ionisation windows around the maxima of the field. As well as a small cluster of saddle point solutions with large imaginary parts, see Fig. 8.2(a). Upon increasing $R$, the additional solutions come further down, clearly as two separate lines. Then however, for a specific value of $R = R_\mathrm{fold}$, the continuous dependence on the momentum is impacted by the coalescence point in Fig. 8.2(b), where two saddle point solutions of different lines coincide for zero momentum. This coalescence marks a fold catastrophe point, which we will discuss in detail below. Increasing $R$ further, now the saddle point solutions move further down towards the real axis and break up into clearly separable sets of solutions, see Fig. 8.2(c). For $R \to \infty$ the solutions will spread equidistantly to form four ionisation windows, similar to the structure at the outset.

The whole set of solutions $t_s$, shown in the complex plane and across a range of parameters $R$ and $p$ clarifies the underlying topology. In Fig. 8.2(d) we show the resulting structure, where solutions are drawn as coloured markers for the various discrete values of $p$. Highlighting solutions for $p = 0$ reveals the symmetry within the apparent 'sheets' of solutions. Firstly, we note a sheet of solutions around $\mathrm{Re}(\omega t) \approx \pi$. These are clearly separated from the remaining structure, throughout the whole switchover. We can therefore confirm what might has been our intuition anyway: the ionisation events around $\mathrm{Re}(\omega t) \approx \pi$ in the $\omega$-dominated field smoothly transition into becoming the ionisation events in the same time period of the $2\omega$ driving field.

[1] This is equivalent to *adding* the second field as $E_2 \cos(2\omega t + \pi)\mathbf{e}_x$. It will become clear in Chapter 9 why we choose this field definition.

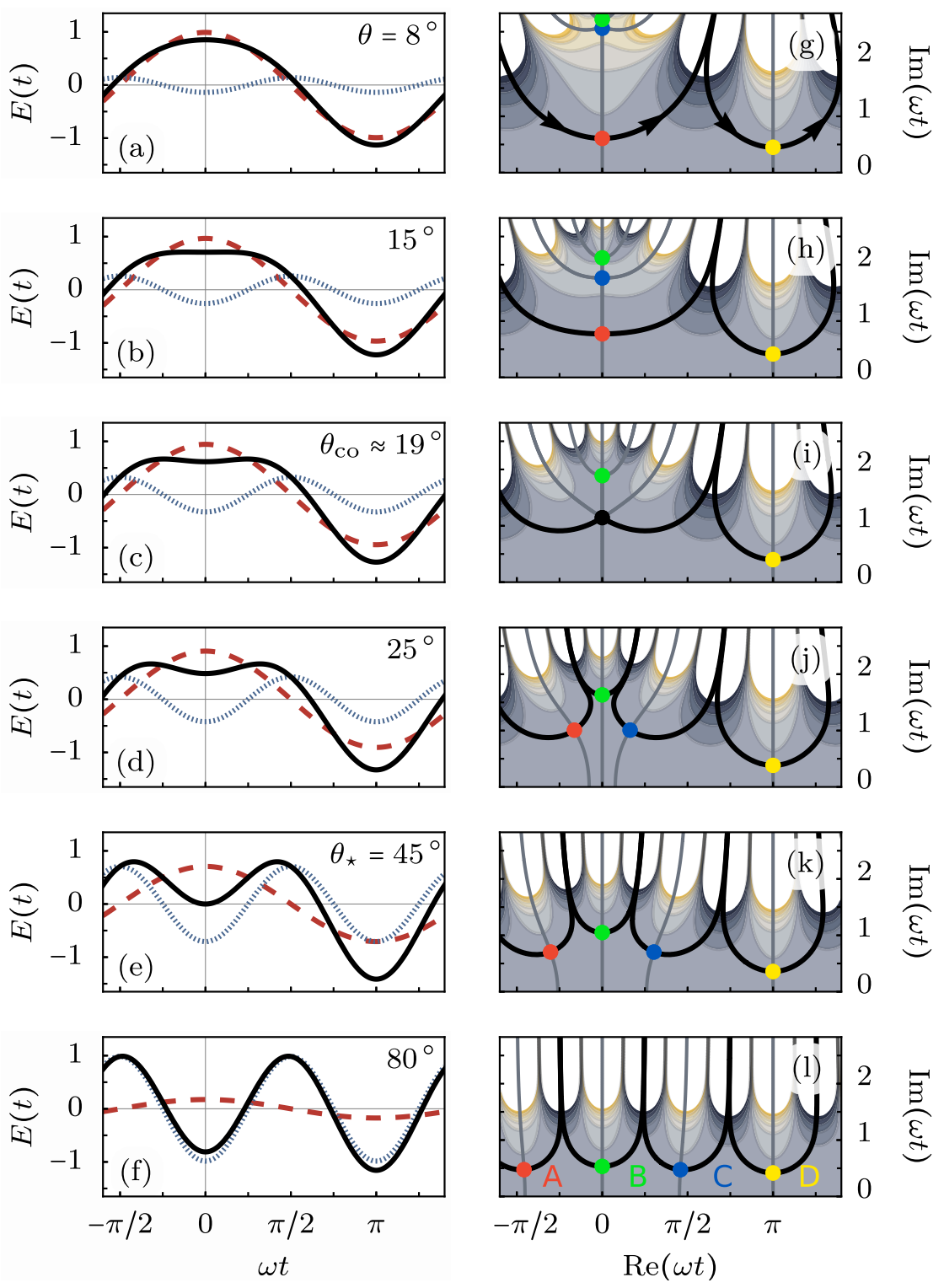

Figure 8.1: Left column (a–f): Total waveform of the bichromatic field Eq. (8.1) (black solid line) and of its components ($\omega$ field: red dashed, $2\omega$ field: blue dotted) for the colour switchover performed by increasing the mixing angle $\theta$ from 8°, through 15° , $\theta_{\mathrm{co}} \approx 19°$ , 25° , $\theta^{\star} = 45°$ and 80°. Right column (g–l): $\mathrm{Im}(S(p,t))$ over the complex $\omega t$ plane for the fields presented on the left, and for $p = 0$. Saddle points $\omega t_s$ are highlighted by coloured dots and can be labelled as shown in panel (l). Their contour lines for constant $\mathrm{Re}(S(p,t))$ are drawn as grey lines, with the resulting integration contour in black. Published similarly in Ref. [1].

The saddle point dynamics within the other half cycle reveals a slightly more intricate structure. From the particular plotting view point in Fig. 8.2(d) it is not so obvious, but the solutions with highest imaginary parts form a sheet that remains separate and disconnected from the rest of the structure throughout the switchover (it will be more obvious in Fig. 8.3 below).

Let us turn to the topologically interesting part of this solutions manifold. For small values of $R$ we can clearly identify the big sheet that corresponds to the first ionisation window of the fundamental $\omega$ field. For increasing values of $R$ they acquire slightly higher imaginary parts and merge with newly incoming saddle point solutions from high imaginary parts. The coalescence point, highlighted in green, acts as an organising centre to this process. Locally, the manifold of saddle points around the fold point looks like a saddle again, as it constitutes a vanishing complex derivative.

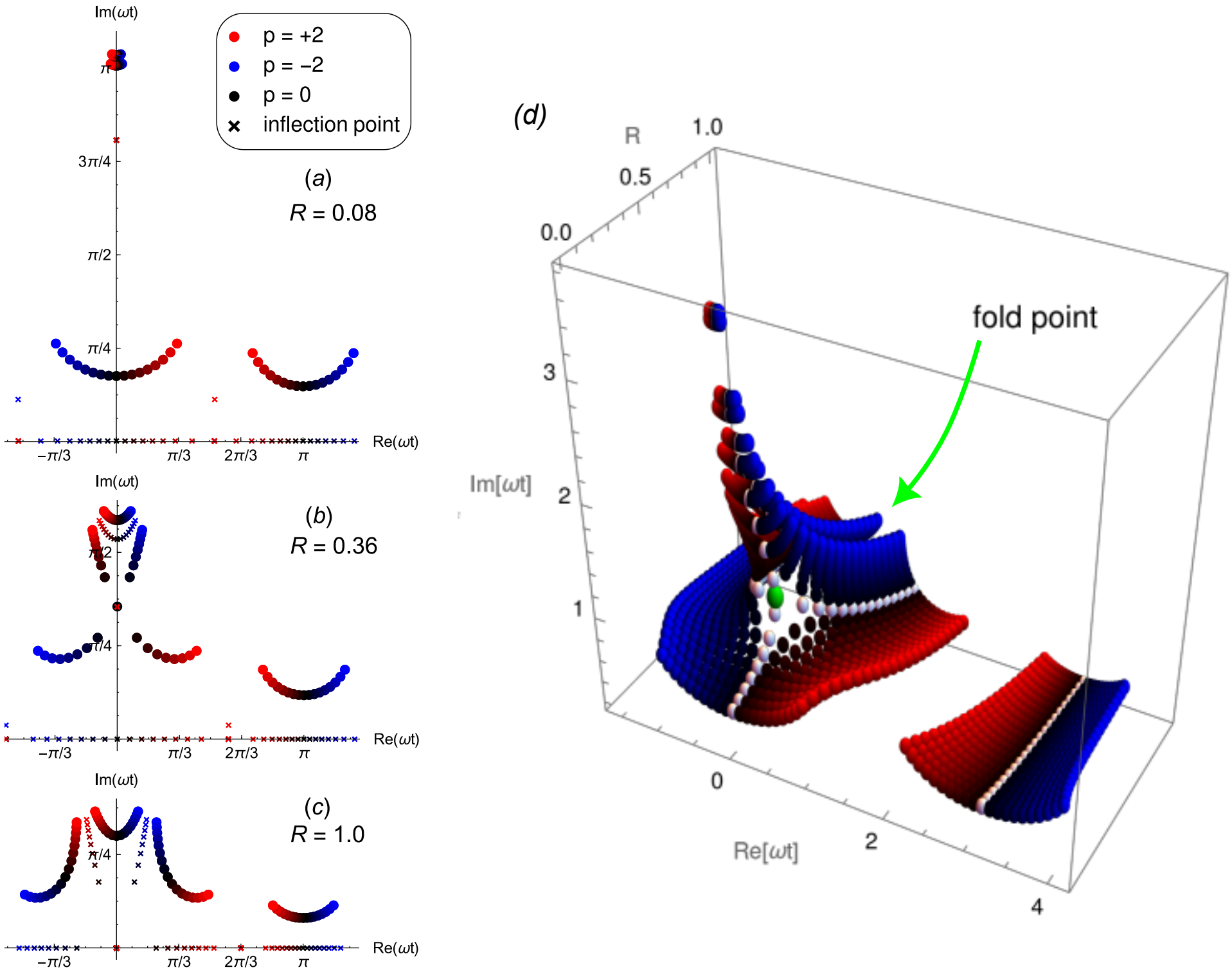

Figure 8.2: Left: Solutions of the first (coloured disks, according to the momentum) and second (coloured crosses) derivative of the action in the complex plane, for fixed values of the amplitude ratio $R$, and a range of momenta between $p = -2$ (blue) and $p = +2$ (red). The amplitude ratios are (a) $R = 0.08$, (b) $R = R_{\text{fold}} \approx 0.36$, and (c) $R = 1.0$. For zero momentum (black), in (b) the solutions for first and second derivative coalesce to a fold catastrophe point. Right: The same solutions for the first derivative, but for a full range of values $R$ values, plotted as a third dimension to the complex plane. Solutions for $p = 0$ are highlighted in white, and the coalescence of two saddle points at $R = 0.36$ is highlighted in green.

The structure Fig. 8.2(d) now somewhat answers the question about the origin of the new ionisation events for a colour switchover. In order to identify ‘where they go’, and furthermore, which of the ‘original’ ionisation events correspond to the ‘new’ ionisation events, we will need to establish a classification of the solutions.

## 8.2 Classifying the solutions

As mentioned and visible from Fig. 8.2(d), the coalescence point acts as a sort of branching point of two sheets. A rigorous classification of the separate surfaces can therefore only be established with knowledge of the exact location of the fold point.

The coalescence of two saddle point solutions constitutes a fold catastrophe point. Hence, to identify the external parameters (in our case $R$ and $p$) for which this coalescence happens we need to simultaneously solve the first and second derivative of the action. As the coalescence point bears interesting features in itself we derive an analytic expression for it below, in Sec. 8.5. Here, let us briefly state that we can identify the coalescence point for the parameters $R_{\text{fold}} \approx 0.36$, $p_{\text{fold}} = 0.$ to be located at $\omega t_{\text{fold}} \approx 0 + 1.1123\text{i}$.

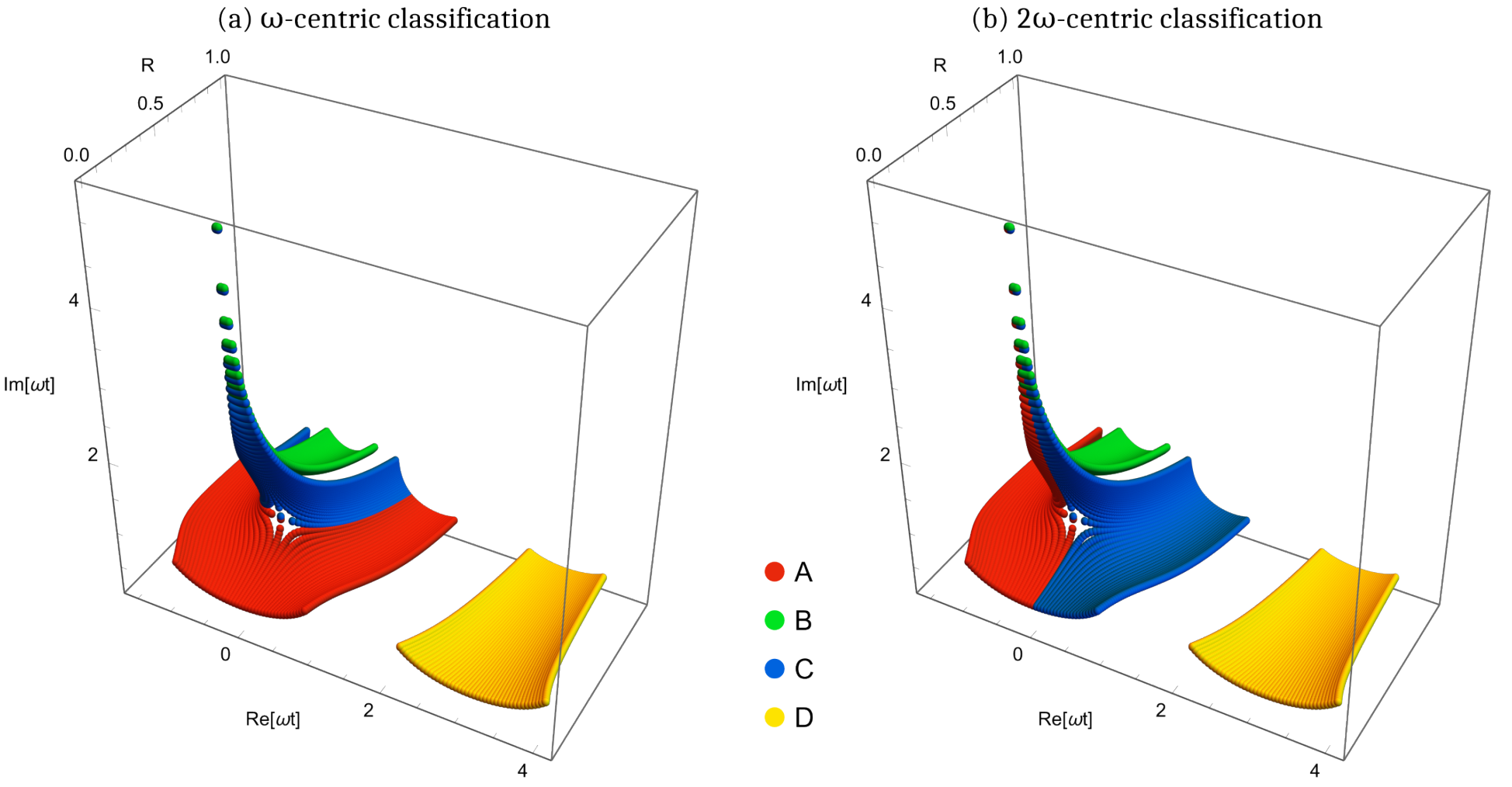

Figure 8.3: The two different classification schemes for the solutions in Fig. 8.2(d).

The coalescence of two saddle point solutions renders the classification across $R$ ambiguous. Clearly, solutions in the later ionisation window can be classified as one separate sheet, say D. Similarly, the surface of solutions with highest imaginary parts can be separated as one class of saddle points, which we call B. For the remaining two solutions A and C we present an $\omega$-centric classification scheme and a $2\omega$-centric classification scheme, depending on where they yield 'consistent' labelling. Both schemes are shown in Fig. 8.3. For the remainder of this chapter we will adhere to the $2\omega$-based scheme, as it proves more meaningful for analysing the contribution of the individual solutions to the total amplitude.

## 8.3 Relevance of saddle points

Now that we have established a consistent labelling scheme for all the saddle point solutions appearing in the colour switchover, let us tackle the next question: *When do the new ionisation events become relevant?* In order to determine whether a saddle point (read: an ionisation event) is a *relevant* contributor to the total integral (read: the total ionisation amplitude), we need to identify whether its steepest-ascent contour intersect the real integration domain (see Sec. 6.1.2).

As explained therein, the steepest-descent and ascent contours are simply the level lines at $\text{Im}(S) = \text{Im}(S(t_s))$ in the contour plot of $\text{Im}(S(t))$ in the complex plane $t \in \mathbb{C}$. In the following we will show those action landscapes, i.e. contour maps of $\text{Im}(S)$, with hills drawn in yellow and valleys in dark grey. We draw the saddle points $t_s$ and their respective level lines at $\text{Im}(S(t_s))$ in grey. The resulting steepest-descent integration contour (i.e., the Lefschetz thimble) is drawn in black.

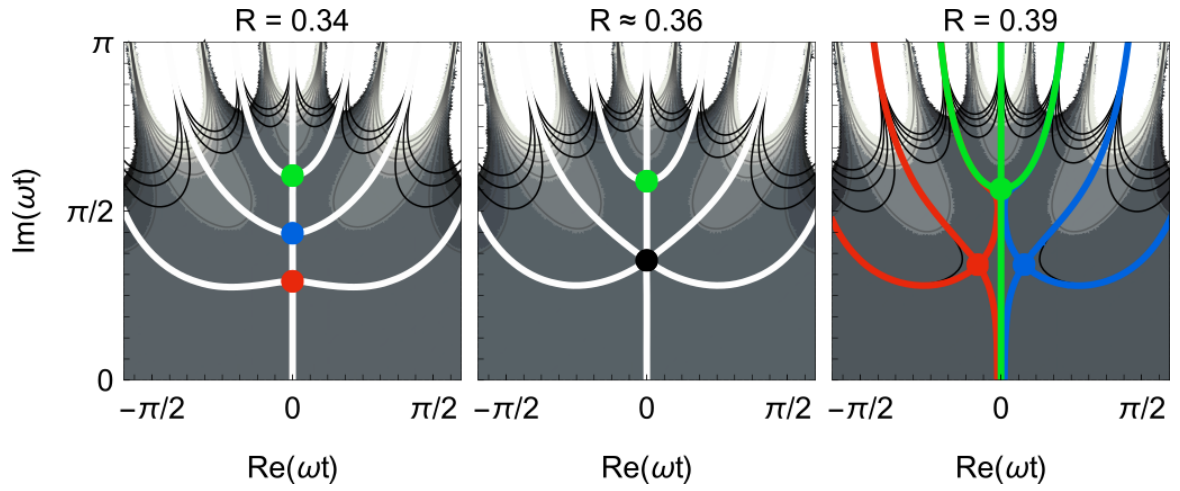

Figure 8.4: Action landscape and contour level lines immediately before (left, $R = 0.34$), at (middle, $R = R_{\text{fold}} \approx 0.36$), and after (right, $R = 0.39$) the coalescence of two saddle points. The contour lines of constant $\text{Re}(S_{\text{ATI}})$ passing through the respective critical points are highlighted.

Let us first discuss this case where $p = 0$. The respective action landscapes across the colour switchover, alongside the electric fields, are shown in Fig. 8.1. At the beginning of the switchover, the newly incoming saddle points are not part of the contour because their steepest-ascent lines do not intersect the real axis. As part of the steepest-descent integration contour saddle point A (red) and D (yellow) are contributors, as expected. For $R = R_{\text{fold}}$ where saddles A and C coalesce, the integration contour goes through this coalescence point, because there is a steepest-ascent route connecting back to real integration domain. In Fig. 8.4 we show a close-up for the topological change of the level lines throughout the fold catastrophe point. It shows that from this stage in the colour switchover (beyond $R = R_{\text{fold}}$), all of the four saddle points are included in the integration contour, and hence relevant contributors.

For the case of non-zero momenta, the action landscapes and hence, the resulting integration contour look different. We show a set of examples taken for $p = \{-1.5, 0, +1.5\}$ (columns) and amplitude ratios $R = \{0.0, 0.15, 0.5, 1.0\}$ (rows) in Fig. 8.5. These landscapes demonstrate an interesting finding: For negative momentum the original saddle point A is pushed away to earlier times (viz. to the left), whereas for positive momenta it is pushed away to later times (to the right). That is, for $p < 0$, the colour switchover changes ionisation events AC to ABCD, whereas for positive momenta the switchover produces CBAD.

Studying the full action landscapes is certainly instructive and not to be dismissed as a tool in understanding saddle-point methods and the deformation of the integration contour. Rather than determining each saddles relevance using the steepest-ascent contours, one could use the steepest-descent paths originating from each saddle point to then construct the suitable integration path. However, as explained in Sec. 4.3.4 this bears additional complications [154]. And, moreover, the full action landscape (in order to determine the contour lines) is not inexpensive to compute, especially when scanning over a whole range of external parameters.

An alternative approach to identifying the relevance of specific saddle points, is to find the Stokes transitions in the respective parameter space. As mentioned in the Sec. 4.2.4, Stokes transitions are topological features of the parameter space, and are linked to their organising centres, the catastrophes. In the case of the fold catastrophe the Stokes lines emanate from Neile's parabola Fig. 3.5. Unfortunately however, there is no sufficient and generic approach to identify Stokes transitions a priori. That means we have to track saddle points across parameter space and check where they are relevant and where they are not. This gives a first estimate of the location of Stokes transitions, which we can then identify as those configurations where two saddles have equal real actions, i.e., $\mathrm{Re}(S(t_{s1})) = \mathrm{Re}(S(t_{s2}))$. The Stokes lines therefore define regions in parameter space in which specific saddle points (or rather, combinations of them) are relevant and we call this a "Stokes map".

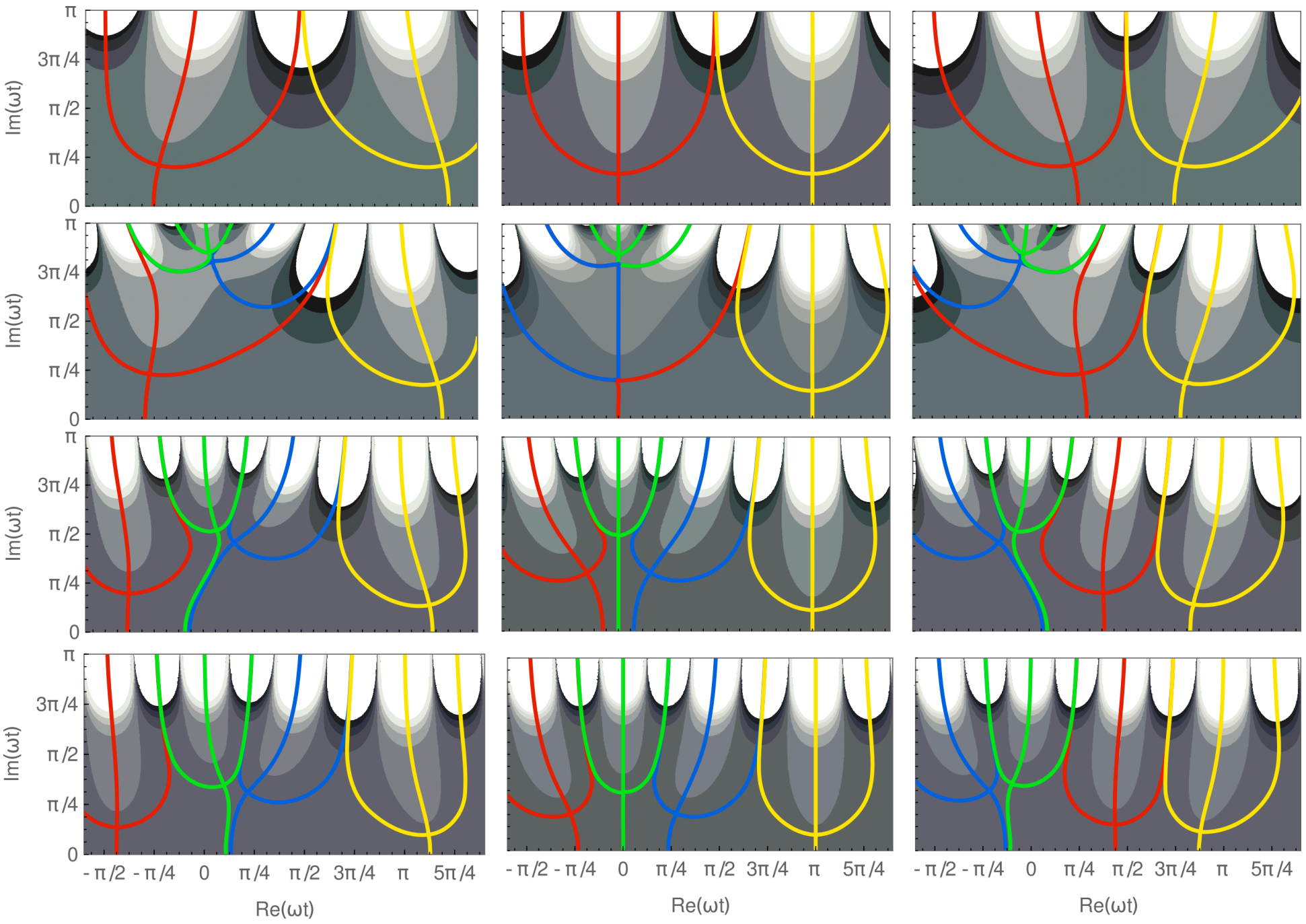

Figure 8.5: For different values of $p$ (columns, $p = \{-1.5, 0, +1.5\}$) and $R$ (rows, $R = \{0.0, 0.15, 0.5, 1.0\}$): Contour map of the action with saddle points and the respective contour level lines of $\mathrm{Re}S_{\mathrm{ATI}}$ highlighted in colours according to the $\omega$-centric classification scheme.

In the case of the colour switchover considered here, the parameter space is spanned by the amplitude ratio $R$ and the momentum $p$. We show the respective Stokes map in Fig. 8.6, where we draw the Stokes lines for both the transition between A and B, as well as between A and C. Unsurprisingly it is symmetric with respect to the sign of momentum. Furthermore, we find the two Stokes lines originate from $R = R_{\mathrm{fold}}$ as the organising centre of this topological structure. For the spectrum at a given amplitude ratio $R$ (viz. a horizontal line across values of $p$) we might cross multiple Stokes transitions and observe multiple cutoffs.

Generally, from both studying the integration paths on the action landscapes, and from looking at the Stokes maps we find that the newly incoming ionisation events start to become relevant contributors for surprisingly low values of the amplitude ratio. The amplitude ratio $R_{\text{fold}}$ is only around 36% of second harmonic, which is lower than we would have expected for the new ionisation events to start contributing.

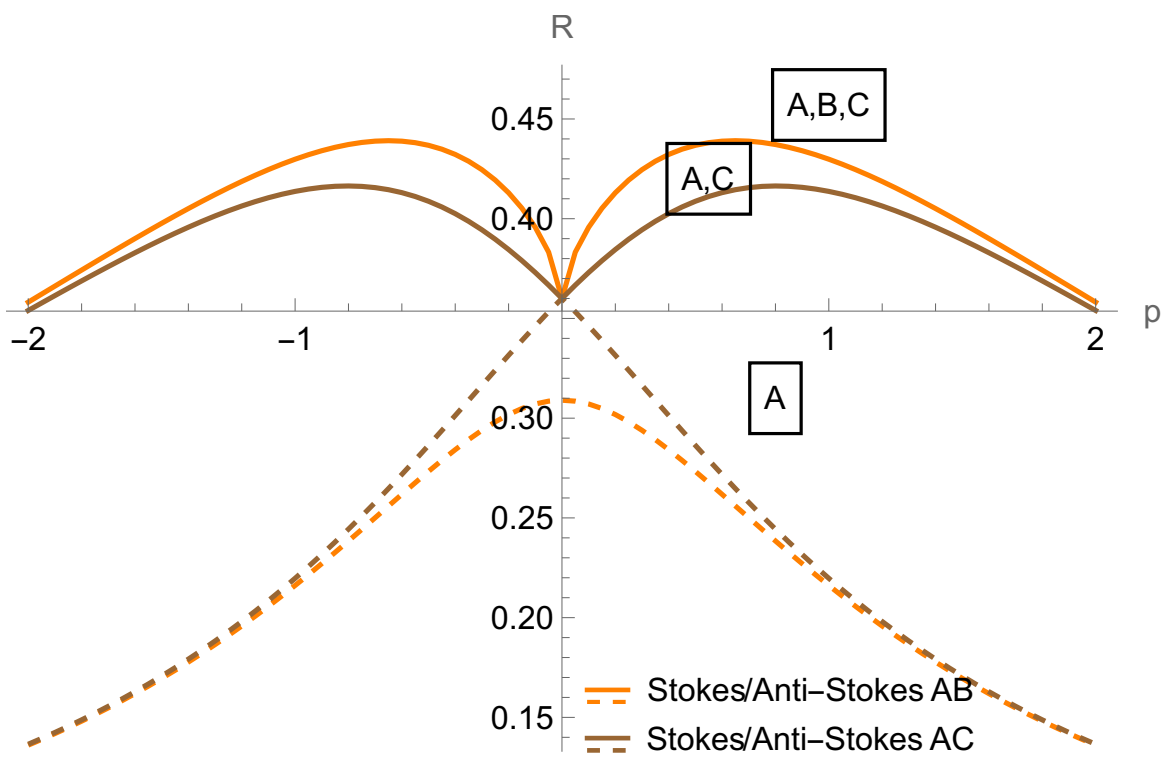

Figure 8.6: Stokes lines (solid) and anti-Stokes lines (dashed) for the colour switchover in parameter space $(p, R)$. The Stokes lines define regions, within which a specific set of saddle points contributes, here denoted with the respective labels in the boxes. The Stokes lines are symmetric in $p$, and coincide at the fold point at $R = R_{\text{fold}}$.

## 8.4 The total spectrum

The newly incoming ionisation events are relevant contributors to the total ionisation yield from a surprisingly early stage of the colour switchover. However, just because they are mathematically relevant to the sum (2.33) that does not mean they contribute a lot. From Eq. (2.33) we understand that the contribution of each individual saddle point is essentially governed by the imaginary part of the action as $|e^{-iS_{\text{ATI}}}| = e^{\text{Re}(-iS_{\text{ATI}})} = e^{\text{Im}S_{\text{ATI}}}$. For a first glimpse into what the (relative) contribution to the spectrum looks like, we therefore plot $\text{Im}(S_{\text{ATI}})$ for each saddle point across the colour switchover for all momenta, and hence in the full parameter space $(p, R)$. Two views of the resulting surfaces are shown in Fig. 8.7, in the $\omega$-centric classification scheme for the saddle points A (red), B (green) and C (blue). We find that the surface for saddle point B is below the one for C, throughout the switchover and for all momenta. Another surface of $\text{Im}(S_{\text{ATI}})$ is curved in similar shape, but a little bit above. Here this is attributed to saddle point C. The highest values of $\text{Im}(S_{\text{ATI}})$ are for saddle point A, which here remains almost flat across the whole switchover. The surface of $\text{Im}(S_{\text{ATI}})$ for saddle point D would be a rather flat sheet above that of A, which we omitted here for visual clarity. We can conclude that D gives the most dominant contribution to the total ionisation yield spectrum, followed by A. B is always be suppressed and C only contributes meaningfully at very low amplitude ratios $R$. However, C does not start contributing below $R_{\text{fold}} \approx 0.36$ anyway, so this can be neglected.

In this $\omega$-centric classification the surfaces for A and C can be identified as on top of each other, touching each other at $p = 0$ for $R > R_{\text{fold}}$. If we chose the $2\omega$-centric classification scheme instead, both surfaces would be symmetric w.r.t. $p = 0$ where they would intersect. The surface of saddle point A would be continuously declining from negative to positive momenta, whereas the surface for B would increase from negative to positive momenta.

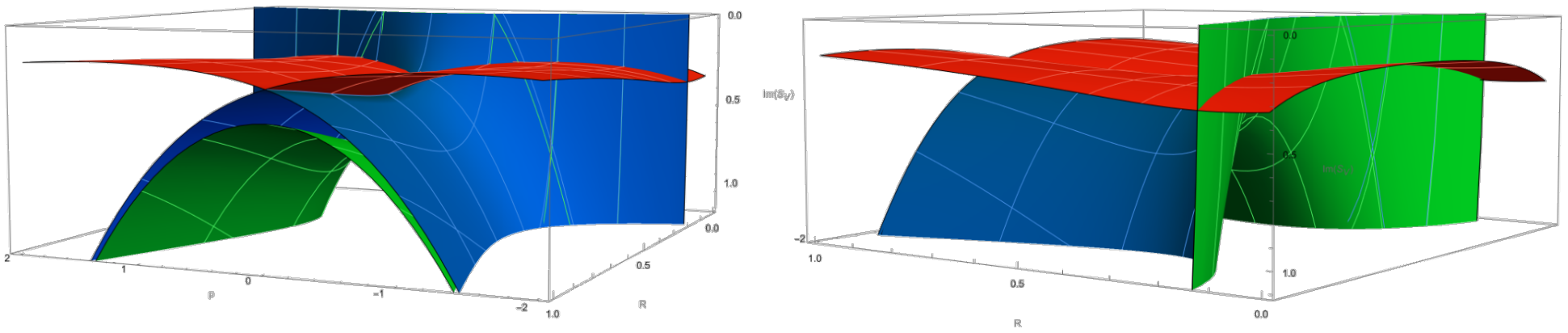

Figure 8.7: Imaginary part of the action $\mathrm{Im}(S_{\mathrm{ATI}})$ for each of the saddle points A (red), B (green) and C (blue) in their dependence of the parameters $R$ and $p$. Views from two different angles shall serve as aid to understand this 3D structure.

Generally, intersections of these $\mathrm{Im}(S_{\mathrm{ATI}})$ surfaces define the Stokes transitions. That is, the projection of the surfaces' intersections in Fig. 8.7 yields the Stokes map Fig. 8.6.

The total spectrum can then be calculated simply as a sum of all the relevant contributions Eq. (2.33), assuming each saddle point's contribution integrates as a Gaussian. We show the resulting spectrum at different stages of the colour switchover in Fig. 8.8. In there, we show the (possible) contributions from each saddle point as coloured markers, according to the $\omega$-based classification scheme. The shapes of the spectral amplitudes for each saddle point clearly resemble lineouts of the surfaces of $\mathrm{Im}(S_{\mathrm{ATI}})$ shown in Fig. 8.7. Generally, we find throughout the full switchover the total spectral amplitude is clearly dominated by the contributions from saddle point D. At the early stage of the switchover the contributions of saddle point A are of similar shape, albeit of less magnitude (see panel a). The contributions of saddles B and C are exponentially decreasing for $|p| > 0$, but at this stage is not included into the total ionisation yield anyway. Upon the further course of the switchover, the contributions of saddle points B and C flatten out more and more. From $R = R_{\mathrm{fold}}$ (panel (b)) onwards they are part of the total amplitude as well, albeit providing only small amplitudes.

The total ionisation amplitude $|\Psi|$, which is the coherent sum of the relevant individual contributions, is shown as a black line. Throughout the colour switchover the order of magnitude of $|\Psi|$ increases, which can be attributed to our particular choice of realisation of the colour switchover. Here, the switchover is performed by *adding* the second colour with gradually increasing amplitude, such that the total field strength gradually increases as well. This of course reflects in the magnitude of the total ionisation amplitude.

At $R = R_{\mathrm{fold}}$ (panel b) there is an artificial peak in the contribution of A for $p = 0$. This is because for this configuration the saddle points A and C coalesce and the second derivative in the denominator vanishes, rendering the total integral to diverge. In order to resolve this artificial discontinuity, we will study this scenario in a dedicated Chapter below.

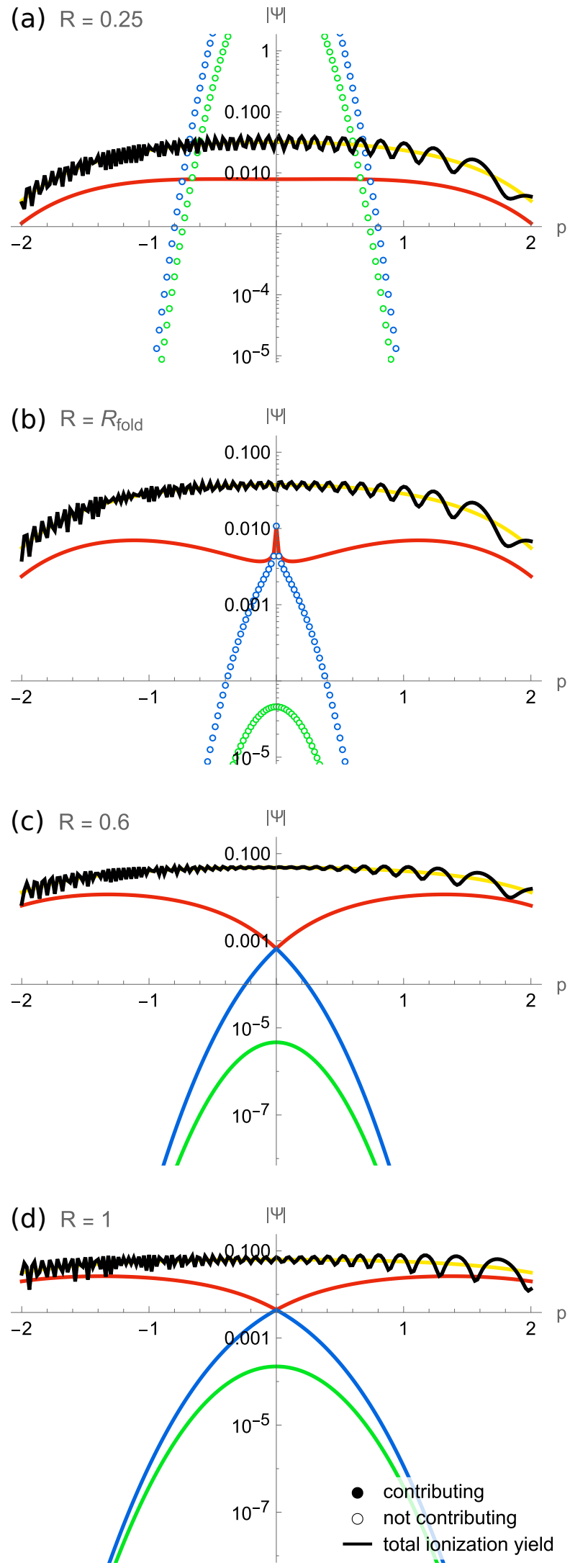

Figure 8.8: Ionisation amplitudes for $R = 0.25, 0.36, 0.6$ and 1 (a through d) for each saddle point, where lines mark relevant contributions and empty dots mark non-relevant contributions, coloured according to the $\omega$-centric classification scheme. The total magnitude of ionisation amplitude $|\Psi|$ is shown in black. Panel (b) shows an artificial discontinuity for the contribution of A (red) at $p = 0$.

# 8.5 The fold catastrophe point

## 8.5.1 Analytic derivation of the fold point

The point where two solutions coalesce acts as an organising centre to the topology of the saddle points and is therefore indispensable not only for the classification but also for determining the relevance to the total integral. The fold point for a one-dimensional function is defined by a vanishing first and second derivative of the action, and a non-vanishing third derivative. To identify the specific mixing angle $\theta$ and momentum $p$ for which this coalescence happens, and the corresponding ionisation time $\omega t$, we can numerically solve this system of equations using a conventional root solver. However, here we want to derive the expression for the specific mixing angle at which the coalescence happens analytically. This will allow us to understand its dependency on external parameters to the system and indicate a link to the canonical forms of catastrophe theory.

For convenience, we use the implementation of the colour switchover in which we *add* the second colour field with a variable amplitude ratio $R$, i.e., we write the field as

$$E(t) = \frac{E_0}{\omega}\cos(\omega t) - R\frac{E_0}{2\omega}\cos(2\omega t)\,. \tag{8.2}$$

We use $\omega t = \phi$ and $A_0 = \frac{E_0}{\omega}$, such that the vector potential[2] reads

$$A(\phi) = -A_0\sin(\phi) + R\frac{A_0}{2}\sin(2\phi)\,. \tag{8.3}$$

Furthermore, we use $\kappa = \sqrt{2\mathcal{I}_\mathrm{p}}$. Our goal is now to find an expression for $R = R_\mathrm{fold}$, such that

$$\frac{\partial S}{\partial \phi} = 0 \qquad \text{and} \qquad \frac{\partial^2 S}{\partial \phi^2} = 0, \tag{8.4}$$

where the action $S(\phi)$ and its derivatives with the above definitions are given as

$$\begin{aligned} S(\phi) &= \frac{1}{2}\kappa^2\phi + \frac{1}{2}\left(p^2\phi + 2p\int A(\phi)\mathrm{d}\phi + \int A^2(\phi)\mathrm{d}\phi\right) \\ &= \frac{1}{2}\kappa^2\phi + \frac{1}{2}\left(p^2\phi + 2p\left(A_0\cos(\phi) - \frac{1}{2}A_0 R\cos^2(\phi)\right)\right. \\ &\quad \left. + \frac{1}{96}A_0^2\left(12R^2\phi - 3R^2\sin(4\phi) - 48R\sin(\phi) + 16R\sin(3\phi) - 24\sin(2\phi) + 48\phi\right)\right) \end{aligned} \tag{8.5}$$

and

$$\begin{aligned} \frac{\partial S}{\partial \phi} &= \frac{1}{2}\kappa^2 + \frac{1}{2}\Big(p^2 + 2p\left(-A_0\sin(\phi) + A_0 R\sin(\phi)\cos(\phi)\right) \\ &\quad + \frac{1}{96}A_0^2\left(+12R^2 - 12R^2\cos(4\phi) - 48R\cos(\phi) + 48R\cos(3\phi) - 48\cos(2\phi) + 48\right)\Big) \end{aligned} \tag{8.6}$$

[2]The vector potential is defined as $A(t) = -\int E(t)\mathrm{d}t$.

as well as

$$\frac{\partial^2 S}{\partial \phi^2} = A_0 p \left(R\cos(2\phi) - \cos(\phi)\right) + \frac{1}{2}A_0^2 \sin(\phi)\left(\left(R^2+2\right)\cos(\phi) + R\left(R\cos(3\phi) - 3\cos(2\phi) - 1\right)\right) \tag{8.7}$$

respectively. We set $p = 0$, because we know the saddle point solutions are symmetric w.r.t. the sign of $p$, and we can assume $R \geq 0$ and $A_0 \geq 0$. Non-trivial solutions to the second derivative (8.7) are then given by

$$\{\phi_i\} = \left\{ -\mathrm{i}\ln\left( \pm\frac{1}{2}\sqrt{\frac{1}{4R^2}+2} \pm \frac{1}{2}\sqrt{\frac{1}{2R^2} - \frac{\frac{1}{R^3}+\frac{8}{R}}{4\sqrt{\frac{1}{4R^2}+2}} - 2 + \frac{1}{4R}} \right), -\mathrm{i}\ln\left(\frac{1\pm\sqrt{1-R^2}}{R}\right) \right\} \tag{8.8}$$

(and the respective solutions including integer multiples of $2\pi\mathrm{i}$ of course). From these, we select the branch with the "+" for the first term and the "-" for the second term, because this is the branch that yields feasible solutions (e.g., positive imaginary parts for $\phi$) within the range of the external parameters. That is, we use

$$\phi_2 = -\mathrm{i}\ln\left( +\frac{1}{2}\sqrt{\frac{1}{4R^2}+2} - \frac{1}{2}\sqrt{\frac{1}{2R^2} - \frac{\frac{1}{R^3}+\frac{8}{R}}{4\sqrt{\frac{1}{4R^2}+2}} - 2 + \frac{1}{4R}} \right). \tag{8.9}$$

in the first derivative and solve

$$\left.\frac{\partial S}{\partial \phi}\right|_{\phi=\phi_2} = 0 \tag{8.10}$$

for $R$. This again yields several branches from which we finally select

$$R_{\text{fold}} = \sqrt{1 + \frac{16\sqrt[3]{2}\kappa^4}{3A_0^4 X} - \frac{8\kappa^2}{3A_0^2} - \frac{36\sqrt[3]{2}\kappa^2}{A_0^2 X} + \frac{X}{3\sqrt[3]{2}}} \tag{8.11}$$

with

$$X = \frac{1}{A_0^4}\left(-729A_0^{10}\kappa^2 + 2160A_0^8\kappa^4 + 128A_0^6\kappa^6 + 3\sqrt{3}\sqrt{A_0^{14}\kappa^4\left(27A_0^2+32\kappa^2\right)^3}\right)^{1/3}.$$

We summarise the external parameters $A_0$ and $\kappa$ to a unified parameter $\Gamma = \frac{\kappa}{A_0}$ such that ultimately we can write

$$R_{\text{fold}} = \frac{\sqrt{-16\Gamma^2 + \frac{8\sqrt[3]{2}(4\Gamma^2-27)\Gamma^2}{X} + 2^{2/3}X + 6}}{\sqrt{6}} \tag{8.12}$$

with

$$X = \sqrt[3]{128\Gamma^6 + 2160\Gamma^4 - 729\Gamma^2 + 3\sqrt{3}\sqrt{\Gamma^4\left(32\Gamma^2+27\right)^3}}. \tag{8.13}$$

This expression gives the amplitude ratio $R_{\text{fold}}$ at which two saddle points coalesce, and which — as explained above — simultaneously marks the amplitude ratio from which on the newly incoming saddle points have to be taken into account. To obtain the location of the actual fold point one can solve the first derivative (8.6) which yields

$$\phi_{\text{fold}} = -\mathrm{i}\ln\left[\frac{1}{2R}\left(1 - Z + \sqrt{\frac{2\left(2R^2 - 4\Gamma R - 1\right)}{Z} - 24\Gamma R - \frac{\sqrt{Z}}{3\sqrt{3}} - 9}\right)\right] \tag{8.14}$$

with

$$\begin{aligned} Z &= \frac{1}{\sqrt{3}}\sqrt{\frac{R\left(8\Gamma^2 - 6R^2 + 6\right)}{Y} + 8\Gamma R + 2RY + 3} \\ Y &= \sqrt[3]{-8\Gamma^3 - 9\Gamma\left(2R^2 + 1\right) + 3\sqrt{3}\sqrt{-\Gamma^2 + R^6 + \left(8\Gamma^2 - 3\right)R^4 + \left(16\Gamma^4 + 20\Gamma^2 + 3\right)R^2 - 1}} \end{aligned} \tag{8.15}$$

and to be evaluated at $R = R_{\text{fold}}$. The mixing angle corresponding to the amplitude ratio is

$$\theta_{\text{fold}} = \operatorname{atan}\left(R_{\text{fold}}\right). \tag{8.16}$$

Note that we have made several assumptions, such as $p = 0$, that were specific to the scenario studied above. We believe, however, that a similar procedure can be followed to identify the coalescence points for more generic setups as well.

## 8.5.2 The spectral amplitude around the fold catastrophe

As mentioned earlier, the saddle-point approximation as such diverges when two saddle points coalesce because the second derivative in the denominator of the prefactor vanishes. This is visible as an artificial peak around $p = 0$ (where the coalescence happens) in Fig. 8.9 for the individual saddles' contribution of saddle A (red), and also reflected in the sum of all contributions (black, SPM). For comparison, we show the value of the integral evaluated along a contour that flew into the complex domain (blue, PLDF). Using the Picard–Lefschetz downward flow removes any artificial discontinuities and the resulting spectrum remains smooth even across momenta for which the saddle points approach each other, and ultimately coalesce.

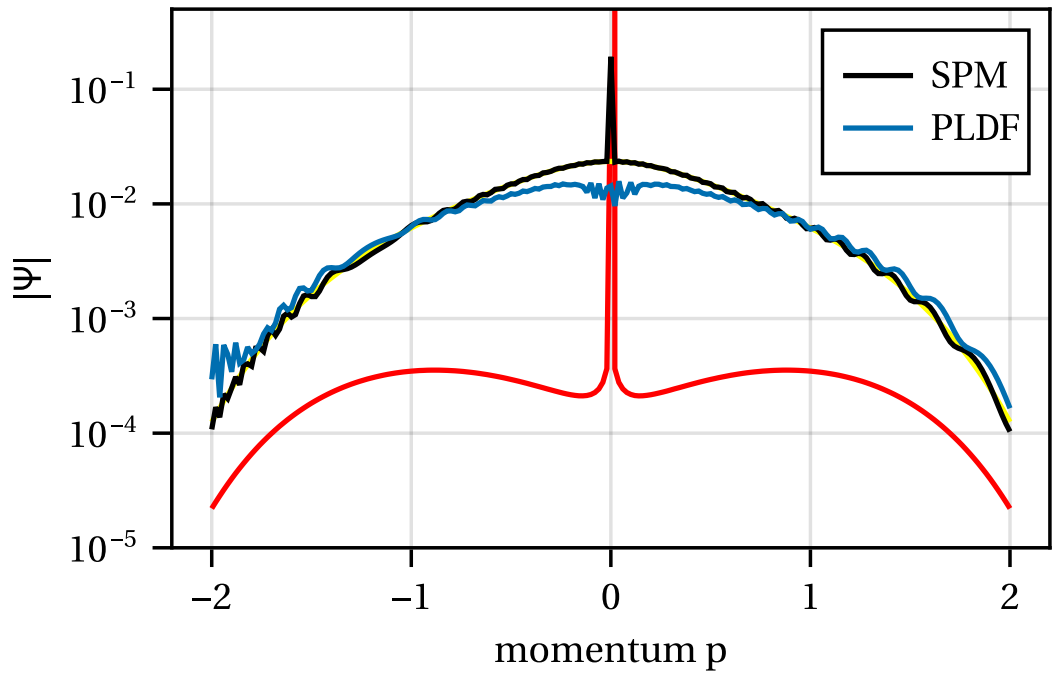

Figure 8.9: ATI spectrum as in Fig. 8.8 at the amplitude ratio $R = R_{\text{fold}}$, with individual saddle points' Gaussian contribution (saddle point D in yellow somewhat obscured by the black line, A in red), their coherent sum (black, SPM), and the integral evaluated along a contour flowed into the complex plane using the Picard–Lefschetz downward flow (blue, PLDF). The latter does not exhibit the artificial discontinuity at $p = 0$.

## 8.6 The colour switchover between $\omega$ and $3\omega$

Above we have studied the specific saddle point dynamics for the gradual replacement of a monochromatic $\omega$ field with its second harmonic — a field of frequency $2\omega$. We identified an interesting topological feature that determine the relevancy of individual saddle points and (spoiler alert) ionisation events that contribute from a surprisingly early stage of the colour switchover, at times when the electric field has not changed sign yet or even vanishes completely. However, in the total spectrum all those dynamics are disguised by the contribution of saddle point D stemming from the second half cycle of the field.

A possibility to "remove" these contributions is to consider a colour switchover from the fundamental field with frequency $\omega$ to its third harmonic — the field of frequency $3\omega$. This introduces a dynamic symmetry to the field shape, an example of which is shown in Fig. 8.10a for the case of equal amplitudes. For the colour switchover from $\omega$ to $3\omega$, the saddle point dynamics in the first half cycle remain unchanged compared to those between saddles A,B and C for the $\omega - 2\omega$ switchover. The second half cycle then follows the same scheme, just with a time-reversal symmetry. The respective saddle points for the colour switchover from $\omega$ to $3\omega$ are shown in Fig. 8.10b, showing analogous features as the structure in the first half cycle of Fig. 8.2.

Generally, the ratio of frequencies between component fields of a colour switchover provides a further external parameter. Exploring the relationship between dynamical symmetries of the driving laser field and the resulting saddle points is done in Ref. [23]. For this thesis, we focus on the colour switchover between $\omega$ and $2\omega$ fields because it is experimentally most applicable. Furthermore, we expect it to serve as a fundamental building block for studying switchovers of other frequency ratios.

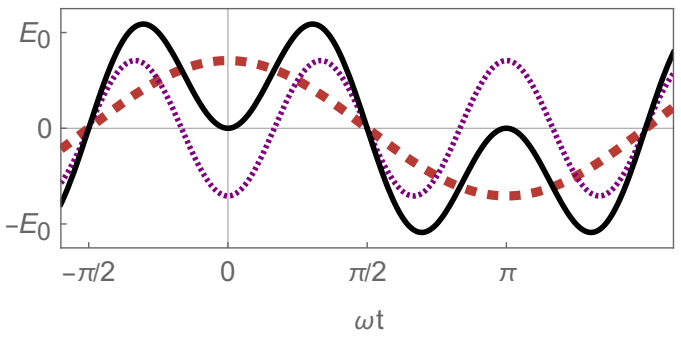

(a) Two-colour field (black solid line) composed of a fundamental field (frequency $\omega$, red) and its third harmonic (frequency $3\omega$, purple), here shown with equal amplitudes.

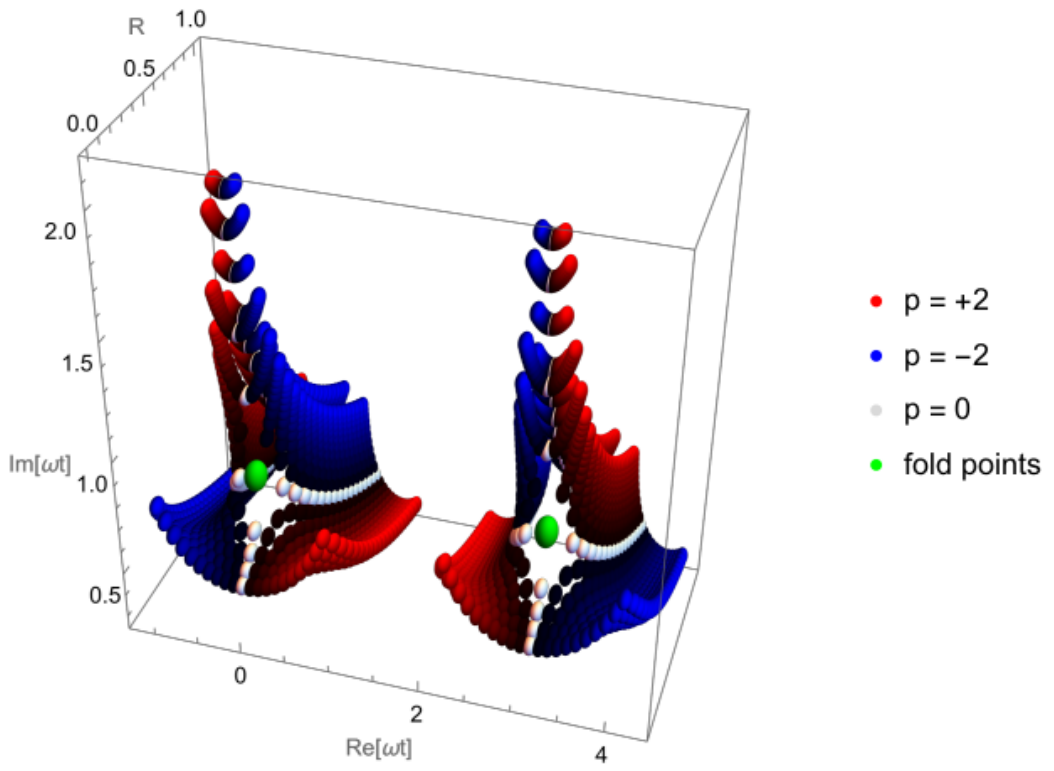

(b) Same as Fig. 8.2(d), but for the transition from a $\omega$ to a $3\omega$ field.

Figure 8.10: The colour switchover between $\omega$ and $3\omega$.

# 9

# Tunnelling without a barrier

During the study of ionisation events throughout the colour switchover we noticed a peculiar feature: as the 'newly incoming' ionisation events start contributing so early, one of them is contributing even in a scenario where both laser fields cancel each other out. At that time, the instantaneous electric field is zero and no tunnelling barrier is created — and yet we seem to have tunnel ionisation?!

In the following Chapter, we want to address some specific features of this intriguing tunnelling event. We dissolve the disconnect by addressing the nonadiabaticity of the tunnelling process itself, discuss the semi-classical features of this particular tunnelling event, and ultimately, explore proposals of how it might be detected in an experiment. The original finding was published in

[1] A. Weber, M. Khokhlova, and E. Pisanty, "Quantum tunneling without a barrier," *Phys. Rev. A* **111**, 043103 (2025).

Some of the results presented here were developed by the undergraduate students Tom Casey, Yann Jannssen, Jude Russell and Xin En (Lyra) So, during their 'third year research project' [189–192], which was supervised by Emilio Pisanty and me. We will credit the students' respective work within each section.

# 9.1 Introduction

In the previous Chapter we studied how new ionisation events emerge in a two-colour switchover scheme. We found that there are four relevant ionisation events in a field configuration where the two component laser fields cancel each other out and the total field amplitude is zero (see Fig. 9.1(a), highlighted with a green arrow). That is, for $R = R_\star = 1$, corresponding to mixing angle $\theta_\star = 45°$, the total field is

$$E_\star(t) = \frac{E_0}{\sqrt{2}} \left(\cos(\omega t) - \cos(2\omega t)\right) . \tag{9.1}$$

Using this as a driving field, there is a saddle point with $\mathrm{Re}(t_s) = 0$, where in fact the instantaneous electric field $E(t) = 0$, and hence no tunnelling barrier is created. This saddle point (labelled 'B') is unambiguously part of the integration path as shown in Fig. 9.1(b). Moreover, ionisation event B contributes even earlier in colour switchover, i.e., between $R_{\mathrm{fold}} < R < R_\star$, where the electric field hasn't changed sign yet, such that the tunnelling happens 'uphill'. This implies that the tunnelling at zero field is no 'exception', but actually a stable feature of the colour switchover scheme. As exciting and counter-intuitive as this is, the contribution to the spectral amplitude of this peculiar tunnelling event is vanishingly low, as can be seen from the spectrum Fig. 9.1(c).

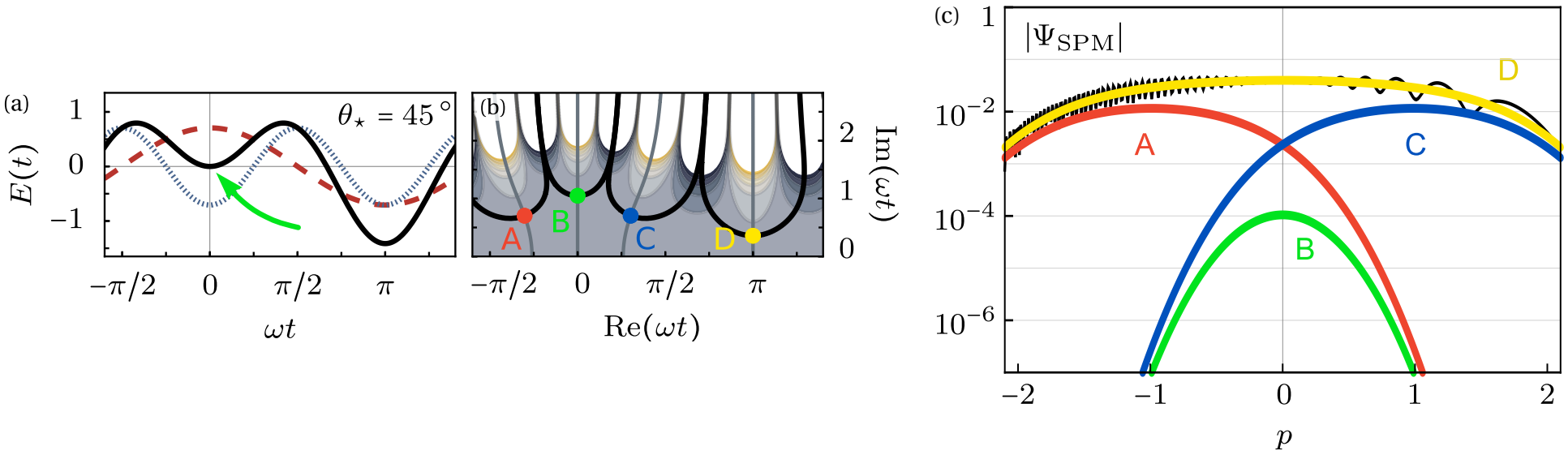

Figure 9.1: Tunnelling-without-barrier ionisation: (a) Two-colour driving field with equal amplitudes (mixing angle $\theta = 45°$, as given in Eq. (9.1)), such that the two field components cancel each other out at $t = 0$ (highlighted by the green arrow). (b) The respective action landscape with relevant saddle points A,B,C,D, same as in Fig. 8.1. Saddle point B clearly is part of the integration path (heavy black line). (c) Magnitude of the spectral ionisation amplitude $|\Psi_{\mathrm{SPM}}(p)|$ (black), and the contribution $|\Psi^s_{\mathrm{SPM}}(p)|$ of each of the four ionisation events shown in (b). For $p = 0$, the contribution of B (green) stems from the ionisation event at zero field. Adapted from [1].

# 9.2 Nonadiabaticity of strong-field tunnelling

This tunnelling event at vanishing electric field is highly counter-intuitive because our understanding of strong-field tunnel ionisation is based on a quasi-static barrier. That is, the ionisation rates that underpin the SFA are derived assuming adiabaticity with a fixed barrier for the duration of the tunnelling process. This phenomenon of tunnelling without barrier can be understood (or rather: the disconnect can be resolved) by remembering that the barrier does in fact change during the tunnelling process, as visualised in the sketch 9.2.

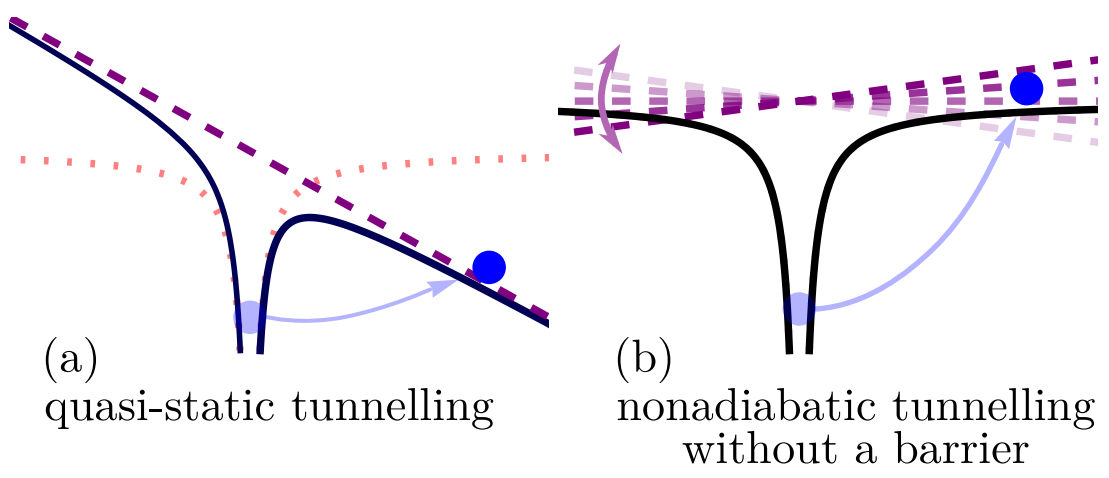

Figure 9.2: Sketch of field-induced tunnelling, with the laser field (purple dashed), the atomic binding potential (pink dotted) and the resulting barrier (black solid) in the (a) quasi-static approach, and (b) nonadiabatic regime in which the laser field changes during the process, from [1].

Within the SFA in the quantum-orbit picture, 'the ionisation time' refers to the saddle point time, which is a complex number. The real part of which is interpreted as the time the electron appears in the continuum, while the imaginary part is the time 'before that', which the electron spends inside the tunnelling barrier [38]. For our peculiar tunnelling with zero field we find the electric field indeed vanishes at the saddle-point time $t_s$. However, the electric field is non-zero at the time 'before' that: In Fig. 9.3 we show the electric field evaluated between $\mathrm{Re}(t_s) + \mathrm{i}\,\mathrm{Im}(t_s) < t < \mathrm{Re}(t_s)$ for saddle point B in the configuration with $\theta = 45°$. Note the plot is to be read from left to right — i.e., the imaginary time axis leads from the tunnel entry (on the left-hand side) towards the tunnel exit (where $\mathrm{Im}(t_s) = 0$) on the right-hand side. We find that the amplitude of the electric field decreases during that time, but is presumingly sufficient to initialise the tunnelling at $t = t_s$.

To illustrate this further we want to showcase the electron trajectories inside this tunnelling barrier. This work was mainly done by student Yann Jannssen. The displacement of the electron is given by Eq. (2.49).

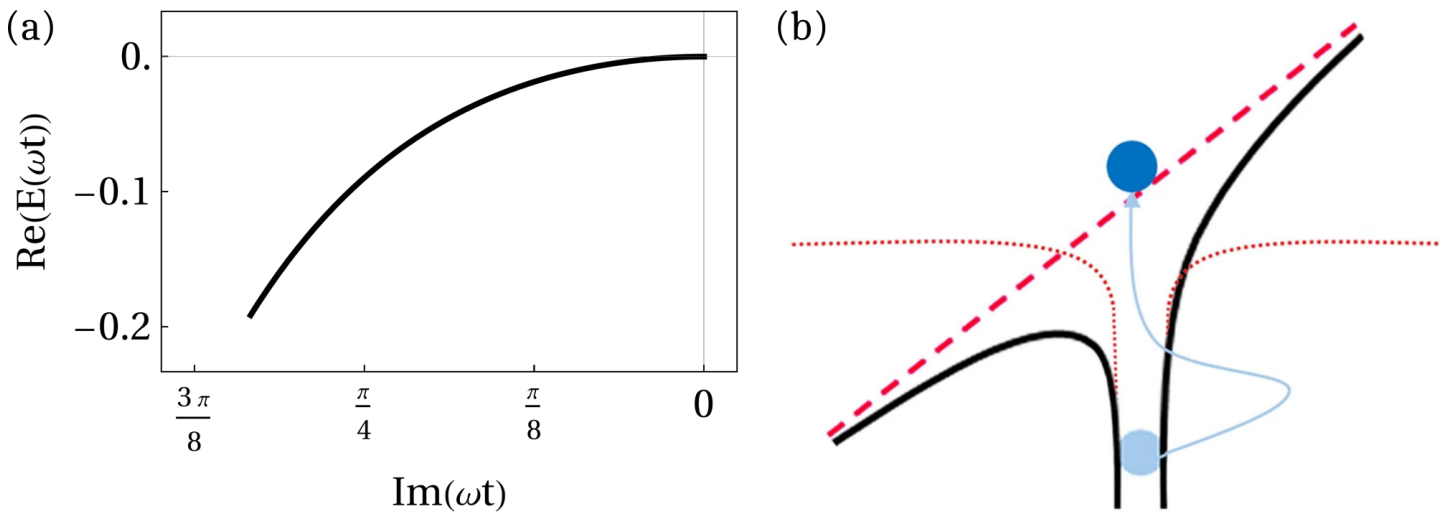

Figure 9.3: (a) Electric field during the tunnelling process for ionisation event B, $\mathrm{Re}(\omega t) = \mathrm{Re}(\omega t_s)$. (b) Sketch of the trajectory inside the tunnel.

As the imaginary part of the saddle point is interpreted as the time the electron spends inside the tunnel, we can calculate the displacement for this part of the time contour and interpret it as the electron's trajectories inside the tunnel. In Fig. 9.4 we show these trajectories inside the tunnel for saddle point B for several values throughout the colour switchover. In each panel, the trajectories are shown for momenta between $|p| = 2$ (red) and $p = 0$ (black). The trajectories are plotted with the imaginary time decreasing from left to right, so that the electron enters the tunnel on the left-hand side of the panel and then appears in the continuum at $\mathrm{Im}(t_s) = 0$ on the right-hand side of the plot. Let us look at the trajectory for a 'normal' ionisation event first, towards the end of the colour switchover where $\theta = 80°$ (panel (e)). Here we find the electron simply gets driven towards one side of the core, namely $\mathrm{Re}(x(\omega t)) > 0$, where it subsequently leaves the tunnel. The trajectories for the other saddle points show similar behaviour (not shown). For earlier stages of the colour switchover (panels (a)-(d)), however, the trajectories lead in one direction and — while still inside the tunnel — slow down and change direction. For a very weak second colour field (panels (a) and (b)) the electron even leaves the tunnel on the other side of the core. These classical trajectories of course should take into account the effects of the Coulomb potential to be more meaningful. For now, they do provide a cute illustration of the process.

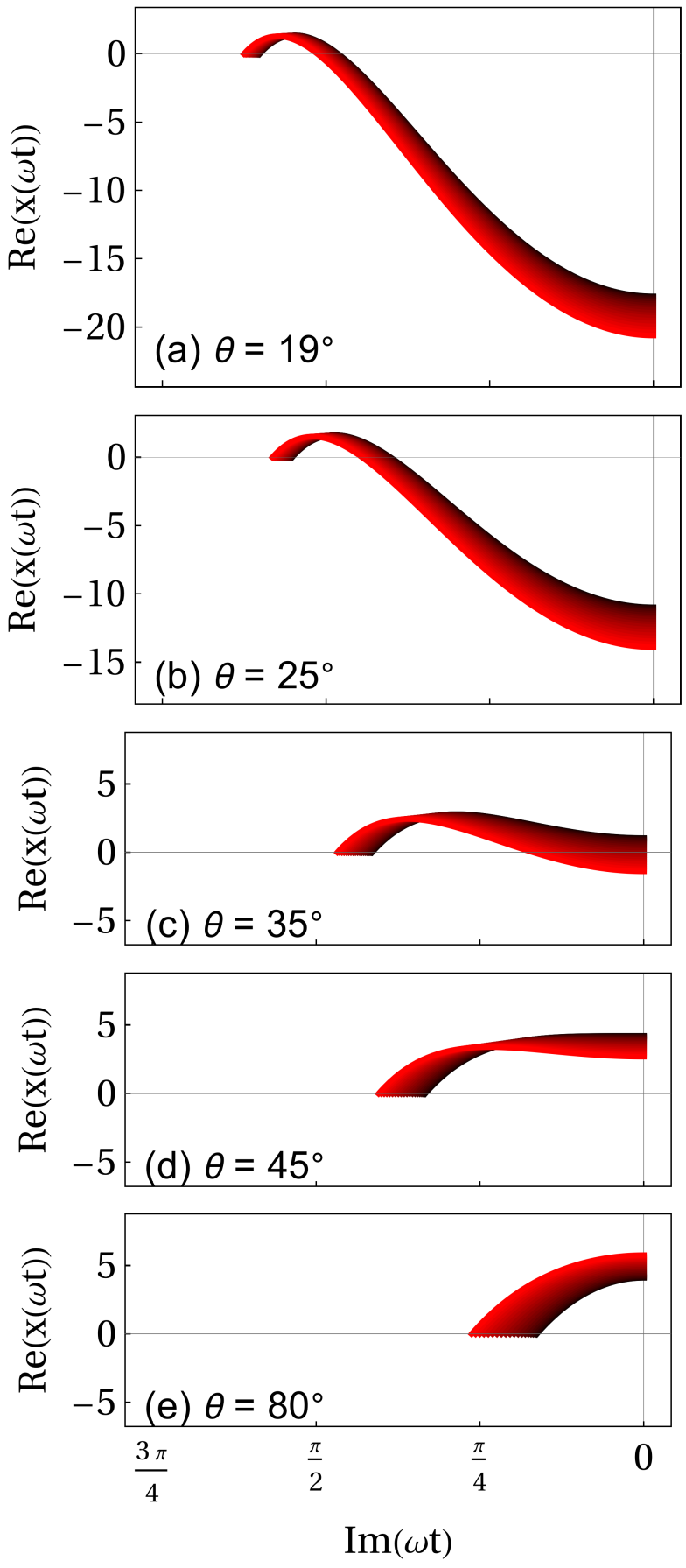

Figure 9.4: Classical electron trajectory during the tunnelling process for ionisation event B throughout the switchover, shown for the mixing angle $\theta$ from 15°, through $\theta_{\mathrm{fold}} \approx 19°$, 25°, 35°, 45° to 80°, (a-f) respectively. The colour indicates momenta between $|p| = 2$ (red) and $p = 0$ (black). Note the $x$-axis has decreasing values of $\mathrm{Im}(\omega t)$ from left to right, so that the electron appears in the continuum at $\mathrm{Im}(\omega t) = 0$ on the right-hand side of the plot.

Within attosecond science, there have been several theoretical and experimental studies to highlight that the process of tunnel ionisation is in fact non-adiabatic. It was found that the change of the laser field during the tunnelling process impacts the electron trajectories and the created harmonic radiation in a measurable way [175]. The tunnelling event at zero field presented here marks the epitome of the nonadiabaticity of strong-field ionisation.

We want to explore the relevance of this tunnelling event depending on the Keldysh adiabaticity parameter $\gamma$. In Fig. 9.5 we show how the total contribution of each of the saddle points changes, as we vary the wavelength of the driving laser field, resulting in a range of Keldysh parameters. We find that, as expected, while the total contribution of ionisation events A,C, and D remain pretty much constant over a large range of wavelengths, the contribution of B decreases for shorter wavelengths.

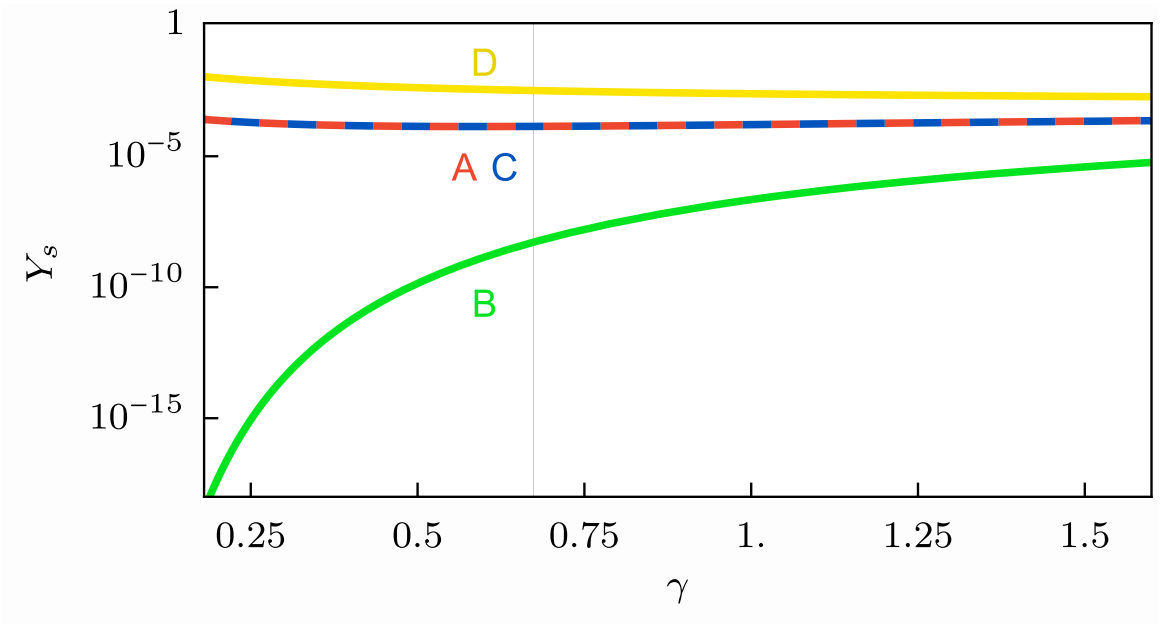

Figure 9.5: Scaling of the total ionisation probability per orbit, for a field shaped like the one shown in Fig. 9.1(a), for a range of Keldysh parameters $\gamma = \sqrt{4\omega\mathcal{I}_\mathrm{p}/5I_0}$, as a function of $\omega$ for constant $\mathcal{I}_\mathrm{p}$ and $I_0$, over the wavelength range $330\,\mathrm{nm} \leq \lambda \leq 3000\,\mathrm{nm}$. The configuration used for Fig. 9.1 yields $\gamma = 0.67$ (grey vertical line). Note that the total contributions of A and C are equal. From [1].

As described in Sec. 8.3 above, the saddle point starts contributing from the coalescence point on. It is therefore illusive to investigate how this coalescence point depends on the adiabaticity parameter. In Fig. 9.6 we show how the amplitude ratio $R_\mathrm{fold}$ scales with the Keldysh parameter, to be deduced from the derivation in Sec. 8.5. The electric field is zero for a two-colour field as defined above, Eq. (9.1)(that is, with the specific phase shift), independently of the Keldysh parameter. To check whether saddle point B remains a contributor to the spectrum in the equal-amplitude configuration therefore means to check whether the coalescence happens before $R = 1$. Fig. 9.6 unambiguously demonstrates that this is indeed the case. The coalescence amplitude ratio $R_\mathrm{fold}$ decreases monotonically with the Keldysh parameter (solid line). The dependence can be well approximated by the asymptotes $1 - \sqrt[3]{\frac{135}{32}}\,\gamma^{\frac{3}{2}}$ in the small-wavelength limit and by $\frac{1}{4\gamma}$ in the long-wavelength limit (dashed and dotted lines respectively). Most importantly, Fig. 9.6 shows that $R_\mathrm{fold} < 1$ for all $\gamma > 0$ (shaded region), and the saddle point B at the equal-amplitudes line $R = 1$ is always part of the integration contour. Hence we conclude that the ionisation event at zero field is a topologically stable feature of the strongly bichromatic driving field.

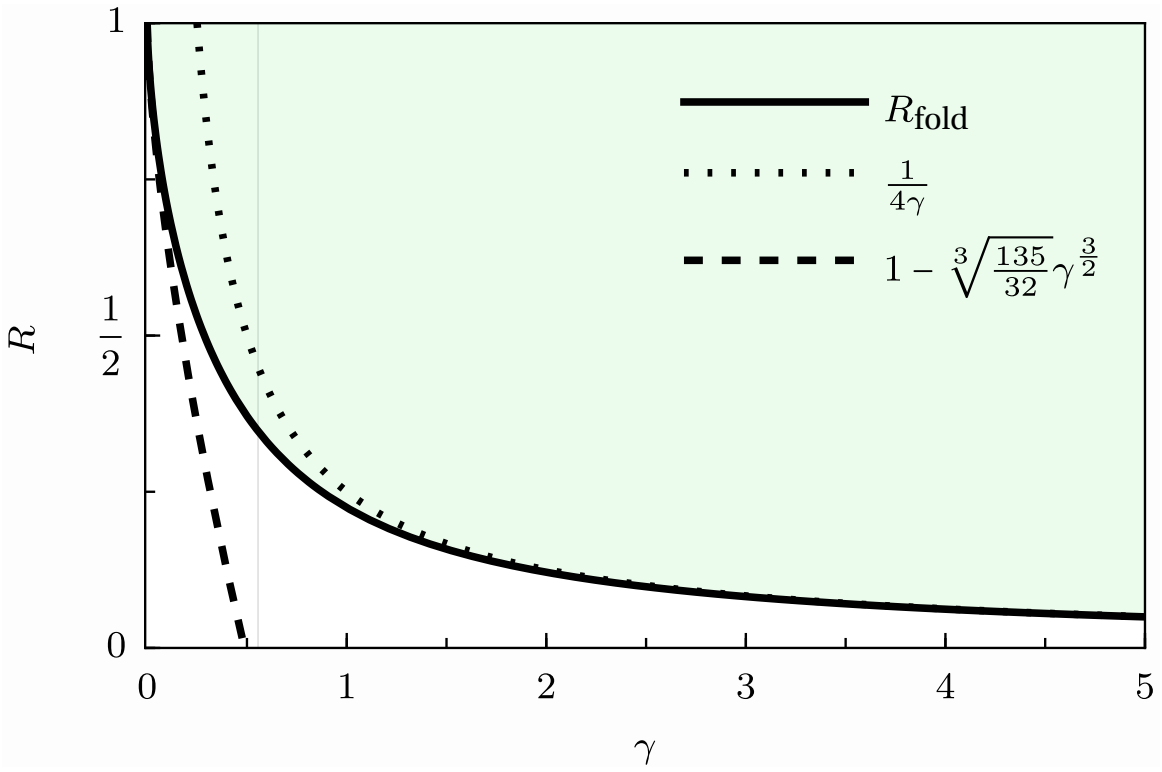

Figure 9.6: Scaling of the amplitude ratio $R_{\text{fold}}$ at which the saddle-point coalescence happens, over a range of Keldysh parameters $\gamma$. The shaded region shows the parameters for which saddles B and C contribute to the integration contour. In dashed lines, asymptotes for the behaviour in the $\gamma \ll 1$ and $\gamma > 1$ regime are shown. The grey line marks the configuration used in Fig. 9.1, for which $\gamma = 0.67$ and the coalescence happens at $R_{\text{fold}} \approx 0.36$, corresponding to $\theta_{\text{fold}} \approx 19°$. From [1].

## 9.3 Semi-classical aspects and avoiding Coulomb effects

Once the electron has tunnelled through the barrier and appears in the continuum we typically treat it as a fully classical particle which is accelerated by the driving laser field. To see whether an electron that escaped out of a non-existing barrier behaves any different once it roams freely in the continuum, we plot those classical trajectories in Fig. 9.7. That is, for the $R = 1$ scenario we show the electron's displacement from the origin outside the tunnel when $t > \text{Re}(t_s)$. The shaded bands in Fig. 9.7 indicate the trajectories with small, but non-zero momentum. We find that there is nothing peculiar about the saddle point B in comparison to A,C, and D. Nevertheless, it reveals that trajectories for A,B and C all spend a long time in the vicinity of the core. This is problematic in so far as that the theoretical model we employ does not consider the effects of the Coulomb potential. In fact, ignoring the effects of the atomic core is very much the key assumption of the SFA. With the electron trajectories hovering in such small distance from the core, however, we expect the Coulomb potential to play a non-negligible role.

Now of course one could refrain from using the SFA and instead solve the TDSE directly. However, this would be computationally more demanding while not giving the same time-resolved insights as the combination of SFA and saddle-point methods. The ionisation amplitude for the tunnelling at zero field would possibly need to be isolated via a Gabor transform, in a time window around the zero field. Alternatively, there exists a variety of 'Coulomb-corrected' approaches to specifically include the Coulomb effects into the SFA [24, 116, 193–195]. For example, they add the Coulomb force explicitly as a driver of the electron's motion in the continuum resulting in Coulomb–Volkov wave functions. This allows to correctly resolve the low-energy structures discussed in the earlier Sec. 3.3.1.

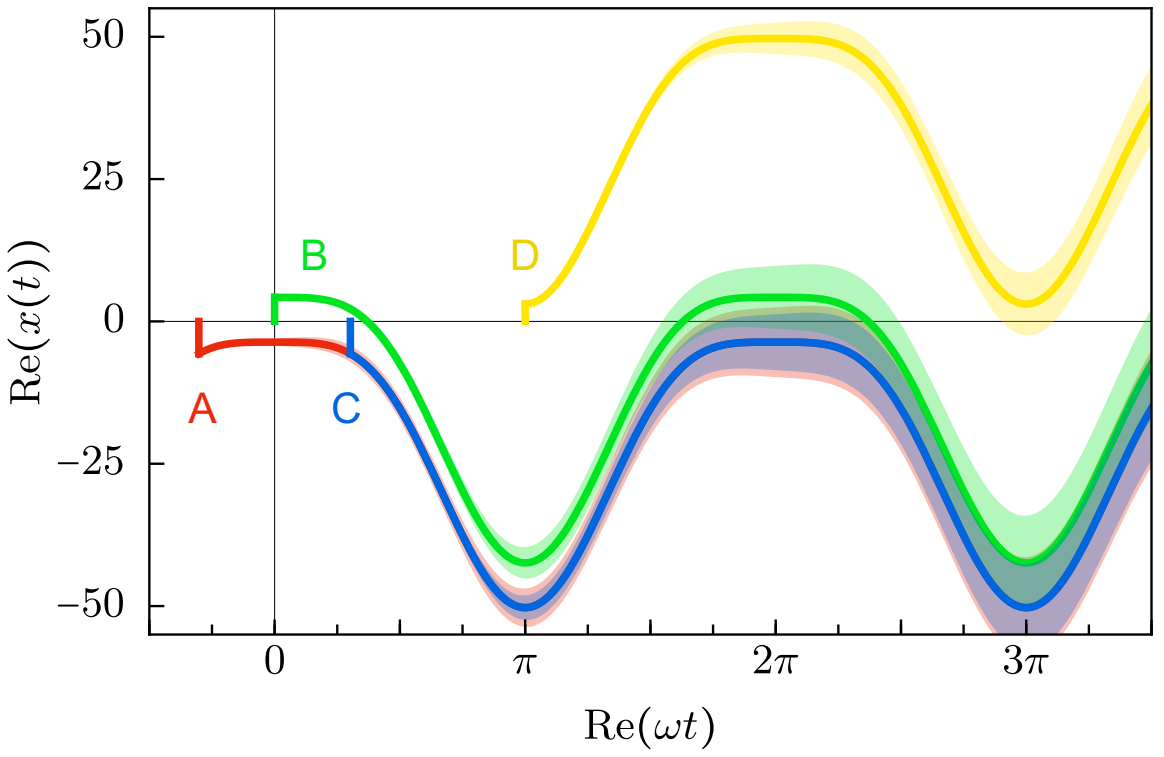

Figure 9.7: Semi-classical trajectories Eq. (2.3) for the field with the four labelled ionisation events shown in Fig. 9.1(b). The semi-transparent bands show the trajectories for solutions with small drift momentum, $|p| < 0.05\,\text{a.u.}$

In our case, however, the more important question is: What would we want from a Coulomb-corrected calculation anyway? As indicated in Fig. 9.1, the ionisation amplitude associated with saddle point B has no noticeable impact on the total spectral amplitude. Even though accounting for the Coulomb force might increase the overall ionisation amplitude, it would do so for all trajectories similarly. Nevertheless, this will be an interesting research project in the future.

For now, we choose a different approach. We discovered the tunnelling-without-barrier event for the case of zero electron momentum and two superimposed fields with equal amplitude that cancel each other out perfectly at the location of the saddle point. The precise location of the saddle points depends smoothly on the field parameters and the electron momentum. Therefore we might as well identify a field configuration that features a tunnelling event at zero field for which the subsequent trajectory *does not* hover nearby the core. By modifying the field configurations w.r.t. amplitude ratio, two-colour phase shift and momentum, our student Lyra So explored whether there would be such a tunnelling event. In Fig. 9.8 we show a selection of her findings. All three configurations of driving field and respective electron momenta show ionisation events that occur at a time when the electric field is zero, and whose resulting trajectories lead away from the core and are therefore less susceptible to Coulomb effects.

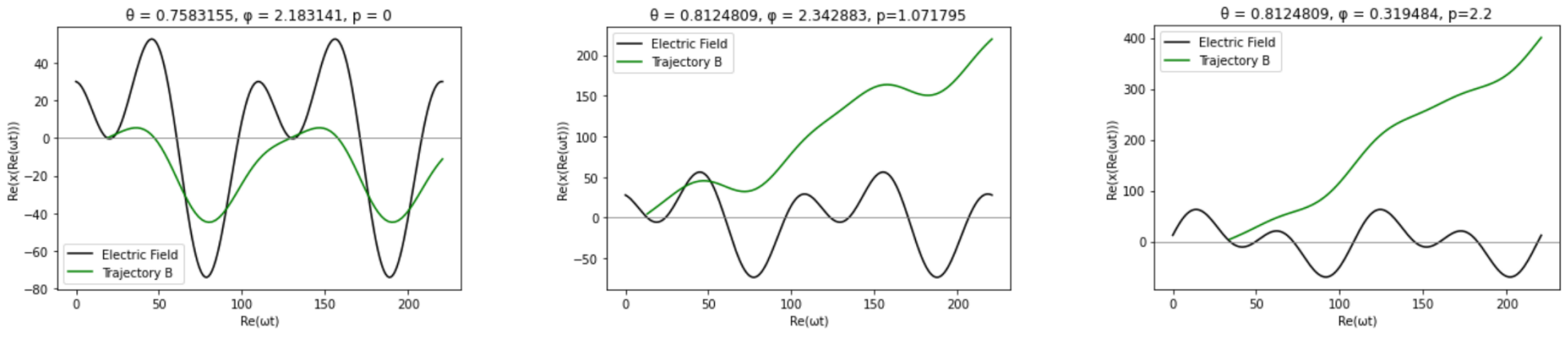

(a) $\theta \approx 43°, \phi \approx 125°, p = 0$ (b) $\theta \approx 46°, \phi \approx 134°, p = 1.07$ (c) $\theta \approx 46°, \phi \approx 18°, p = 2.2$

Figure 9.8: Several configurations of electric fields (black line) that, for a specific momentum $p$, feature an ionisation event at a time, when the instantaneous field amplitude is zero. The resulting trajectories originating from these events are shown in green. Taken from Lyra So's report [192].

## 9.4 Observation using phase-of-the-phase spectroscopy?

For the driving field parameters and target chosen above (and in our publication [1]), the contribution of the saddle point corresponding to tunnelling without a barrier is a few orders of magnitude below the contribution of the other saddle points, see Fig. 9.1(c). Directly observing tunnelling without a barrier solely through measuring a spectrum is therefore not feasible. Luckily however, attosecond science offers a variety of specialised measurement schemes that have been developed precisely to isolate and identify such subtle contributions. Our student Tom Casey explored several of these approaches in order to identify a suitable scheme to detect a significant contribution of the tunnelling without a barrier. A common technique is to add a perturbative field in orthogonal polarisation direction. Analysis of the resulting holographic photoelectron momentum distributions can allow to extract information about contributions from individual orbits, as briefly alluded to in Sec. 2.1.2. For our specific setup, however, this approach does not provide any detectable signature of saddle point B.

A more indirect way of separating out the contribution from several contributing trajectories is phase-of-the-phase spectroscopy [196–198]. The idea is the following, also shown in Fig. 9.9. Upon a scan of the phase shift $\varphi$ between the driving field and an added perturbation, the ionisation yield for a given electron momentum changes periodically. This change itself might not be sufficiently large to be detectable itself. E.g., adding a perturbative $4\omega$ field did not lead to any noticeable changes in the total yield structure. However, we may analyse the change in terms of a Fourier series, where in fact the first order terms might be sufficient. That is, we may write the total yield for a given momentum $p$ and at a given phase shift $\varphi$ as a sum over contributions from several relevant trajectories $s$ as usual, and express this in terms of a single Fourier component:

$$\Psi(p,\varphi) \approx \sum_s \sqrt{\frac{2\pi}{\mathrm{i}S''}}\,\mathrm{e}^{-\mathrm{i}S(p,\varphi,t_s)} \approx A_1 \mathrm{e}^{\mathrm{i}\phi}\,. \tag{9.2}$$

The phase $\phi$ and amplitude $A_1$ of this Fourier component can now be analysed for the several trajectories, or summations thereof.

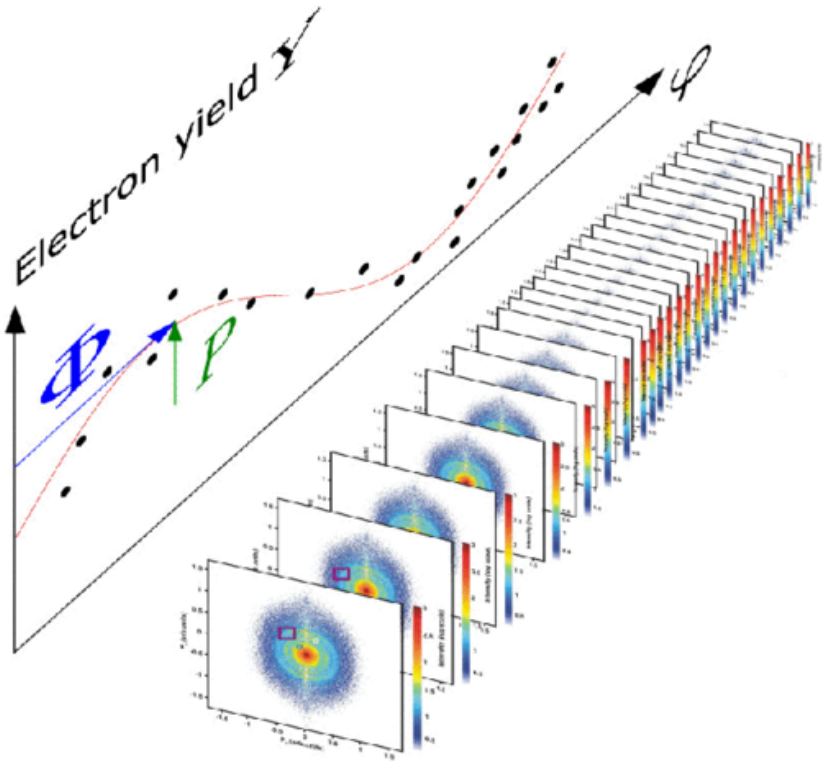

Figure 9.9: Schematic illustration of the phase-of-the-phase spectroscopy: A series of PEMDs is taken upon a scan of the relative phase $\phi$. The electron yield of a given momentum (small purple square) is fitted with the first order of a Fourier series, i.e., a trigonometric function with amplitude $P$. From [196].

In Fig. 9.10 we show the results for the case of the tunnelling without barrier, as done by our student Tom. In order to identify a possible signature of the ionisation event that happens at a time without barrier, the phase-of-the-phase spectroscopy is performed for the total ionisation yield $\Psi$ including this peculiar ionisation event ('with B') and neglecting it ('without B'). The added perturbative field that is used with a phase shift against our usual equal-amplitudes two-colour field Eq. (9.1) is $E_y(t) = E_3 \cos\big(4\omega t + \phi\big)$, where $E_3$ is $0.03E_0$ and the field is polarised along the orthogonal direction. Fig. 9.10(a) shows the complex phase $\phi$ (hue shows the phase, brightness shows the amplitude) of the Fourier component of the total ionisation yield across the momentum plane $(p_x, p_y)$. In the top panels the ionisation event B is taken into account when summing over the several contributions, in the bottom panel it is neglected. For comparison, the left-hand side panels show the case of equal amplitudes ($\theta = \pi/4$) of the driving field, whereas on the left-hand side the second field is stronger ($\theta = 2\pi/5$). While indeed for an increased mixing angle the contribution of saddle point B becomes more distinguishable, the electric field at the corresponding ionisation event in this case is not zero. And hence, it is no situation of zero-field tunnelling any more. However, the zero-field case on the left-hand side does actually yield a noticeable difference: In panel (b) we show a line out of the complex phase plot at $p_y = 0.2$ and plot the amplitudes $A_1$ for the Fourier components for the total ionisation yield with (red line) and without (blue line) ionisation event B. Now this line out shows a miniscule difference in amplitude between the two ionisation amplitudes. Note that these results are for a slightly modified Keldysh parameters compared to the original publication. As shown in Fig. 9.5, the (relative) contribution of the individual saddle points depends largely depends on the Keldysh parameter. Tom Casey therefore optimised the setup parameters such as to yield a noticeable difference between 'with B' and 'without B' at all. The used parameters are $E_0 = 0.106\,\text{a.u.}$, $\omega = 0.082\,\text{a.u.}$ and $\mathcal{I}_\text{p} = 0.53\,\text{a.u.}$, where the decreased frequency seemed to be key in identifying an observable difference.

In conclusion, this method provides a signature of the tunnelling event that happens at a time without barrier in the total spectrum. While this signature is arguably tiny, phase-of-the-phase spectroscopy seems to be a promising tool to potentially make out a distinguishable contribution of this tunnelling event without barrier.

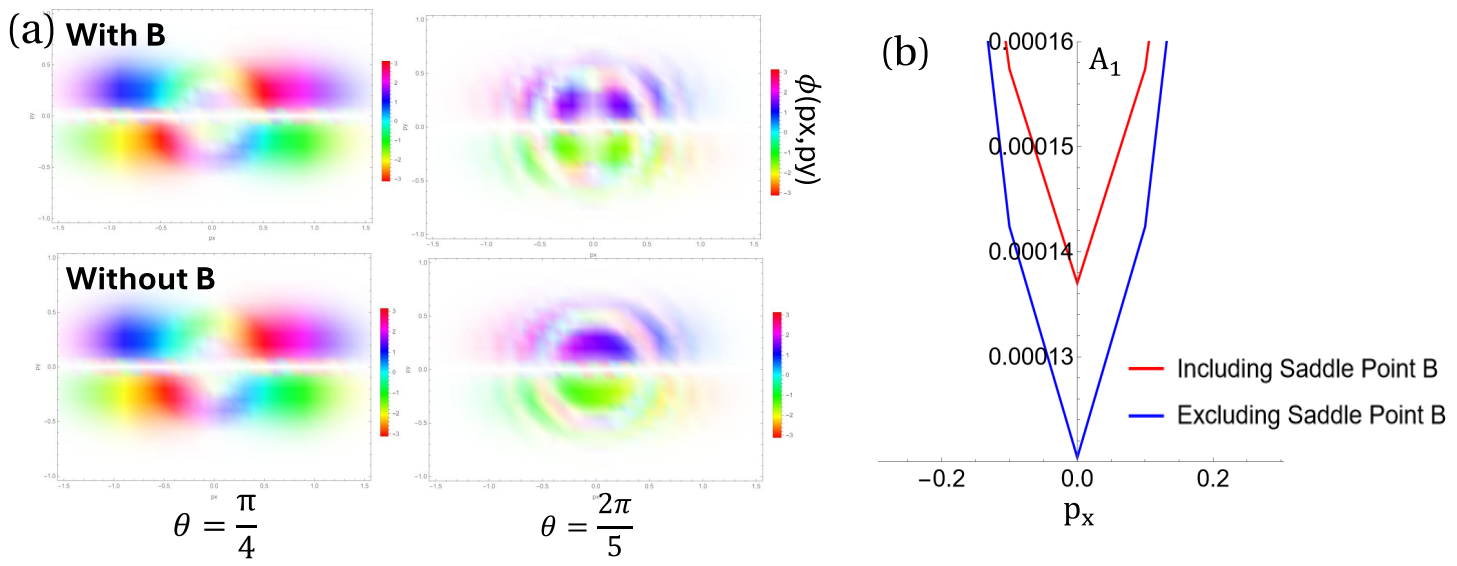

Figure 9.10: Results of the phase-of-the-phase spectroscopy for the tunnelling-without-barrier scenario. Explanation in the main text. From Tom Casey's report [189].

## 9.5 Intensity scans in HHG — a route to observe tunnelling without barrier?

In the previous sections we investigated the phenomenon of tunnelling without a barrier in the context of strong-field ionisation. A natural next step is to examine whether electrons released at times when the instantaneous electric field vanishes can return to the parent ion and contribute to the generation of high-order harmonics. Owing to its intrinsic three-step nature — ionisation, propagation, and recombination — HHG involves a considerably richer dynamic structure than ionisation alone. This additional complexity makes HHG a particularly sensitive probe of subtle strong-field and subcycle effects.

In the following we briefly review intensity-scan techniques in HHG experiments, which have proven particularly useful to identify the interference of distinct quantum paths. Building on this approach, we then apply such scans to HHG driven by a two-colour field, with the specific aim of isolating the contributions of electron trajectories that tunnel into the continuum at times when no potential barrier is present.

### 9.5.1 Intensity scans in HHG

#### Intensity scans and quantum-path interference

The fundamental intensity of the driving field, denoted as $I_0$, has a well-understood impact on the HHG spectrum. The intensity determines the ponderomotive energy of the driver and so directly enters the estimate of the high-order harmonic cutoff Eq. (2.5). With that, the overall shape of an HHG spectrum — namely strong harmonics at the ionisation threshold $\mathcal{I}_p$, a plateau spanning a long range of harmonic orders, followed by a sharp drop in intensity at the cutoff — is 'stretched' horizontally by the fundamental intensity of the driver.

For harmonics within the plateau and otherwise fixed configuration, a scan of the fundamental intensity therefore effectively corresponds to a scan over several harmonic orders. Consequently, just as interference fringes appear across harmonic orders in a conventional HHG spectrum, analogous interference patterns emerge in the signal of a fixed harmonic order when the driving-field intensity is varied. Intensity scans have therefore become an established and insightful tool for disentangling the contributions of different electron trajectories and for demonstrating their interference [184].

In practice, the intensity can be controlled through various parameters of the laser setup. Experimentally, measuring the yield of a given harmonic order as a function of the driving field intensity reveals characteristic oscillations, which can be directly attributed to quantum-path interference. This is what we refer to as an intensity scan. Although such measurements are macroscopic — since they include the propagation of the harmonic radiation into the far field — the observed behaviour can be consistently explained in terms of the microscopic strong-field approximation. In particular, the individual electron trajectories exhibit distinct intensity dependences, an aspect that will be discussed in detail below.

Within the microscopic SFA picture, each trajectory contributes with the (real part of the) semi-classical action as a dipole phase Eq. (2.50). For monochromatic driving fields, the dipole phase is often assumed to depend linearly on the intensity, such that the dipole for a given saddle point can be written as $\mathbf{D}_s(q\omega) \approx \mathrm{e}^{-\mathrm{i}\alpha I_0}$, where $\alpha$ is the phase coefficient of the respective trajectory. Considering the semi-classical action $S_{\mathrm{HHG}}$ in its dependency on the fundamental intensity, $S_{\mathrm{HHG}}(t_{\mathrm{i}}, t_{\mathrm{r}}) = S(I)$, we can derive $\alpha$ from a Taylor expansion about $I_0$:

$$S(I) \approx S(I_0) + \left.\frac{\partial S}{\partial I}\right|_{I=I_0} \cdot (I - I_0) \tag{9.3}$$

By writing $\left.\frac{\partial S}{\partial I}\right|_{I=I_0} = \alpha$, and absorbing constants into $\tilde{S}$, we may simplify this to

$$S(I) \approx \tilde{S}(I_0) + \alpha I \tag{9.4}$$

such that in the integrand of the dipole we obtain the term

$$\mathrm{e}^{-\mathrm{i}S(I)} \approx \mathrm{e}^{\mathrm{i}\tilde{S}(I_0)}\mathrm{e}^{\mathrm{i}\alpha I}\,. \tag{9.5}$$

For the case of a monochromatic driving field, the dominant contribution to the harmonic signal stems from a pair of short and long trajectories. The corresponding phase coefficients are shown in Fig. 9.11.

For increasing intensity, the $\alpha$ coefficients exhibit a similar shape to the imaginary parts of the return time of the corresponding saddle points as $\mathrm{Im}(t_{\mathrm{r}})$ enters the semi-classical action directly. The phase-matching condition for the harmonic response in the far-field then ultimately refers to the idea that the harmonic signal is stronger if the phase coefficients of the respective dipoles match, such that the signal constructively interferes. The phase coefficients of the individual trajectories can therefore be used as an indicator for intensity regimes in which the HHG will be strong.

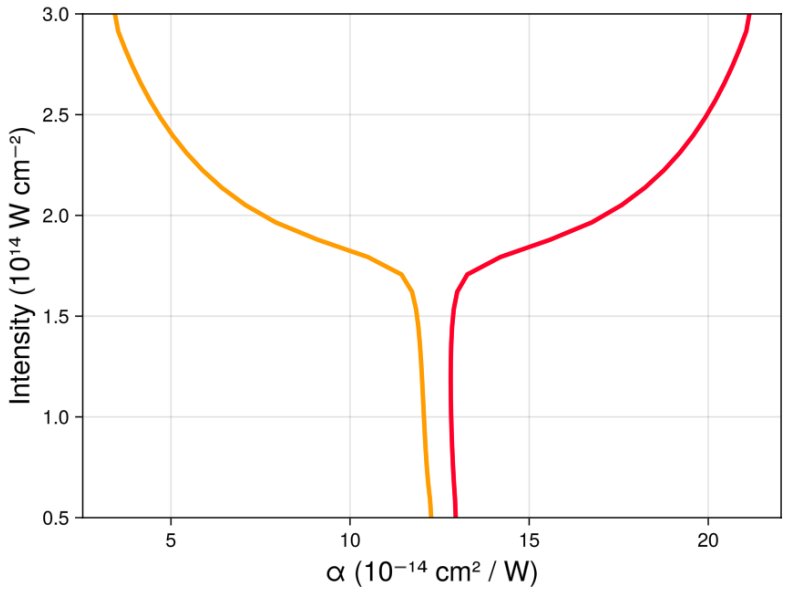

Figure 9.11: Phase coefficient $\alpha$ for the short (orange) and the long (red) trajectories of a monochromatic driving field contributing to harmonic order 25. From Jude Russell's report [190].

**Link to scaling parameters for Lefschetz thimbles**
Mathematically, the interference fringes of the various contributing trajectories are simply the Airy oscillations of the total integral stemming from the contributions of the individual relevant saddle points. As explained in Sec. 4.2.1, for an integral of the shape $\int e^{ik\phi} = e^{ikH} \int e^{kh}$, the thimble associated with a certain saddle point can be 're-used' for different values of the large prefactor $k$. The methods of Picard–Lefschetz theory are hence particularly useful in cases where the integral needs to be evaluated across a large range of values of $k$, for example for optical lensing problems where $k$ is the radiation frequency. Not only can the integration contour be re-used, but from the variation of a thimble's contribution across $k$ it is possible to determine whether the given saddle point is relevant to the integral at all.

For the semi-classical action of the SFA-based HHG integral the fundamental acts *almost* as such a scaling prefactor to the phase. Unfortunately it does not factor out perfectly, but we can get some inspiration of the methods that are based on the integrals' dependency on this large parameter.

**Intensity variations in experiments**
Another motivation to study intensity scans in the microscopic HHG dipole response stems from the comparison to experimental measurements. Macroscopically, a lot of atomic targets extend over a spatial domain where they interact with the driving laser beam. On this given target plane, the laser beam has a certain spatial intensity profile across which the intensity 'seen' by an individual atom differs by orders of magnitude. Performing an intensity scan therefore correspond to measuring harmonics emitted from a limited spatial region of the generating plane.

The generation for high-order harmonics requires driving laser field intensities on the order of $10^{14}\mathrm{W/cm}^2$. These intensities are challenging to achieve in experimental setups, and impossible to measure directly. Typically, a beam is focussed into the gas jet of atomic targets to reach intensities high enough for HHG to happen. The intensity that is reached at the focus is typically estimated from measurements outside the focus and the focussing geometry itself. Those estimates, however, can be wrong by orders of magnitudes. Not only is it impossible to measure the intensity at the focus, but also the intensity might fluctuate. Being able to provide an intensity scan therefore yields a valuable asset to improve the comparison with experimental data.

## 9.5.2 Intensity scans to identify signatures of tunnelling without barrier

The technique of distinguishing contributions from the several trajectories was applied to the case of a bichromatic field that features a counter-intuitive tunnelling event at a time where the electric field is zero, and hence no barrier is created. This work was undertaken by our student Jude Russell with the goal of identifying a potential observable signature of this tunnelling event in the context of HHG.

**HHG in for the scenario of tunnelling without barrier**
The field shape that features this curious tunnelling event is a $\omega - 2\omega$ co-linear driving field, where the two fields have equal amplitude and a phase shift of $\pi/2$. The respective saddle points, limited to travel times up to one optical cycle, are shown in Fig. 9.12 in the complex time planes of ionisation and recombination (panel (a)). In panel (b) we show the complex plane of the corresponding travel times, as well as the range of harmonic orders. Analogously to the ionisation-only discussion above, we identify four distinct ionisation bursts. This can be made out from the 'clusters' of saddle point times in the complex ionisation time plane on the left-hand side of Fig. 9.12(a). Each of those ionisation bursts gives rise to several trajectories. As we have seen from considering ionisation during the full switchover, the attribution of ionisation events to a specific burst might change throughout the switchover for the various final momenta. In analogy, for a range of harmonic orders, the classification of trajectories to changes as we perform the colour switchover. This is detailed in a chapter below, where we perform a colour switchover for a slightly different scenario.

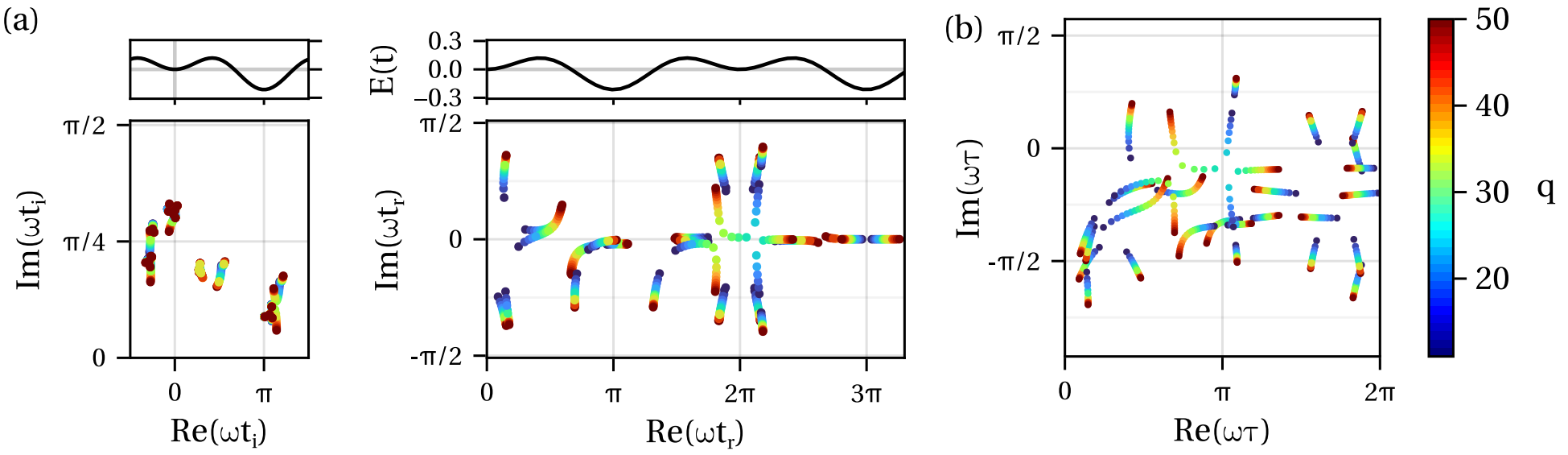

Figure 9.12: (a) All saddle points ionised within one cycle of the two-colour field configuration shown on top in the complex planes of ionisation (left) and recombination (right) time. (b) Travel times and the range of harmonic orders for those saddle points.

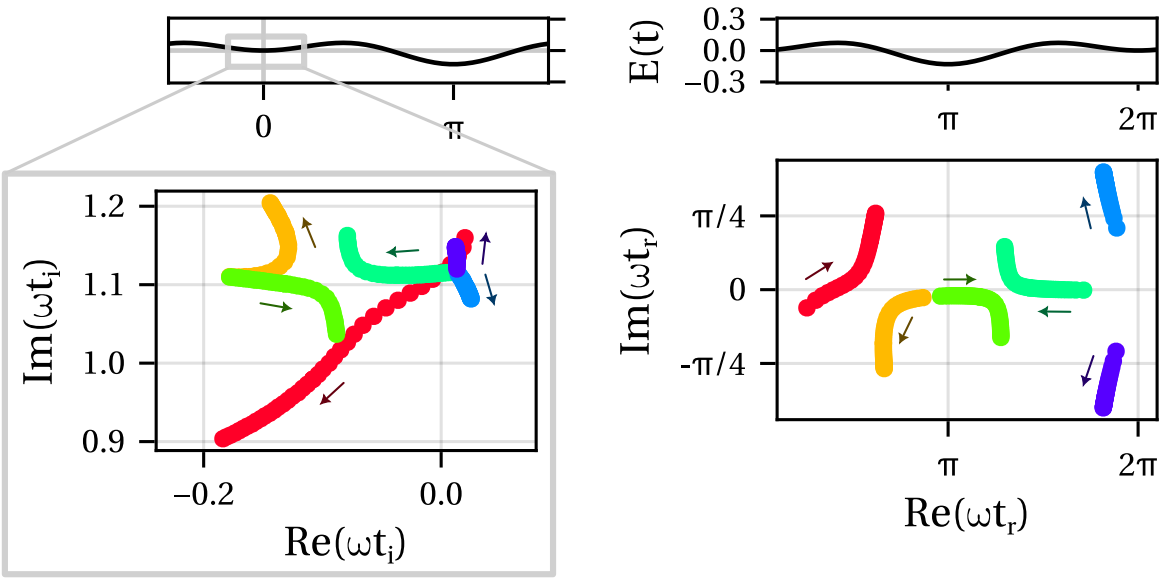

Figure 9.13: Saddle points ionised around $\mathrm{Re}(\omega t) \approx 0$, for a range of harmonic orders. Shown in the complex plane of ionisation (left) and recombination (right) and classified (colour), small arrows denote the direction of increasing harmonic order.

Here, however, we only consider this particular field shape at the fixed amplitude ratio $E_2/E_1 = 1$ and do not worry about possible classification issues arising from changing it. Even upon a small perturbation around the respective field amplitudes, the number of relevant trajectories as well as their classification does not change. For each of the four ionisation bursts we consider the shortest six relevant trajectories, limited in travel time to less than one cycle. For the ionisation burst B, at which the instantaneous electric field vanishes, they are shown in Fig. 9.13 where the separate branches of trajectories are colour-coded and the direction of increasing harmonic orders within each branch is indicated by a small arrow. The corresponding semi-classical electron trajectories are shown in Fig. 9.14 for the harmonic order 20.

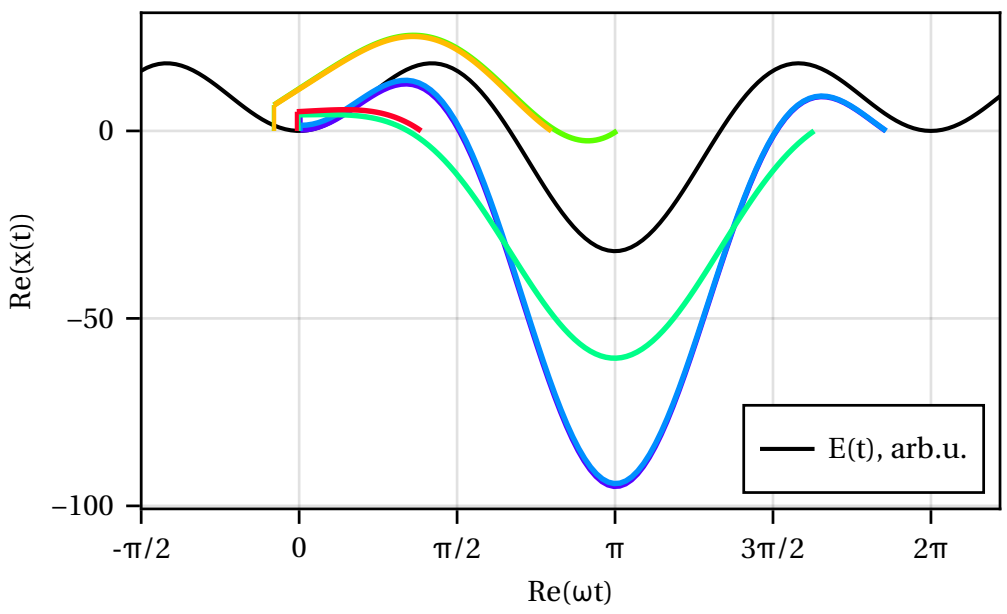

Figure 9.14: Semi-classical electron trajectories for the saddle points shown in Fig. 9.13, for harmonic order 20, as well as the electric field (black).

In Fig. 9.15 we show the total spectrum of the considered driving field, where we show the trajectories' individual contributions, coloured according to their ionisation burst, as well as the total harmonic intensity (in black). As expected, the contributions originating from ionisation burst B (green) are orders of magnitude below the contributions stemming from the other ionisation bursts. In the following sections, we will explore two techniques that potentially allow to identify signatures of trajectories from ionisation burst B.

**Intensity-dependent phase-coefficients**

Once the trajectories have been classified for a given field configuration, this classification can be used to track the corresponding saddle-point solutions as the fundamental intensity is varied. This enables a systematic analysis of how each trajectory evolves throughout the intensity scan. In particular, we can extract and compare the associated dipole phase coefficients. The intensity dependence of these coefficients for all trajectories stemming from one laser cycle, as well as only those from around $\mathrm{Re}(\omega t_i) \approx 0$ is shown in Fig. 9.16. In the above example of a monochromatic field, Fig. 9.11, the dependency is clearly different between the short and long trajectory and would henceforth allow a distinction in, e.g., an experimental setup. However, for this two-colour field configuration there are many more trajectories at play. Throughout the intensity scan there are several pairwise divergences, corresponding to missed approaches of saddle points in the complex plane (see Fig. 9.13) and indicating cutoff-like features in the spectrum.

In order to isolate the contribution of the second ionisation burst, or at least to potentially identify a signature of it, we examine how the $\alpha$ coefficient depends on the individual field components' amplitudes $E_1$ and $E_2$. For that we write a Taylor expansion

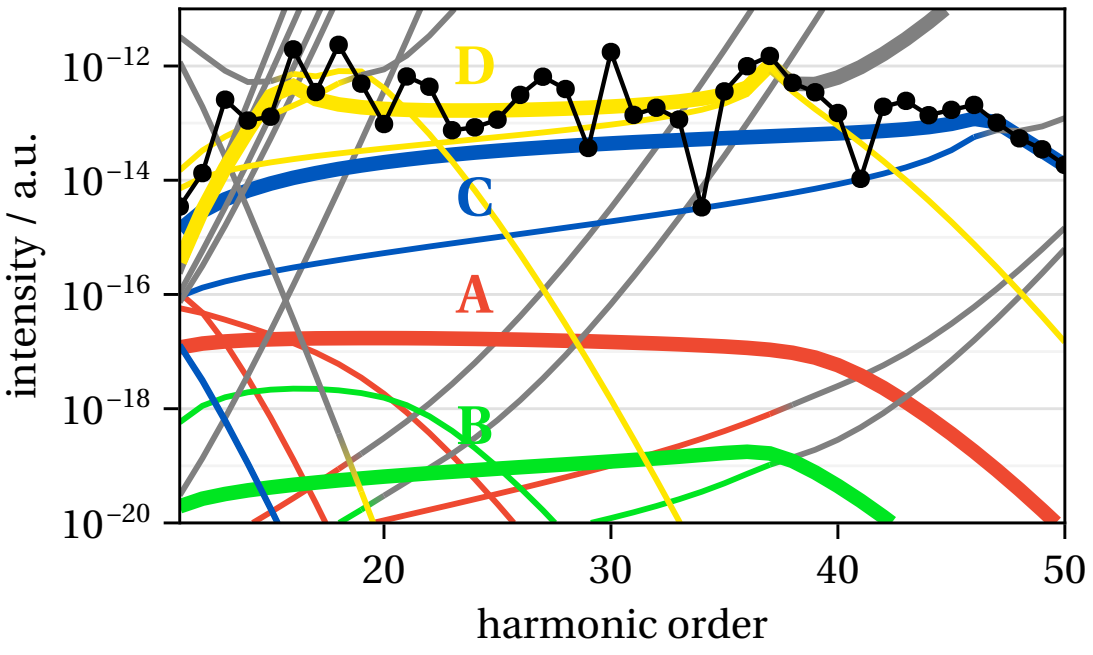

Figure 9.15: Harmonic spectrum with contributions from the individual trajectories, relevant contributions coloured according to their ionisation burst A,B,C,D. Non-relevant contributions are drawn grey and the dominant contribution from each ionisation burst is highlighted as a heavy line. The total harmonic intensity is shown in black, calculated as the sum of Gaussian contributions.

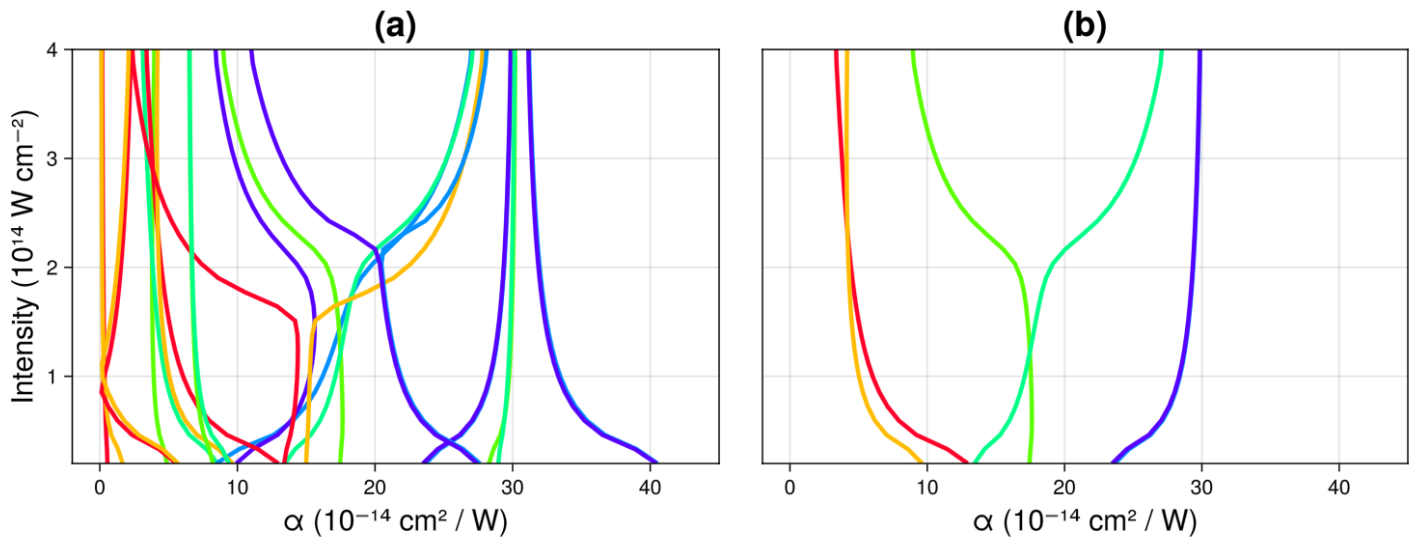

Figure 9.16: Intensity-dependent quantum-phase coefficients $\alpha$ for harmonic order 40 for (a) all trajectories within one laser cycle, and (b) only those from the second ionisation window. From Jude Russell's report [190].

w.r.t. those two quantities:

$$S(E_1, E_2) \approx S_0 + \frac{\partial S}{\partial E_1}\Delta E_1 + \frac{\partial S}{\partial E_2}\Delta E_2 + \frac{1}{2}\left(\frac{\partial^2 S}{\partial E_1^2}\Delta E_1^2 + 2\frac{\partial^2 S}{\partial E_1 \partial E_2}\Delta E_1 \Delta E_2 + \frac{\partial^2 S}{\partial E_2^2}\Delta E_2^2\right) \quad (9.6)$$

about the point $(E_{10}, E_{20})$ and hence $\Delta E_1 = E_1 - E_{10}$ and $\Delta E_2 = E_2 - E_{20}$. From that, we capture the non-linear dependency of the phase coefficient on the individual field amplitudes in terms of the phase coefficient matrix:

$$\alpha = \begin{pmatrix} \alpha_{11} & \alpha_{12} \\ \alpha_{21} & \alpha_{22} \end{pmatrix} = \begin{pmatrix} \frac{\partial S}{\partial E_1} & \frac{\partial S}{\partial E_1 \partial E_2} \\ \frac{\partial S}{\partial E_2 \partial E_1} & \frac{\partial S}{\partial E_2} \end{pmatrix} \quad (9.7)$$

This allows to compare the intensity sensitivity upon changes in the field amplitudes for the various ionisation bursts. In Fig. 9.17 we show the coefficients of the alpha matrix for the dominant trajectory within each ionisation burst — coloured according to the scheme in the previous sections and as highlighted in Fig. 9.15. We are still on the quest to identify features that distinguish the trajectories that enter the continuum at a time when the electric field is zero, i.e., the second ionisation burst (shown in green). We find the corresponding trajectory has the strongest sensitivity in coefficient $\alpha_{11}$. For the remaining coefficients this trajectory is less sensitive, or as sensitive as trajectories from other ionisation bursts. These findings appear consistent across the range of harmonic orders in the plateau.

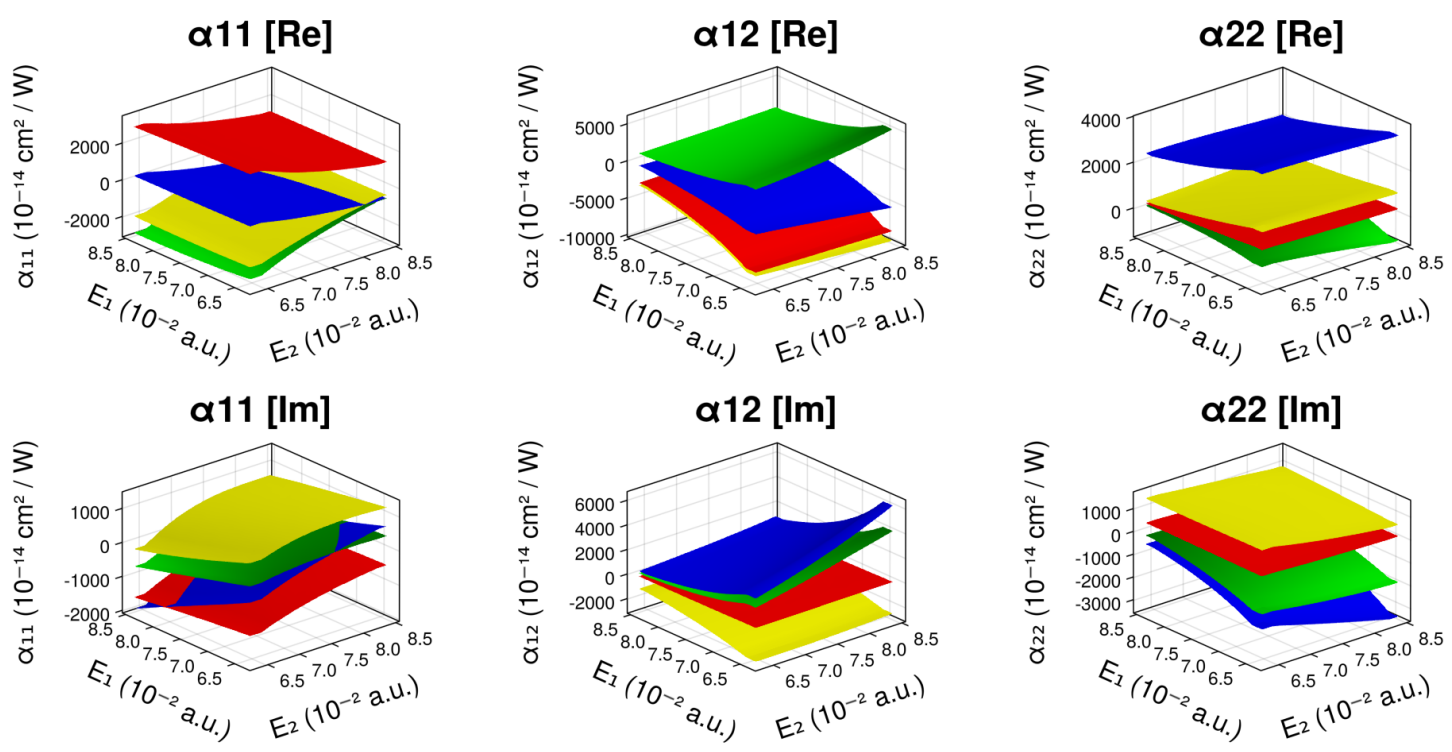

Figure 9.17: Surfaces of the elements in the quantum-path phase coefficient matrix for the dominant trajectory for each of the four ionisation bursts in one cycle for a strongly bichromatic driving laser field. The tunnelling without a barrier event is shown in green. The top row is calculated as the second derivative of the real part of the action and the bottom row is for the same calculation with the imaginary part. From Jude Russell's report [190].

Similarly, but a more quantitative approach, the dependency on the driving field parameters can be evaluated in terms of the expansion accuracy. In Fig. 9.18 we show the real part of the semi-classical action (solid sheets), as well as its Taylor expansion Eq. (9.6) (semi-transparent sheet) for two different trajectories (top row: ionised around zero field, bottom row: other, arbitrary trajectory). For the trajectory that begins when the field is vanishingly small, the linear approximation (panel (a)) yields minimal deviation from $\mathrm{Re}(S)$, and adding the second-order term (panel (b)) does not increase the accuracy. That is, the dipole phase can be well approximated by a linear dependency on the field amplitudes. In contrast to that, for the trajectory from ionisation burst four (yellow, bottom row), the linear approximation (panel (d)) does not capture the curvature across the direction of $E_2$. The second order term adds significantly to the accuracy of the approximation (panel (e)).

These results indicate that the dominant trajectories originating from the time at which the electric field vanishes depend predominantly linearly on the driving field parameters, with the strongest sensitivity in the $\alpha_{11}$ coefficient. In contrast, trajectories from the other ionisation bursts, exhibit markedly more complex intensity dependencies. Taken together with the scaling behaviour of the contribution from ionisation event B with the Keldysh parameter Fig. 9.5, this points towards a potentially observable signature of tunnelling-without-barrier trajectories when varying the driving-field intensity.

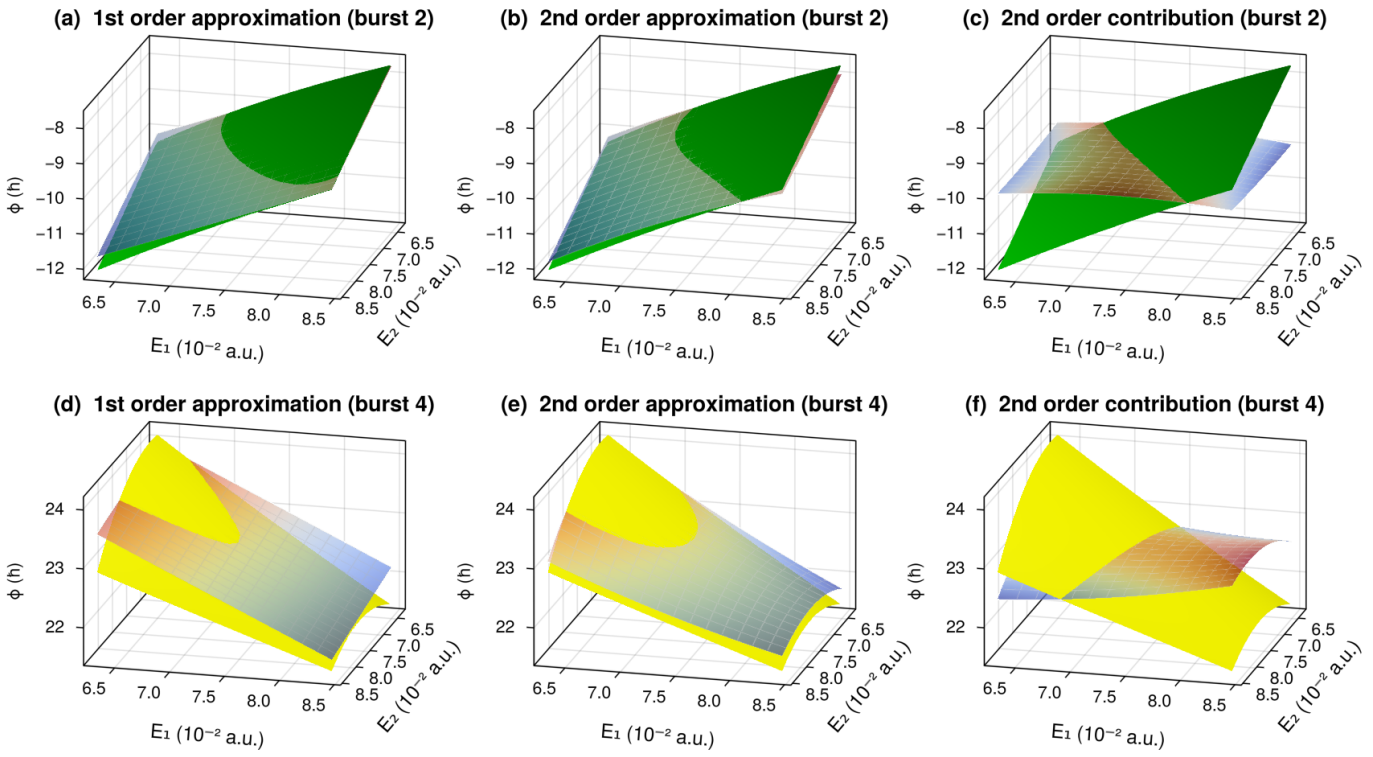

Figure 9.18: Taylor series approximations of Re($S$) compared with its computed surfaces for the second ionisation burst (top row) and the fourth burst (bottom row) over a scan of independent $E_1$ and $E_2$. Computed surfaces are in solid green and yellow, while the approximation surfaces are semi-transparent. The first-order approximations (a, d) are compared with the second-order approximations (b, e) and the pure second-order contributions (c, f). From Jude Russell's report [190].

## 9.5.3 Conclusion

We have demonstrated that Picard–Lefschetz methods can readily be used to perform intensity scans, where there are no further 'complications', i.e., no Stokes transitions and no coalescing saddle points. We showed results for intensity scans for a two-colour field configuration that features a scenario at which ionisation occurs when the electric field is (almost) zero, indicating tunnelling without a barrier. For those scenarios, we examined the respective phase coefficients $\alpha$ of the harmonic dipole in their dependency on the fundamental intensity of the driving field. We found the corresponding trajectories have the highest sensitivity in the $\alpha_{11}$ coefficient, capturing the second-order dependency on the fundamental driving field's amplitude, compared to the trajectories stemming from the other ionisation bursts. While we have not presented an immediate course of action to observe signature of tunnelling without a barrier in HHG, these results clearly show the way towards a possible observable.

# 10

# HHG during the colour switchover

The tools of Picard–Lefschetz theory allow us to link the regime of perturbative second-colour drivers with the regime of fully bichromatic fields. As a technique to probe this transition in Chapter 7 we introduced the colour switchover scheme where we gradually replace a driving field with its second harmonic, via two-colour fields of increasing intensity ratio. The therein posed questions about the origin, the dynamics and the relevance of saddle points implicitly assume the ability to track the saddle point solutions upon the colour switchover. This immediately links to the problem of finding a classification of solutions into different 'types' of trajectories that holds upon a changing number of ionisation bursts.

For a colour switchover scheme we present the answers to those questions in the following sections. We disentangle the saddle point dynamics in the complex plane and demonstrate methods to classify the multitude of saddle point solutions consistently across the colour switchover. Based on this classification, we showcase examples of saddle point solutions, their respective electron trajectories, and their relevance to the total harmonic dipole. Ultimately, we examine the resulting harmonic spectra across the colour switchover and attribute specific features to the individual trajectories.

A much more concise version of this Chapter is published in Ref. [2].

## 10.1 The colour switchover with constant $U_p$

Within this Chapter, we consider a colour switchover with a constant ponderomotive potential $U_p$, such that the total energy of the driving field and the Keldysh parameter remain constant throughout. With that, the *classical* harmonic cutoff Eq. (2.5) for the spectrum of a monochromatic driver is the same at the beginning and after completion of the colour switchover. We perform the colour switchover between the fundamental $\omega$-field and its colinear second-harmonic $2\omega$ field, and restrict our discussion to the case of zero phase delay between the fields. The corresponding vector potential reads

$$A(t) = A_1 \cos(\omega t) + A_2 \cos(2\omega t) \tag{10.1}$$
$$\text{where} \quad A_1 = \cos(\theta)\frac{E_0}{\omega} \quad \text{and} \quad A_2 = \sin(\theta)\frac{E_0}{2\omega}\,.$$

The electric field is hence given by

$$E(t) = E_1 \sin(\omega t) + E_2 \sin(2\omega t) \tag{10.2}$$
$$\text{where} \quad E_1 = E_0 \cos(\theta) \quad \text{and} \quad E_2 = 2E_0 \sin(\theta)\,.$$

Here, we have used the *mixing angle* $\theta$ as a parameter to tune the amplitude ratio of the two constituent fields as $E_2/E_1 = 2\tan\theta$. For $\theta = 0°$ the field Eq. (10.2) corresponds to a monochromatic field of frequency $\omega$, and $\theta = 90°$ corresponds to a monochromatic field of frequency $2\omega$. For all intermediate values of $\theta$, Eq. (10.2) forms a two-colour field with varying amplitude ratio.

The results of this chapter are obtained for the following parameters: The wavelength $\lambda = 800\,\text{nm}$ ($\omega = 0.057\,\text{a.u.}$), total intensity $I_0 = 4\times10^{14}\,\text{W/cm}^2$ ($E_0 = 0.107\,\text{a.u.}$), and $\mathcal{I}_p = 13.7\,\text{eV} = 0.5\,\text{a.u.}$ The ponderomotive energy is then $U_p \approx 0.88\,\text{a.u.}$, and the resulting Keldysh parameter is $\gamma = 0.534$ throughout the whole colour switchover. The classical harmonic cutoff order is $q_{max} \approx 58$ and the ionisation threshold order is $\mathcal{I}_p/\omega \approx 9$. To cover the spectrum in its entirety and study the behaviour well beyond the cutoff, we consider harmonic orders $10 \le q \le 100$.

In the small top subpanels of Fig. 10.1 we show the total electric field $E(t)$ for the values $\theta = 1°, 5°, 13°, 22°, 33°$ and $66°$ (panels (a)–(f)). This demonstrates the key intrigue of the scheme, which we have motivated in Chapter 7: At the initial stage of the switchover ($\theta = 1°$) we have two (distinct and clearly separated) ionisation bursts within one cycle of the fundamental $T$, just after the maxima of the field at $\omega t = \pi/2$ and $3\pi/2$ (panel (a)). After the *continuous* transition to $\theta = 90°$ however, we have four ionisation bursts in that same time frame. This can be seen in panel (f), where we show the saddle points at $\theta = 66°$, which is structurally just before completion of the cutoff. Especially with the understanding that ionisation events (and the associated recombination times, making them trajectories) correspond to saddle points, and saddle points are topologically stable features of analytic functions, this 'jump' in number of ionisation events raises questions about the origin and classification of trajectories. The Picard–Lefschetz methods developed in Sec. 6.2 allow us to answer these questions.

## 10.2 Saddle-point chaos in the complex plane

Upon performing the colour switchover, the saddle points in the complex planes of ionisation and recombination time reveal rather complicated dynamics, which makes a consistent classification of trajectories tedious and non-unique, but still possible. Before we delve into the details of how to track and classify the solutions, however, let us first showcase the problem. In Fig. 10.1 we show the saddle points for some intermediate steps of the colour switchover. For each mixing angle (row) we show the complex plane of ionisation and recombination times on the left-hand side (panels (a) – (f)), with the respective electric field above. On the right-hand side panels (g)–(l) we show the complex travel time. Within each panel we show saddle point solutions for harmonic orders between 10 and 100. In contrast to earlier visualisations the colour of the markers is now used to distinguish the separate 'types' of trajectories. Before we explain the strategies employed to establish such a classification, let us first discuss step by step the respective saddle point dynamics in the complex planes. For that, we limit our considerations to saddle points with travel times less than one cycle of the laser field, $\mathrm{Re}(t_r - t_i) = \mathrm{Re}(\tau) < T$, and we neglect those solutions that are never relevant to the harmonic spectrum.

At the very initial stage of the colour switchover (not shown here) the field is monochromatic and the saddle points are as shown in earlier chapters (see Fig. 2.11). There are two distinct ionisation bursts, centred around the maxima of the electric field. The corresponding recombination times subsequently extend across the whole cycle. The travel times show the characteristic missed approach between short and long trajectories. As soon as we increase the amplitude of the $2\omega$ field, additional saddle points come in from high imaginary values. This is shown in Fig. 10.1(a), where $\theta = 1°$. In the complex plane of ionisation (left) we find new saddle points entering from high imaginary parts. Just after, in Fig. 10.1(b), we find two 'clusters' of saddle points around $\mathrm{Re}(\omega t_i) \approx \pi/2$ and $3\pi/2$, with $\mathrm{Im}(\omega t_i) \lesssim \pi/4$. Additionally, at high imaginary values ($\mathrm{Im}(\omega t_i) \approx 3\pi/4$) there is a new cluster of saddle points. In the complex recombination time plane (right) the new saddle points come in from both highly positive and highly negative imaginary parts. In panel (b) and (c), the complex travel time plane shows how the two half cycles of the field loose their symmetry. That is, for the monochromatic driving field both half cycles map to the same points on the travel time plane. As soon as the two half cycles are not symmetric any more, the previously simple shape will unfold into two distinct structures.

Upon a further increase of the mixing angle, the new saddle points move more and more towards the real axis, compare Fig. 10.1(c)–(e). In the ionisation time panels we find the cluster of new saddle points moves downwards. They subsequently spread out such as to form the equidistant four ionisation bursts that we expect for the monochromatic $2\omega$ driving field at the final stage of the colour switchover. On the recombination time panels some of the interesting dynamics is slightly more visible: the newly incoming trajectories 'repel' the old trajectories and slowly squeeze themselves into the periodic structure of short and long trajectories. Interestingly, focusing on the ionisation times, we find that the typical structure around the first ionisation burst (around time $\omega t_i = \pi/2$, structure as seen in Fig. 10.1(b) is 'ripped apart' by the newly incoming saddle points, which subsequently push some solutions to earlier times and some solutions to later times. In contrast, the ionisation window of the second half cycle remains largely undisturbed.

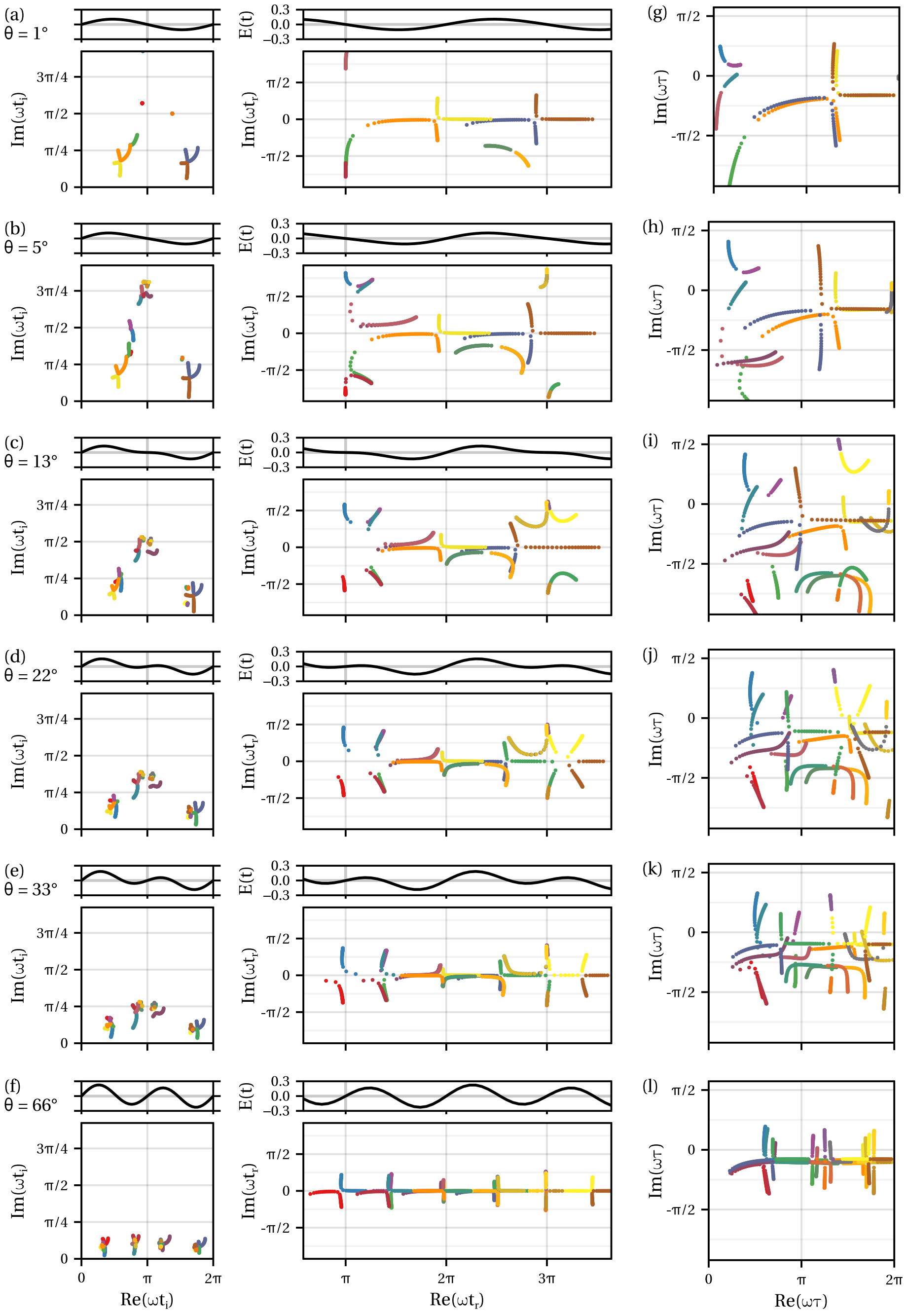

Figure 10.1: Saddle points in the complex plane for several steps ($\theta = 1°, 5°, 13°, 22°, 33°$ and $66°$ along the rows) of the colour switchover. Each row shows the complex-time panels for ionisation and recombination (a)–(f), as well as their difference, i.e., the travel time complex plane (g)–(l). For the range of harmonic orders ($10 \leq q \leq 100$) the solutions trace lines in the complex planes, here coloured according to the classification scheme.

The travel time panels are included here for conceptual clarity. They indicate that both at the beginning and the end of the colour switchover the driving field has a dynamical symmetry which results in a nicely structured set of saddle point solutions. The transition between the two, however, reveals rather chaotic structures. To obtain a consistent classification of all saddle point solutions we require some strategies, such as to resolve scenarios like the one in panel (d), where multiple types of trajectories perform missed approaches.

## 10.3 Classifying the saddle-point solutions

Whether a given saddle point contributes to the total spectrum can be decided by using the necklace algorithm introduced in Sec. 5.2.3. This procedure is entirely independent of the classification of the saddle point. However, if we want to study the resulting physical phenomena of adding a second colour driver and increasing its relative amplitude, then 'tracking' and classifying the individual saddle point solutions becomes necessary. In the following we lay out the procedure we employed to classify the saddle point solutions during the colour switchover.

### 10.3.1 General outline of the strategy

We begin by classifying the saddle point solutions at the 'final state' of the colour switchover, i.e., the monochromatic $2\omega$ driving field, because this provides the maximum available number of saddle points. This is furthermore motivated by the earlier finding that the '$2\omega$-centric' classification scheme was preferred in the classification of ionisation (only) saddle points in the previous Chapter 8.

For this configuration, the classification of solutions is rather trivial as the saddle points for subsequent harmonic orders trace clearly distinguishable 'lines' in the complex plane, see Fig. 10.2. In the top row (panel (a)) we show all saddle point solutions for harmonic orders $10 \leq q \leq 100$ which will eventually contribute to the harmonic spectrum, with the complex plane of ionisation times on the left, and complex plane of recombination times on the right. In Fig. 10.2(b) we show only the solutions for the first ionisation burst of the $2\omega$ driver, stemming from around $\mathrm{Re}(\omega t_\mathrm{i}) \approx \pi/2$. As a (somewhat arbitrarily chosen) naming convention, we labelled these solutions according to their travel time (in letters A through F) and the respective ionisation burst (in numbers 1 through 4). The respective labels for the trajectories are then annotated as A1, B1, C1, D1, E1 and F1 for the separate lines of saddle point solutions. The saddle point structure and the resulting labelling scheme follows analogously for the other three ionisation bursts. This establishes a classification scheme for the colour switchover at the final stage where $\theta = 90°$.

From there on, as a general strategy, we now decrease the mixing angle $\theta$ in small steps and simply infer each saddle points' classification from the previous step. This *mostly* works, as for every small change in $\theta$ the saddle points only move slightly. At several instances during this routine for the full switchover procedure, however, we will encounter complications and ambiguities which we address in the following.

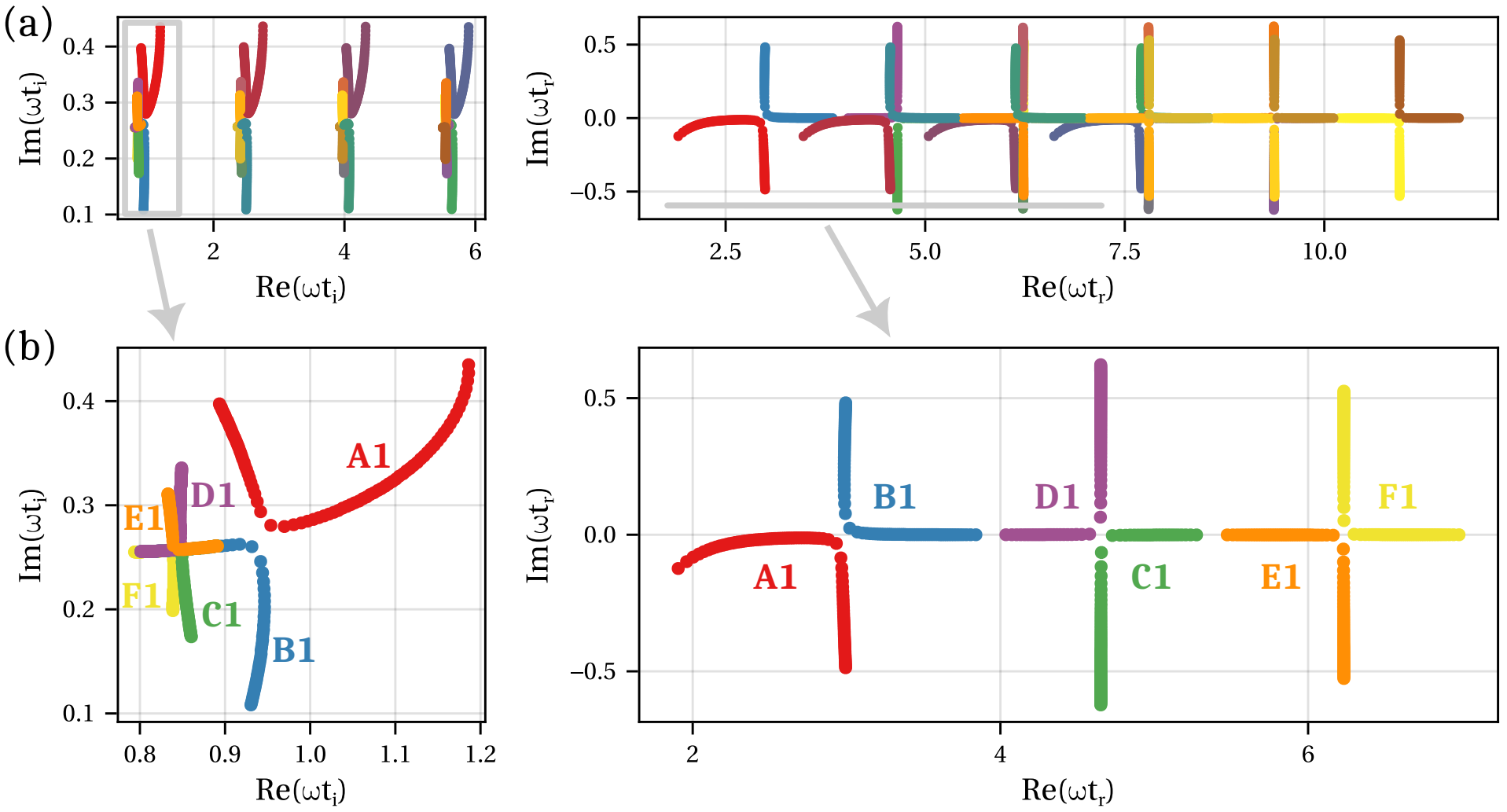

Figure 10.2: Saddle points for harmonic orders $10 \leq q \leq 100$ in the complex time planes of ionisation (left) and recombination (right), for the final stage of the colour switchover — a monochromatic driver with frequency $2\omega$. In (a) we show all the relevant saddle points, and in (b) only those stemming from the first ionisation window at around $\mathrm{Re}(\omega t_\mathrm{i}) \approx \pi/2$. Solutions are coloured and annotated according to their classification tags, here A1, B1, C1, D1, E1 and F1.

## 10.3.2 Resolving missed approaches

When saddle points of different types perform a missed approach this might lead to a misclassification due to the chosen discretisation in both the mixing angle $\theta$ and the harmonic order $q$. An example of such a situation is shown in Fig. 10.3, where we show a set of saddle points for several discrete values of $q$, constituting two types of trajectories. Across a scan of the external parameter $\theta$ the two trajectories approach each other closely (shown in the top panels). This induces difficulties for the classification which is usually based on a simple 'nearest neighbour' approach. At $\theta = \theta_\star$ shown in the top centre panel the distance $\Delta q$ between two subsequent orders of the same type is larger than the distance $\Delta\theta$ to that of the other type, $\Delta q > \Delta\theta$. A classification approach that is solely based on the vicinity to surrounding saddle points therefore gives the wrong result. In situations like this there are two possible solutions: first, we can use the 'organising centre' of the missed approach. This is a fold catastrophe point, at which the determinant of the Hessian vanishes. To identify this point we assume a complex-valued harmonic order $q \in \mathbb{C}$ and identify whether a the saddle point (that wants to be classified) lies on the same side of a line of it, as sketched in the bottom panel of Fig. 10.3. This strategy is detailed in the later section Sec. 12.1, where we discuss fold points that constitute the HHG cutoff in particular. Alternatively however, as a second method, we can increase the discretisation in $q$ and/or $\theta$ to resolve the dynamics of the saddle points in the complex plane. The former solution of actually identifying folds, however, proved to be more robust and also more insightful.

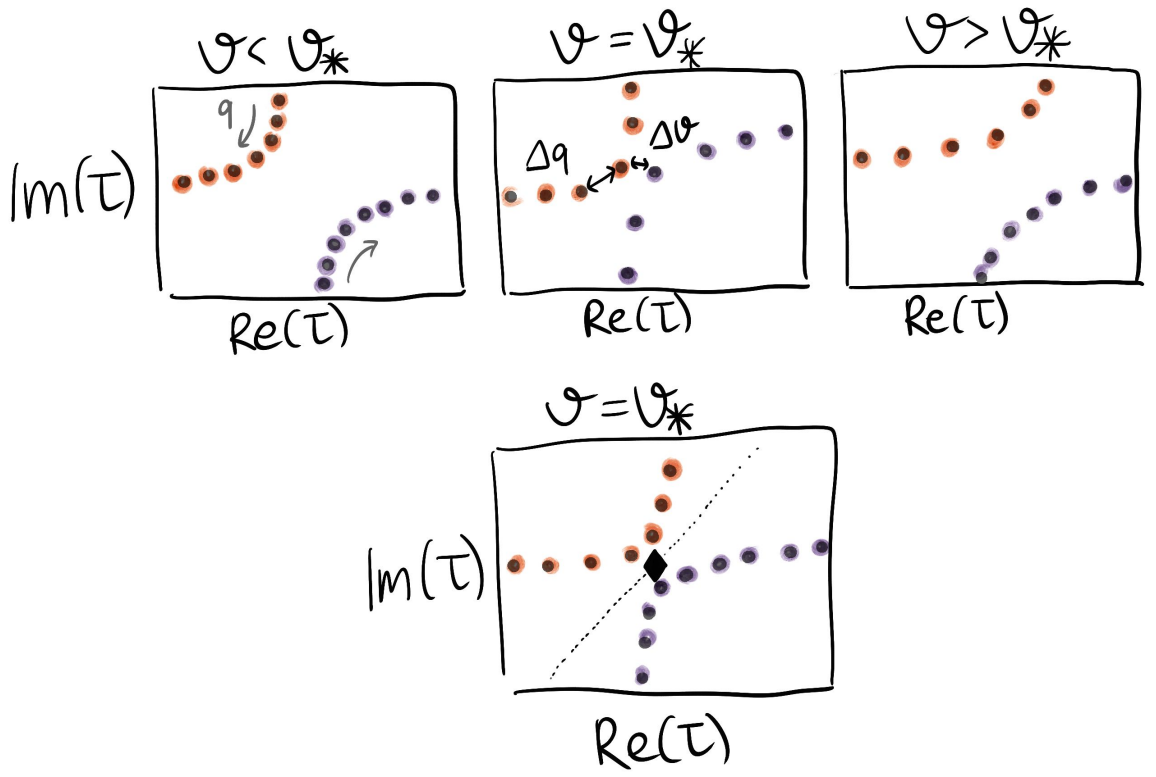

Figure 10.3: Difficulties and strategies for classifying saddle points that perform a missed approach. Top panels: Across a scan of $\theta$, the two branches of saddle points may approach each other so close that two solutions of subsequent $q$ within one branch are further apart than the two solutions of the separate branches. Bottom panel: Identifying the fold point in between the two branches allows to rigorously separate the two branches, independent of the discretisation step sizes.

### 10.3.3 Resolving actual fold coalescences

For our specific implementation of the colour switchover with zero two-colour phase delay there are rare cases for which two saddle point solutions actually coincide to a fold catastrophe point. That is, even though we can find a fold catastrophe point sat in between every missed approach by assuming $q \in \mathbb{C}$, for some specific cases the imaginary part of $q$ vanishes, such that the fold occurs for fully real-valued parameters. In this case, the classification is truly ambiguous and we simply have to make a choice. An example of a situation like this is visualised in Fig. 10.4. Both before (left panel, $\theta < \theta_f$) and after (right panels, $\theta > \theta_f$) the classification of saddle points into two separate branches is trivial. At the exact coalescence (centre panel, $\theta = \theta_f$), however, the two branches coincide at the fold catastrophe point (marked by a black diamond).

Upon the scan of $\theta$ across this fold catastrophe, we can either choose the classification tag to remain fixed for low harmonic orders (i.e., ‘anchoring’ the classification at low $q$, see top right panel), or to remain the same for high harmonic orders (anchoring at high $q$, see bottom right panel). For the few actual folds that appear throughout the performed colour switchover, we have decided to anchor the classification at low harmonic orders for consistency with the relevance of the respective saddle points.

### 10.3.4 Resolving missed approaches and actual coalescence between three trajectories

As part of the colour switchover we have identified a cusp catastrophe point, at which three saddle point solutions are expected to coalesce. For the case of zero phase shift considered here, the exact cusp occurs only at complex-valued $q$ and $\theta$, even though the imaginary parts remain very small. The details of this, as well as its consequences for the observable spectrum are discussed in the later Sec. 10.5. For the moment, we focus on the problem of consistently[1] classifying the saddle-point solutions.

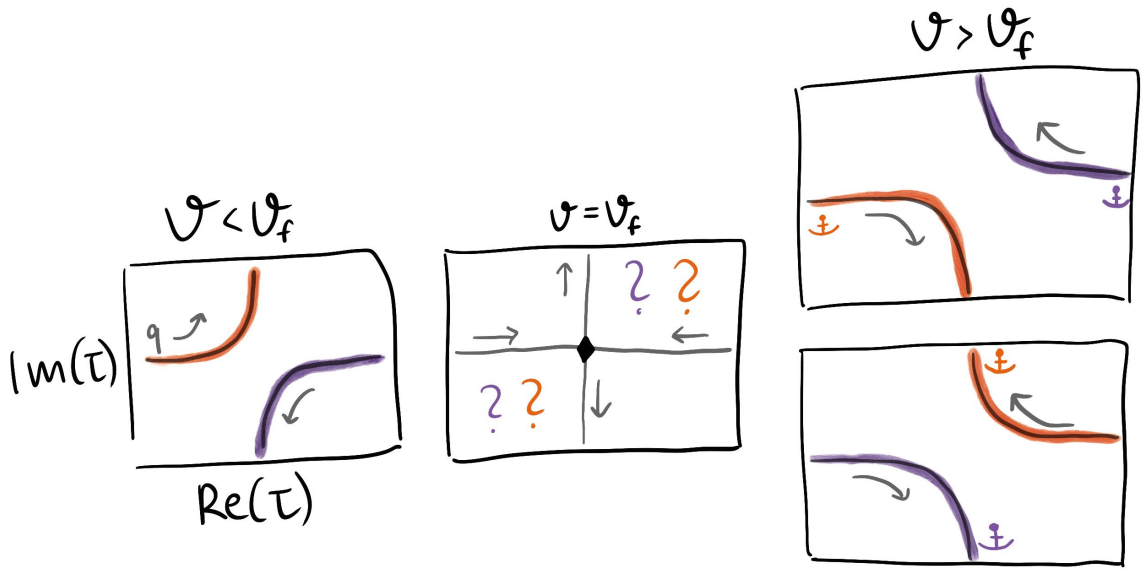

Figure 10.4: Strategy for classifying saddle points into branches across an actual fold. For the fold configuration (centre panel), the two branches coincide and the classification is truly ambiguous. We propose to anchor the classification at either low harmonic orders (top panel on the right), or at higher harmonic orders (below that).

Within the colour switchover for real-valued parameters, the three saddle-point branches undergo a missed approach rather than a true coalescence, as can (hardly) be seen in the recombination time panel of Fig. 10.1(d) at $\mathrm{Re}(t_\mathrm{r}) \gtrsim 3\pi$, and as sketched schematically in Fig. 10.5. To maintain a consistent labelling of the saddle points throughout this missed approach we employ the second of the two classification strategies previously introduced for around the fold catastrophe. That is, increasing the discretisation step size in $q$ and $\theta$ in order to clarify the labelling of each saddle point solution.

A more rigorous classification strategy that explicitly incorporates the actual coalescence point of the three solutions as an organising centre has yet to be developed. For the missed approach of two trajectories, we identified the fold point in order to create a clear separatrix between the two traversing saddle point solutions. The approach is not directly applicable here. While fold points can indeed be identified between each pair of trajectories, the separatrix criterion is only locally valid and therefore insufficient for a global classification.

In a configuration where three saddle points fully coalesce into a cusp catastrophe the classification would be truly ambiguous. In such cases we could invoke a similar approach as for the fold above, where we anchored the classification at low (or high) harmonic orders, as schematically shown in Fig. 10.5. Although this situation does not arise for our configuration here at $\varphi = 0$, where only a missed approach occurs, we found it useful to keep this perspective in mind, as it motivates practical computational strategies. In particular, we iteratively infer the classification from the lower-$q$ neighbouring saddle points, starting from the smallest value of $q$.

[1] It is impossible to classify the saddle points across a true coalescence, because at the coalescence point all identity is lost. But for the practical application of scanning across a specific parameter we can try to find a meaningful classification.

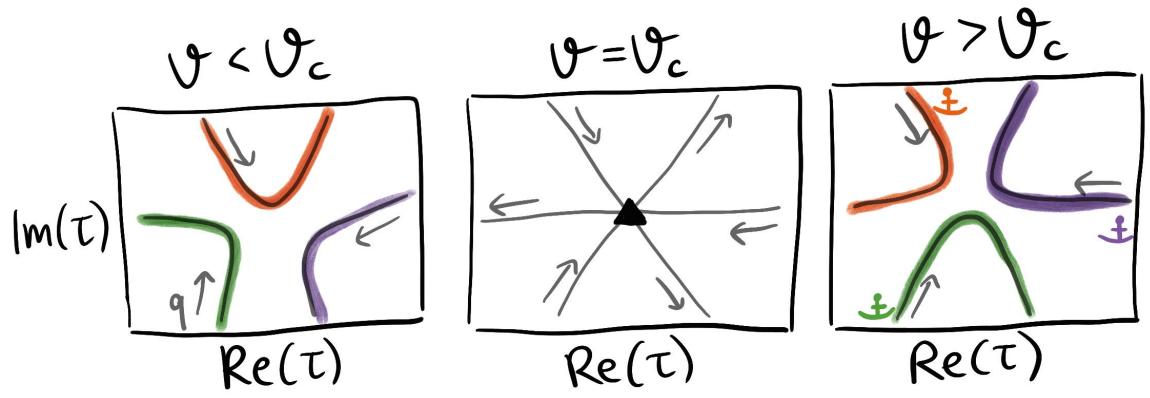

Figure 10.5: Strategy for classifying saddle points into branches across an actual cusp. Similar to the scenario of a fold, we choose to anchor the classification at the low-order harmonics.

### 10.3.5 Tracking individual trajectories' saddle points throughout the colour switchover

Once we have classified the saddle points into separate trajectories we can study individual trajectories' dynamics upon the colour switchover. For example, this enables us to answer the question 'where do the ionisation events go?', posed by the colour switchover scheme. In Fig. 10.6(a) we therefore show the saddle points in the complex planes for the initial stage of the colour switchover, i.e., the monochromatic driving field where $\theta = 0°$. We find the familiar structure of an ultra-short (green, labelled C1, and pink, labelled M2), a short (orange, E1, and dark blue, A4) and a long trajectory (yellow and brown) within each half cycle of the driving field, where the short and long trajectory perform missed approaches in the complex planes. The saddle point solutions are coloured according to the (arbitrary) classification system established above. Note that the ultra-short trajectory M2 was not part of our earlier considerations (e.g. in Fig. 10.1), because the respective saddle points remain irrelevant to the total integral for all considered harmonic orders and throughout the full colour switchover.

In Fig. 10.6(b)-(e) we show how the saddle points of the four trajectories E1, A4, M2 and C1, respectively, move in the complex plane upon performing the colour switchover, i.e., upon increasing $\theta$. Within each panel, we show the ionisation and recombination complex time planes as usual, but the markers are coloured according to harmonic orders as earlier (from $q = 10$ in blue, to $q = 100$ in red). Panels (a) and (b) show the evolution of the first short trajectory from the two half cycles of the $\omega$-driver. For clarity, we have coloured the non-relevant saddle points in grey. We recognise the familiar feature: For both of these short trajectories the saddle points for high-harmonic orders are not relevant, i.e., they are discarded beyond the cutoff. Interestingly, however, upon the switchover, these two trajectories evolve quite differently.

The short trajectory of the first half cycle (E1, panel (a)) moves closer towards the real axis and towards *earlier* ionisation times and *later* recombination times. That is, upon the colour switchover the trajectories' travel time increases. At the final stage of the colour switchover, this trajectory will eventually become the third shortest trajectory of the first ionisation burst of the $2\omega$-driver, see the orange markers labelled E1 in Fig. 10.2. In contrast, the short trajectory of the second half cycle (A4, panel (b)) also moves closer to the real axis, but towards *later* ionisation times and *earlier* recombination times. That is, upon the colour switchover this trajectory decreases its travel time. Upon completion of the switchover, we find this trajectory remains the first shortest trajectory of the latest (fourth) ionisation burst of the $2\omega$ driver. That is, the equivalent of trajectory A1 marked red in Fig. 10.2, but for ionisation burst 4.

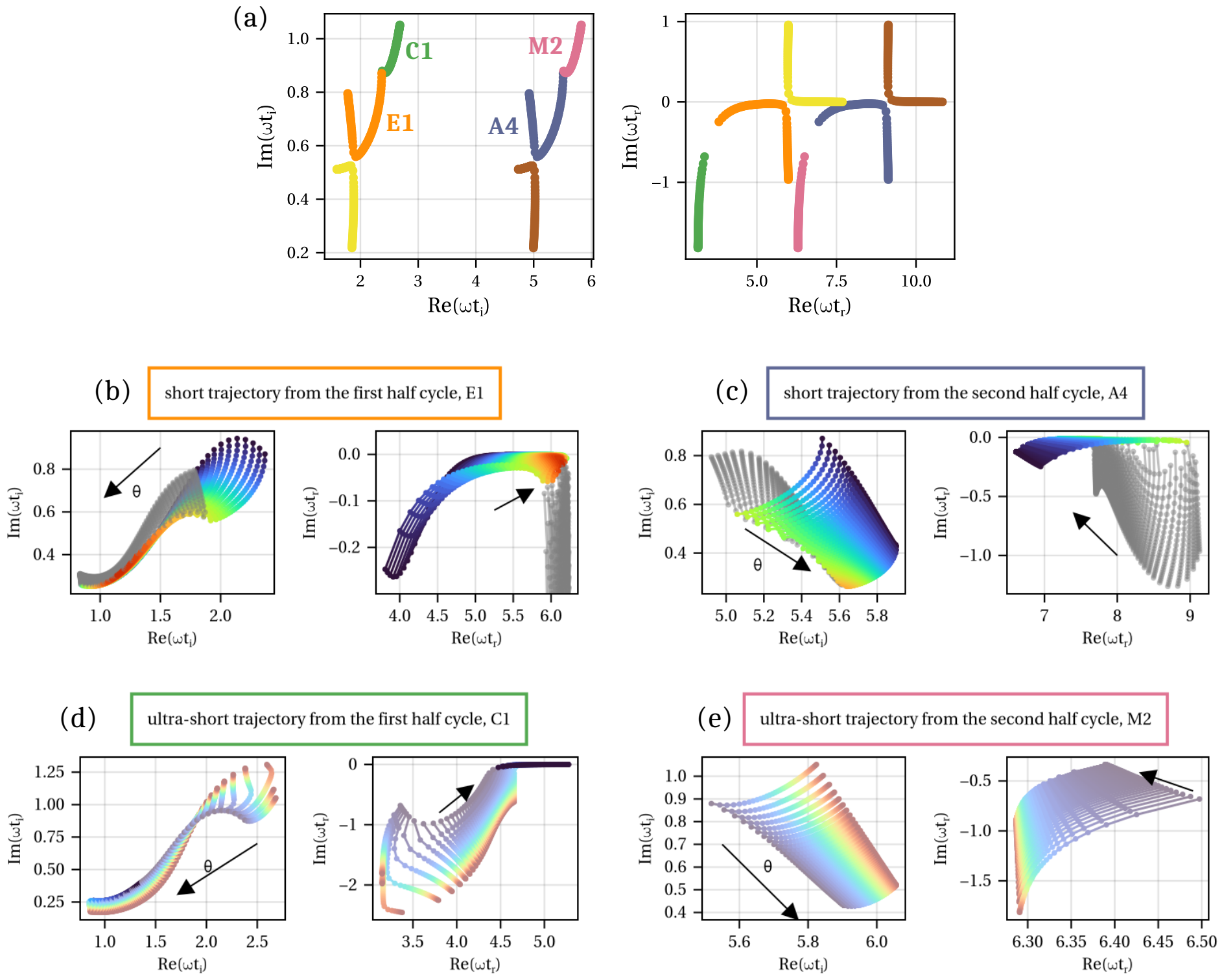

Figure 10.6: (a) Classified saddle points in the complex planes of ionisation and recombination for the monochromatic driving field at the initial stage of the colour switchover where $\theta = 0°$. Panels (b) – (d) show how the saddle points of the four trajectories move upon performing the colour switchover. The direction in which they move upon increasing $\theta$ is indicated by the black arrow. The saddle points are coloured according to the harmonic order, and grey (for (b) and (c)) or in a brighter shade (for (d) and (e)) for non-relevant saddle points.

A similar analysis follows for the two ultra-short trajectories, C1 and M2 in panels (d) and (e) respectively. We draw the markers for non-relevant saddle points in slightly less bright colours, still denoting the harmonic colours (in contrast to panels (b) and (c) where they were just drawn grey). This is simply because saddle points of trajectory C1 only contribute at the very end of the colour switchover and only for low harmonic orders. Saddle points for trajectory M2 never contribute to the total integral. Note that this is also the reason why trajectory M2 did not appear in our earlier discussion, e.g., in Fig. 10.2. Let us nevertheless compare the behaviour of the associated saddle points in the complex planes. Similarly to the short trajectory, also the ultra-short trajectory of the first half cycle moves towards lower imaginary times for both ionisation and recombination upon the switchover (panel (d)). Ionisation drifts towards earlier times and recombination towards later times, such that the travel time increases. At the final stage of the colour switchover, the ultra-short trajectory of the first half-cycle of the $\omega$-field has become the third short trajectory of the first ionisation burst, see Fig. 10.2. The ultra-short trajectory of the second half cycle (M2, panel (e)) has a different behaviour. Upon increasing $\theta$ the saddle points move only slightly towards the real axis without notable changes in their behaviour across the harmonic orders. The saddle points remain non-relevant and eventually constitute the ultra-short trajectory of the last (fourth) ionisation burst of the $2\omega$-driver.

The saddle points that constitute the 'long' trajectories at the beginning of the colour switchover (yellow and brown in panel (a) of Fig. 10.6) undergo more complicated dynamics. This makes depictions like the above less clear, hence we omit them here. But we may briefly state that the long trajectory of the first half cycle ultimately remains the long(est) trajectory of the first ionisation burst of the $2\omega$-driver. It remains relevant throughout the entire colour switchover. The long trajectory of the second half cycle, however, is involved in (several) actual coalescences. Loosely speaking, upon the coalescence the clear 'line' of saddle points is discontinuously broken up into parts and connected with parts of a line of saddle points of a different trajectory, as sketched in Fig. 10.4. As a result, some of the saddle points end up to constitute the long trajectory of the last ionisation burst of the $2\omega$-driver, while others do something else.

In conclusion, this showcased how the ability to classify saddle points consistently allows to track individual trajectories throughout the switchover. It furthermore demonstrates that the rigorous classification is impossible as each of the involved coalescences renders the classification ambiguous and we had to make a choice how to continue the classification between two neighbouring steps of the switchover.

## 10.4 Energy-time relations

Having classified the saddle points and identified how individual trajectories behave upon the colour switchover, let us move one step further towards the observable harmonic spectrum. The harmonic intensity is determined by the radiation dipole created from the interference of several quantum paths. Even without knowing any catastrophe theory one can reason that the total intensity is enhanced if the several quantum paths constructively interfere and their respective dipoles are in phase. As the dipole phase is the semi-classical action and oscillates with $\mathrm{Re}(\omega t_\mathrm{r})$, it is instructive to plot the energy upon return (given by the harmonic order) over the recombination time. Regions where the quantum paths are more 'flat' indicate regions of enhanced harmonic intensity. This can be understood intuitively, as several trajectories recollide shortly after each other and all produce radiation of similar energy. It can be utilised to explain harmonic enhancements even in a classical picture [37]. For a monochromatic driving field as the very initial step of the colour switchover, those are shown earlier, in Fig. 2.11(b). Here, we show the energy-time relations for two intermediate steps of the colour switchover in Fig. 10.7 by plotting the harmonic orders $q$ against the real part of the recombination times $\mathrm{Re}(\omega t_\mathrm{r})$ for the respective saddle points. Empty circles denote non-relevant saddle points, and the colour follows the classification scheme introduced above. In panel (a) we find a rather flat region around $\mathrm{Re}(\omega t_\mathrm{r}) \approx 3\pi$ which may cause an increase in harmonic response around the corresponding harmonic order 40, as well as the region of a harmonic cutoff for slightly earlier recombination times at $q \approx 50$. Furthermore this visualisation allows to generally identify how many trajectories will interplay to produce a given harmonic order. For example, in panel (b), we find many more trajectories play into the generation of a given harmonic order, which may lead to increased signal.

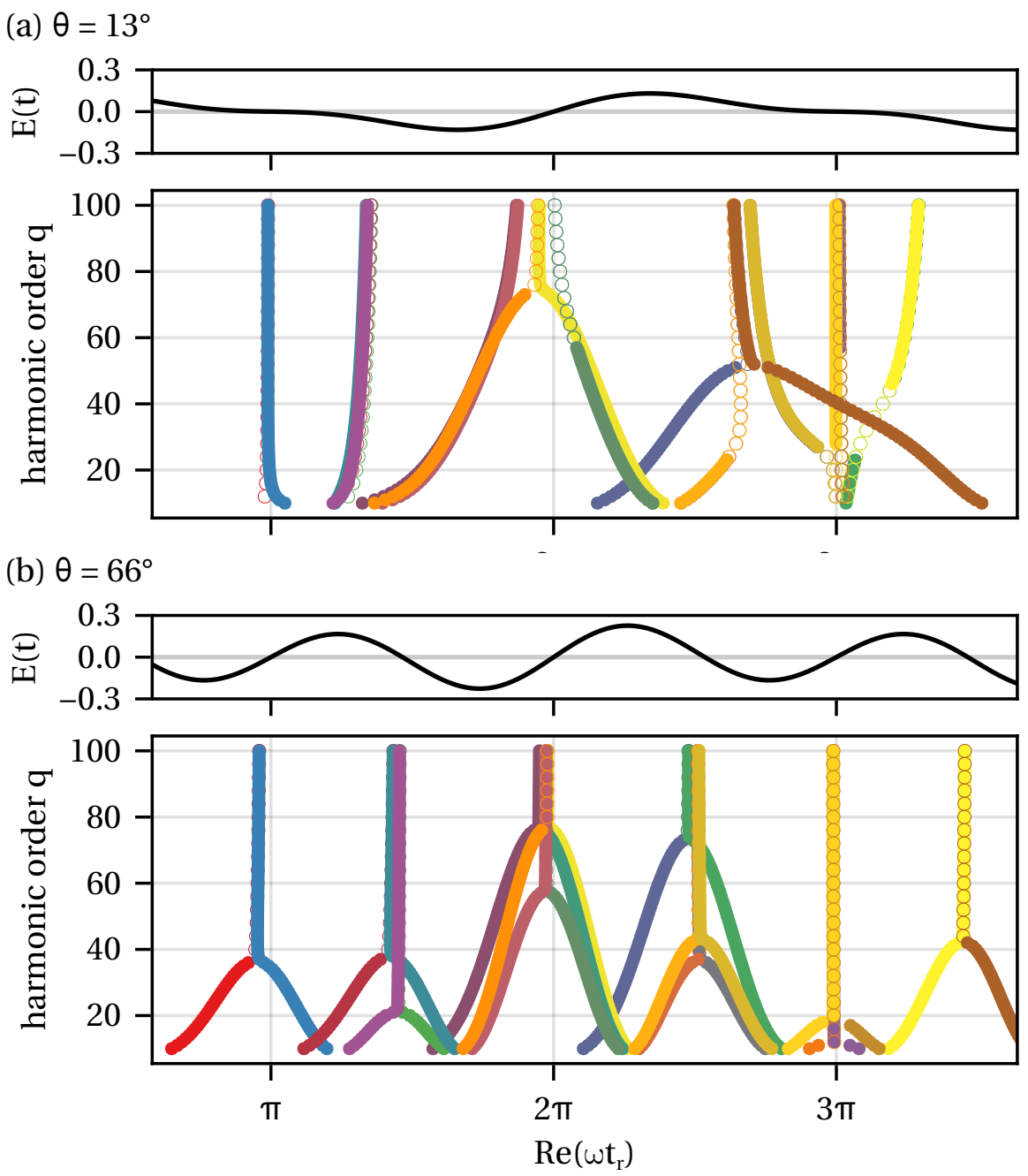

Figure 10.7: Energy-time relations for the colour switchover at (a) $\theta = 13°$ and (b) $\theta = 66°$, shown by plotting the harmonic order over the real time of recombination. For reference, the respective electric field is shown on top.

## 10.5 Harmonic response — spectra

Having developed a consistent labelling for all saddle points throughout the colour switchover, we can now turn to study the harmonic spectra, including the contributions from the several saddle points. Let us highlight a few curious features that impact the observable harmonic spectrum. On the right-hand side of Fig. 10.8 we present the harmonic spectra with their contributions from the individual saddle points for the respective stages of the colour switchover, which are shown on the left-hand side. In the spectra, we coloured the individual saddle points' contributions according to their ionisation window (indicated on the left-hand side), and we show the resulting total intensity (SPM, from Eq. (6.4)) in black. The contributions of non-relevant saddle points are shown as faint lines.

For the initial stage of the colour switchover, i.e., a purely monochromatic field with frequency $\omega$, the HHG spectrum looks like the one shown in Fig. 2.17. The contributions from short and long trajectories stemming from the same half cycle interfere with each other and, due to the dynamical symmetry of the driving field, result in the cancellation of the odd-order total intensities. A slight perturbation to the driving field breaks this behaviour. In panels (a) and (f) of Fig. 10.8 we show the field and the resulting spectrum shortly after the initial stage of the colour switchover ($\theta = 1°$), i.e., a two-colour driving field with a weak $2\omega$ component. We see that the comb-like structure of Fig. 2.17(b) is broken. Moreover, the contributions of short-long pairs from within each half cycle start to separate, visible in particular at the harmonic cutoff which is here around order 60.

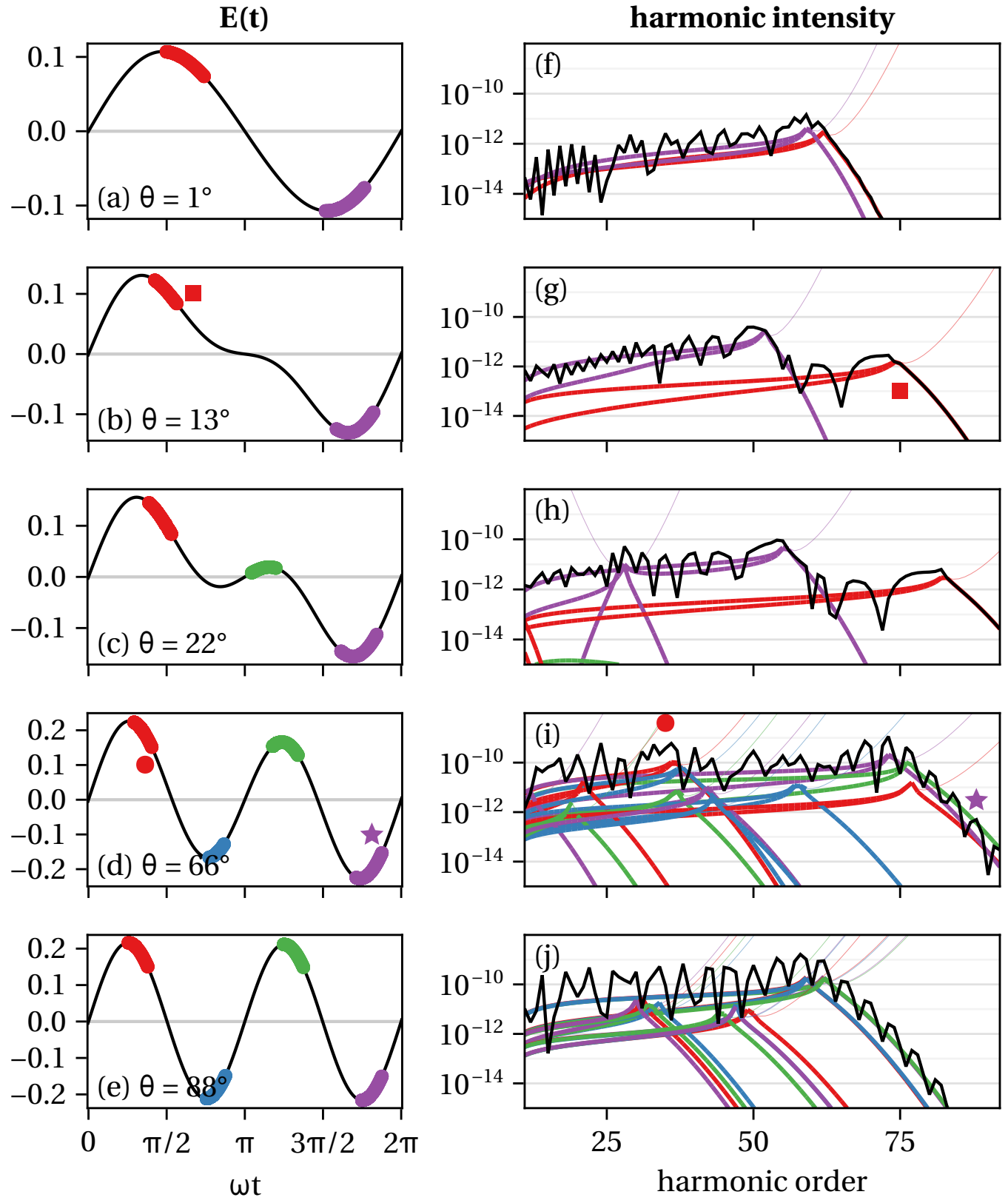

Figure 10.8: The colour switchover scheme: Electric fields $E(t)$ (in a.u. , left column) and respective HHG spectra $I(q\omega)$ (in a.u. , right column). The contributions of the various trajectories are colour-coded based on their ionisation times marked in the electric field. The markers in (g) and (i) indicate the specific contributions corresponding to the trajectories in Fig. 10.10.

When increasing the strength of the second-colour field (panels (g)–(j)), this spreading becomes more pronounced. The newly incoming saddle point solutions ultimately cause the harmonic cutoff in the spectrum to break up into two visible cutoffs (around $q = 50$ and $q = 75$ in (g)). For the later stage of the colour switchover the newly emergent trajectories produce multiple cutoffs, in panel (j) at $q \approx 32$, 47 and 60. The former, however, do not impact the total shape of the harmonic spectrum as they stem from higher-order return pairs of saddle points, i.e., trajectories with longer travel times and hence weaker contribution [172].

Upon the full completion of the colour switchover we restore the expected suppression of odd harmonic orders (of the $2\omega$ driver) due to the symmetry of the driving field.

## 10.6 Tracking trajectories throughout the switchover

Apart from looking at the shape of the total harmonic response and how the several ionisation windows interfere, the necklace algorithm gives us the unique capability to calculate the intersection number for any individual saddle point, and hence, to follow individual quantum orbits' contribution throughout the colour switchover. This has been inaccessible within the existing understanding of saddle-point methods or their extension to uniform approximations. In particular, tracking saddle points allows to examine how the individual trajectories' impact on the spectrum, as well as the corresponding electron trajectories, change.

### 10.6.1 Relevance of trajectories

In Fig. 10.9 we show the relevance to the total spectrum for all the 24 trajectories involved in the colour switchover, classified as above. Each panel shows whether the corresponding saddle points are relevant (coloured region) or not (grey) for the range of harmonic orders (horizontal axis) and across the colour switchover (vertical axis). For example, trajectories A1 and A2 (top left panels) are irrelevant at the beginning of the colour switchover. From about $\theta = 30°$, low harmonic orders start to become relevant until eventually all harmonic order up to $q \approx 60$ are relevant at completion of the switchover (at $\theta = 90°$). We will discuss the behaviour of selected trajectories below.

For now, the overview in Fig. 10.9 is shown to demonstrate some preliminary insights, as well as to present what we used as a 'tool' to conduct the classification. One main observation is that there are certain distinct behaviours that are followed by several trajectories. For example A1, A2, C1 share similar features, as well as A4, B3, C2, D3 and E2. Furthermore, the aforementioned coalescences and ambiguities in the classification show up in this visualisation as clear branch cuts, e.g. for trajectory F4. Generally, the boundary between the relevant and non-relevant regions are the Stokes lines or the branch cuts in this parameter space. This visualisation therefore demonstrates what we mentioned in the earlier Sec. 4.2.4: knowing the Stokes lines *a priori* and, hence, being able to identify these regions of contribution, would make the usage of the necklace algorithm for every single saddle point superfluous. This would be a great advantage, as the necklace algorithm is theoretically robust, but its numerical implementation sometimes suffers from choosing appropriate discretisation parameters and gives the wrong result. This can be seen in some of the panels of Fig. 10.9, where individual pixels are grey (non-relevant) within a region of relevance. For example for trajectories C4 or D1. Re-running the necklace algorithm with slightly different discretisation parameters would certainly remove these artefacts. We have chosen to keep them here, in order to demonstrate this behaviour. To further proceed with the calculation of the harmonic spectrum, of course, we correct for these artificial mis-decisions. The white regions for very low mixing angles, for example for labels C4 and E4, denote regions where no corresponding saddle points were found within our generously chosen domain of $(t_i, t_r) \in \mathbb{C}^2$. This is because for these low mixing angles the respective saddle points are coming in from asymptotically high imaginary ionisation times, as can be seen in Fig. 10.1(a).

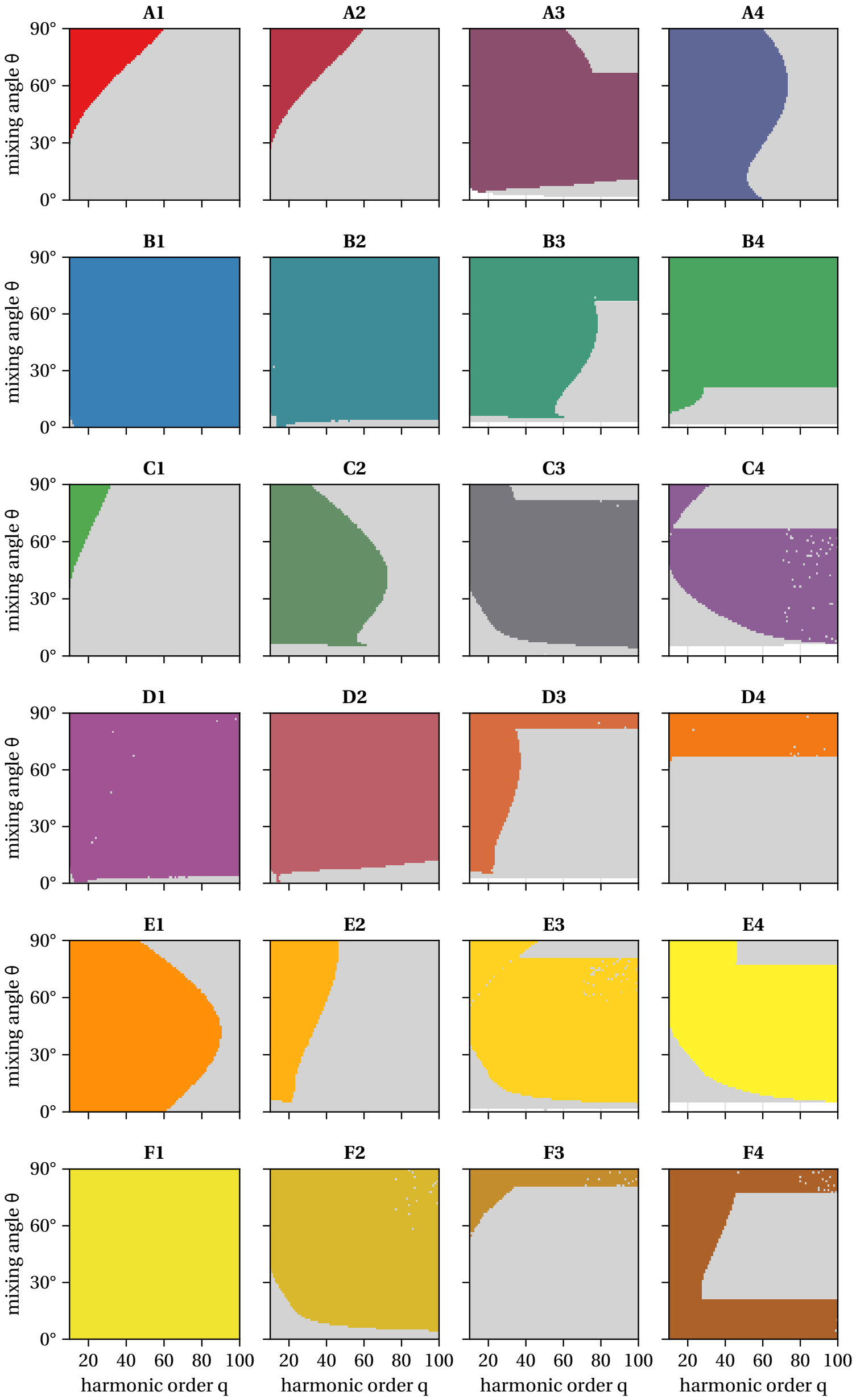

Figure 10.9: Relevance regions of all the classified saddle points. Each panel shows where in the parameter space of harmonic order $q$ and mixing angle $\theta$ the saddle points are relevant to the total spectrum (coloured area) or where they are not relevant (grey).

## 10.6.2 Electron trajectories

In Fig. 10.10 we showcase four ('types' of) trajectories in detail, as representative examples of common behaviours. In the left column we show whether the respective saddle point is relevant (coloured) or not (grey), as it is tracked for the range of harmonic orders and throughout the colour switchover. Hence, the boundary of the coloured region of contributions are the Stokes lines in parameter space, drawn as a black line. In the second column we show the specific semi-classical electron trajectories, as introduced in Fig. 2.15, for a fixed mixing angle (indicated with the rainbow-coloured horizontal line in the left panel) and a range of harmonic orders denoted with the respective colour. Similarly, the third column shows the trajectories for a fixed harmonic order (indicated by the vertical bar in the left panel) and across the colour switchover denoted with the colour gradient. For non-relevant trajectories the lines are drawn faint where it does not lead to confusion.

The first row in Fig. 10.10 shows the trajectory labelled A1, which is eventually the most dominant short trajectory for the first ionisation window of the $2\omega$ field. That is, from panel (a) we find that for $\theta = 90°$ this saddle point is relevant up to harmonic order $q \approx 60$, which constitutes the cutoff of Fig. 10.8(j). Prior to that (for $\theta < 90°$), this saddle point only contributes for lower harmonic orders, or not at all. We find this behaviour particularly interesting, as this saddle point only starts contributing quite late in the colour switchover, but then in fact plays a prominent role for the spectrum of the fully $2\omega$ driving field. The trajectories shown in the centre panel (for all harmonic orders) correspond to the contribution marked with a circle in the harmonic spectrum in Fig. 10.8(i).

In contrast to that, the trajectory A4 showcased in the second row contributes to (at least) the early plateau throughout the full colour switchover. It is the first, and hence most dominant, short trajectory starting from ionisation burst four, marked in Fig. 10.8(i) with a star. This trajectory remains one of the most dominant contributors to the spectrum throughout the colour switchover, so that its Stokes line marks a noticeable cutoff in the spectrum. The Stokes line in the left panel explains the shift in the high-order harmonic cutoff observed from orders $q = 60$ for $\theta = 0°$ to 75, 82, 78 and 60 for 13°, 22°, 67° and 88° as seen in Fig. 10.8(f)-(j) respectively.

In the third row we show trajectory F1, the first long trajectory of the second ionisation burst of the $2\omega$ driver around $\omega t_i = 2.22$, marked in Fig. 10.8(b) and (g) with a square. As a long trajectory it is relevant for all harmonic orders and remains so throughout the entire colour switchover.

The trajectory shown in the last row is B4, which is involved in the caustic mentioned to explain the enhancement around $q = 28$ in Fig. 10.8(h). Around the caustic there is a saddle-point coalescence that introduces ambiguity in the classification and shows up as a clear discontinuity in panel Fig. 10.10(d). For the beginning of the colour switchover this saddle point was a higher order return. After this branch cut, however, this trajectory eventually becomes the long trajectory of the fourth ionisation burst of the $2\omega$ field.

To conclude, Fig. 10.10 demonstrates how the electron trajectories change smoothly upon parameter scans, but their relevance to the total spectrum may change abruptly. In turn, tracking contributions from distinct quantum orbits throughout a parameter scan allows us to attribute the observable features of the harmonic spectra to these specific saddle-point dynamics.

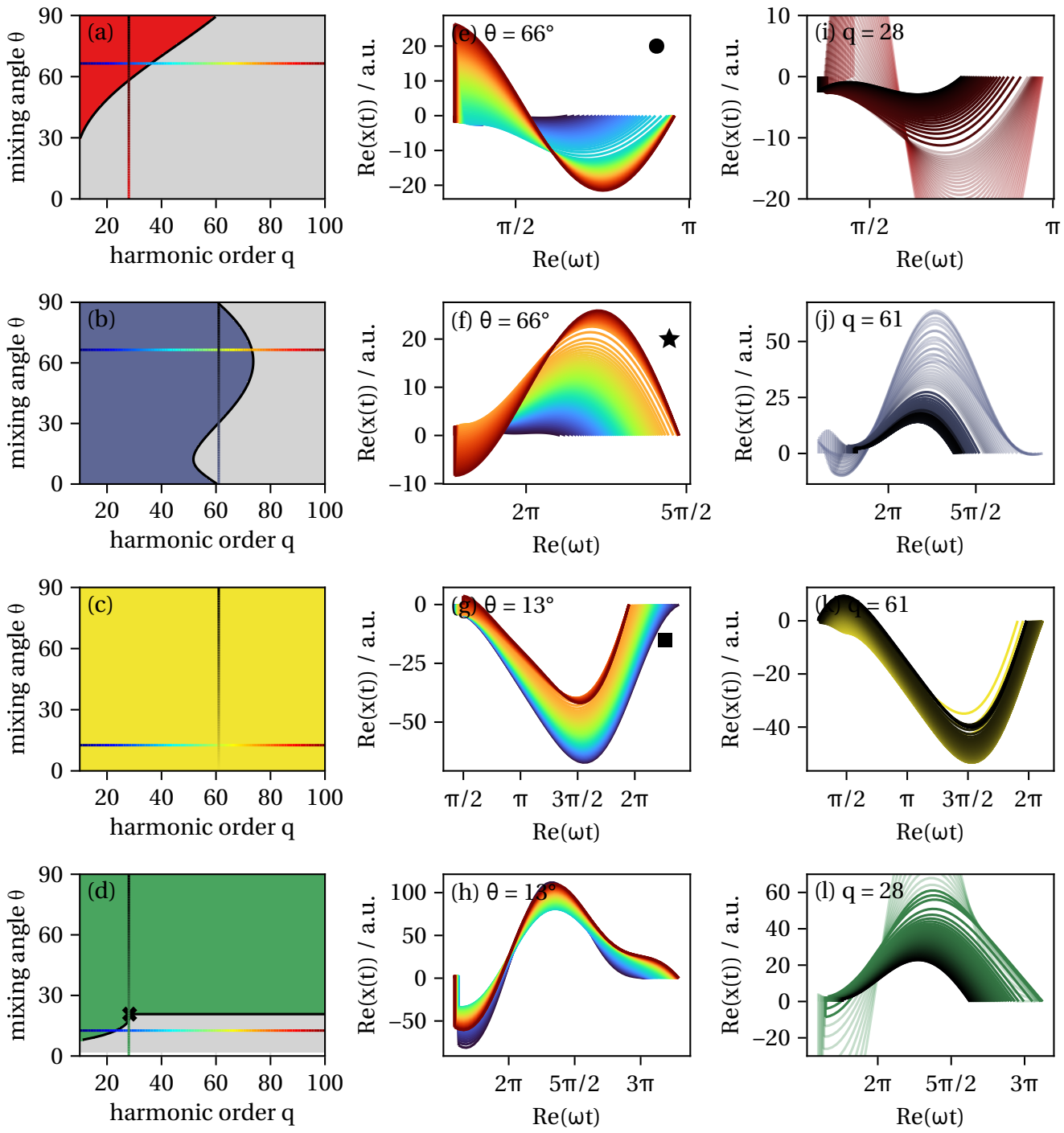

Figure 10.10: Tracking four different saddle points (rows) throughout the colour switchover and harmonic orders. They are the trajectories labelled A1, A4, F1 and B4 according to the classification scheme introduced earlier. Left column: showing whether the saddle point is relevant (coloured) or not (grey). Centre column: Trajectories according to Eq. (2.49) for the indicated $\theta$ value (horizontal line in the left panel), for all harmonic orders (lines coloured respectively). Right column: Trajectories for the indicated harmonic order (vertical line in the left panel), throughout the colour switchover (i.e., all values of $\theta$, colour shaded respectively). Markers in (e), (f) and (g) attribute contributions to the spectra in Fig. 10.8(j) and (g).

## 10.7 The interplay of contributions to the spectrum

With a detailed understanding of the relevance of the individual trajectories within the two-dimensional parameter space spanned by the harmonic order and the mixing angle we are now in a position to assume a more explicitly catastrophe-theory-inspired perspective. Rather than analysing the individual trajectories across the colour switchover, we focus on the total harmonic intensity, which is composed from the coherent summation of contributions from all relevant trajectories, and examine its behaviour in the two-dimensional parameter space.

As a first step, Fig. 10.11 illustrates the number of relevant saddle-point contributions that enter the total harmonic signal. This number varies between four (at the very beginning of the colour switchover, $\theta = 0°$) and 24 (upon completion of the switchover, $\theta = 90°$). In between, we observe an intricate structure, whereby every change in the number of contributing saddles marks a Stokes transition. For the central region of the colour switchover (when $30° \lesssim \theta \lesssim 60°$) the number of contributing saddle points remains constant throughout the harmonic plateau. As expected, around the harmonic cutoff order there are many changes in the number of contributors. As Stokes transitions are linked to catastrophe points, they may result in observable caustics within the total signal. Therefore, Fig. 10.11 alludes to the complexity of the underlying interference structure.

In Fig. 10.12 we show the total harmonic intensity across the same parameter space, in a logarithmic colour scale. The most obvious interference pattern are the fringes between harmonic orders, which can be attributed to the half cycle symmetry. Furthermore, however, we find two big lobes of increased intensities. One signifies the (main) harmonic cutoff, making an inverted S-shape around harmonic order $q \approx 58$ across the colour switchover. The other lobe emerges at $\theta \approx 40°$ from low harmonic orders towards the cutoff harmonic order $q \approx 58$ at the end of the colour switchover, where it coincides with the former cutoff-lobe. This one can be attributed to the cusp catastrophe originating in the fourth ionisation burst, which we study in more detail in the later Sec. 12.2.

Overall, Fig. 10.12 clearly reveals rich interference structures that closely resemble those characteristic for the canonical diffraction integrals shown in Sec. 3.2.4. This observation provides strong motivation to further use catastrophe theory as a framework to analyse and interpret the underlying dynamics. In this sense, the present chapter establishes a direct link between the individual trajectory picture and the wave patterns that emerge from their interference.

## 10.8 Conclusions

The versatility of the Picard–Lefschetz methods we developed in Chapters 5 and 6 allows us to address questions of a new class of parameter scans embodied by the colour switchover. The gradual replacement of a monochromatic driving field with its second harmonic, via two-colour configurations of increasing amplitude ratio, connects the perturbative second-colour regime to fully bichromatic driving fields. Using the necklace algorithm, we are able to identify which of the many quantum orbits are relevant to the total spectrum, for any intermediate driving-field configuration. In the perturbative case, we can attribute the unfolding of the high-harmonic cutoff to the dominant pairs of trajectories from the respective half cycles. Increasing the relative strength of the second harmonic eventually leads to newly emerging ionisation bursts that produce topologically stable enhancements in the spectrum. These arise due to the unavoidable proximity to a three-fold saddle-point coalescence, i.e., a cusp catastrophe. Moreover, tracking individual saddle point solutions throughout the colour switchover allows us to show how the electron trajectories react to the change of driving field.

We find an interesting similarity between the colour switchover considered for strong-field ionisation (in Chapter 8) and that for the generation of harmonics. In both cases, new ionisation (and recombination) saddle points enter the complex planes from high imaginary values. In order for them to become relevant, however, ionisation events have to undergo a fold coalescence. For HHG, it appears the emergence of new trajectories is tightly linked to the underlying cusp catastrophe. The specifics of this particular cusp point will be detailed in Sec. 12.2 further below.

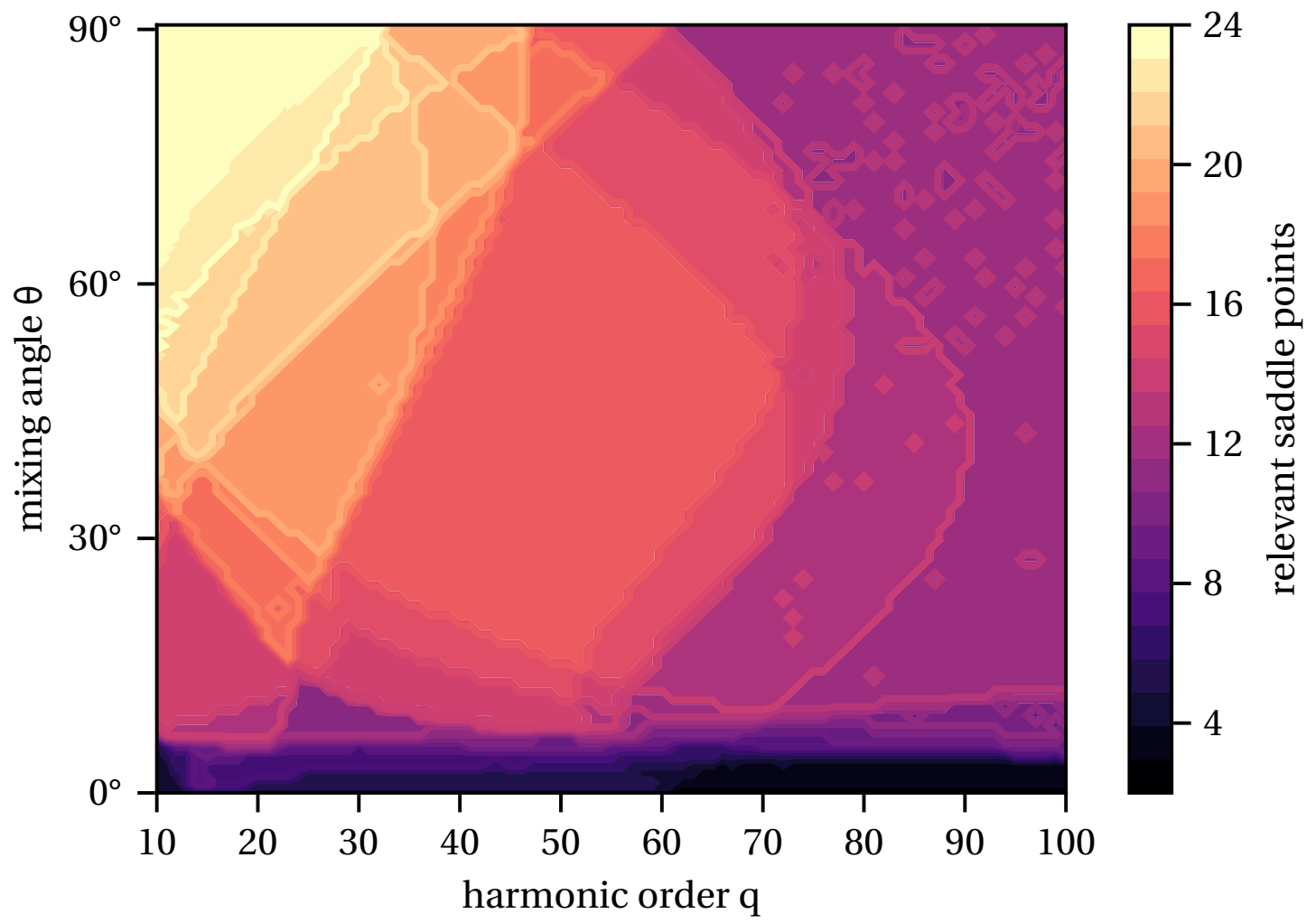

Figure 10.11: Number of contributing saddle points (colour bar) to the integral for the total harmonic response, across harmonic orders (horizontal) and the colour switchover (i.e., for the mixing angle, vertical).

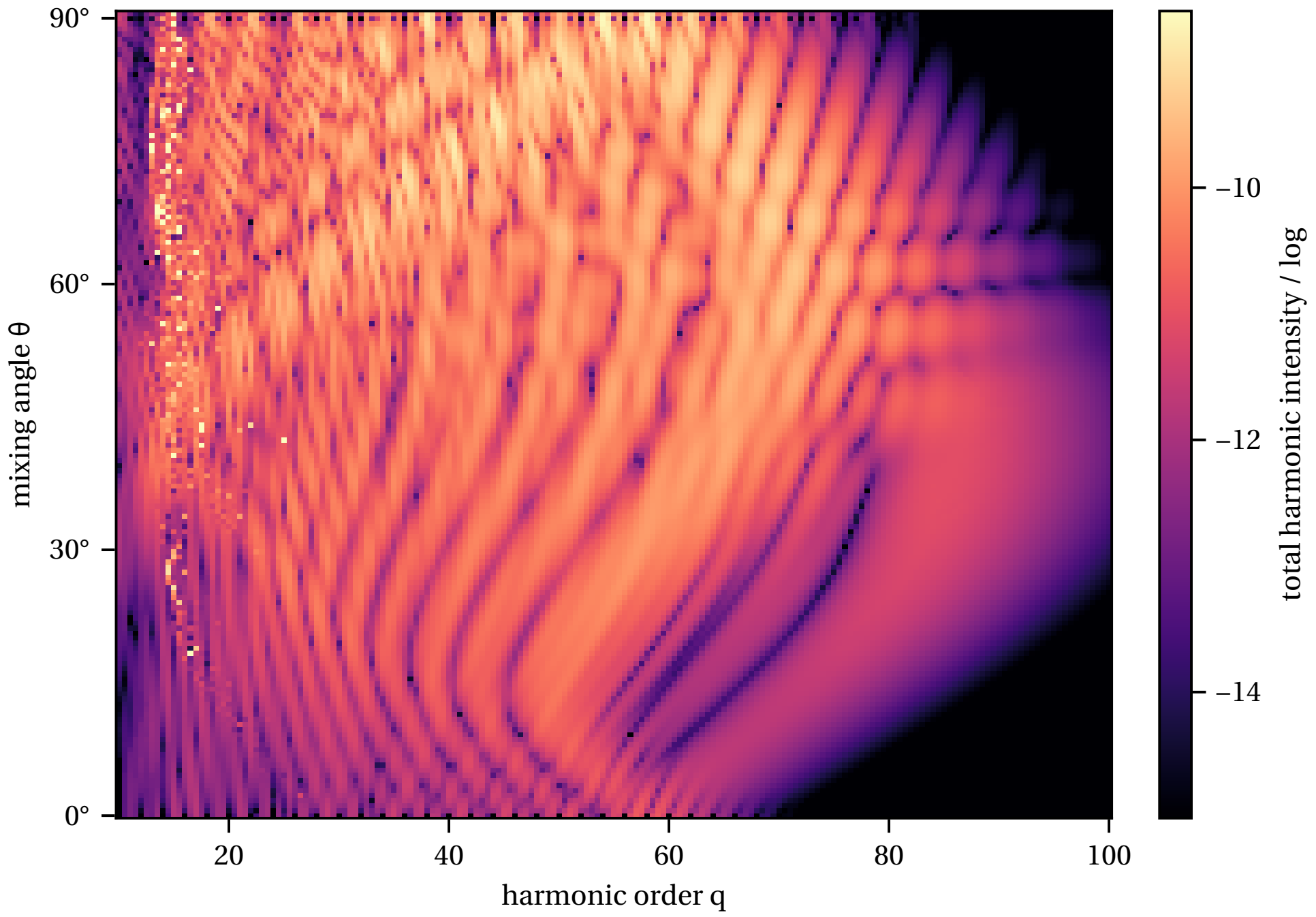

Figure 10.12: Total harmonic intensity (colour bar) across harmonic orders and colour switchover, calculated using the downward flow algorithm.

# 11

# Phase scans in co-orthogonal two-colour HHG

While the previous chapters examined the variation of the amplitude ratio in a two-colour driver, this chapter focuses on a different control parameter: the relative phase between the fundamental field and its second harmonic. We specifically consider regimes where the second-colour field is no longer strictly perturbative, but reaches intensity ratios in the order of ten percent of the fundamental field. Upon a scan of the two-colour relative phase, oscillations of the harmonics' intensity at the frequencies of the driving field have been observed in many experimental and theoretical studies. Their interpretation in this strongly polychromatic regime, however, remains understood less well.

The results presented here follow from a fruitful collaboration within the XUV Attosecond Source Wavefront Optimization (XAWO) project at the ELI ERIC Beamlines facility. Experimental measurements were carried out by Xiaozhou Zou and Lucie Jurkoviča. Here, we present the theoretical analysis of saddle-point dynamics and the intensity modulations upon a two-colour phase scan. Following that, we demonstrate how the observed intensity modulations can be attributed to individual electron trajectories and ultimately signifies quantum-path interference. This was partially developed in close collaboration with Xiaozhou Zou, and has been presented in his PhD thesis [199] and accepted for publication [3]. Ongoing work continues to build on these results, approaching the saddle-point analysis in a more systematic way, to further disentangle the interplay of several orbits across a phase delay scan in a macroscopic target range.

## 11.1 Introduction

When combining multiple light fields, their relative phase plays a central role in shaping the resulting electromagnetic waveform and its nonlinear interaction with matter. In attosecond science, this sensitivity is commonly exploited by superimposing a fundamental driving field with its second harmonic. A relative phase delay can thereby be precisely tuned through controlled modifications of the optical path length, for example using a rotatable nonlinear crystal. Scans of this phase delay therefore constitute a powerful interferometric tool for probing strong-field electron dynamics on a subcycle timescale.

A common scheme is the co-propagating superposition of a fundamental field with its second harmonic. Varying the relative phase difference[1] between the two components allows to measure ionisation time [182], selectively suppress or enhance harmonic output [176], shape the spectral properties of the attosecond pulse [178], and ultimately demonstrates the control over the taken trajectories [187]. Similarly, adding a second colour field in orthogonal polarisation direction offers increased control over the electron dynamics and the generated attosecond pulse, both theoretically [200], and experimentally [201]. As the second colour is added in orthogonal polarisation direction, the two half cycles of the field are not symmetric any more and hence, both even and odd harmonic orders are generated. Furthermore, the polarisation (or rather, emission angle) of the harmonic radiation offers yet another observable to encode information about the electron dynamics. This has been exploited to retrieve the ionisation and recombination times [173, 182] and tunnelling times [175]. Previous studies have been conducted with *weak* commensurate fields, where the second harmonic field would act as a perturbation to the processes triggered by the fundamental driving field — we will refer to these as "perturbative" two-colour fields. The goal of the experimental XAWO campaign was to go beyond this perturbative regime and add a strong second harmonic field to the fundamental driver. In those scenarios we expect the second colour component to not only 'slightly modify' the electron trajectories but to actually add new, relevant trajectories, at least for some phase shifts.

On the theoretical side, previous studies would use full TDSE calculations, the numerical integration of the SFA integral — possibly windowing the time integration to achieve a trajectory selection — or, a fully classical approach to analyse the processes in this strongly bichromatic regime. Using Picard–Lefschetz theory, as introduced in Chapter 6, now provides the tools to study two-colour phase scans in the quantum-orbit approach even for two-colour fields with a strong second colour component.

Let us briefly comment on some aspects relevant to the experimental realisation of phase-delay scans. High-order harmonic generation driven by two-colour fields can be implemented either in an interferometer-like configuration, where the properties of each colour are controlled independently in separate arms, or in a colinear 'in-line' setup, schematically shown in Fig. 11.1. A detailed description of the experimental implementations used in this project can be found in Refs. [199, 202]. The second harmonic field is generated using a beta-barium borate (BBO) crystal. Rotating the calcite plate modifies the optical path lengths experienced by the fundamental and second harmonic components, thereby changing their relative phase delay. A calibration procedure is required to relate the crystal rotation angle to the corresponding phase shift. This procedure, however, yields only the relative phase between the two colours and provides no access to their absolute optical phases. Further details of this calibration are given in the aforementioned theses. Overall, such phase-delay scans are experimentally straightforward to perform. At the same time, certain limitations must be kept in mind. In particular, the laser intensity 'at the atom' is challenging to measure directly and must instead be inferred indirectly. As a consequence, the intensity ratio between the $\omega$ and the $2\omega$ components can only be given as rough estimates, typically based on the focal beam sizes together with constraints imposed by the optical layout and beam-propagation simulations. Furthermore, the rotation of the calcite plate modifies the phase delay between the two incoming *pulses* of fundamental and second harmonic. In the following sections we study the influence of the relative two-colour phase as the phase difference between the two field components, neglecting any shape of the pulse envelope. As a continuation of this enjoyable collaboration with ELI Beamlines we aim to extend our approach to include the effects of the pulse envelopes.

[1] Note that here we consider solely the phase difference $\varphi$, which we refer to as the "two-colour phase", between the two field components, as introduced in Sec. 2.2. This is related to (but not equivalent) to the phase delay between the envelope of two *pulses* of different frequencies.

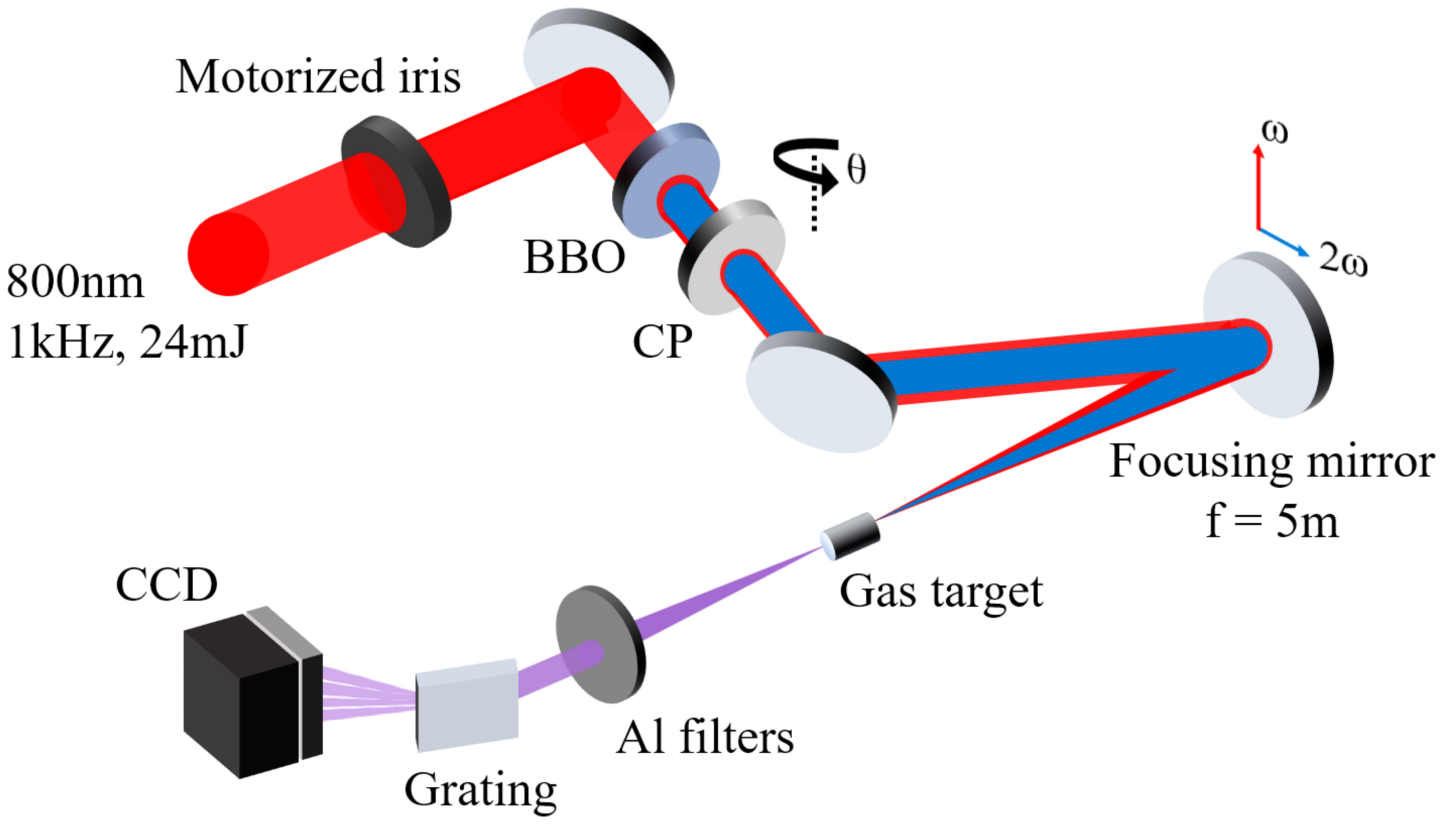

Figure 11.1: Schematic of the experimental setup. The fundamental beam (red) generates the second harmonic (blue) in a BBO crystal. The two-colour relative phase is controlled by the fine rotation of the calcite plate (CP) $\theta$ angle. The bichromatic field is then focused into an argon gas cell by a spherical mirror with a 5 m focal length, and the high harmonics are characterised by a flat-field spectrometer. Taken from [199].

## 11.2 Saddle-point-based approach to the phase scan

### 11.2.1 Theoretical methods

In the following, for the calculation of the saddle points, their relevance to the total HHG integral, as well as the resulting per-orbit dipoles we use the methods described in great detail in Chapter 6 above. The total harmonic dipole will be calculated as a sum over Gaussian contributions of the relevant saddle points, Eq. (2.50). We consider a driving field comprised of two linearly polarised fields with frequencies $\omega$ and $2\omega$, with co-orthogonal polarisation axes as introduced in Sec. 2.2. We assume the fundamental field in $x$-direction and the second harmonic field in the $y$-direction, with the two-colour relative phase $\varphi$, such that the total electric field reads:

$$\mathbf{E}(t) = E_{01}\sin(\omega t)\mathbf{e}_x + E_{02}\sin\big(2\omega t + \varphi\big)\mathbf{e}_y\,. \tag{11.1}$$

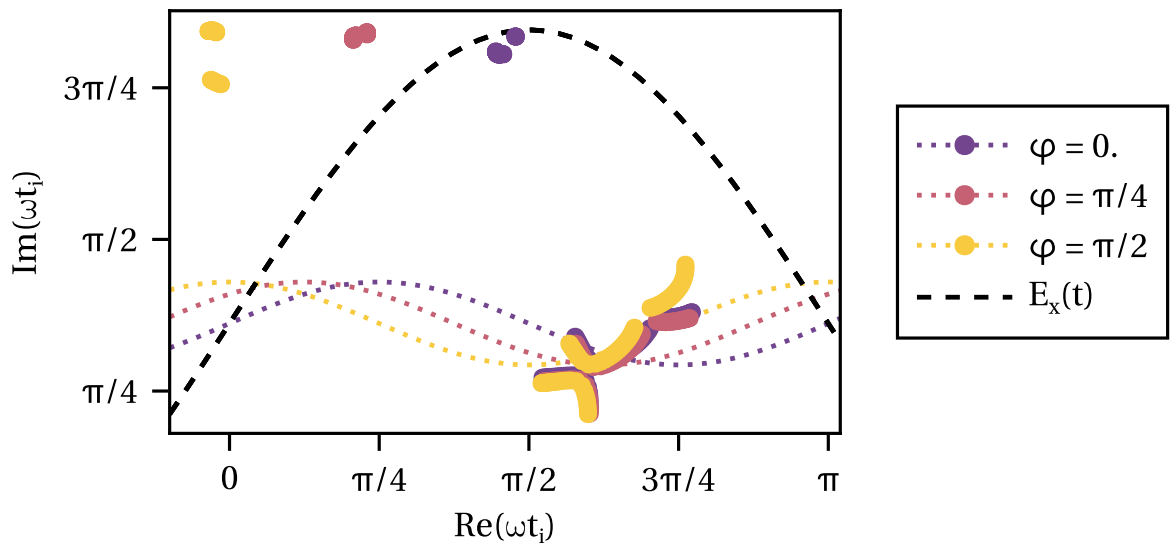

Figure 11.2: Complex ionisation times of the saddle points (coloured markers) for a range of harmonic orders (not colour-coded) stemming from the field components (fundamental $x$-component in black dashed, weak second colour component in coloured dotted lines) of a perturbative two-colour field ($R_I = 2\%$), for three different relative phases of the second colour field. The different two-colour phases $\varphi = \{0, \pi/4, \pi/2\}$ introduce new saddle points at high $\mathrm{Im}(\omega t_\mathrm{i})$, at different $\mathrm{Re}(\omega t_\mathrm{i})$, and slightly modify the saddle points stemming from the fundamental driver at $\mathrm{Im}(\omega t_\mathrm{i}) \approx \pi/4$.

The amplitudes of the two constituent fields are expressed in terms of the *intensity ratio* as $R_I = I_{2\omega}/I_\omega$, with the intensities $I_{2\omega}$ and $I_\omega$ of fundamental and second-harmonic driving field, respectively. That is, $E_{01} = E_0$ and $E_{02} = \sqrt{R_I}E_0$, where $E_0 = \sqrt{I_\omega}$. This convention is used in accordance with estimates from the experiments.[2] Throughout this chapter we will state the intensity ratio in percentages, just to differentiate it from other conventions for amplitude ratios in this thesis. Note that, even though the electric field $\mathbf{E}(t)$ itself is now a vector, the action $S_\mathrm{HHG}(t_\mathrm{i}, t_\mathrm{r})$ remains a scalar. The HHG dipole $\mathbf{D}(q\omega)$, however, is a vector because it includes the dipole matrix element at recombination time, which ultimately includes the vector potential.

For the calculations in the upcoming sections we use the following configuration parameters: $\mathcal{I}_\mathrm{p} = 15.76\,\mathrm{eV} = 0.58\,\mathrm{a.u.}$ (Argon), $I_\omega = 1.5\times 10^{14}\mathrm{W/cm^2}$ ($E_0 = 0.065\,\mathrm{a.u.}$), and $\lambda = 800\,\mathrm{nm}$ ($\omega = 0.057\,\mathrm{a.u.}$). The classical harmonic cutoff order for this configuration is around 29 (for intensity ratios up to $\approx 12\%$), and the threshold harmonic order is at $q \approx 10$, such that we will typically consider harmonic orders in the range of 15 to 40, if not mentioned otherwise.

### 11.2.2 Saddle-point dynamics for a perturbative two-colour field

As thoroughly discussed in the previous Chapter 10 for a co-linear two-colour laser field, the added second-colour driver introduces new saddle points, coming in from high imaginary parts for the ionisation times. This holds true for a co-orthogonal field as well. The real part of the newly introduced saddle points depends on the two-colour phase of the second-colour field. That is because the ionisation saddle points are around the maxima of the respective field component. This is shown in Fig. 11.2, where we show the ionisation times of the saddle points for harmonic orders (not colour-coded) for the case of a perturbatively added second colour field, compared for three different two-colour phases. The fundamental $\omega$-field (black dashed $E_x(t)$) causes the ionisation shortly after its maximum at $\mathrm{Re}(\omega t) \approx \pi/2$, resembling the familiar saddle point structure of a monochromatic field at around $\mathrm{Im}(\omega t_\mathrm{i}) \approx \pi/4$. The weak second colour field, with $R_I = 12\%$, is shown in dotted lines for three different two-colour phases $\varphi = \{0., \pi/4, \pi/2\}$ (purple, burgundy, yellow, respectively). At around the (real) time of the respective field maxima, we find saddle points at high imaginary parts, $\mathrm{Im}(\omega t_\mathrm{i}) \approx 3\pi/4$. For an intensity ratio as low as this, the saddle points stemming from the second-colour field are not yet relevant contributors to the total dipole. However, the perturbative field slightly modifies the saddle points from the fundamental driving field, as can be seen by taking a closer look at the respective saddle point structure in Fig. 11.2. This change in ionisation time can be observed experimentally and can furthermore be used to actually reconstruct the ionisation times, as shall be detailed below.

[2] Note that an increase of $R_I$ will also increase the total intensity experienced by a single atom, which would make this convention unsuitable for a colour switchover.

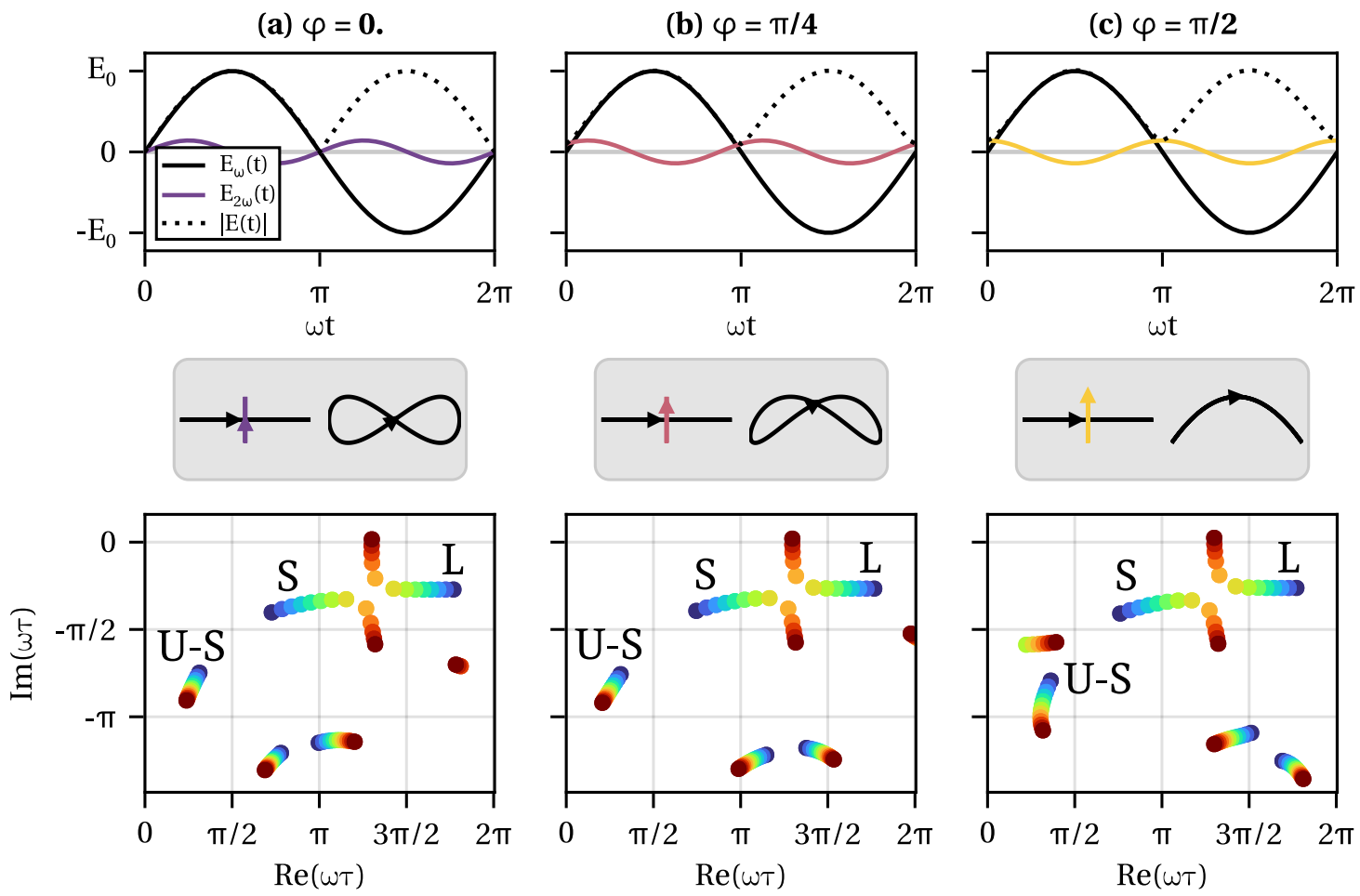

Figure 11.3: Driving fields (top and middle row) and complex-valued travel times (bottom row) for a weak second-colour driver ($R_I$ = 2%) for three different two-colour phases $\varphi$ (columns).

For a quantum-orbit-based analysis of two-colour phase scans in HHG we wish to classify the saddle point solutions, such that, for example, we can study individual trajectories' contributions to the total harmonic dipole later on. For the case of a weak second-colour field we therefore show the complex travel times for the three different two-colour phases in Fig. 11.3. For each phase (columns (a) $\varphi = 0$, (b) $\varphi = \pi/4$, (c) $\varphi = \pi/2$), we show the electric field components in the top panel where the solid line shows the fundamental driver, the coloured dotted line denotes the weak second harmonic, and the total field's amplitude is drawn in black dashed. Below, we show the Lissajous figures of the fields in the $x-y$-plane, demonstrating how they create different polarisation shapes. The real and imaginary parts of the travel times for the respective saddle points within one cycle of the fundamental are shown in the bottom row, coloured according to the harmonic orders ranging from 15 (blue) to 40 (red). For all three two-colour phases we find the familiar structure of the missed approach of two branches, at around $\omega\tau \approx +\frac{5\pi}{4} - \frac{\pi}{4}\mathrm{i}$. These are the short (S) and long (L) trajectories of the fundamental driving field, and we find they remain largely unchanged across the three exemplified two-colour phases. A ultra-short (U-S) trajectory is apparent (at around $\omega\tau \approx \frac{\pi}{4} - \frac{3\pi}{4}\mathrm{i}$), but remains seemingly unaffected by the second colour field as well. At very low imaginary travel times ($\mathrm{Im}(\omega\tau) \lesssim -\pi$), we find the new trajectories, associated with the second colour field.[3] These indeed move towards longer (real) travel times for an increased two-colour phase.

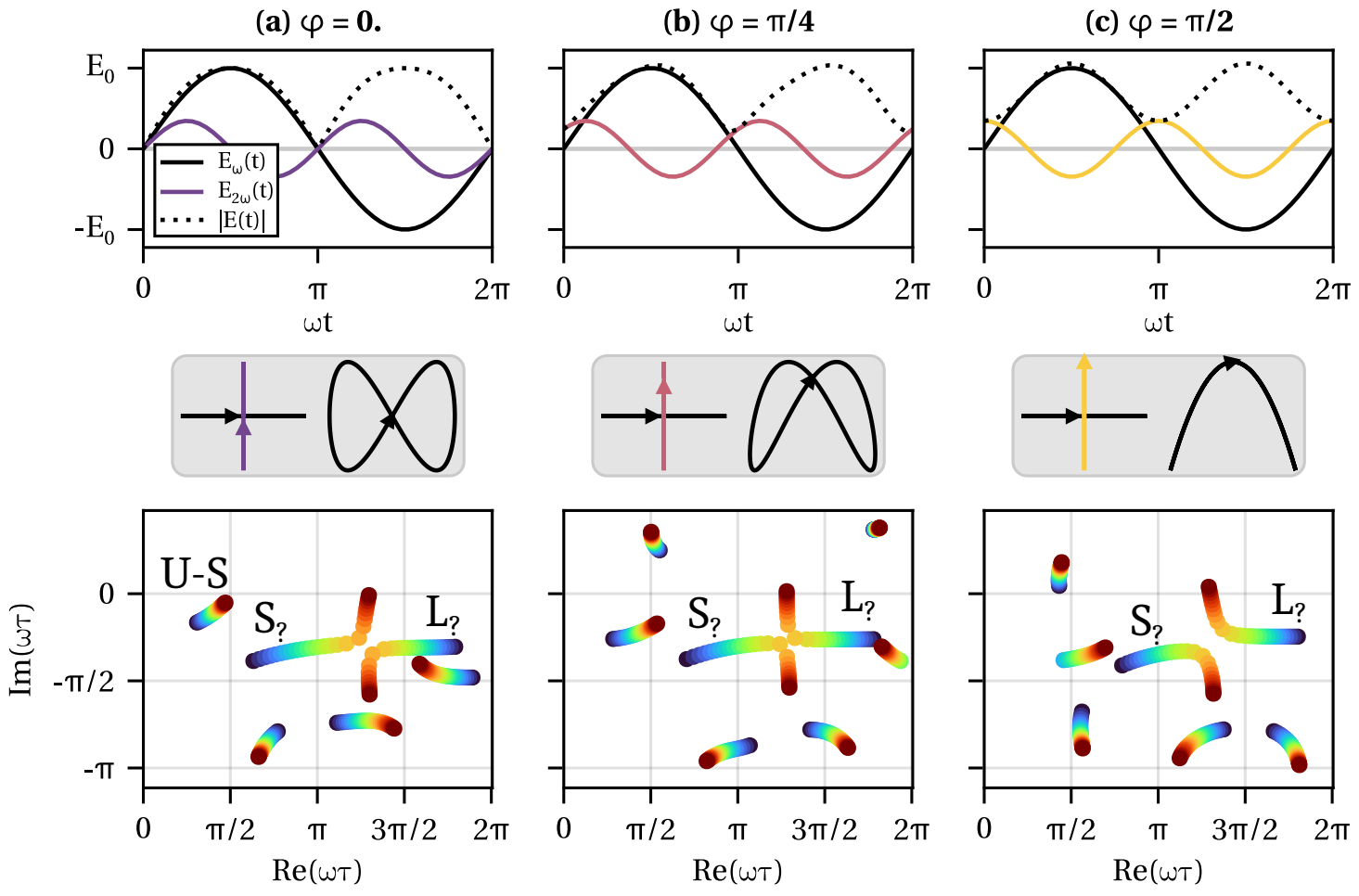

Figure 11.4: Driving fields (top and middle row) and complex-valued travel times (bottom row) for a strong second-colour driver ($R_I = 12\%$) for three different two-colour phases $\varphi$ (columns).

For a classification of saddle point solutions throughout a full two-colour phase scan, we can simply track the individual branches of solutions across the two-colour phase scan. This is possible because the solutions associated with the weak second-colour driver do not 'interfere' (in the non-quantum-mechanical sense of the word) with the fundamental's pair of short and long trajectories.

### 11.2.3 Saddle-point dynamics for a strongly bichromatic driving field

If the second-colour component is added with a higher relative amplitude, however, the saddle point landscape changes significantly. In Fig. 11.4 we show the corresponding driving fields and respective saddle point travel times as above. Firstly, let us remark on the driving fields, shown in the top and centre rows. As showcased in the introduction

[3] Note that $\tau = t_r - t_i$, and with $\mathrm{Im}(t_r) \approx 0$, the high magnitudes of $\mathrm{Im}(t_i)$ seen earlier result in low values of $\mathrm{Im}(\tau)$.

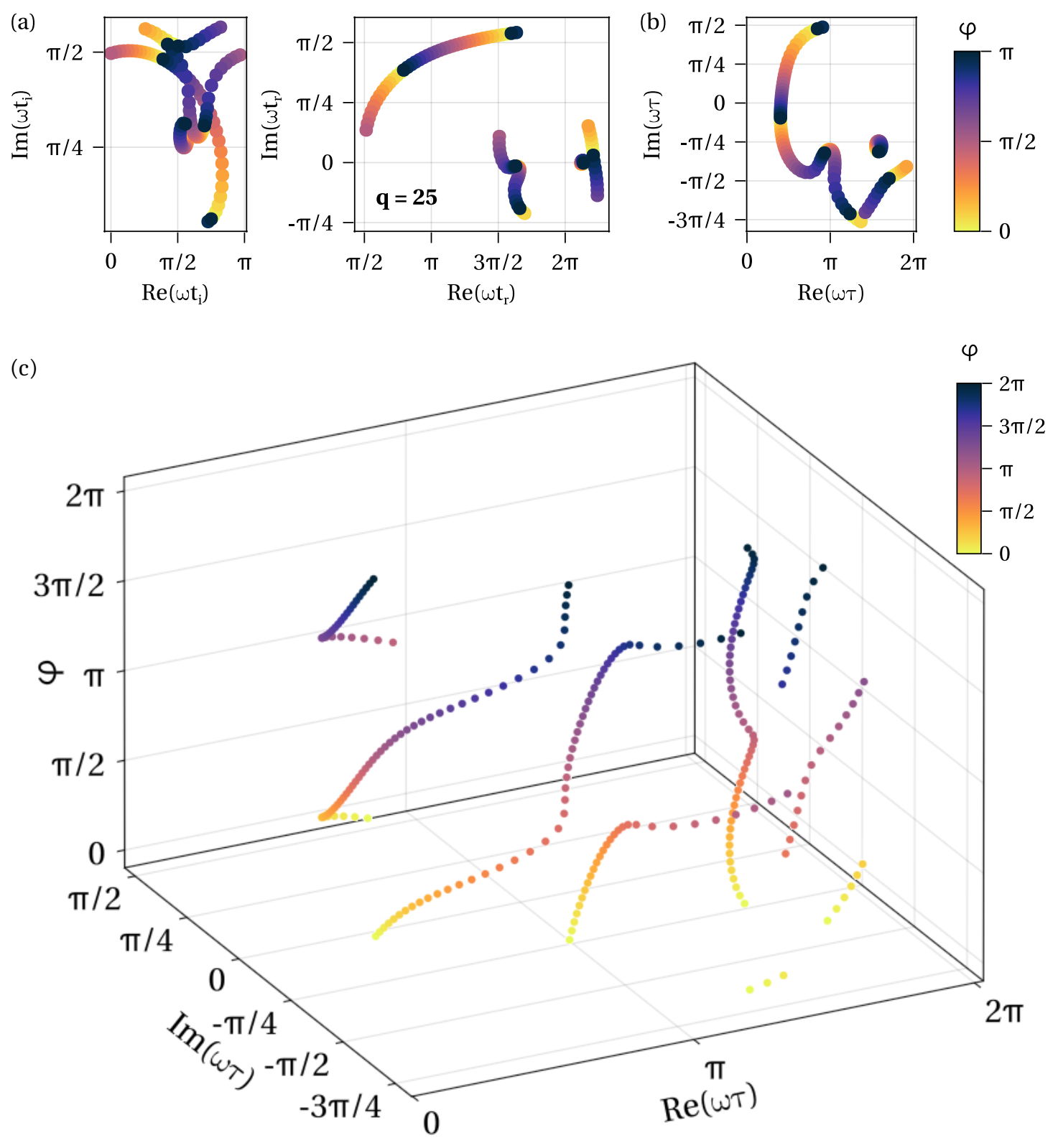

Figure 11.5: Selected saddle points in the complex planes of (a) ionisation and recombination, and (b) travel time for harmonic order $q = 25$, for two-colour phases $0 \leq \varphi \leq \pi$ (colour bar). The driving field is a strong second-colour driving field ($R_I = 12\%$). The travel times from (b) are shown for the full period of $0 \leq \varphi \leq 2\pi$ as a third vertical axis in (c).

Sec. 2.2, for a strong second-colour field the relative phase has a substantial effect on the total driving field. Correspondingly, the saddle point structure remains less stable compared to the case of a weak second-colour field. For each two-colour phase scan we may identify the dominant missed approach stemming from the fundamental driver, as well as other trajectories that may be associated with the additional driving field. Attempting a rigorous classification of the solutions across a performed two-colour phase scan, however, is challenging as the saddle points undergo some intricate dynamics. In particular, we find the new trajectories start to interfere with the fundamentals' short and long trajectory solutions in a way that makes a rigorous classification non-trivial. This is further complicated as for the strong second-colour field the new trajectories start to become relevant contributors to the harmonic dipole (at least for some harmonic orders).

To further visualise this in Fig. 11.5 we show how the saddle points for a fixed harmonic order, here $q = 25$, move in the complex planes upon a two-colour phase scan. As usual, in (a) we show the complex planes of ionisation and recombination time, and in (b) the complex travel time plane. We show a selection of saddle points, ionised within the first half cycle of the driving field, and they are coloured according to the two-colour phase $\varphi$ (see colour bar), which (for panels (a) and (b)) varies between 0 and $\pi$. In both, the recombination and the travel time plane we find saddle points at $\varphi = \pi$ overlap with those from $\varphi = 0$. That is, upon the two-colour phase scan some saddle points take other saddles' place in the complex plane. To clarify this, in Fig. 11.5(c) we show the travel times in the complex plane and show the two-colour phase as a third, vertical axis. Note, that in this, we show the full revelation of the two-colour phase, going from $\varphi = 0$ to $2\pi$. This demonstrates that a classification scheme that holds consistently across the two-colour phase cannot be based on absolute values within the complex plane but has to take dynamics features into account.

One particular situation that furthermore makes the consistent classification of saddle point solutions throughout a two-colour phase scan challenging is visualised in Fig. 11.6. We show a selection of saddle points for a range of harmonic orders (coloured), for the case of a strong second-colour field ($R_I = 12\%$), and across a small range of two-colour phases in panels (b). The saddle points may be classified as four separate trajectories, involved in a complicated interplay which has yet to be disentangled. Panels (a), (c) and (d) show different views of the real and imaginary parts of the travel times across a small range of two-colour phases in the third dimension. This corresponds to a situation between column (a) and (b) of the previous figure Fig. 11.4,[4] and hints at an underlying catastrophe and branch cut that complicates the consistent classification across a two-colour phase scan.

[4] Note that the temporal structure of the problem is $\pi/2$-periodic in the two-colour phase, such that the range $\varphi = 2.0$ to 3.0 shown in the 3D structure in Fig. 11.6 appears in the same fashion between $\varphi = 0.43$ to 1.43.

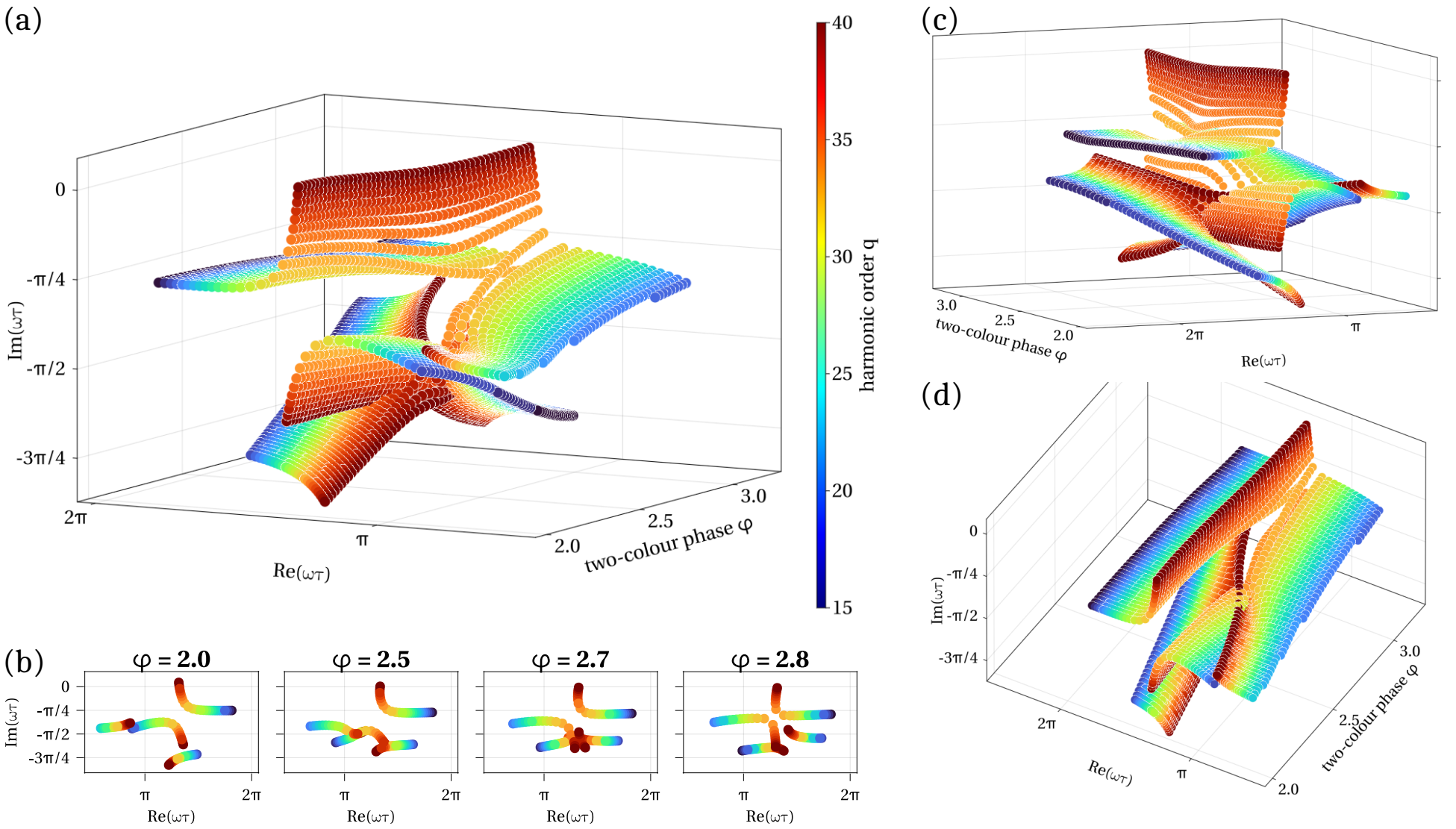

Figure 11.6: Intricate structure of saddle points hinting an underlying catastrophe for $R_I = 12\%$, shown as real and imaginary part of the travel time across the two-colour phase, $(\mathrm{Re}(\tau), \mathrm{Im}(\tau), \varphi)$, in several views (a), (c), (d). Slices of it at fixed $\varphi = \{2.0, 2.5, 2.7, 2.8\}$ are shown in (b).

## 11.3 Oscillations of the harmonic intensity

The two-colour phase scan bears some intriguing challenges for the classification of saddle points and the rigorous and consistent definition of 'types' of trajectories, especially for two-colour fields with a strong second-colour component. Experimentally, however, performing the two-colour scan is rather simple, independently of the amplitude ratio of the two field components. Here, we therefore do not dwell on developing a cunning classification scheme to disentangle the intricate structure of saddle points, but move swiftly on towards observables that may be compared with experimental measurements.

Let us first examine the case of a perturbative second-colour field which we consider with $R_I = 2\%$. For the case of such a perturbative second-colour field we can identify the short and long trajectory throughout the two-colour scan, as the respective saddle points are only slightly modified by it. The saddle points from the second-colour field are not yet relevant to the total dipole. For both short and long trajectories the harmonic intensity smoothly oscillates over the two-colour phase scan. In Fig. 11.7 we show those oscillations separately for short (top) and long (bottom) trajectories, and normalised within each harmonic order. Across all harmonic orders the oscillations follow the fundamental driving field's frequency, which can be proven analytically [203]. The short trajectories, however, show a significant drift.[5] This drift will be relevant in the subsequent section. For now, note that the long trajectories also exhibit such a drift, but not as prominent. The 'gap' around harmonic order 32 (yellow) can be attributed to the high-order harmonic cutoff. Around the harmonic cutoff the saddle points move 'fast' in the complex plane, such that for two subsequent harmonic orders in the cutoff region the saddle points are further apart than, say, for two subsequent orders in the plateau region.

[5]We want to avoid the term phase shift, just to avoid confusion with all the other phases relevant in this context.

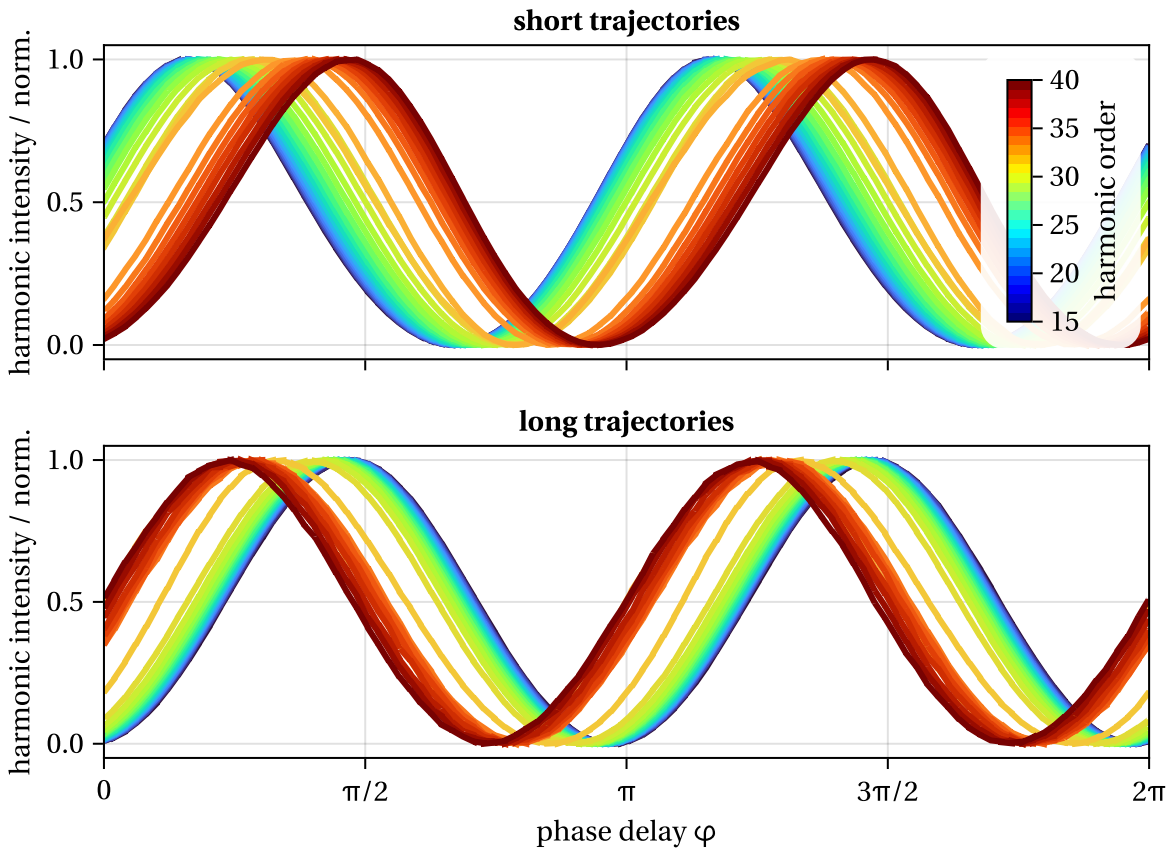

Figure 11.7: Intensity oscillations of short (top) and long (bottom) trajectories upon a two-colour scan of a weak second-harmonic driving field ($R_I = 2\%$), for a range of harmonic orders.

Let us turn to the oscillations of the total harmonic intensity. These are shown in Fig. 11.8 for harmonic orders of the plateau region, odd orders in the top panel, even orders in the bottom panel. We show the total intensity as a solid line across the two-colour scan. Individual saddles' intensities are drawn as markers. Note that in this case the saddles' intensities exhibit clear oscillations, in the same orders of magnitude for even and odd harmonic orders. They can be attributed to short (S) and long (L) trajectories. The total intensities, however, show signatures of quantum path interference as even harmonic orders (bottom panel) are significantly suppressed compared to the odd orders (top panel). This observation will be discussed in more detail below.

For the case of a strong second-colour field component the situation becomes more interesting. As outlined above, a thorough classification of saddle point solutions across the phase-delay scan is not as straight-forward. Thanks to Picard–Lefschetz theory we can calculate the individual saddles' relevance and obtain the individual intensities as well as the total intensity, regardless of the classification. In Fig. 11.9 we show the intensity oscillations in the same fashion as above, but for a strongly polychromatic driving field with $R_I = 12\%$. For odd harmonic orders (top panel) the total intensity exhibits smooth oscillations, similarly to the case of a perturbative driver. The individual saddles' intensities, however, now draw a different picture: For each harmonic order we find one 'line' of contributions that oscillates smoothly as in the case of the weak second-colour driver. These can be associated with the long trajectories and are marked with L in Fig. 11.9. The other set of solutions, however, do not oscillate smoothly, but are in fact different, individual 'humps'. These can be attributed to short(er) trajectories (marked as $S_1$ and $S_2$). However, it becomes now clear that the short trajectory at, say $\varphi = 0$ is not the same object as the short trajectory at, say, $\varphi = \pi$. In turn, for a small region around $\varphi = 3\pi/4$ (and $7\pi/4$) there are in fact two short trajectories that contribute in comparable orders of magnitude. This is a central finding that complicates further analysis as we shall see below. For the even harmonic orders the individual saddles' contributions show the same behaviour. The total intensity of each harmonic order has a less uniform behaviour. It follows a 'double hump' structure that is more pronounced for higher harmonic orders.

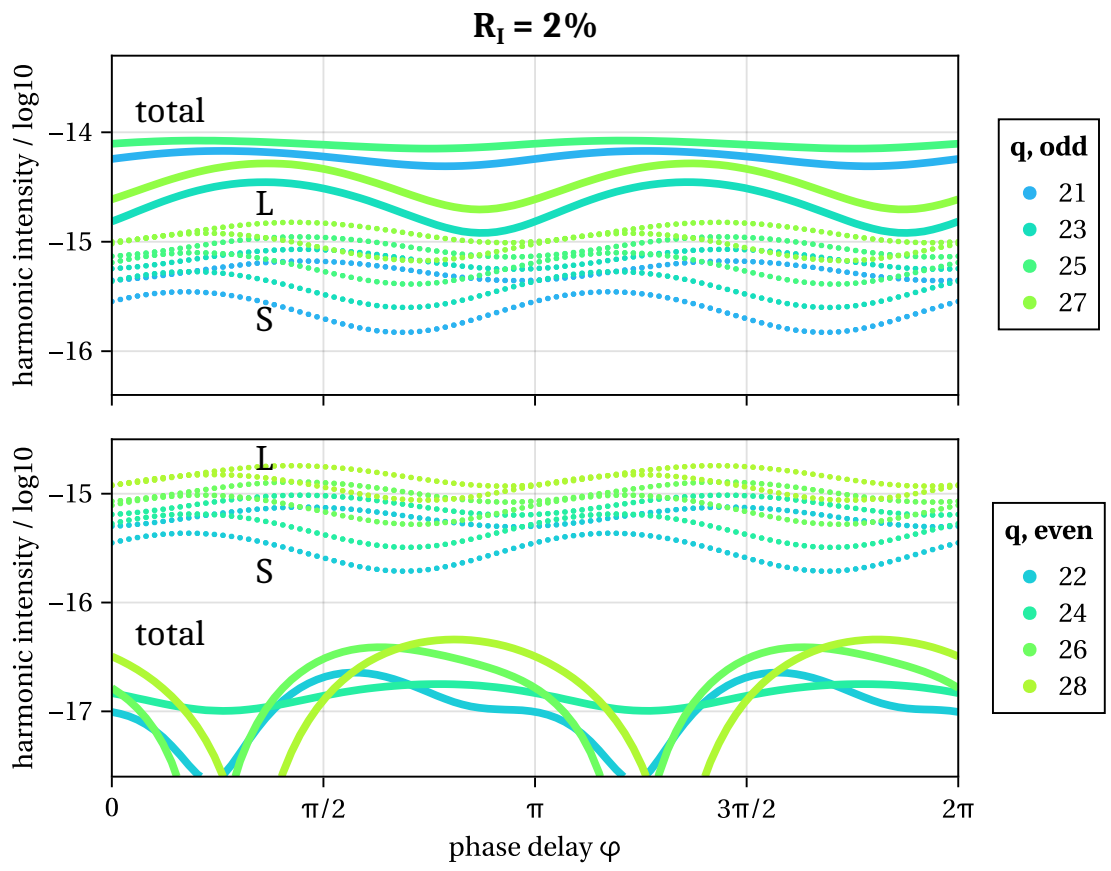

Figure 11.8: Intensity oscillations in the total intensity (solid lines), as well as intensities from individual saddle points (markers) for odd (top panel) and even (bottom panel) harmonic orders of the plateau region, for a perturbative second-colour field. Note the sections of the $y$-axis are different for both panels.

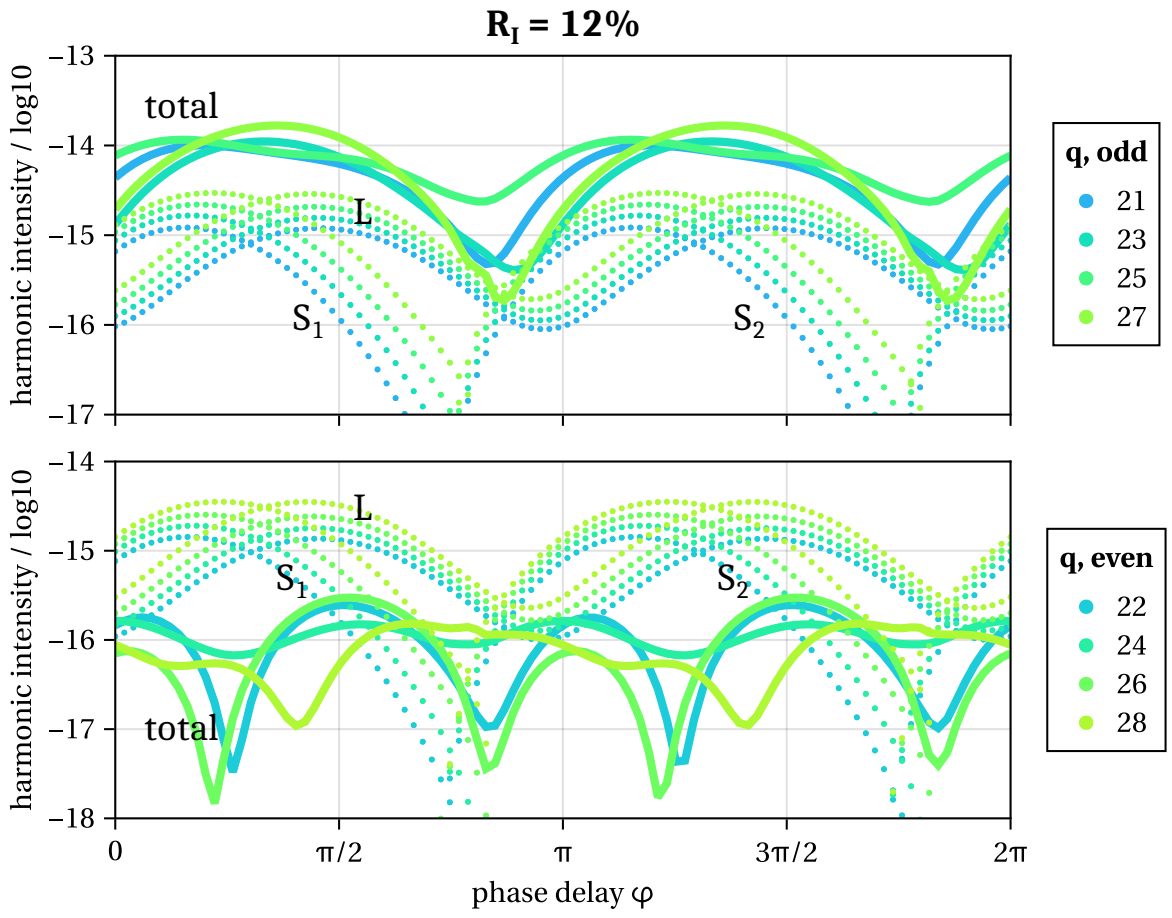

Figure 11.9: Intensity oscillations in the total intensity (solid lines), as well as intensities from individual saddle points (markers) for odd (top panel) and even (bottom panel) harmonic orders of the plateau region, for a strong second-colour component. Note the sections of the $y$-axis are different for both panels.

## 11.4 Interference of harmonic dipoles

In the previous section we analysed the intensity oscillations as calculated from the individual electron trajectories. While this provides valuable insights into the underlying dynamics, an even more compelling goal is of course the interpretation and analysis of actual experimental measurements. In the experiment considered here, the intensity ratio was estimated to be about $R_I = 12\%$, a regime in which the observed oscillations in the harmonic signal can be interpreted as clear signatures of interference between individual quantum orbits. In the following we show the corresponding results and provide physical interpretation of the observed features. These results are the product of close collaboration and scientific discussion with Xiaozhou Zou, who performed the experimental campaign, although the theory development is my own. A summary of the theory has been previously presented in Ref. [199]. Furthermore, a concise version of the following results has been accepted for publication [3].

### 11.4.1 Experimental observation of the oscillations

The oscillation upon phase-delay scan can be observed experimentally, and the divergence of the measured signal allows to distinguish the contributions from short and long trajectories, at least for perturbative driving fields. In Fig. 11.10(a) we show the experimentally measured and normalised harmonic intensity for a phase-delay scan for a series of harmonic orders. Firstly, we clearly see that odd harmonic orders are produced more intense than even orders. We identify two major bright 'blobs' for each harmonic order. The line out for harmonic order 24 is shown as circles in panel (c), where we find the signal oscillates with a single frequency. These oscillations are compared to the theoretical calculations, as shown previously. For even harmonic orders, the signal is generally rather weak. Suitable post-processing of the data (details can be found in Ref. [199]) allows the identification of a bimodal oscillation, shown for harmonic order 25 in panel (b).

This behaviour is relatively consistent across harmonic orders of the plateau. As explained above, the oscillations are ultimately a result of the symmetries of the driving field. The two-colour field $\mathbf{E}(t)$ composed of a $\omega$ component linearly polarised along $x$ direction, and a $2\omega$ component linearly polarised along $y$ direction obeys a half-cycle symmetry upon rotation about the $x$ axis such that $R_x\mathbf{E}(t+\frac{\pi}{\omega}) = \mathbf{E}(t)$, for the rotation about the $x$ axis $R_x$. The electric field of the emitted harmonic $\mathbf{E}_{q\omega}(t) = \mathbf{D}_s(q\omega)e^{iq\omega t}$ inherits this symmetry. Consequently, the field is linearly polarised in the $x$ direction for odd harmonic orders $q$, and linearly polarised in the $y$ direction for even $q$ [204]. From an alternative perspective, these oscillations can also be interpreted as clear signatures of interfering quantum paths, which become evident when analysing the individual electron trajectories, as discussed in the following.

### 11.4.2 Explanation in terms of quantum path interference

Let us now deconstruct the total intensity into contributions stemming from the several quantum orbits, each creating a respective harmonic dipole $\mathbf{D}_s(q\omega)$. Generally, the harmonic radiation emitted from HHG driven by strongly polychromatic driving fields may be arbitrarily polarised. The polarisation of the emitted radiation is therefore best described in terms of the polarisation ellipse introduced in Fig. 2.19. That is, rather than analysing the vector components of the created harmonic dipole $\mathbf{D}_s(qw)$, we will focus on the vector components of the polarisation ellipse drawn by the harmonic electric field $\mathbf{E}_{q\omega}(t)$. For the driving field considered here, the emitted harmonic radiation is linearly polarised such that it is sufficient to examine the components of the major polarisation axis, $M_x$ and $M_y$, respectively. The intensity along the minor axis $N$ is negligible, with $|\mathbf{N}|/|\mathbf{M}|$ in the order of 5%.

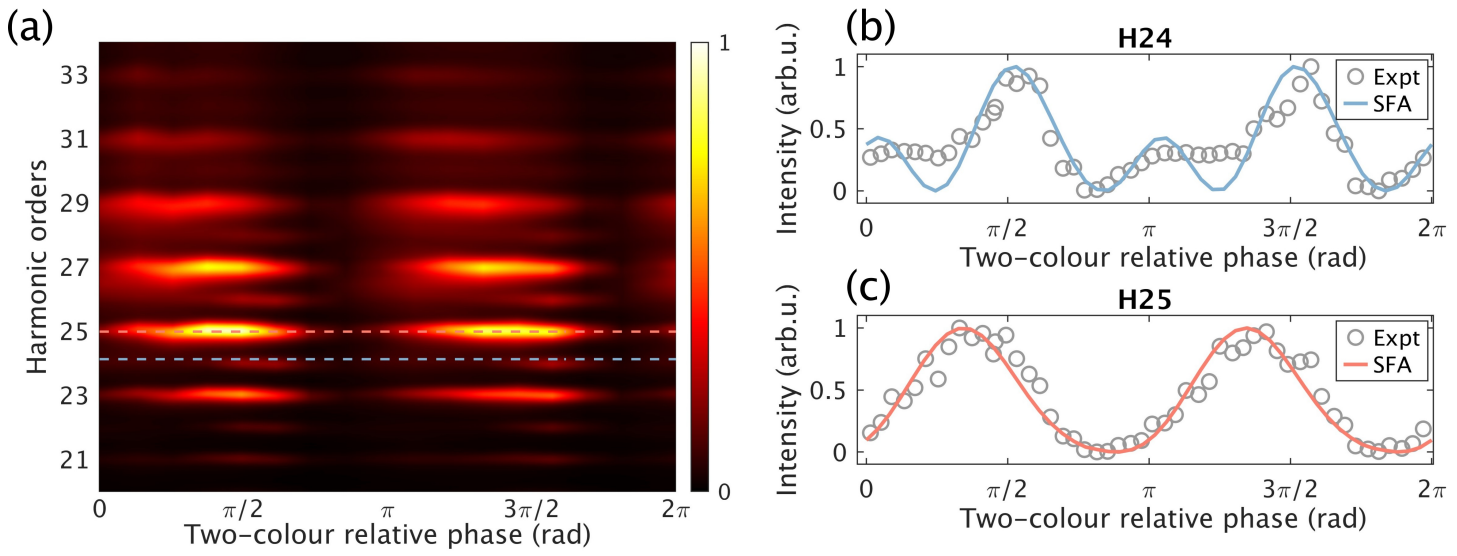

Figure 11.10: (a) Experimental data of the two-colour phase-delay scan. (b,c) Experimental data (circles) and theoretical simulation (lines) of the total intensity yield modulation for harmonic order (b) 24 and (c) 25. Taken from [199].

In Fig. 11.11 we show the modulation of the harmonic signal in the $x$ and $y$ directions (left and right column respectively) for the two subsequent harmonic orders 24 and 25 (top and bottom row respectively). Within each panel we show the total intensity modulation across a two-colour phase scan in the top subplot. In the bottom subplot, we show the contributions associated with the individual quantum paths in several shades of violet. The contributions are shown in terms of the components ($x$ components in (a) and (c), $y$ components in (b) and (d)) of the major polarisation axis of the harmonic radiation associated with the harmonic dipole from the respective saddle point.

From Fig. 11.11 we can now understand the following: For harmonic order 24 (top row) the trajectories from within the two subsequent half-cycles (solid and dashed lines respectively) produce dipoles with opposite orientation, and their polarisation axes are symmetric along the $x$ axis (see lower subplot in panel (a)). As a result, their contributions interfere destructively in the $x$ direction such that the $x$ component of the total intensity vanishes (see upper subplot in panel (a)). The $y$ components of the trajectories, however, are oriented in the same direction for the two half-cycles (dash-dotted lines in the bottom figure of panel (b)). Hence, their contributions interfere constructively to produce an oscillating, non-zero harmonic signal along the $y$ direction (panel (b)).

For harmonic order 25 the situation is inverse: The $x$ components of the individual trajectories' major dipole axes are oriented in the same direction for the two half-cycles and interfere constructively. This produces an oscillatory total signal (panel (c)). Conversely, the $y$ components have opposite orientation for the two half-cycles and therefore interfere destructively such that the total signal is zero (panel (d)).

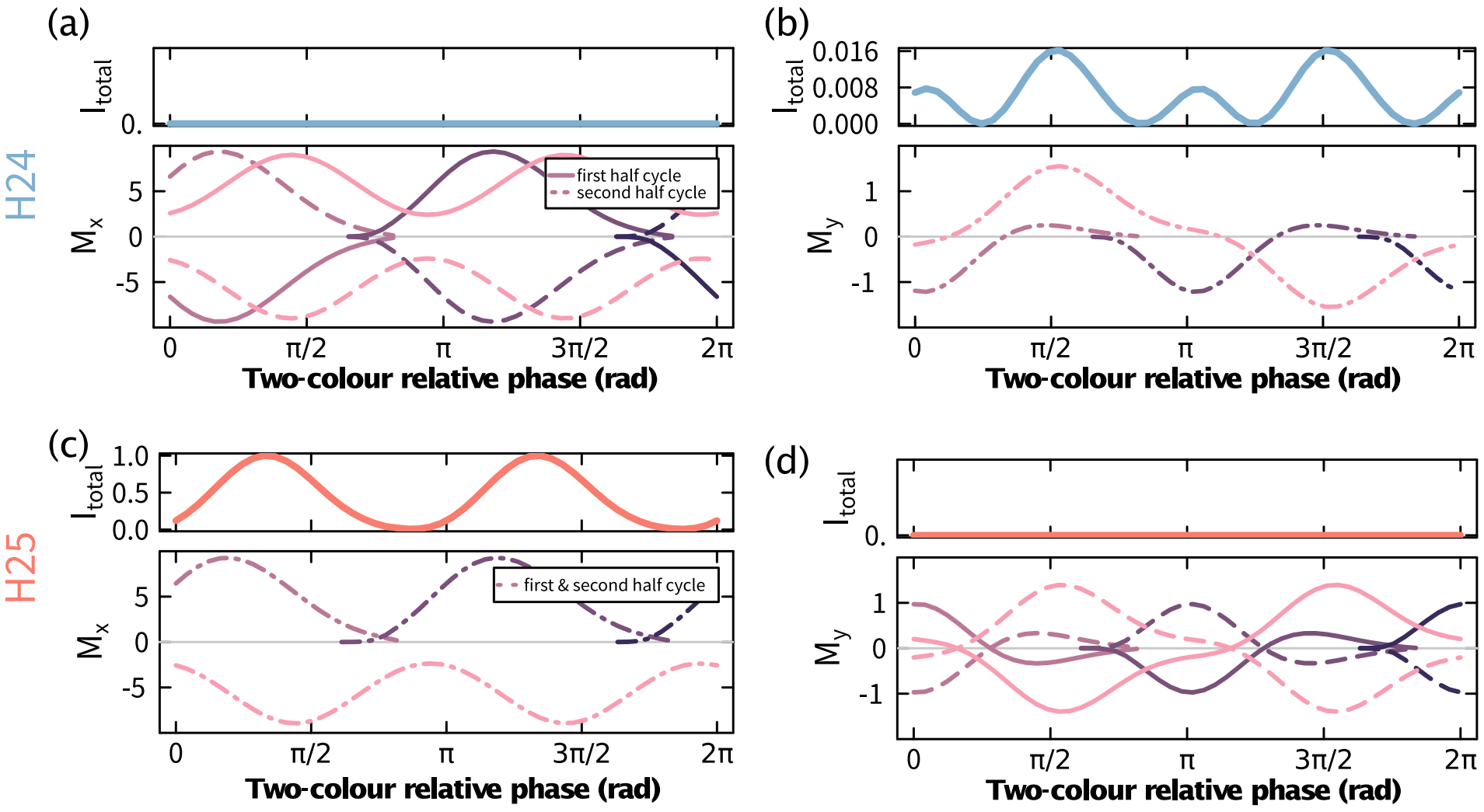

Figure 11.11: Total harmonic intensity modulation (top subpanels) as well as the main contributions from the most dominant trajectories (bottom subpanels) throughout a scan of the two-colour phase $\varphi$. Those contributions are given in terms of the components of the major polarisation axes of the harmonic field ellipse, $M_x$ and $M_y$. In the top row, we show (a) the $x$ (parallel to the $\omega$ polarisation) and (b) the $y$ (parallel to the $2\omega$ polarisation) components for harmonic order 24, and in the bottom row (c,d) the same components for harmonic order 25. Taken from [199].

To visualise this further, in Fig. 11.12 we show the electron's displacement during the propagation in the laser field, given by Eq. (2.49). For the two most dominant trajectories for harmonic order 24 (red) and 25 (blue) generated from a field with $\varphi = 0$. That is, these orbits correspond to the two short trajectories whose dipole polarisation axes are depicted in Fig. 11.11(a) and (b) in violet. In Fig. 11.12(a) we show the components of the displacement in $x$ and $y$ direction. We find that the $x$ component is opposite for the two half-cycles, i.e. negative for $t \lesssim -20$ (heavy solid lines) and positive for $t \gtrsim -20$ (heavy dashed lines), while the $y$ direction is identical in both half-cycles (thin dotted lines). In Fig. 11.12(b) we plot exactly the same orbits, but in $x$-$y$ space. As expected, we find the trajectories start away from the core, where the electron appears in the continuum, are subsequently driven further away and then return in a slingshot-movement back to the core (at $(0,0)$) where they recombine. Most importantly, we can clearly see that the trajectories in the first (solid line) and second (dashed line) half-cycles appear symmetric with respect to the $x$ axis. Generally, while the trajectories only change slightly from harmonic order 24 to harmonic order 25, Fig. 11.12(a) and (b) both clearly demonstrate the symmetry imposed by the driving field. That is, the trajectories in the second half-cycle are flipped along the $x$ axis with respect to the trajectory in the first half-cycle. Moreover, both figures show that the displacements in the $y$ direction (up to $\approx 5\,$a.u.) are smaller than those in the $x$ direction (up to $\approx \pm 13\,$a.u.). This can readily be explained as the $y$ component of the driving field is weaker than the $x$ component.

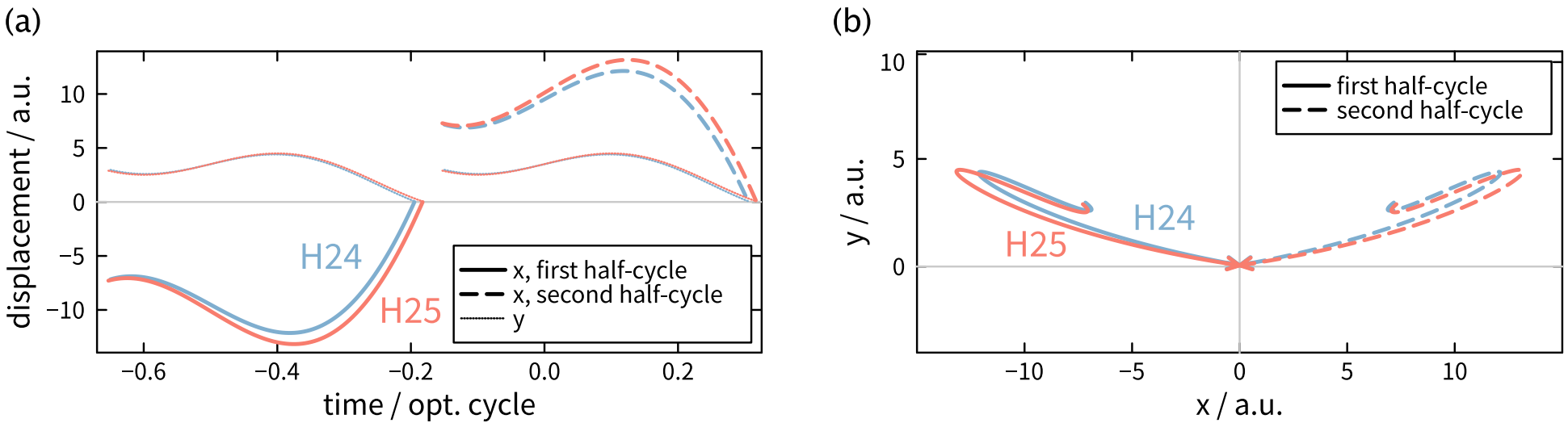

Figure 11.12: Displacements for two short trajectories within one optical cycle for a two-colour field with $R_I = 12\%$ as above, with $\varphi = 0$, for the two subsequent harmonic orders 24 and 25 shown (a) over time, and (b) in space. Taken from [199].

Somewhat independently of the two-colour phase scan, at first, it may be surprising that the behaviour of two subsequent harmonic orders is so different even though the saddle points themselves depend smoothly on the harmonic orders. To resolve this disconnect, let us show yet one more visualisation to demonstrate the interplay of several quantum orbits. In Fig. 11.13 we present the created harmonic field $\mathbf{E}_{q\omega}(t)$ from the relevant quantum orbits for the two harmonic orders 24 and 25, for a driving field with $R_I = 12\%$ and at $\varphi = 0$. In panel (a) we show the total harmonic field for these two subsequent harmonic orders in black. Because they are both linearly polarised they appear as straight lines along the $x$ and $y$ axis, respectively. Note that the maximum field amplitude of harmonic order 25 is much larger, corresponding to a higher intensity — shown in panel (b) and (c) of Fig. 11.11 at $\varphi = 0$.

The harmonic fields created from the individual quantum orbits' dipoles $\mathbf{D}_s(q\omega)$ are drawn in colours and are of elliptic shape. For further clarity, we show these ellipses in a non-natural aspect ratio for harmonic orders 24 and 25 separately, in panels (b) and (c). We find the shape of the harmonic ellipses does not change much between the two subsequent orders. How come then their superposition, i.e., the total harmonic field, changes so drastically between one order and the next? For that, we have indicated the harmonic field at emission time $t = \mathrm{Re}(t_{\mathrm{r},s})$ as a dotted line, corresponding to the spectral phase of the dipole. Intuitively, the total harmonic field is created as the superposition of the individual fields at this time. Hence, the advancing phases of the individual field ellipses for subsequent harmonic orders lead to drastic changes of this coherent superposition.

### 11.4.3 Conclusions

The above analysis reproduces the experimentally observed intensity modulations in phase-delay scans, as well as the characteristic polarisation properties of the emitted harmonics, with even orders polarised along the $y$ direction and odd orders along $x$, as required by the symmetry of the driving field. In particular, the harmonic signal in the $x$ direction oscillates with the fundamental frequency, while the emission along $y$ oscillates with the second-harmonic frequency. This behaviour can be understood in terms of the interference of individual quantum orbits, whose contributions add constructively or destructively depending on the emission direction. Owing to the spatio-temporal symmetry of the driving field, trajectories originating in the second half cycle are mirrored with respect to the $y$ axis compared to those from the first half cycle. As a consequence, their contributions cancel in the $x$ direction for even harmonic orders, while for odd orders the cancellation occurs in the $y$ direction. The observed intensity modulations in the phase-delay scan can therefore be interpreted as a clear signature of interference of quantum orbits.

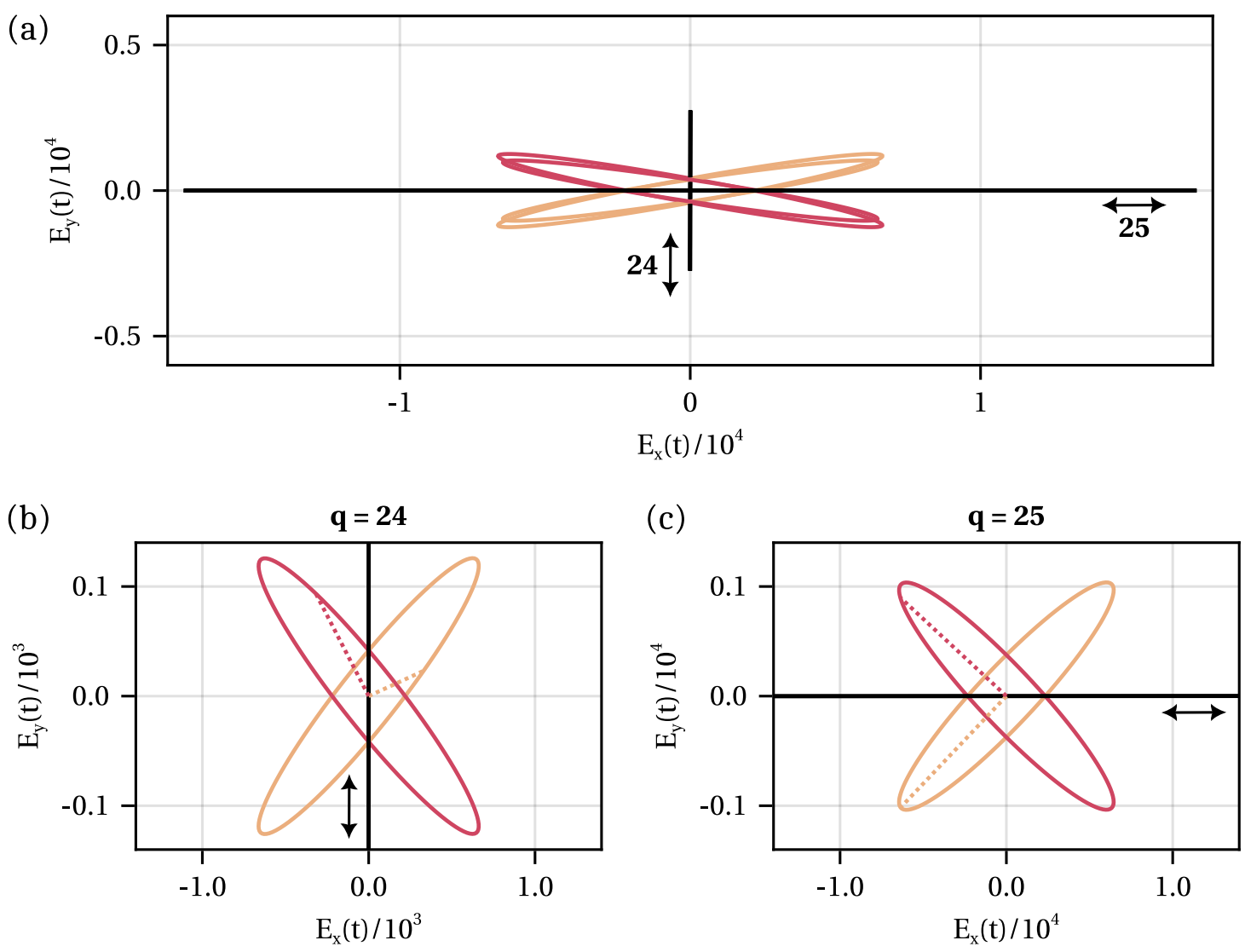

Figure 11.13: Harmonic fields from individual saddle points' dipoles (coloured ellipses) and the total harmonic field (black) for harmonic orders 24 and 25. (a) Harmonic fields in natural aspect ratio such that the total fields appear as straight lines along the $x$ and the $y$ direction for harmonic orders 24 and 25 respectively. Panels (b) and (c) are shown in distorted aspect ratios, and the electric field at $t = 0 = \mathrm{Re}(t_{r,s})$ is indicated by a dotted line.

## 11.5 Outlook

In conclusion, the analysis presented in this chapter demonstrates the broader utility of two-colour phase scans as a means to study the underlying electron dynamics in HHG driven by two-colour fields on a subcycle time scale. For perturbative drivers, the observed intensity modulations upon the phase delay scan can be used to retrieve ionisation and recombination times. Going beyond the perturbative regime allows us to probe the electron dynamics even further. With the decomposition of the harmonic dipoles into contributions from separate quantum paths, we can provide valuable insights into the interpretation of the total intensity modulations as clear signatures of quantum path interference.

Looking ahead, a central challenge in quantitatively comparing experimental observations with the theoretical calculations lies in obtaining a reliable estimate of the laser intensities of the two field components, and hence, their ratio. For the majority of the calculations presented here an intensity ratio of $R_I = 12\%$ for the case of a ‘strong’ second colour field was assumed. However, comparison between the experimentally observed oscillations and the simulations shown in Fig. 11.9 suggests that a higher intensity ratio may provide a better description. In particular, for larger ratios the characteristic ‘double-hump’ structure becomes more pronounced, potentially yielding improved agreement with the measured data.

This uncertainty is not the only limitation of the present comparison. The calculations discussed in this chapter are based on simplified single-atom SFA simulations, whereas experimentally HHG is generated in an extended gas medium. In such a macroscopic environment the emitted radiation interferes and propagates before reaching the detector, thereby modifying the observable signal. Moreover, the finite focussing geometry of the driving laser beams implies that individual atoms experience different local intensities and intensity ratios across the interaction volume.

In a follow-up project, these effects are being addressed more carefully. Ongoing work combines spatial and temporal beam-propagation simulations with the saddle-point-based microscopic response, followed by a propagation of the individual quantum orbits’ harmonic dipoles to the far field. This aims at achieving a more complete and quantitative description of the experimental observables and disentangling the contributions of the various involved quantum paths.

# 12
# Caustics and catastrophes

In this Chapter, we finally bring together the concepts from catastrophe theory and the methods developed from Picard–Lefschetz theory, and apply them to scenarios in attosecond science which exhibit caustics. We illustrate how these complementary frameworks provide a powerful language for analysing strong-field dynamics governed by interfering quantum trajectories.

Three representative examples of catastrophes in the context of HHG are discussed. We begin with a review of the fold catastrophe at the HHG cutoff, based on Ref. [48]. Although well established, this example is included here because the associated classification and analysis techniques form the foundation for the treatment of higher-order catastrophes.

We then focus on a cusp catastrophe that emerges during a smooth colour switchover from an $\omega$ to a $2\omega$ driver, via two-colour fields of increasing amplitude ratio, in a colinear configuration. For this case, we analyse the underlying saddle-point structure, identify its organising centre, and demonstrate how its signatures manifest in the harmonic response. Parts of this analysis are mentioned in our recently published Ref. [2].

Finally, we briefly discuss the appearance of a swallowtail catastrophe in a co-orthogonal $\omega - 2\omega$ driving-field configuration, thereby completing the natural hierarchy of catastrophes relevant to this system and highlighting the broader applicability of the presented approach.

## 12.1 The fold, at the high-harmonic cutoff

In Sec. 3.2.4, we introduced the fold catastrophe in a general mathematical setting. Here, we specialise this discussion to fold catastrophes that appear in the context of HHG, demonstrating how they can be identified and utilised.

The most prominent example of a (hidden) fold catastrophe in HHG is the high-harmonic cutoff of the spectrum. In the simplest case of a monochromatic driving field this is where the short and long electron trajectories undergo a missed approach, which is a characteristic signature of an underlying fold catastrophe that lies outside the accessible parameter range (e.g., real-valued parameters). For the case of more complex, polychromatic driving fields, the increased number of contributing trajectories allows similar fold-like interactions to occur within the high-harmonic plateau itself. In particular, for the introduction of additional frequency components and the systematic increase of their relative amplitudes — as done within the colour switchover scheme discussed in Chapter 10 — we can expect new pairwise trajectory interactions. Moreover, saddle-point coalescences also occur at the ionisation threshold, $q\omega = I_p$, such that low-order harmonics can likewise be interpreted in terms of a fold catastrophe.

Within the context of HHG, the fold catastrophe has been shown to play a central role in the enhancement of harmonic emission near the cutoff. As discussed in Chapter 12, the mathematically rigorous link between the missed approach of short and long trajectories, and a fold catastrophe hidden in complex-valued parameter space has only been formalised relatively recently in Ref. [48]. That work demonstrated that the fold point uniquely determines the Stokes transition at which one of the contributing trajectories must be discarded. Furthermore, it established a consistent classification scheme for saddle-point solutions in the vicinity of the fold. Finally, it showed that the harmonic intensity across the entire spectral range can be approximated using an interference signature that is uniform around the cutoff, with the fold point itself serving as the sole required input. In the following we want to revise the most relevant aspects of this discussion, largely based on Ref. [48].

### 12.1.1 Definition of the fold catastrophe point

As introduced in Sec. 3.2.4, fold points are critical points at which the determinant of the deformation tensor vanishes but its eigenvalues are non-zero. For a univariate function of one variable this simply means that first and second derivative vanish simultaneously while the third derivative remains non-zero.

The action of the two-dimensional HHG integral in the SFA framework is a function of both $t_\mathrm{i}$ and $t_\mathrm{r}$. In Ref. [48] this two-dimensional function is reduced to a one-dimensional problem by exploiting the idea that the saddle-point equation of ionisation can be understood as an implicit definition of the (saddle point) ionisation time. That is, the two equations that define saddle points are (same as in the introduction of this thesis)

$$\frac{\partial S_{\mathrm{HHG}}}{\partial t_{\mathrm{i}}} = \frac{1}{2}\left(\mathbf{p}_s(t_{\mathrm{i}}, t_{\mathrm{r}}) + \mathbf{A}(t_{\mathrm{i}})\right)^2 + \mathcal{I}_{\mathrm{p}} = 0 \tag{12.1a}$$

$$\frac{\partial S_{\mathrm{HHG}}}{\partial t_{\mathrm{r}}} = \frac{1}{2}\left(\mathbf{p}_s(t_{\mathrm{i}}, t_{\mathrm{r}}) + \mathbf{A}(t_{\mathrm{r}})\right)^2 + \mathcal{I}_{\mathrm{p}} - q\omega = 0\,. \tag{12.1b}$$

The former of which yields the implicit definition $t_{\mathrm{i}} = t_{\mathrm{i}}(t_{\mathrm{r}})$, which can then be inserted into the latter. As a result, the saddle points can be found by solving a single equation, where the right-hand side is identified with the derivative of the Volkov part of the full semi-classical action:

$$q\omega = \frac{1}{2}\left(\mathbf{p}_s(t_{\mathrm{i}}(t_{\mathrm{r}}), t_{\mathrm{r}}) + \mathbf{A}(t_{\mathrm{r}})\right)^2 + \mathcal{I}_{\mathrm{p}} = \frac{\partial S_{\mathrm{V}}}{\partial t_{\mathrm{r}}}(t_{\mathrm{i}}(t_{\mathrm{r}}), t_{\mathrm{r}})\,. \tag{12.2}$$

Solutions to this describe a Riemann surface in the space of complex recombination time and harmonic order:

$$\mathcal{S} = \left\{(t_{\mathrm{r}}, q) \in \mathbb{C}^2 : q\omega = \frac{\partial S_{\mathrm{V}}}{\partial t_{\mathrm{r}}}(t_{\mathrm{i}}(t_{\mathrm{r}}), t_{\mathrm{r}})\right\}. \tag{12.3}$$

This surface is topologically equivalent to that defined by the total derivative of the full action $\frac{\mathrm{d}S_{\mathrm{HHG}}}{\mathrm{d}t_{\mathrm{r}}}(t_{\mathrm{i}}(t_{\mathrm{r}}), t_{\mathrm{r}})$. In the following, let us therefore consider the Volkov action only, using the implicit definition of the ionisation time at the saddle point, i.e., the univariate function $S_{\mathrm{V}}(t_{\mathrm{r}}) = S_{\mathrm{V}}(t_{\mathrm{i},s}(t_{\mathrm{r}}), t_{\mathrm{r}})$. The total derivative then reads

$$\frac{\mathrm{d}S_{\mathrm{V}}}{\mathrm{d}t_{\mathrm{r}}}(t_{\mathrm{r}}) = \frac{\partial S_{\mathrm{V}}}{\partial t_{\mathrm{r}}}(t_{\mathrm{i},s}(t_{\mathrm{r}}), t_{\mathrm{r}}) + \frac{\mathrm{d}t_{\mathrm{i},s}(t_{\mathrm{r}})}{\mathrm{d}t_{\mathrm{r}}}(t_{\mathrm{r}})\,\frac{\partial S_{\mathrm{V}}}{\partial t_{\mathrm{i}}}(t_{\mathrm{i},s}(t_{\mathrm{r}}), t_{\mathrm{r}})\,, \tag{12.4}$$

where the second term vanishes by definition of the saddle point $t_{\mathrm{i},s}(t_{\mathrm{r}})$ such that

$$\frac{\mathrm{d}S_{\mathrm{V}}}{\mathrm{d}t_{\mathrm{r}}}(t_{\mathrm{i},s}(t_{\mathrm{r}}), t_{\mathrm{r}}) = \frac{\partial S_{\mathrm{V}}}{\partial t_{\mathrm{r}}}(t_{\mathrm{i},s}(t_{\mathrm{r}}), t_{\mathrm{r}})\,. \tag{12.5}$$

Subsequently, the second derivative reads

$$\frac{\mathrm{d}S_{\mathrm{V}}^2}{\mathrm{d}t_{\mathrm{r}}^2}(t_{\mathrm{i},s}(t_{\mathrm{r}}), t_{\mathrm{r}}) = \frac{\mathrm{d}}{\mathrm{d}t_{\mathrm{r}}}\frac{\partial S_{\mathrm{V}}}{\partial t_{\mathrm{r}}}(t_{\mathrm{i},s}(t_{\mathrm{r}}), t_{\mathrm{r}}) = \frac{\partial S_{\mathrm{V}}^2}{\partial t_{\mathrm{r}}^2}(t_{\mathrm{i},s}(t_{\mathrm{r}}), t_{\mathrm{r}}) + \frac{\mathrm{d}t_{\mathrm{i},s}}{\mathrm{d}t_{\mathrm{r}}}(t_{\mathrm{r}})\frac{\partial S_{\mathrm{V}}^2}{\partial t_{\mathrm{i}} t_{\mathrm{r}}}(t_{\mathrm{i},s}(t_{\mathrm{r}}), t_{\mathrm{r}})\,. \tag{12.6}$$

Unfortunately, the term $\frac{\mathrm{d}t_{\mathrm{i},s}}{\mathrm{d}t_{\mathrm{r}}}$ cannot be evaluated explicitly, as $t_{\mathrm{i},s}(t_{\mathrm{r}})$ is only given implicitly via Eq. (12.1a). Luckily, differentiating that definition yields this term as well:

$$\frac{\mathrm{d}}{\mathrm{d}t_{\mathrm{r}}}\frac{\partial S_{\mathrm{V}}}{\partial t_{\mathrm{i}}}(t_{\mathrm{i},s}(t_{\mathrm{r}}), t_{\mathrm{r}}) = \frac{\partial^2 S_{\mathrm{V}}}{\partial t_{\mathrm{i}} t_{\mathrm{r}}}(t_{\mathrm{i},s}(t_{\mathrm{r}}), t_{\mathrm{r}}) + \frac{\mathrm{d}t_{\mathrm{i},s}}{\mathrm{d}t_{\mathrm{r}}}(t_{\mathrm{r}})\frac{\partial^2 S_{\mathrm{V}}}{\partial t_{\mathrm{i}}^2}(t_{\mathrm{i},s}(t_{\mathrm{r}}), t_{\mathrm{r}}) = 0\,. \tag{12.7}$$

This can be solved for $\frac{\mathrm{d}t_{\mathrm{i},s}}{\mathrm{d}t_{\mathrm{r}}}$ to read

$$\frac{\mathrm{d}t_{\mathrm{i},s}}{\mathrm{d}t_{\mathrm{r}}} = -\left(\frac{\partial^2 S_{\mathrm{V}}}{\partial t_{\mathrm{i}} t_{\mathrm{r}}}(t_{\mathrm{i},s}(t_{\mathrm{r}}), t_{\mathrm{r}})\right) \Big/ \frac{\partial^2 S_{\mathrm{V}}}{\partial t_{\mathrm{i}}^2}(t_{\mathrm{i},s}(t_{\mathrm{r}}), t_{\mathrm{r}})\,, \tag{12.8}$$

which we can insert into the second derivative Eq. (12.6), yielding

$$\frac{\mathrm{d}S_{\mathrm{V}}^{2}}{\mathrm{d}t_{\mathrm{r}}^{2}}(t_{\mathrm{i},s}(t_{\mathrm{r}}),t_{\mathrm{r}})=\frac{\partial S_{\mathrm{V}}^{2}}{\partial t_{\mathrm{r}}^{2}}-\left(\frac{\partial^{2}S_{\mathrm{V}}}{\partial t_{\mathrm{i}}t_{\mathrm{r}}}\right)^{2}\Big/\frac{\partial^{2}S_{\mathrm{V}}}{\partial t_{\mathrm{i}}^{2}}\,, \tag{12.9}$$

where we have omitted the dependencies $(t_{\mathrm{i},s}(t_{\mathrm{r}}),t_{\mathrm{r}})$ on the right-hand side for clarity. A fold point can now be found by simultaneously solving Eqs. (12.1a) and (12.9) to be zero.

The same result can be obtained by requiring the determinant of the deformation tensor to vanish. The deformation tensor (i.e., the Hessian matrix) for the full HHG action reads

$$\mathcal{H}=\begin{pmatrix}\frac{\partial^{2}S_{\mathrm{HHG}}}{\partial t_{\mathrm{i}}^{2}} & \frac{\partial^{2}S_{\mathrm{HHG}}}{\partial t_{\mathrm{i}}t_{\mathrm{r}}}\\ \frac{\partial^{2}S_{\mathrm{HHG}}}{\partial t_{\mathrm{i}}t_{\mathrm{r}}} & \frac{\partial^{2}S_{\mathrm{HHG}}}{\partial t_{\mathrm{r}}^{2}}\end{pmatrix} \tag{12.10}$$

with the determinant

$$\det(\mathcal{H})=\frac{\partial^{2}S_{\mathrm{HHG}}}{\partial t_{\mathrm{i}}^{2}}\frac{\partial^{2}S_{\mathrm{HHG}}}{\partial t_{\mathrm{r}}^{2}}-\left(\frac{\partial^{2}S_{\mathrm{HHG}}}{\partial t_{\mathrm{i}}t_{\mathrm{r}}}\right)^{2}, \tag{12.11}$$

where the condition for a fold is

$$\det(\mathcal{H})\big|_{(t_{\mathrm{i},f},t_{\mathrm{r},f})}=0\,. \tag{12.12}$$

These second partial derivatives are equivalent to the derivatives of the Volkov action above:

$$\begin{aligned}\frac{\partial^{2}S_{\mathrm{HHG}}}{\partial t_{\mathrm{i}}^{2}}&=\frac{\partial^{2}S_{\mathrm{V}}}{\partial t_{\mathrm{i}}^{2}}\\ \frac{\partial^{2}S_{\mathrm{HHG}}}{\partial t_{\mathrm{r}}^{2}}&=\frac{\partial}{\partial t_{\mathrm{r}}}\left(\frac{\partial S_{\mathrm{V}}}{\partial t_{\mathrm{r}}}+q\omega\right)=\frac{\partial^{2}S_{\mathrm{V}}}{\partial t_{\mathrm{r}}^{2}}\\ \frac{\partial^{2}S_{\mathrm{HHG}}}{\partial t_{\mathrm{i}}t_{\mathrm{r}}}&=\frac{\partial}{\partial t_{\mathrm{r}}}\frac{\partial S_{\mathrm{V}}}{\partial t_{\mathrm{i}}}\,.\end{aligned} \tag{12.13}$$

Hence, the vanishing of Eq. (12.9) is equivalent to the vanishing of Eq. (12.11).

Notably, this second condition for the fold is independent of the harmonic order $q$, as well the first condition, i.e., the vanishing Volkov action at ionisation, Eq. (12.1a). The saddle-point equation for recombination (12.1b), evaluated at the fold times $(t_{\mathrm{i},f},t_{\mathrm{r},f})$ can therefore be understood as a definition for the harmonic energy $E_f$ at the fold:

$$\frac{\partial S_{\mathrm{V}}}{\partial t_{\mathrm{r}}}(t_{\mathrm{i},f},t_{\mathrm{r},f})=q_f\omega=E_f \tag{12.14}$$

with the harmonic order $q_f$, which turns out to be a complex-valued number.

This unexpected and unintuitive finding is a breakthrough in the application of catastrophe theory to attosecond science problems. In Ref. [48] this was developed in application to the high-order harmonic cutoff. For a monochromatic driving field, the identified fold point sits in between the saddle points of the short and the long trajectory, and the harmonic order $q_f$ provides an exact value for the harmonic cutoff. It is furthermore shown how the imaginary part of $q_f$ corresponds to the interference strength between the short and long trajectories throughout the HHG spectrum.

Technically, a missed approach of two ('lines' of) saddle point solutions may appear for other parameter scans as well, not only across the harmonic order. For example, continuously varying the amplitude ratio between the two component fields of a bicircular field yields similar behaviour [48]. Beyond that, for a phase delay scan in a two-colour driver, the saddle points of a fixed harmonic order similarly perform a missed approach. In all these cases, it can be helpful (and hence, desired) to identify the (possibly complex) values of a certain parameter, say $x$, at which the two trajectories fully coalesce. To numerically identify this coalescence point for a given harmonic order $q$, we can solve the above fold conditions Eqs. (12.1a) and (12.9) (or Eq. (12.12)), and simultaneously $\frac{\partial S_V}{\partial t_r}(t_{i,f}, t_{r,f}) = q\omega$ for $x$. In practice, this can be done using a multi-variate Newton's method root solver. For that we define $S(x) = S_V(t_{i,f}, t_{r,f})$ with $S_V$ evaluated at parameter $x$ and at the fold $(t_{i,f}, t_{r,f})$ which defines a harmonic order $q_f$. We then iteratively solve $\frac{dS(x)}{dt_r} = q$ for $x$.

### 12.1.2 Classification of saddle points around a fold

In most practical examples a parameter scan will not cause two saddle-point solutions to fully coalesce but instead cause a 'missed approach'. The most common example, which has been mentioned many, many times in this thesis already, is the high-order harmonic cutoff (e.g., seen in Fig. 2.11). In those cases it is often numerically challenging to correctly classify the solutions. Mostly, this can be solved by decreasing the step size of the respective parameter scan, for example, by choosing smaller discretisation steps for the harmonic order. However, as the two branches can become arbitrarily close — and in fact fully coalesce even — this approach is not always effective (or efficient). Nonetheless, any such missed approach is a hint for a fold catastrophe. Sometimes (in our case, often) this fold point is inaccessible to the conventional parameter space. In the example of the HHG cutoff this means that for real-valued harmonic orders the two saddle points will (almost) never coincide. But, if the parameter space is extended to permit complex-valued harmonic orders, a coalescence of the two trajectories can be found. The fold point then sits in the centre of the two branches[1] of saddle point solutions. Hence, it can be used as a natural 'separator' between the two sets of saddle-point solutions in order to simplify their classification. The rigorous procedure for this classification was developed in Ref. [48] and we want to briefly review it here, because we used it in Chapter 10.

The lines that the saddle-point solutions trace for a range of real-valued harmonic orders can be interpreted as a contour of the surface $\mathrm{Im}\left(\frac{dS_V}{dt_r}\right)$, at the level $\mathrm{Im}\left(\frac{dS_V}{dt_r}\right) = 0$. As a potentially helpful but certainly not rigorous analogy, this follows from the idea that a fold catastrophe is a saddle point 'on top of' another saddle point, such that by viewing the problem within the frame of the first saddle point, we observe the second one. At the fold harmonic order $q = q_f$, the level lines at $\mathrm{Im}\left(\frac{dS_V}{dt_r}\right) = 0$ coincide and form an orthogonal cross as it is characteristic for saddle points. For values of $q$ further away from $q_f$, these contour lines behave like level lines around a plain saddle point, viz. forming a missed approach. That the contour lines are orthogonal at the actual fold point is what informs the approach of the classification scheme developed in Ref. [48], and which is visualised in Fig. 12.1, in the complex plane of travel time $\tau$.

[1]As it turns out they are in fact *branches* of the same Riemann surface, as shown in Ref. [48].

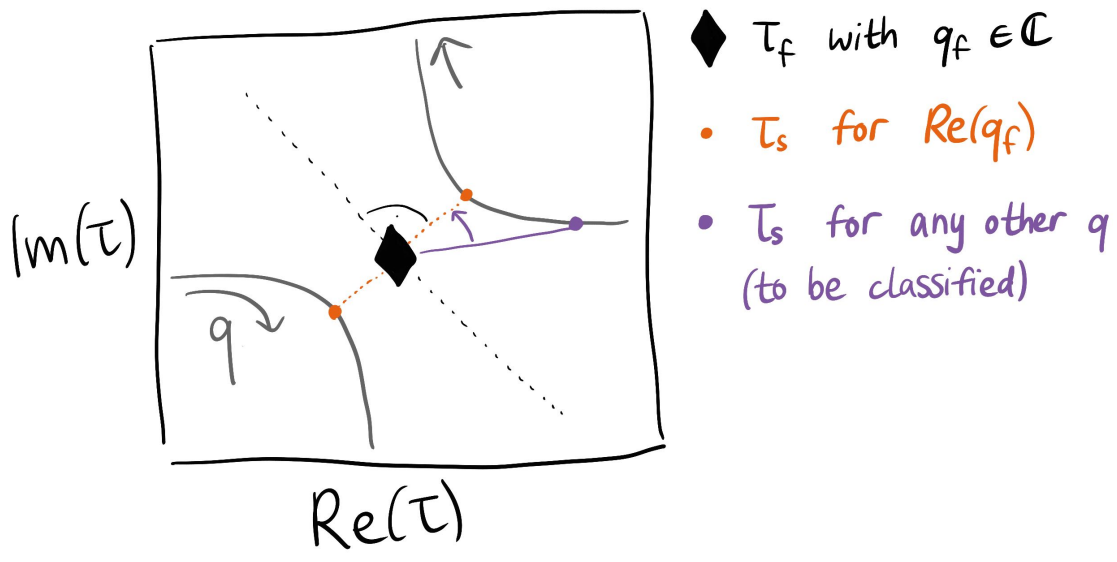

Figure 12.1: Rigorous classification of saddle points performing a missed approach by using the fold point (diamond, for $q_f \in \mathbb{C}$) in between. Any saddle point $\tau_s$ (purple) can be classified to belong to either trajectory by evaluating the inner product of its distance vector to the fold (purple line) with the distance vector of the saddle points for $\mathrm{Re}(q_f)$ to the fold (orange).

As a first step, we identify the fold point $(t_{\mathrm{i},f}, t_{\mathrm{r},f})$ (drawn as the black diamond $\tau_f$) and the corresponding complex-valued fold harmonic order $q_f \in \mathbb{C}$. Then, we find the saddle points for $q = \mathrm{Re}(q_f)$. In Fig. 12.1 they are highlighted in orange, and they are the saddle points that are in closest distance[2] from the fold point. For each of these two saddle points $t_{\mathrm{sep}}$, we define the vector to the fold point:

$$\delta t_{\mathrm{sep}} = t_{\mathrm{sep}} - (t_{\mathrm{i},f}, t_{\mathrm{r},f}) \,. \tag{12.15}$$

For any given saddle point $(t_{\mathrm{i},s}, t_{\mathrm{r},s})$ we can now check 'on which side' of the fold point it lies. For that we use the sign of the inner product between this separation vector $\delta t_{\mathrm{sep}}$ and the vector from the saddle point to the fold point as a criterion:

$$\mathrm{Re}\left(((t_{\mathrm{i},s}, t_{\mathrm{r},s}) - (t_{\mathrm{i},f}, t_{\mathrm{r},f})^* \cdot \delta t_{\mathrm{sep}})\right) \gtrless 0 \,. \tag{12.16}$$

This is visualised in the sketch Fig. 12.1, where the vector to an arbitrary saddle point $\tau_s$ (in purple) is shown in relation to that separation vector (in orange).

When applying this classification strategy, it is important to keep in mind that it relies on the assumption that the fold point locally behaves like a canonical fold catastrophe. Consequently, the procedure might become less accurate for harmonic orders far away from the fold. Moreover, in the context of HHG, typically multiple fold points arise, corresponding to missed approaches between short and long trajectories associated with higher-order returns. A rigorous and globally consistent classification of all saddle-point solutions therefore requires partitioning the complex time plane into appropriate regions, each containing only a single fold. A rigorous execution of this is shown in the earlier

[2] We define the distance in the double complex plane $\mathbb{C}^2$ between two points $(t_{\mathrm{i}1}, t_{\mathrm{r}1})$ and $(t_{\mathrm{i}2}, t_{\mathrm{r}2})$ as the Euclidean norm $\|(t_{\mathrm{i}2} - t_{\mathrm{i}1}, t_{\mathrm{r}2} - t_{\mathrm{r}1})\| = \sqrt{|t_{\mathrm{i}2} - t_{\mathrm{i}1}|^2 + |t_{\mathrm{r}2} - t_{\mathrm{r}1}|^2}$.

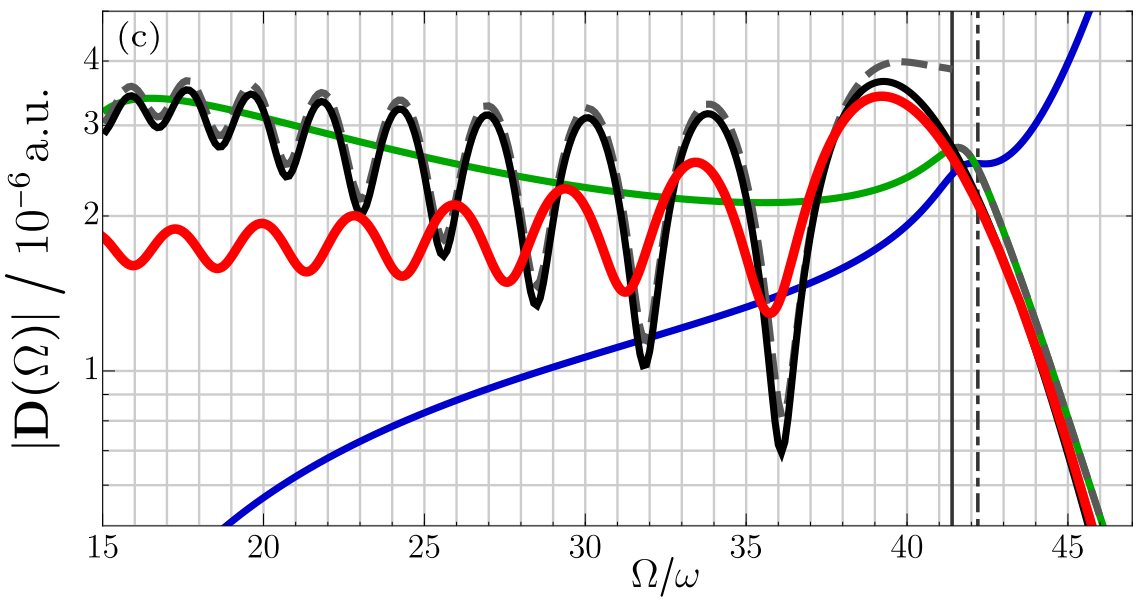

Figure 12.2: The HCA (red) in comparison to the uniform approximation (black) and the saddle-point approximation (grey dashed). Individual saddles' contributions are shown in green and blue. From [48].

Fig. 3.13, where an unstructured set of saddle point solutions (in panel (b)) is classified using the fold points in between each pairwise missed approach (panel (c)).

### 12.1.3 An approximation for the full spectrum

When saddle points are in close vicinity or even coalesce the approximation of their contribution as Gaussians becomes inaccurate. This becomes apparent when evaluating the total integral across a range of external parameters. Thereupon the discrepancy between the exact integral and its saddle-point approximation is largest at parameters where the saddles are close. This can be seen in Fig. 2.17.

For canonical diffraction integrals with coalescing saddles, uniform approximations are developed to provide integral values that are smooth across a large range of scaling factors of the phase function in the exponent, see Sec. 3.2.3. In the case of HHG, there is no such explicit scaling factor. However, the harmonic order enters the phase of the integral as a factor. Hence, methods of uniform approximation may be applied to the analysis of the structure of the whole spectrum.

For example, nearby the high-order harmonic cutoff of a HHG spectrum driven by a monochromatic laser field, the dipole integral can be expressed in terms of an Airy function around the two saddle points [46, 119]. With the knowledge of the exact coalescence point in terms of a fold catastrophe, it is possible to construct an approximation of the integral across the entire harmonic spectrum. This was developed in Ref. [48] and is called the "harmonic cutoff approximation" (HCA). The main idea is to identify the fold point $(t_{i,f}, t_{r,f})$ for the high-order harmonic cutoff and to expand the action around it to third order. As explained above, the fold point may only be found for complex-valued harmonic orders, at $q = q_f \in \mathbb{C}$. Including scaling factors to account for this offset into the complex plane, the HCA for the harmonic dipole of a given real-valued harmonic order

reads:

$$\mathbf{D}_f(q\omega) = \sqrt{\frac{2\pi}{\mathrm{i}\frac{\partial^2 S_\mathrm{V}}{\partial t_\mathrm{i}^2}(t_{\mathrm{i},f}, t_{\mathrm{r},f})}} \frac{2\pi}{\mathrm{e}^{2\pi\mathrm{i}\frac{k}{3}} A_f^{1/3}} \mathbf{f}(t_{\mathrm{i},f}, t_{\mathrm{r},f}) \mathrm{e}^{-\mathrm{i}S_\mathrm{V}(t_{\mathrm{i},f}, t_{\mathrm{r},f}) + \mathrm{i}q\omega t_{\mathrm{r},f}} \mathrm{Ai}\left(\frac{(q_f - q)\omega}{\mathrm{e}^{2\pi\mathrm{i}\frac{k}{3}} A_f^{1/3}}\right) \tag{12.17}$$

with the coefficient

$$A_f = \frac{1}{2} \frac{\mathrm{d}^3 S_\mathrm{V}}{\mathrm{d}t_\mathrm{r}^3}(t_{\mathrm{i},f}, t_{\mathrm{r},f}) .$$

Note that — similarly to when totalling contributions from several saddle points — the expression Eq. (12.17) provides the contribution from a single fold point, encapsulating the contribution from a single pair of short and long trajectories. For a more accurate description of the full spectrum we need to sum these contributions from several fold points, e.g., from cutoffs of the higher-order return trajectories as well. In Fig. 12.2 we show the resulting dipole compared to the uniform approximation [46, 119] (black), individual saddles' contribution (green and blue lines), their coherent summation with Gaussian contributions (saddle-point approximation, grey dashed), and the HCA (red) for a monochromatic driving field. At the cutoff (here around harmonic order 42), the HCA is quantitatively accurate, and it provides a reasonable estimate throughout the entire spectrum. Note however, that in order to obtain the HCA we require to solve only for a single critical point — namely the cutoff fold. In contrast, the remaining methods require solving for saddle points on a fine grid of harmonic orders $q$.

Apart from at the high-order harmonic cutoff exemplified here, the determinant of the deformation tensor vanishes for the threshold harmonic orders as well. Hence, in practice, solving for fold points in the complex plane often returns these too. For a purely monochromatic driving field they constitute an exact coalescence for the real-valued threshold order $q = \mathcal{I}_\mathrm{p}/\omega$.

## 12.2 The cusp, a tunable intensity enhancement in the colour switchover

In the previous Section, we examined the fold catastrophe as it appears in HHG, when two electron trajectories either coalesce or when it forms the organising centre of a missed approach. In that context, only a single external control parameter is required. Following the natural hierarchy of elementary catastrophes, the next structurally stable critical point is the cusp catastrophe, which has codimension $K = 2$ and therefore requires two independent control parameters. Since the codimension directly determines the degree of intensity enhancement associated with the resulting caustic, the cusp emerges naturally as the next interesting object of study following the fold point. Notably, to the best of our knowledge, cusp catastrophes have not been explicitly identified or discussed in the attosecond-science literature to date.

To identify a cusp within the context of HHG, we have to introduce control over an additional parameter. In this Section, we use the amplitude ratio of the two components of a collinear two-colour driving field as this second parameter. The general features of the smooth transition between a monochromatic $\omega$ field and its second harmonic have already been discussed in the earlier Chapter 10. Within this colour switchover, we encountered a missed approach between three electron trajectories — a feature that indubitably hints at a hidden cusp catastrophe. This observation provides an excellent and physically well-motivated example, and therefore forms the central focus of the present Section.

### 12.2.1 Definition and identification of a cusp

The 'textbook' definition of a cusp catastrophe for a univariate polynomial function is a vanishing first, second and third derivative with a non-zero fourth derivative. In most cases, however, the considered function does not appear in this canonical form. Therefore we will have to employ the conditions introduced in Sec. 3.2.4, Eq. (3.24). That is, a cusp point of a function is defined by fulfilling the following three conditions:

1. the first derivative in all directions vanishes
2. the determinant of the deformation tensor (the Hessian matrix) vanishes
3. the derivative of one of the eigenvalues of the deformation tensor vanishes in its canonical direction.

The former two conditions are what defines a fold point as detailed above. The third condition requires some further elaboration. Let $\mathcal{H}$ be the deformation tensor with the eigenvalues $\pm\epsilon_1$ and $\pm\epsilon_2$ with the corresponding eigenvalues $\mathbf{v}_1$ and $\mathbf{v}_2$. The precise formulation of condition (3) is then that the dot product between the eigenvector and the gradient of its eigenvalue vanishes [71, 205], for $i = 1$ or 2:

$$\mathbf{v}_i \cdot \nabla\epsilon_i = 0\,. \tag{12.18}$$

In practice, the conditions can be solved numerically with a finite-differences scheme for the differentiation of the eigenvalue fields. Analogously to the calculations for the fold, to solve these conditions we require the external parameter to assume complex values. In particular, because the cusp catastrophe has codimension $K = 2$ we require two complex-valued parameters. One is the harmonic order, and the other is a parameter that modifies the driving laser field shape. To identify a cusp that occurs for real-valued parameters we proceed analogously to the fold case, and iteratively try to minimise the imaginary part of the obtained solution by adjusting a third free parameter.

In the following we will exemplify the appearance of a cusp within the context of the colour switchover in two-colour HHG. Considering the colour switchover with fixed zero phase delay between the two component fields, $\varphi = 0$, we solve the above conditions by choosing the mixing angle $\theta$ and the harmonic order $q$ as external parameters. We identify a cusp point for:

$$\begin{aligned}\theta_c &= (21.28 + 0.03\text{i})^\circ \\ q_c &= 27.95 - 0.1\text{i} \\ (t_{i,c},\, t_{r,c}) &= (90.87 + 6.31\text{i},\ 174.34 + 0.03\text{i})\,.\end{aligned}$$

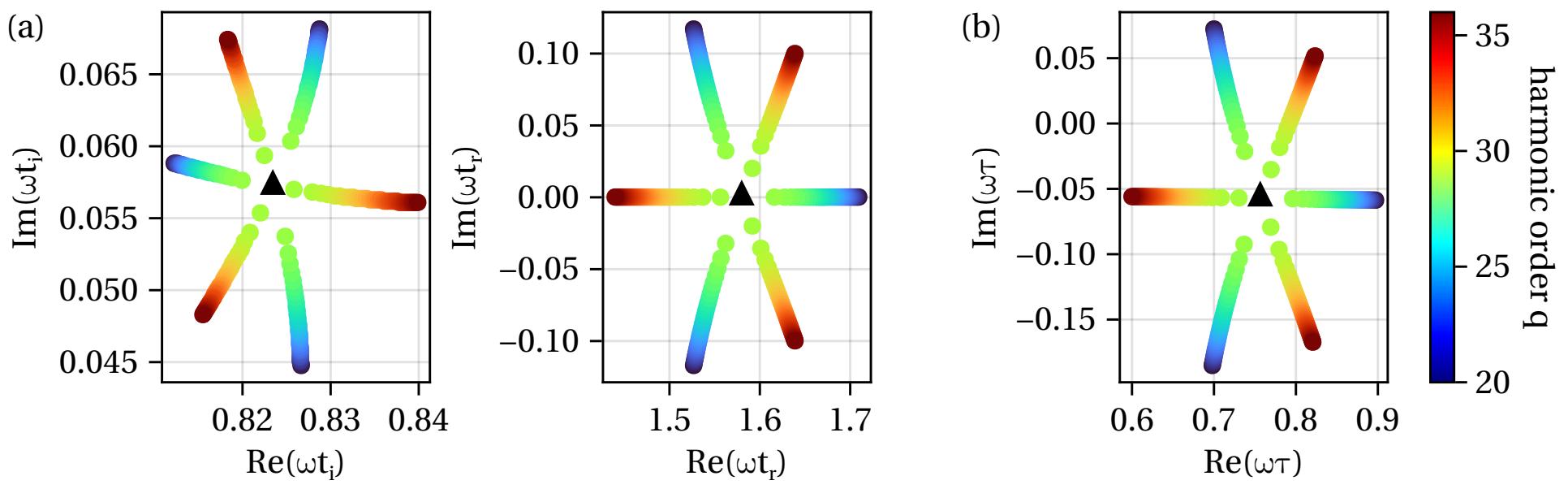

Figure 12.3: Saddle points in (a) the complex ionisation and recombination plane, as well as (b) the complex travel time plane (panel (b)), for $\theta = \theta_c \in \mathbb{C}$ and a range of harmonic orders with fixed imaginary part $\mathrm{Im}(q) = 0.1\mathrm{i}$. Note the colour bar for the harmonic orders on the right has a smaller range of harmonic orders than usual. The black triangle marker indicates the cusp point.

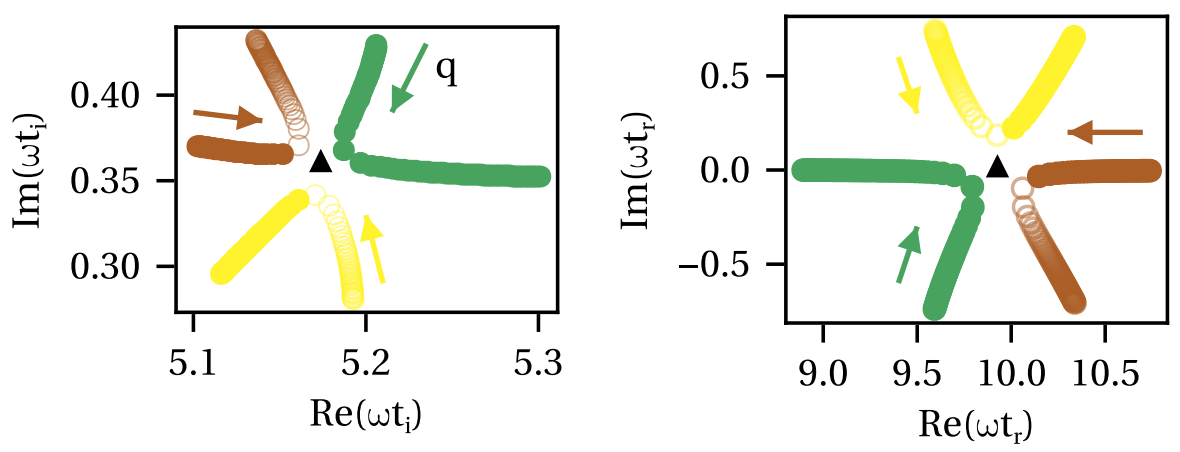

Figure 12.4: Saddle points in the complex plane for the spectrum shown in Fig. 10.8(h), taken at mixing angle $\mathrm{Re}(\theta_c) = 21.28°$ indicating a cusp catastrophe (cusp point is drawn as a triangle), and with the trajectories classified as explained in Chapter 10 and denoted by colours.

For this complex-valued mixing angle $\theta = \theta_c$, three saddle point solutions coalesce into a cusp point, shown in Fig. 12.3 where we have plotted saddle points for $q = q_r + \mathrm{i}\,\mathrm{Im}(q_c)$, for a range of $q_r \in \mathbb{R}$ (colour of the saddle points), as well as the cusp point (triangle).

When performing the colour switchover ‘in real life’, of course, we would only assume real-valued mixing angles. Therefore, we only see a missed approach of the three trajectories, shown in Fig. 12.4, where the cusp point (drawn as a triangle) now clearly serves as a centre point in between the different trajectories. We have employed the classification of saddle points into trajectories as introduced in Chapter 10. The fact that the exact coalescence is ‘hidden’ in the complex parameter space means that a consistent classification of the saddle-point solution can be achieved by taking smaller discretisation steps in $q$ (and eventually $\theta$), as explained for the case of a missed approach between two trajectories. Note that saddle points ‘move faster’ the closer they approach the catastrophe point. This can be seen as in Fig. 12.3 the saddle points shown are for steps $\Delta q = 0.25$, and yet — near the coalescence point — this is not fine enough to reliably capture the geometry.

As mentioned earlier, for the case of zero two-colour phase delay $\varphi$ we can only identify an exact coalescence of three saddle points (i.e., fulfilling the three conditions above) if the mixing angle $\theta$ and the harmonic order $q$ assume complex values, even though their imaginary parts might be very small. We find that for a different values of $\varphi$ we can still identify a cusp point, however, the imaginary parts to $\theta$ and $q$ will increase. We therefore conclude that the cusp — even if hidden — is a stable feature of the colour switchover, as it an integral part of the parameter space $(\varphi, \theta, q)$ that defines the setup. In Fig. 12.5 we show the parameters $(\varphi \in \mathbb{R}, \theta_c \in \mathbb{C}, q_c \in \mathbb{C})$ for which we identify the cusp, in several projections. The specific parameters for $\varphi = 0$ are marked in black and the marker size denotes imaginary parts. The $(\mathrm{Re}(q), \mathrm{Re}(\theta))$ projection in panel (b) indicates how the specific harmonic order which is enhanced by the coalescence changes as we perform the colour switchover. Panel (c) confirms that for colour switchovers with a different two-colour phase delay $\varphi$ there is still a cusp catastrophe point, which we will demonstrate below. However, the increased imaginary parts of the external parameters (indicated by the marker size) suggest that it plays a less dominant role for the total spectrum.

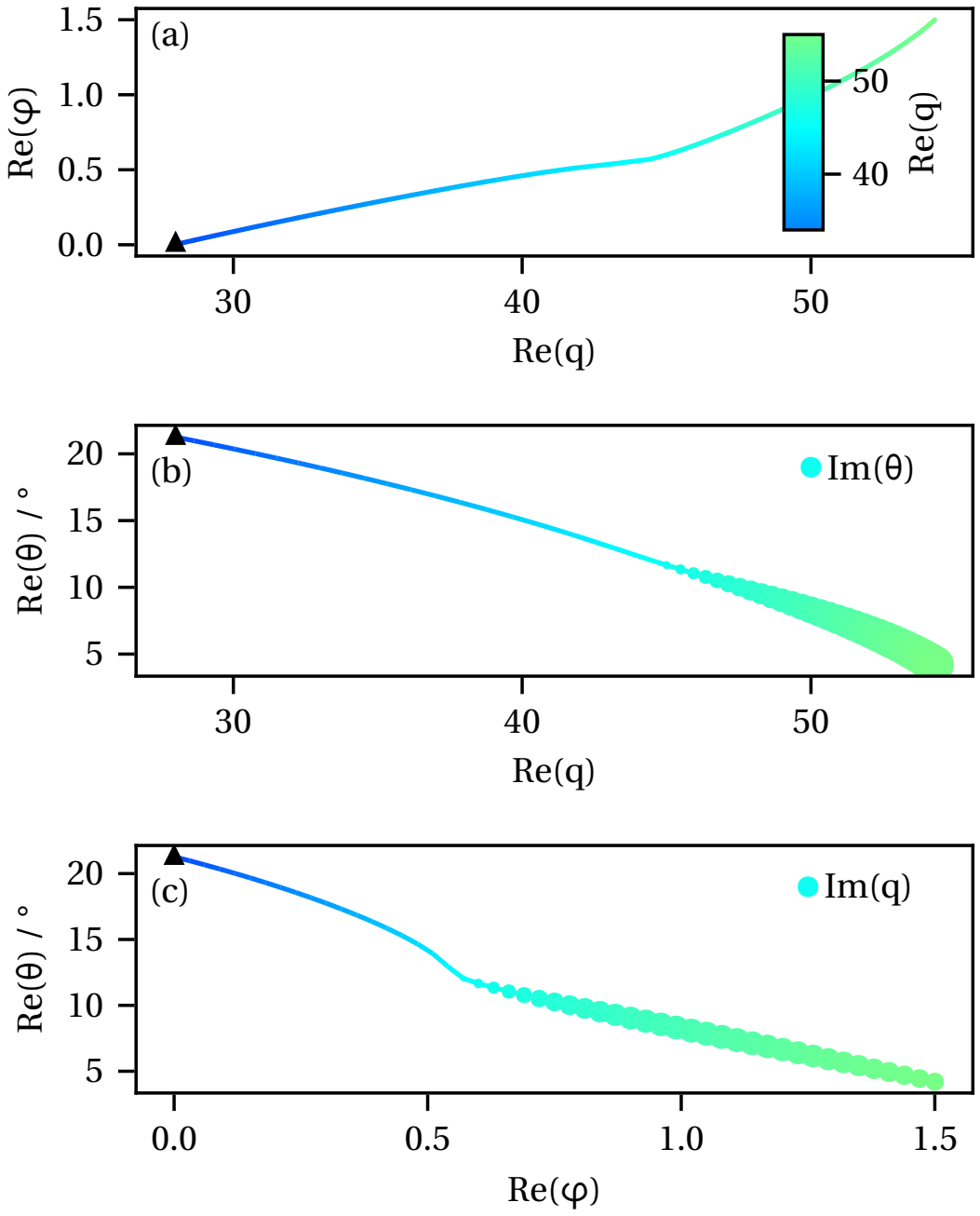

Figure 12.5: Projections of parameter combinations $(\theta, \varphi, q)$ for a cusp catastrophe point at which three saddle points coalesce. For the exact coalescence we assume $\theta \in \mathbb{C}$ and $q \in \mathbb{C}$ and denote the respective imaginary parts as marker size in panels (b) and (c). In all three panels the colour indicates the real part of the harmonic order $q$. The cusp point for $\varphi = 0$ is drawn as a triangle.

As the cusp causes a caustic enhancement in the spectrum, Fig. 12.5(a) suggests that, for example, harmonic order $q = 40$ may be enhanced by tuning the two-colour field to a phase shift of $\varphi = 0.5$. However, larger imaginary parts will diminish the effect of the exact coalescence point on the spectrum, as the spectrum is always calculated for real-valued external parameters only.

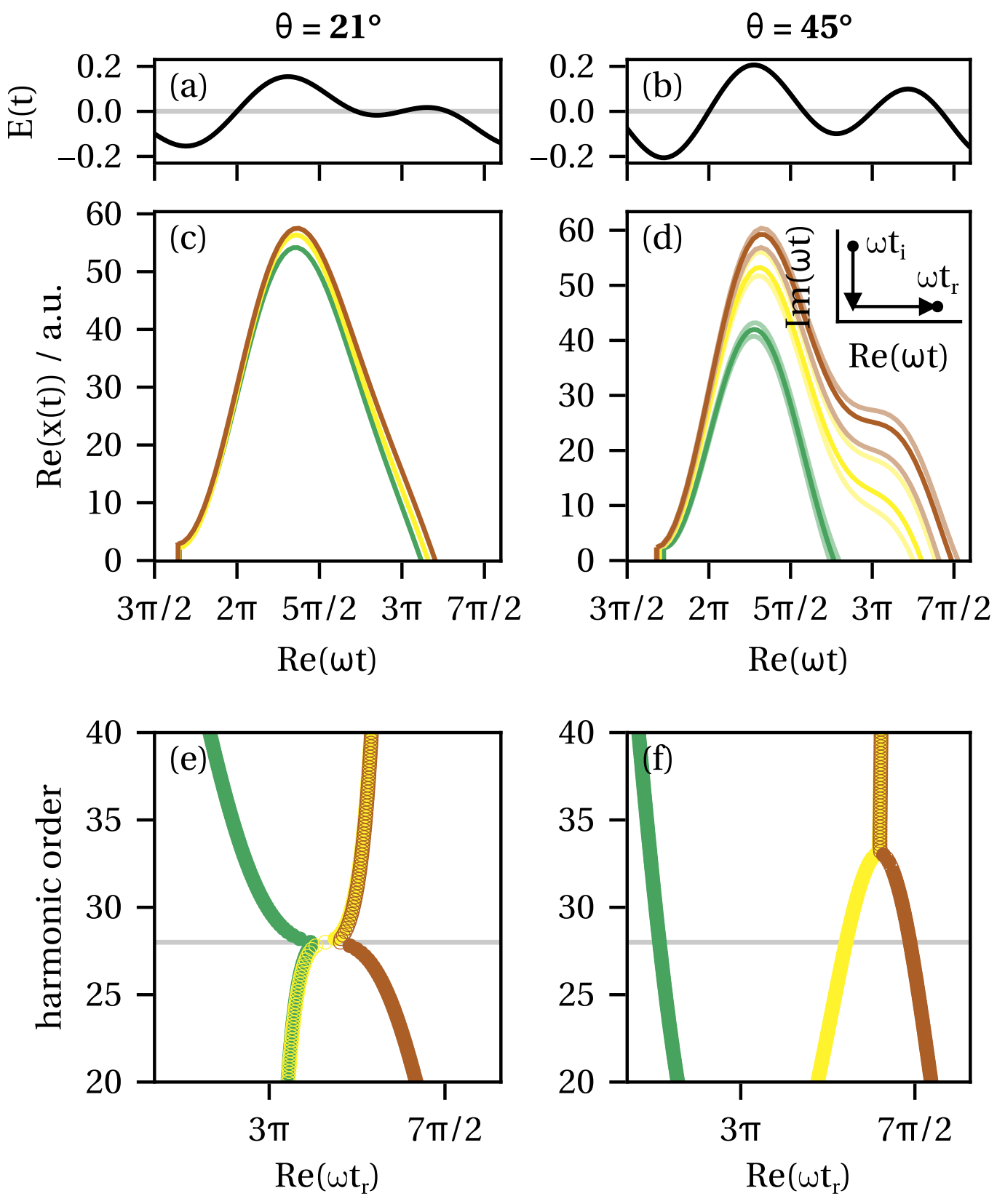

Figure 12.6: Electric fields (panels (a) and (b)) and three according semi-classical electron trajectories (panels (c) and (d)) for three saddle points contributing to harmonic order $q = 28$. The respective energy-time relations are shown in panel (e) and (f). The three assimilating trajectories for $\theta = 21°$ ($E_2/E_1 = 0.78$) on the left-hand side cause the enhancement seen in the respective spectrum Fig. 10.8(h). The temporal contour along which the integral Eq. (2.49) is evaluated is shown in the inset of panel (d).

### 12.2.2 Semi-classical picture of colliding trajectories

For a more intuitive explanation of the origin of the enhanced harmonic intensity we can look at the semi-classical electron trajectories around the cusp point. In Fig. 12.6 we show the trajectories for the saddle points involved in the missed approach in comparison to the saddle points of the same trajectory, but for a slightly different mixing angle. Clearly, for $\theta = \theta_c$ (panel (c)) the three trajectories almost coincide. In particular, they recombine at the same time with the same velocity (think gradient at time of recombination) which allows their dipole phases to interfere constructively. In contrast, for a slightly different mixing angle, the three trajectories (panel (d)) are of different shape and recollide at different times. This becomes even more apparent from the energy-time relations shown in the panels below. In the vicinity of the cusp (panel (e)), the energy-time curves are almost flat, for all three trajectories, corresponding to a similar dipole phase and hence, constructive interference of the signals. In comparison, at a different mixing angle the energy-time relations (panel (f)) are steep, such that the dipole phase changes rapidly over recombination (emission) time.

### 12.2.3 Caustic features in the harmonic response

We motivated the study of catastrophe points by promising caustic enhancement features in the observable harmonic spectrum. In Fig. 12.7 we show the spectrum at $\theta = \theta_c \approx 21°$ with the contributions from the individual saddle points, calculated by the usual saddle-point approximation Eq. (6.4) (in colours), as well as the total intensity (in black). Around harmonic order $q = \mathrm{Re}(q_c) \approx 28$ we clearly identify an enhancement by half an order of magnitude. Furthermore, note that the total intensity as calculated by the standard saddle-point approximation technically breaks down at/in the vicinity of the cusp, which causes the artificial discontinuity around this harmonic order 28. The total intensity calculated using the downward flow procedure (PLF, blue line) on the other hand is smooth across the spectrum and the enhancement feature.

To hone in on the aforementioned tunability of the cusp within the parameter space, in Fig. 12.8 we show the spectrum around a cusp for a different two-colour phase delay. For $\varphi = \pi/8$ we identify the cusp at $\theta = 16.37 + 0.06\mathrm{i}$ (as can be read off from Fig. 12.5(c)) and the harmonic order $q = 37.85 - 0.08\mathrm{i}$. The spectrum Fig. 12.8 is calculated for the real parts of those parameters. This demonstrates that by tuning the phase delay we gain control over the enhanced harmonic order within the harmonic spectrum. In the subsequent Section, which details the swallowtail catastrophe within the context of HHG, this feature will be discussed further. Note that the artificial discontinuity in Fig. 12.8 due to the cusp appears as an even more pronounced peak here in the total intensity as calculated by the standard saddle-point method. Again, the downward flow behaves smoothly across the caustic.

### 12.2.4 Conclusions

We conclude that the cusp catastrophe is a stable feature of the colour switchover. That is, even if we chose to perform a colour switchover for a different two-colour phase shift we would be able to find a cusp catastrophe point, 'hidden' in the complex-valued parameter space for $\theta$ and $q$. The more hidden the exact cusp catastrophe point is in the complex-valued parameter space, the less enhancement we will observe in the spectrum. This was found for the fold point at the high-harmonic cutoff as well, where the imaginary part of the harmonic-cutoff order determines the depth of the interference fringes between the short and long trajectories throughout the spectrum.

The cusp catastrophe discussed in this Section hints towards a further line of research associated with the colour switchover in the 'ionisation only' setup, examined in Chapter 8, where we had identified a fold catastrophe as a necessary feature for the onset of the contribution of the newly incoming ionisation events. The final momentum $p$ for direct photoelectrons in ATI can be seen analogously to the harmonic orders $q$ of the emitted radiation in HHG. Therefore we expect to find a cusp point within the colour switchover of ATI, possibly by considering complex-valued momenta and/or amplitude ratios. This could explain particular enhancement features in photoelectron momentum distributions and serve as a good starting point to understand, for example, how to analytically model the harmonic response around the cusp points — in analogy to the previously discussed Harmonic Cutoff Approximation around the fold point at the cutoff.

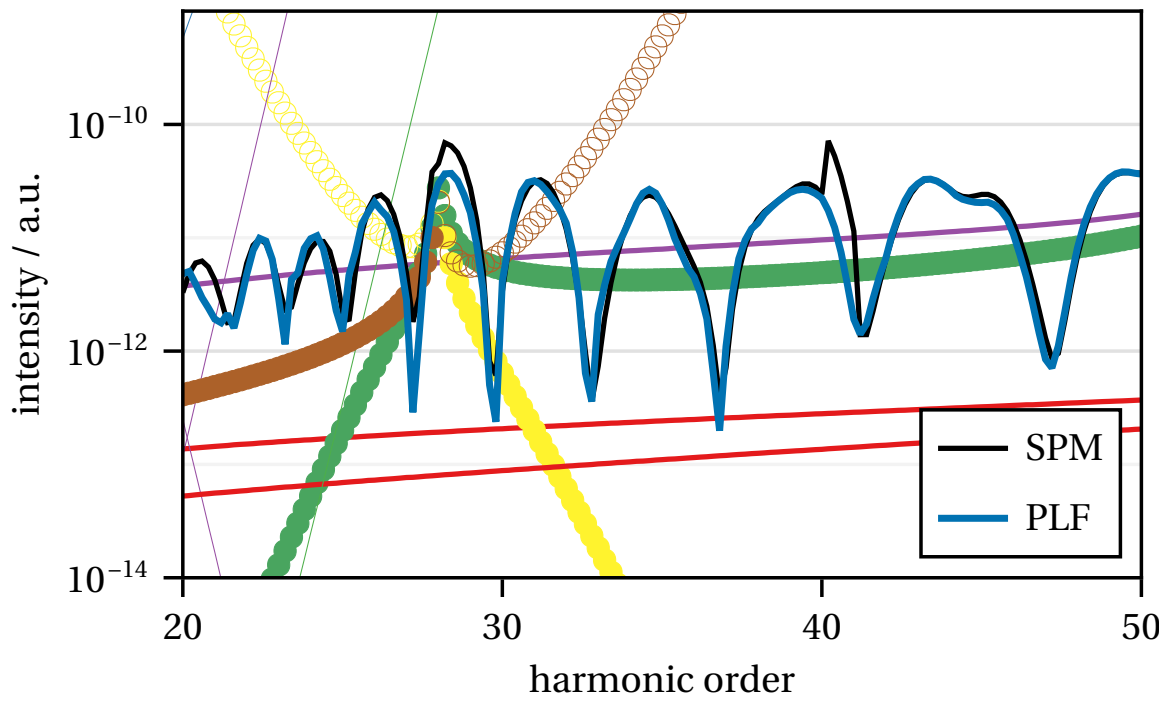

Figure 12.7: Zoom in on the spectrum of Fig. 10.8(h) (at $\theta = \theta_c \approx 21°, \varphi = 0$), highlighting the three different contributions (as brown, yellow and green markers) that cause the enhancement of harmonic order 28 due to the nearby cusp catastrophe point.

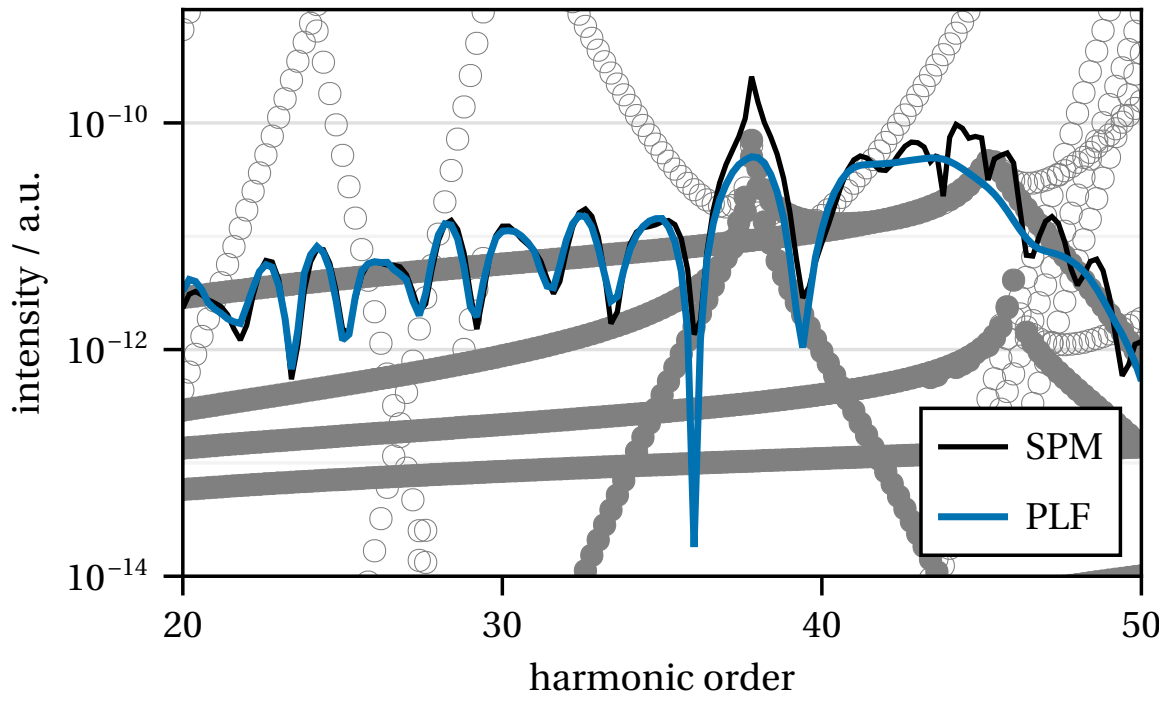

Figure 12.8: Harmonic spectrum for the two-colour field with $\varphi = \pi/8$ and $\theta = 16.37$ with individual (unclassified, hence grey) contributions from separate saddle points, and an enhancement of harmonic order 38 due to the nearby cusp catastrophe point. The total intensity is calculated using a sum of Gaussian contributions (SPM, black) and the downward flow (PLF, blue).

## 12.3 The swallowtail, an observed caustic in two-colour HHG

The swallowtail catastrophe is, to date, the only catastrophe that has been both experimentally observed and explicitly named in the context of attosecond science. Its first observation was reported in Ref. [72] in HHG driven by a co-linear two-colour field. This finding was revisited more recently in Ref. [37]. Both studies report a pronounced enhancement of a specific harmonic order when HHG is driven by a $\omega - 2\omega$ field where the second harmonic component is rather strong and has a certain phase delay to the fundamental driver. Beyond the experimental observation, Ref. [72] explicitly established the connection to catastrophe theory by modelling the relevant diffraction integrals as polynomial functions incorporating the laser-field parameters, and demonstrated qualitative agreement with the measured spectra.

In contrast, Ref. [37] focusses primarily on the enhancement of a (tunable) harmonic order. The explanation for the stronger signal was given in terms of coinciding classical trajectories. Here, we briefly illustrate how the tools of Picard–Lefschetz methods can provide additional and complementary insights into this phenomenon.

### 12.3.1 The (attempted) identification of a swallowtail

At a swallowtail catastrophe, four saddle points coalesce into a single higher-order critical point. In the vicinity of this point, the set of critical (saddle-point) solutions unfolds into four distinct branches, or sheets, in parameter space. In principle, the external parameters defining the swallowtail point can be identified using the conditions outlined in Sec. 3.2.4. For the cusp we required a vanishing gradient, determinant of the deformation tensor, and directional derivative of one of the eigenvalue fields. In addition to that, for a swallowtail we require the dot product between the eigenvector and the gradient of the directional derivative of the eigenvalue field to vanish. That is,

$$\mathbf{v}_i \cdot \nabla (\mathbf{v}_i \cdot \nabla \epsilon_i) = 0 \tag{12.19}$$

The implementation of these conditions within the application of the semi-classical HHG action is an ongoing challenge. The exact identification of a swallowtail catastrophe at which four quantum orbits coalesce is therefore the focus of future research.

The definition of the swallowtail catastrophe in the publications [72] and [37] is based on simple classical models. In Ref. [72] the swallowtail is identified to be at $\tilde{R} = 0.44$, $\varphi = 0.535\,\text{rad}$ for the two-colour driver $E(t) = E_1 \sin(\omega t) + E_2 \sin\big(2\omega t + \varphi\big)$, where $\tilde{R} = E_2/E_1$. In Ref. [37] the harmonic intensity for order 23 was maximised at $R \approx 0.2$, $\varphi = -1\,\text{rad}$ for the definition of the two-colour driving field

$$E(t) = \sqrt{(1-R)}E_0 \cos(\omega t) + \sqrt{R}E_0 \cos\big(2\omega t + \varphi\big). \tag{12.20}$$

Even though experiments were conducted within different experimental parameters (e.g., different fundamental intensity, different gases), resulting in different harmonic orders to be enhanced, topologically, these two setups represent the same structure. Around these reported 'classical' swallowtail points the two driving field definitions are equivalent.

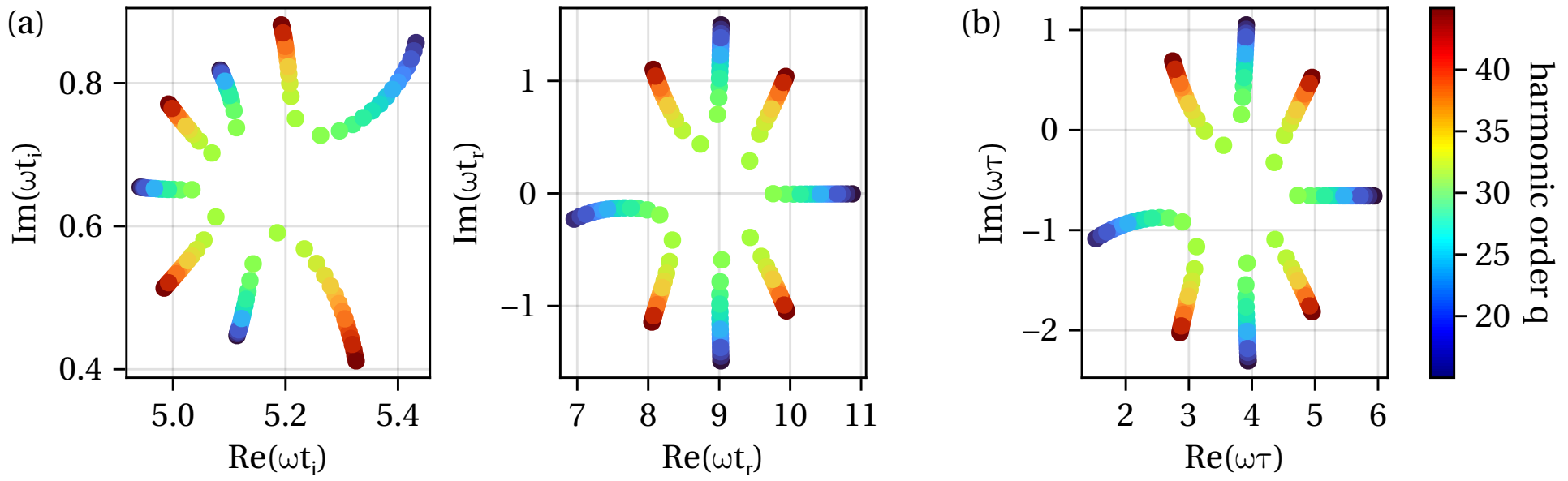

Figure 12.9: Saddle points in the complex ionisation and recombination time planes (panel (a)), as well as the complex travel time plane (panel (b)), in the vicinity of the expected swallowtail with $R = 0.16$ and $\varphi = 0.59$, for the range of real-valued harmonic orders (coloured).

For the following considerations we adapt the definition of the two-colour field Eq. (12.20), used in [37]. We use $\mathcal{I}_p = 15.76\,\mathrm{eV} = 0.58\,\mathrm{a.u.}$, $\lambda = 1030\,\mathrm{nm}$, corresponding to $\omega = 0.044\,\mathrm{a.u.}$ and $I_{\mathrm{total}} = 0.92 \times 10^{14}\mathrm{W/cm^2}$, making $E_0 = 0.051\,\mathrm{a.u.}$ For the three external parameters that the swallowtail requires, we take the two-colour phase delay $\varphi$, the amplitude ratio $R$, and the harmonic order $q$. For the *exact* coalescence of four saddle points we expect these to acquire imaginary parts, in analogy to the procedures for the fold and the cusp catastrophe. The evaluation of the HHG integral and the corresponding spectral intensity, however, is only meaningful for real-valued parameters. We therefore elegantly ignore our current inability to rigorously identify the exact swallowtail point and instead focus on considering the 'classical' swallowtail parameters, derived from the above publications. For our definition here they are $\varphi = 0.59$, $R = 0.16$, $q = 30.$, with an expected swallowtail point around $(t_{\mathrm{i},s}, t_{\mathrm{r},s}) \approx (5.1 + 0.66\mathrm{i}, 9. - 0.04\mathrm{i})$.

## 12.3.2 The missed approach of four trajectories

While the exact identification of the swallowtail parameters is yet to be determined, we present several signatures — and clear indicators — of its existence. The first, and most mundane signature is the close approach of four saddle points. Therefore, the saddle points for $(\varphi, R) = (0.59, 0.16)$ are shown in Fig. 12.9, for a range of real-valued harmonic orders $q$. Along the harmonic orders the saddle points trace clear branches in the complex plane that are seemingly repelled from a common centre point. Analogously to the treatise of the earlier catastrophes, this missed approach of four trajectories hints a swallowtail catastrophe point that sits exactly at this centre point. As the saddle points do not fully coalesce upon varying $\theta$ and/or $\varphi$, the consistent classification of saddle points into separate trajectories can simply be achieved by decreasing the discretisation step size. Any arising doubt (most likely by my future self) regarding the nature of the higher-order critical point — and, in particular, the existence of an actual swallowtail point and not a higher-ranked catastrophe — shall be removed by examining the saddle points themselves. Ultimately, catastrophes are defined by the coalescence of saddle points, and in the present case the problem (even if allowing for complex-valued parameters) only admits four such solutions. Consequently, the highest-order critical point can be formed by the coalescence of exactly these four saddle points, consistent with the defining structure of a swallowtail catastrophe.

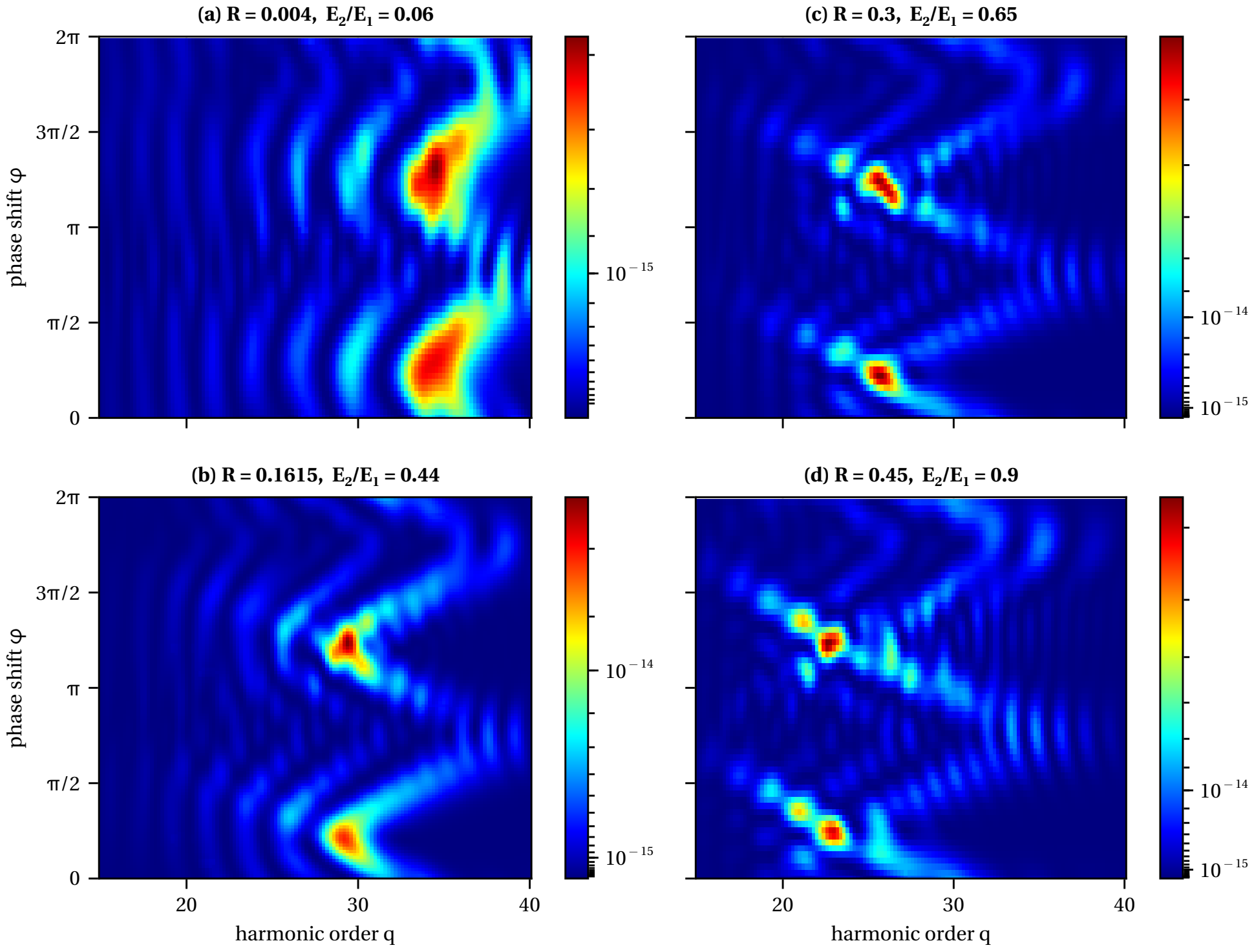

Figure 12.10: Harmonic intensities across a range of harmonic orders $q$ and two-colour phase delays $\varphi$, for the fixed amplitude ratios (a) $R = 0.004$, (b) $R = 0.1615$, (c) $R = 0.36$ and (d) $R = 0.45$. Note the colour ranges vary across the configurations to highlight the diffraction structure within each panel. The most intense point is at the expected swallowtail configuration (b).

### 12.3.3 Observable caustics

In order to observe the diffraction pattern characteristic for the swallowtail, we have to look at the intensity across two-dimensional projections of the parameter space. In Ref. [72] this is elegantly achieved by showing the harmonic intensity across a range of harmonic orders and phase shifts, for three distinct amplitude ratios. In Fig. 12.10 we show the same type of projection, for several amplitude ratios, ehibiting similarity with the canonical diffraction patterns from Sec. 3.2.4. The amplitude ratio for which we expect the swallowtail catastrophe is $R = 0.1615$, where $E_2/E_1 = 0.44$, shown in panel (b).

At the swallowtail point, the standard saddle-point approximation breaks down as the Hessian in the denominator vanishes due to the coalescence of four saddle points. We therefore evaluate the SFA integral exactly, using the downward flow procedure. The resulting harmonic intensity shows a clear qualitative agreement with the canonical diffraction pattern of the swallowtail catastrophe, including the emergence of a caustic structure in parameter space. By contrast, in Ref. [72] the harmonic response was modelled phenomenologically by directly evaluating a swallowtail diffraction integral constructed from a polynomial of swallowtail degree. Our approach complements the established results by providing an un-biased approach to the same universal wave features.

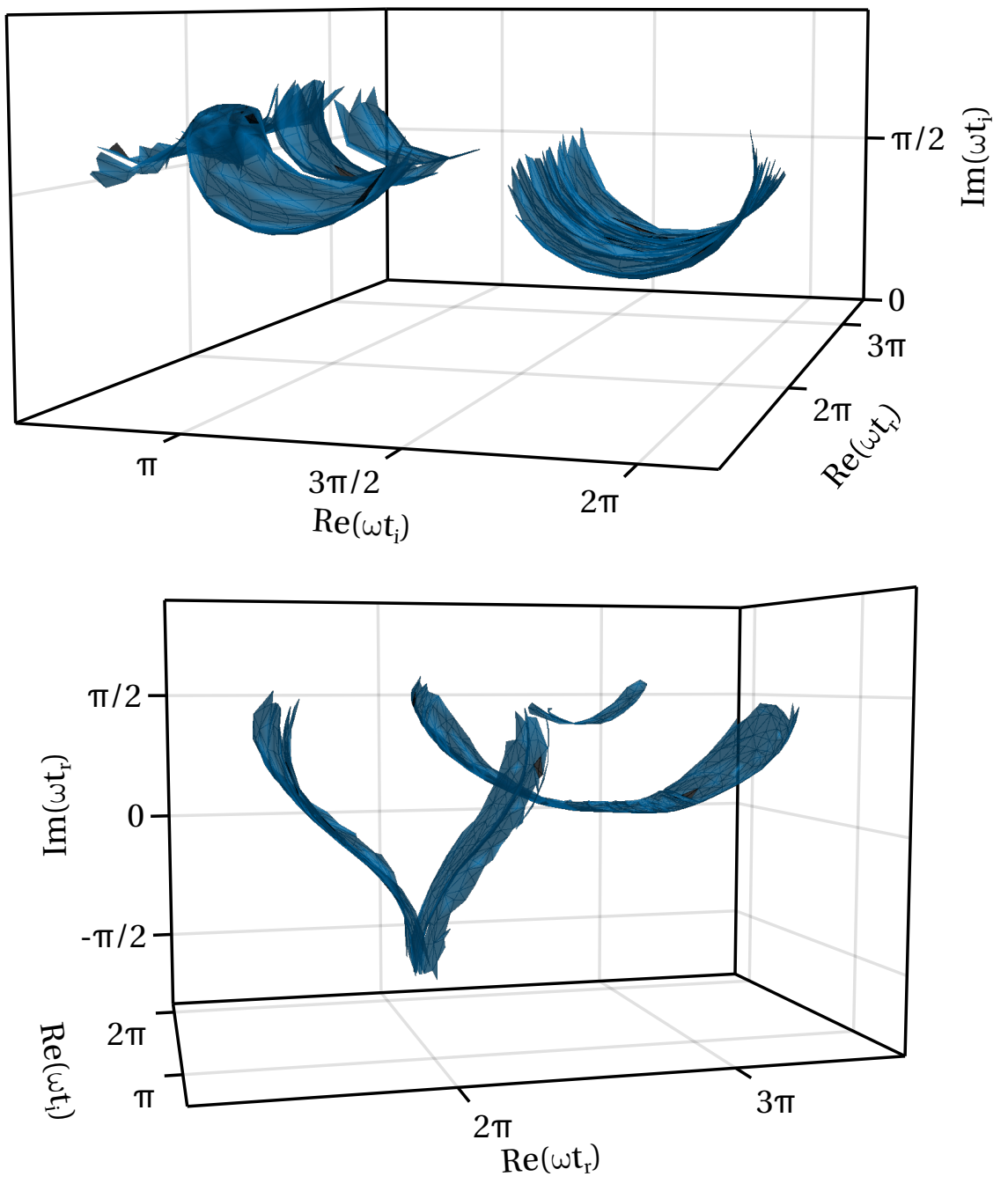

Figure 12.11: Lefschetz thimble in the ionisation-time projection $(\mathrm{Re}(t_\mathrm{i}), \mathrm{Im}(t_\mathrm{i}), \mathrm{Re}(t_\mathrm{r}))$ (top) and recombination-time projection $(\mathrm{Re}(t_\mathrm{r}), \mathrm{Im}(t_\mathrm{r}), \mathrm{Re}(t_\mathrm{i}))$ (bottom), for the expected swallowtail configuration $(\varphi, R, q) = (0.59, 0.16, 30.)$.

In Fig. 12.11 we show the respective Lefschetz thimble for the expected swallowtail configuration at $(\varphi, R, q) = (0.59, 0.16, 30.)$. Notably, the surface is rather flat over recombination time compared to the thimbles attached to 'normal' saddle points (e.g., the two surfaces seen around $\mathrm{Re}(\omega t_\mathrm{r}) \approx 2\pi$ in the bottom panel).

## 12.3.4 The bifurcation set around a swallowtail point

To obtain more clarity about the structure of critical points it can be insightful to visualise the bifurcation set, i.e., the (surface of) critical points for which the determinant of the deformation tensor is zero. This surface of fold points is what is typically visualised in catastrophe theory, see Fig. 3.7. In this section we show some preliminary results for the case of the expected swallowtail. In order to solve the fold conditions we had to assume complex-valued parameters. Typically, we would complexify the harmonic order $q$. To further solve the cusp and the swallowtail conditions, we assume the other parameters to assume non-zero imaginary parts as well. As a result, the bifurcation set is embedded in a 6-dimensional parameter space of $(\theta \in \mathbb{C}, \varphi \in \mathbb{C}, q \in \mathbb{C})$. In Fig. 12.12 we plot this bifurcation set in a projection, and only consider $\mathrm{Im}(\varphi) = 0$. In panels (a) and (d) we show the fold points in the projection $(\mathrm{Re}(\varphi), \mathrm{Re}(R), \mathrm{Re}(q))$ from two different view angles, respectively, where we have chosen $\mathrm{Im}(R) = -0.01$. In panel (b) we show 'slices' through this structure for fixed values of $\mathrm{Re}(R)$. The colour of the markers denotes the real part of the largest eigenvalue of the deformation tensor, with the respective colour bar shown in panel (c). We find this projection shows structural similarity with the bifurcation set of the swallowtail catastrophe with four surfaces that merge at a point in the centre, as in Fig. 3.7. Hence, we should be able to identify the two ridges towards the presumed centre of the structure as the solutions to the cusp conditions. For different values of $\mathrm{Im}(R)$, the whole structure is sheared and skewed. It is intriguing and encouraging to identify the exact point (which we assume exists) that acts as an organising centre to this structure. Furthermore, we have yet to understand the implications of the complexification of the parameter space in more detail.

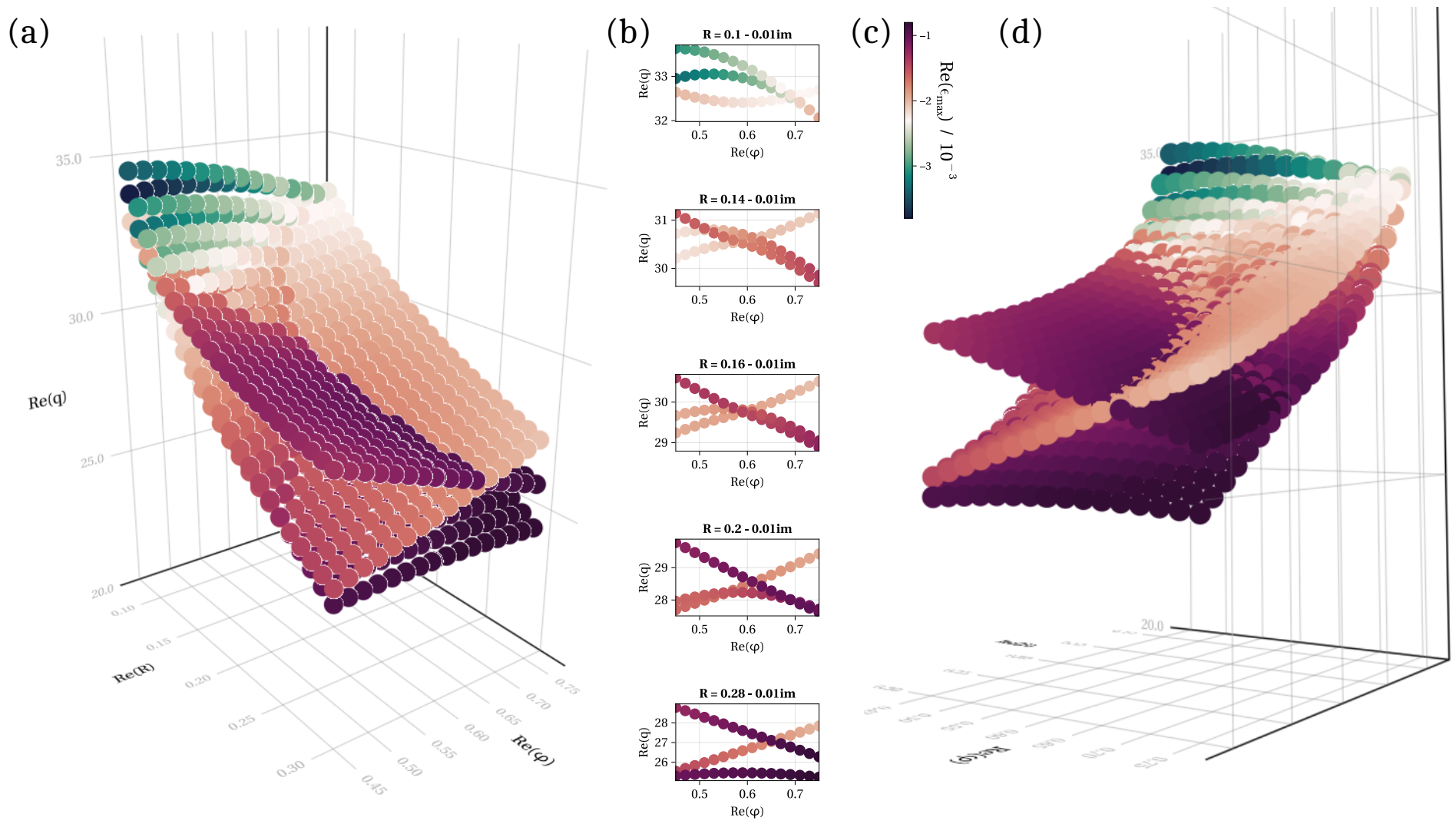

Figure 12.12: The bifurcation set in $(\varphi, R, q) \in \mathbb{C}^3$, in the vicinity of the expected swallowtail point. Shown are folds for $\mathrm{Im}(\varphi) = 0.$ and $\mathrm{Im}(R) = -0.01$. Panels (a) and (d) show the projection in $(\mathrm{Re}(\varphi), \mathrm{Re}(R), \mathrm{Re}(q))$ space from two different view angles. Panel (b) shows the projections in the $(\mathrm{Re}(\varphi), \mathrm{Re}(q))$ plane for the fixed values $\mathrm{Re}(R) = 0.1, 0.14, 0.16, 0.2, 0.28$, respectively. Markers are coloured according to the real value of the largest eigenvalue of the deformation tensor (colour scale in panel (c)).

### 12.3.5 Conclusions

The saddle points, the diffraction pattern and the bifurcation set all provide strong evidence for the existence of a swallowtail catastrophe point within the setup of HHG driven by a co-linear two-colour field. From the harmonic intensities shown in $(q,\varphi) \in \mathbb{R}^2$ parameter space we find that the caustic yields a substantial enhancement in comparison to its vicinity in parameter space. Further work aims to identify the exact swallowtail parameters at which four semi-classical trajectories coalesce, and to disentangle the bifurcation set in its 6-dimensional space.

# Part IV

## Conclusions and Back Matter

# 13

# Conclusions

## Summary

The creation of light pulses of attosecond duration via the generation of high-order harmonics (HHG) emitted from atomic targets is typically understood in terms of a semi-classical model involving several interfering electron trajectories. In this picture, electrons are ionised, accelerated by the laser field, and eventually recombine with their parent ion, emitting radiation at harmonic frequencies. While this interpretation has proven remarkably successful, it is fundamentally rooted in intuitive arguments derived from saddle-point approximations. In this thesis, we provided a rigorous mathematical link between the fully quantum-mechanical description of these processes and their intuitive explanation in terms of distinct semi-classical trajectories.

For that, we introduced the key concepts of Picard–Lefschetz theory and develop corresponding numerical methods for the evaluation of one- and two-dimensional highly-oscillatory integrals. Central to this framework is the downward flow algorithm, which deforms the original integration contour to the complex domain into so-called Lefschetz thimbles — contours along which the oscillations of the integrand are minimised. The original integral can then be rewritten as a sum of contributions from selected thimbles, while still providing an exact representation of the original expression. Upon convergence of the downward flow, each thimble is attached to a critical point of the integrand, and whether a given critical point contributes to the total integral is determined by its intersection number with the original integration contour.

For one-dimensional integrals, this intersection number can be identified straightforwardly. For two-dimensional integrals, however, no established general method previously existed. To overcome this limitation we developed the necklace algorithm which allows us to reliably identify the intersection number of a given critical point. In turn, this approach enables a complementary perspective on highly-oscillatory integrals: assuming the critical points are known, their individual relevance can be assessed directly and the total integral reconstructed as a sum over the contributing saddle points. If the critical points are simply saddles in the complex plane, their local contributions are approximately Gaussian, yielding the familiar saddle-point approximation. The numerical methods developed in this thesis are readily applicable to integrals of the form $\int e^{i\phi(\mathbf{x})}\,d\mathbf{x}$ for (somewhat) arbitrary phase functions $\phi(\mathbf{x})$. They are available as an open-source julia package.

We demonstrated that the integrals describing both the HHG response and the direct photoelectron signal from tunnel ionisation within the strong-field approximation (SFA) framework can be evaluated using these techniques. Historically, the interpretation of these processes in terms of semi-classical electron trajectories emerged from applying the saddle-point approximation to the SFA integrals for monochromatic driving fields, where the saddle points define the relevant ionisation and recombination events. In recent years, however, attosecond dynamics have increasingly been studied using polychromatic laser drivers, which allow the waveform driving the strong-field process to be deliberately shaped. In particular, when fields of commensurate frequency are added to the fundamental driver with more-than-perturbative amplitudes, it becomes unclear which electron trajectories contribute to the dynamics. Using the necklace algorithm, we were able to identify the relevant contributors to the total integral independently of driving-field symmetries, trajectory classifications, or heuristic arguments.

Furthermore, the use of the downward flow algorithm allows the evaluation of integrals in parameter regimes where multiple saddle points lie in close proximity, the traditional saddle-point approximation breaks down, and caustics emerge in the observable signal. We therefore presented a series of results that apply the methods of Picard–Lefschetz theory to such scenarios in attosecond science.

As a central unifying framework of this thesis, we introduced the concept of the colour switchover, which describes the gradual transition from monochromatic driving fields to strongly polychromatic waveforms. This approach is applied to both strong-field ionisation and HHG. We analysed how the addition of a second-colour field introduces new ionisation events and determine the amplitude ratios at which these newly emerging trajectories begin to influence the dynamics. As part of this analysis, we identified a particularly striking feature in which ionisation occurs at times when the instantaneous electric field vanishes and no tunnelling barrier is present, providing a clear manifestation of the nonadiabatic nature of strong-field ionisation.

Building on that, we showcased the colour switchover in the context of HHG to demonstrate the unique capabilities of Picard–Lefschetz methods in identifying the relevant contributors to the harmonic spectrum. We found that newly emerging trajectories can become relevant already at comparatively low amplitude ratios. As in the case of ionisation alone, the colour switchover in HHG is characterised by the near-coalescence of saddle points, which appears to be a necessary condition for new trajectories to contribute.

A further application, developed in collaboration with experimentalists, concerns scans of the relative phase delay between the two components of an orthogonally polarised driving field. In this setting, analysing the individual electron trajectories enabled the observed intensity oscillations across the phase-delay scan to be disentangled and interpreted in terms of quantum-path interference.

Finally, we combined the methods of Picard–Lefschetz theory with concepts from catastrophe theory to investigate the appearance of caustics in attosecond science. We revised how the HHG cutoff can be treated as a fold catastrophe, analysed in detail the cusp catastrophe emerging during the colour switchover, and alluded to the appearance of a swallowtail catastrophe which was observed experimentally in phase-delay scans of HHG driven by two-colour fields of varying amplitude ratios.

Ultimately, this thesis develops a unified framework to evaluate highly oscillatory integrals across extended parameter scans, both with numerical accuracy sufficient to resolve caustics and with an unambiguous decomposition into semi-classical contributions. We apply this framework in particular to attosecond science, such as to solve and *understand* the quantum dynamics governing strong-field processes.

## Future Work

This thesis serves as a proof of principle, demonstrating that the ideas of Picard–Lefschetz theory can be applied to the SFA integrals that compute the ionisation amplitude and the HHG response. Below, we want to list a subset of the many promising ideas and possible future research directions following from this thesis.

In this thesis, we have implemented the downward flow and the necklace algorithm. Of course, there are several ideas to improve the functionality of these specific numerical tools. For example, evaluating the necklace algorithm is currently rather computationally expensive when applied across a large parameter scan where no information on the Stokes transition is available. We would like to explore whether the relation between the two dimensions of the integral can be exploited to improve the performance of the procedure. Furthermore, we currently have no means of identifying the sign of the intersection number, nor the possibility of intersection numbers larger than one.

Beyond that, the ideas of Picard–Lefschetz theory allow for a wider range of numerical and conceptual methods that we look forward to employing. For example, we would like to understand caustic features in terms of trajectories, by identifying the Lefschetz thimbles that span those temporal regions. Tracking the contribution of a particular thimble across parameter scans will allow us to predict the emergence of caustic features in observable HHG spectra. A step tightly linked to it, is the identification of the actual catastrophes. As shown in Sec. 12.3, in the HHG response of a two-colour driving field, we find four trajectories in close vicinity undergoing a missed approach. In the near future, we want to identify the (conditions for the) canonical parameters that allow the semi-classical solutions to actually coalesce. Motivated by this example, we aim to develop — in continued collaboration with Job Feldbrugge — a solver that identifies swallowtail catastrophes for generic functions.

On the application side, there are several outlooks for this thesis. A currently ongoing project is the further study of HHG in co-orthogonal two-colour fields, in collaboration with the team at ELI Beamlines, providing experimental results and fruitful scientific discussions. This project is aimed at disentangling the contributions from different trajectories that create harmonic dipoles across the full macroscopic target. Identifying which trajectories are relevant for which field configurations, and where — in a suitable parameter space — trajectories interfere in a 'catastrophic' way, will allow us to attribute features of the measured spectra to specific caustic effects.

The methods outlined in this thesis are by no means restricted to specific driving-field configurations. As such, in collaboration with Nicola Mayer we have already obtained preliminary results for calculating the HHG response from chiral molecules driven by synthetically chiral light fields. In such systems, the harmonic response is expected to differ for opposite molecular enantiomers, opening a pathway towards chiral discrimination. In particular, we find promising early results for calculating the harmonic emission from chiral molecules interacting with vector-vortex beams with azimuthal and radial polarisations. Accurately describing these interactions requires a more careful treatment of the dipole matrix transition elements appearing in the prefactor of the HHG dipole integral. Since these matrix elements typically originate from computationally expensive quantum-chemistry calculations, a direct numerical evaluation of the full integral becomes prohibitively costly. Restricting the evaluation to the physically relevant saddle points therefore constitutes a highly valuable simplification, offering both computational efficiency and deeper insight into the underlying dynamics, and ultimately informing the identification of viable schemes for chiral-sensitive measurements.

## Outlook

Within attosecond science, our methods open the door to quantum-orbit approaches for studying strong-field processes with a broader class of light sources — for example, structured light or quantum light. In a way, we have shown the application of Picard–Lefschetz methods for the integration over ionisation (and recombination) time. However, introducing the rich toolbox of Picard–Lefschetz to attosecond science hopefully inspires evaluation techniques for integrals over other variables as well. For instance, describing HHG in solids requires integration over lattice momenta. We are intrigued to explore whether Picard–Lefschetz theory can prove useful for evaluating such integrals more efficiently, and for understanding the corresponding strong-field processes in terms of semi-classical trajectories.

Beyond attosecond science, we found that the application of Picard–Lefschetz theory remains surprisingly limited, despite the ubiquitous role of path integrals throughout physics. We hope that the work presented in this thesis contributes to making Picard–Lefschetz theory more widely accessible and encourages its adoption in research areas both adjacent to and beyond attosecond science. Moreover, during our exploration of catastrophe theory, we encountered numerous mathematical and physical (research) settings in which caustics naturally arise. Many of these problems could benefit from the application of Picard–Lefschetz methods as a systematic and robust means of resolving the appearing caustic structures and studying the underlying semi-classical constituents.

# List of Figures

# Acknowledgements

This thesis would not have been possible without the guidance, support, and supervision I received throughout my PhD. First and foremost and above and beyond all, I want to thank my incomparable supervisor Emilio Pisanty. His rigorous scientific approach, inexhaustible enthusiasm for complex analysis and the SFA, and his uncompromising standards were challenging at times but ultimately deeply rewarding. I am especially grateful that he encouraged me to travel to conferences, supported me in doing public outreach, teaching, supervising, and many other research-related activities. I feel immensely honoured to have been his first PhD student, and greatly enjoyed discovering that path together, along which we learnt a lot, and I would not have wanted it any other way.

For being my second supervisor, I thank Amelle Zaïr — for her continued efforts in establishing the AttoKing's research group and for providing supportive space for discussion. Whenever an extra pinch of precision was needed, Margarita Khokhlova was there, offering additional clarity and scientific thoroughness, which I truly appreciate.

From the very beginning of writing this thesis, I dreaded the moment of writing these acknowledgements — mainly because I do not know how to acknowledge Job Feldbrugge in a truthful way that does not embarrass myself. With every meeting we had he removed the clouds of confusion and pointed me towards the sunshine, reminding me of the joy of doing scientific research. His calm and genuine way of teaching always left me inspired, motivated, and full of new ideas to pursue. I very much look forward to the continuation of this collaboration.

My PhD was carried out within the Photonics & Nanotechnologies group, and I want to express my sincere admiration for Anatoly Zayats for leading the research group with such powerful calmness and elegance. I am also grateful to everyone in the office who contributed to building the Social Community of the seventh floor. Most of all, I deeply appreciate the company of Andrew, who provided me with intellectual, emotional, nutritional, and linguistic support in unreasonable amounts. From hiding emergency snack stashes for me in the office, to secret nightly dance-offs in the seminar room, seeing questionable shows at the Edinburgh Fringe and a freezing dip in the ocean — Thank you for being such an excellent partner in crime and for all the Firlefanz!

Beyond that, I am happy to have shared parts of this PhD journey with Xiaozhou as a friend and collaborator. Further thanks goes to Ben, for extensive conversations of all kinds; to Anna, for the best hugs; to Nihal, for nightly juggling sessions in the office; to Zakaria, for bringing in some fresh perspectives; and to Nicola for bringing the chill vibes and for giving my research a little more purpose.

Thanks (again!) to Emilio's encouragement, I was able to travel to many conferences, workshops, and schools that introduced to me a wonderfully welcoming international scientific community and made me see places I would otherwise have never travelled to. Experiencing science in different parts of the (academic) world allowed to get fresh perspectives on, well, everything. I am very grateful for everyone I met for delightfully sharing so much more than only our scientific research.

The past four years would not have been nearly as enjoyable without the people around me, and I have drawn a great deal of motivation from my friends. I thank Natalie for indulging my enjoyment of caustics through an intense artistic collaboration, for sharing insights into her work and sparking introspection about my own, for hours on loud dance floors and silent walks through hidden nature in London, and for showing me the genuinely nice bits of the city. Thank you for being interested in my work and seeing it with entirely different eyes!

Willie, Hannah, and Jacob — thank you for making me feel at home, for the laughter, silliness, food, sweat, and for quite literally lifting me up and down so many times.

Kerttuli, for being the most inspiring little bundle of creativity in the truest sense of the word.

Jaqui, without whom I would likely have procrastinated this thesis indefinitely: thank you for actually making me write it, and for your inspiring diligence and perseverance.

Generally, I want to thank all of my friends who provided joy, grounding, and upside-down perspectives, and regularly reminded me of the sheer irrelevance of my research.

I am beyond thankful to Peter and Siw♡, for sharing their home with me, and together with Manny and Christine for turning it into a peaceful retreat within this big and hasty city.

By moving to London, I was incredibly lucky to meet Sean Gandini and Kati Ylä-Hokkala. Your life energy, creativity, enthusiasm, and your healthy balance between joy and ambition have been truly inspiring. With the times of almost daily juggling sessions, you became like a family to me. I very much enjoy shepherding a zoo of mini juggling patterns, taming some of the beastly gorillas, and occasionally confronting "the monster". Thank you for bringing so much joy — to me and to the world!

Mark and Adrian, thank you for all the count-less hours of juggling, the long drives through the British countryside and beyond, and for the mutually supportive friendship with the full range of (e)motions. Thank you for being so ambitious and rigorous in this silly hobby and finding so many clever and sweet ways of incentivising my (and our) practice. My apologies for all the bad passes.

In fact, I would actually also like to thank the entire juggling community, and the PassOut convention in particular. If it wasn't for a random chat over breakfast with Iain, who encouraged me to get in touch with Job, my PhD might not have progressed beyond what is now only two Chapters of it. Especially PassOut 2025/26 — right in the final throes of thesis writing — was a wonderfully wholesome and restoratively silly week 🐙

I am grateful to Andrew, Ben, Danylo, Iain, Ismaël, Mark and Nihal for proofreading parts of this thesis. And Job again! Just because I can't thank him enough!

If I had to dedicate this thesis to someone, it would probably be Julia — not only the coding language, but foremostly my sister. Julia, thank you for giving me purpose, for reminding me what actually matters in life, and for always being there for me. For being my outer inner voice, for sharing your wisdom and experience, for being my best friend, and for doing all the things I would never dare to do. I am incredibly proud of you.

Finally, I want to thank my family. Opa, for sharing your experiences, your unjustified belief in me, and giving (mostly outdated) advice on how to pursue a PhD. To my dad, to whom I have explained attosecond science about as many times as there are harmonics in a spectrum — thank you for encouraging me to pursue things properly and to not be boring, but to follow my own interests.

The last words I want to dedicate in deep gratitude to my mum Heike♡. Thank you for living life full power, for sparkling with *corianduli di luce*, for the ability to find joy in every situation and for leaving us with an omnipresent trust.

*Anne*
*London, January 2026*